\documentclass[12pt]{article}
\usepackage{setspace}
\usepackage{epsfig}
\usepackage{amsmath}
\usepackage{amssymb}
\usepackage{graphicx}
\def\be{\begin{equation}}
\def\ee{\end{equation}}
\def\ba{\begin{array}}
\def\ea{\end{array}}
\def\beqn{\begin{eqnarray}}
\def\eeqn{\end{eqnarray}}

\def\bt{\begin{tabular}}
\def\et{\end{tabular}}
\def\bc{\begin{center}}
\def\ec{\end{center}}

\begin{document}
\title{Implications of non minimal lepton mass textures for Dirac neutrinos}
\author{Samandeep Sharma, Priyanka Fakay, Gulsheen Ahuja$^*$, Manmohan
Gupta\\ {\it Department of Physics, Centre of Advanced Study,
P.U.,
 Chandigarh, India.}\\
\\{\it $^*$gulsheenahuja@yahoo.co.in}}
\date{}
\maketitle

\renewcommand{\baselinestretch}{1.50}\normalsize
\begin{abstract}
 In the light of the recent measurement of the leptonic mixing
angle $\theta_{13}$, implications of the latest mixing data have
been investigated for non-minimal textures of lepton mass matrices
pertaining to Dirac neutrinos. All these texture specific lepton
mass matrices have been examined for their compatibility with the
latest data in the cases of normal hierarchy, inverted hierarchy
and degenerate scenario of neutrino masses. The implications of
all the three lepton mixing angles have been investigated on the
lightest neutrino mass as well as the Jarlskog's CP violating
parameter in the leptonic sector.
\end{abstract}

\section{Introduction}
Ever since being proposed by Pauli, neutrinos have been a sort of
fascinating puzzle for the physicists. The recent observation of
non zero leptonic mixing angle $\theta_{13}$
\cite{t2k}-\cite{reno} has provided significant boost to the
sharpening of implications of the neutrino oscillations and has
added another dimension to neutrino physics by implying the
possibility of CP violation in the leptonic sector, hence
deepening the flavor puzzle further. The non zero value of
$\theta_{13}$, on the one hand, restores the parallelism between
the mixings of quarks and leptons, on the other hand, its
unexpectedly `large' value signifies the differences between
these, the leptonic mixing angles being large as compared to the
quark counterparts.

\par In the absence of any viable theory of flavor dynamics for explaining the fermion masses and mixings, approaches
followed on the theoretical front can broadly be catagorized into
`top-down' and `bottom-up'. Despite large number of attempts using
the `top-down' perspective \cite{topdown} , we have yet to arrive
at a viable approach which accounts for the vast amount of data
related to flavor mixings. Therefore, in the present work, we
follow the `bottom-up' approach consisting of finding the
phenomenological fermion mass matrices which are in tune with the
latest low energy data. In this context, texture specific mass
matrices have got a good deal of attention \cite{tex} and play an
important role in explaining the flavor mixing phenomena. In
particular, texture specific mass matrices provide valuable
information to explain the quark and lepton masses and mixings,
for details we refer the reader to \cite{singreview}. In the
context of leptons, it needs to be mentioned that with the texture
approach, minimal textures for Dirac neutrinos have recently
\cite{ptep} been ruled out, however, a detailed and comprehensive
analysis for the non-minimal textures in light of the recent
measurement of $\theta_{13}$ is yet to be carried out.

\par   The purpose of the present work is to carry out a
systematic and comprehensive study of  non-minimal lepton mass
matrices  using the texture zero approach. In the context of
neutrinos, the issue of these being Dirac like or Majorana
particles is still an open question for the physicists since the
Dirac neutrinos have not yet been ruled by experimental data. In
the present work, considering the neutrinos to be Dirac like
particles, we have made an attempt to examine the compatibility of
 non-minimal lepton mass matrix textures with the
recent neutrino mixing data for all the three neutrino mass
hierarchies i.e. normal, inverted and degenerate scenario of
neutrino masses. It may be noted that while carrying out the
analysis, we consider the charged lepton mass matrices and the
effective neutrino mass matrices having parallel structures i.e.
texture zeroes being located at identical positions, in consonance
with some classes of family symmetries and grand unified theories
\cite{so10} wherein such parallel structures emerge naturally.
\par The detailed plan of the paper is as follows. In Section (2) we discuss the
 general lepton mass matrices in the Standard Model (SM). Inputs
used in the  analysis are given in Section (3). Results and
discussion pertaining to normal hierarchy, inverted hierarchy and
degenerate scenario of neutrino masses for texture 2 zero, texture
4 zero and texture 5 zero lepton mass matrices are presented in
Section (4). Finally, Section (5) summarizes our conclusions.

\section{General lepton mass matrices in the Standard Model} \label{metho}
In the Standard Model (SM) \cite{st}-\cite{sm} of particle
physics, the lepton mass matrices are arbitrary $3\times3$ complex
matrices, thus containing a total of 36 real parameters. Using the
`Polar Decomposition Theorem', the lepton mass matrices in SM can
be considered to be hermitian without any loss of generality
bringing down the number of free parameters from 36 to 18. To
facilitate the formulation of phenomenological mass matrices,
which perhaps are compatible with the GUT scale mass matrices, it
has been suggested \cite{nmm} that in order to avoid fine tuning
amongst the elements of the mass matrices, these should follow a
`natural hierarchy' i.e. $ (1,1),(1,2),(1,3)\lesssim (2,2), (2,3)
\lesssim (3,3)$. The number of free parameters in these mass
matrices can further be reduced using the facility of Weak Basis
(WB) transformations. These WB approaches broadly lead to two
possibilities for the texture zero lepton mass matrices. In the
first possibility observed by Branco {\it{et al.}} \cite{branco},
one ends up with a structure wherein one of the matrices is a
texture two zero type and the other is a one zero type, i.e.
\beqn
(M_l)_{(1,1)}=(M_l)_{(1,3)}=(M_l)_{(3,1)}=(M_{\nu D})_{(1,1)}=0,\\
 or~(M_l)_{(1,1)}=(M_{\nu D})_{(1,1)}=(M_{\nu D})_{(1,3)}=(M_{\nu D})_{(3,1)}=0,
\label{branc}\eeqn
$M_l$ and $M_{\nu D}$ correspond to the charged
lepton and the Dirac neutrino mass matrix respectively. In the
second possibility, given by Fritzsch and Xing \cite{frxing}, one
ends up with a texture two zero structure for the lepton mass
matrices, wherein both the lepton mass matrices assume a texture
one zero structure, viz.,
\be
(M_l)_{(1,3)}=(M_l)_{(3,1)}=(M_{\nu D})_{(1,3)}=(M_{\nu
D})_{(3,1)}=0.
\label{branc}
\ee 
Although the two possibilities are equivalent,
however for the present work we have followed the Fritzsch-Xing
approach, this choice can be justified as the mass matrices
obtained by Fritzsch-Xing approach not only exhibit parallel
texture structures but can also be diagonalized exactly making the
construction of the corresponding lepton mixing matrix easier.
Therefore, the general lepton mass matrices within the framework
of SM can be given as
\be
 M_{l}=\left( \ba{ccc}
C_{l} & A _{l} & 0      \\
A_{l}^{*} & D_{l} &  B_{l}     \\
 0 &     B_{l}^{*}  &  E_{l} \ea \right), \qquad
M_{\nu D}=\left( \ba{ccc} C_{\nu } & A _{\nu } & 0\\
A_{\nu }^{*} & D_{\nu } & B_{\nu }     \\
 0 &     B_{\nu }^{*}  &  E_{\nu } \ea \right),
\label{t20}\ee
with $A_{l(\nu)}=
|A_{l(\nu)}|e^{i \alpha_{l,\nu}}$ and $B_{l(\nu)}=|B_{l(\nu)}|e^{i \beta_{l,\nu}}$.

 \par Following the methodology connecting the lepton mass matrices to the mixing matrix detailed
 in \cite{singreview}, one can carry out diagonalization of a general
 mass matrix $M_k$ by expressing it as
\be M_k=Q_k M_k^r P_k, \ee where $Q_k$, $P_{k}$ are diagonal phase
matrices given as Diag$(e^{i \alpha_k}, 1, e^{-i \beta_k})$ and
Diag $(e^{-i \alpha_k}, 1, e^{i \beta_k})$ respectively and
$M_k^r$ is a real symmetric matrix. $M_k^r$ can be diagonalized by
an orthogonal transformation $O_k$, e.g.,
\be M_k^{diag}=O_k^T M_k^r O_k \ee
which can be rewritten as 
\be M_k^{diag}=O_k^T Q_k^\dagger M_k P_k^\dagger O_k. \ee

The elements of the general diagonalizing transformation can
figure with different phase possibilities, however, these
possibilities are related to each other through the phase matrices
\cite{singreview}. For the present work, we have chosen the
possibility

\begin{equation}
O_k=\left(\ba{ccc} O_k(11) & O_k(12) & O_k(13)\\ O_k(21) &
-O_k(22) & O_k(23)\\ -O_k(31) & O_k(32) & O_k(33) \ea \right),
\end{equation}
where

\begin{equation*} O_k(11) =  \sqrt{\frac{(E_k -m_1)(D_k + E_k - m_1 -
m_2)(D_k + E_k -m_1 -m_3)}{(D_k +2 E_k -m_1 -m_2 -m_3) (m_1 -m_2)
(m_1-m_3)}}
\end{equation*}

\begin{equation*}
 O_k(21) = \sqrt{\frac{(m_1-C_k)(m_1-E_k)}{(m_1-m_2)(m_1-m_3)}}
\end{equation*}

\begin{equation*} O_k(31)= \sqrt{\frac{(E_k
-m_2)(E_k-m_3)(m_1-C_k)}{(m_1-m_2)(m_1-m_3)(E_k-C_k)}}
\end{equation*}

\begin{equation*}
O_k(12) = \sqrt{\frac{(E_k-m_2)(m_3-C_k)(m_1-C_k)}{(E_k -
C_k)(m_1-m_2)(m_3-m_2)}}
\end{equation*}

\begin{equation*}
O_k(22) = \sqrt{\frac{(E_k-m_2)(C_k -m_2)}{(m_1-m_2)(m_3-m_2)}}
\end{equation*}

\begin{equation*}
O_k(32)= \sqrt{\frac{(-E_k + m_1)(C_k -m_2)(E_k
-m_3)}{(m_1-m_2)(m_2-m_3)(E_k-C_k)}}
\end{equation*}

\begin{equation*}
O_k(13)= \sqrt{\frac{(-C_k +m_1)(-E_k + m_3)(C_k
-m_2)}{(m_1-m_3)(m_3-m_2)(C_k -E_k)}}
\end{equation*}

\begin{equation*}
O_k(23)=\sqrt{\frac{(-m_3-C_k)(E_k -m_3)}{(m_1-m_3)(m_3-m_2)}}
\end{equation*}

\be
O_k(33)=\sqrt{\frac{(E_k-m_1)(E_k - m_2)(m_3-C_k)}{(C_k
-E_k)(m_1-m_3)(m_3-m_2)}}, \, \label{diat20} \ee

 $m_1$, -$m_2$, $m_3$ being the eigen values of $M_k$.

 \par Using the above expressions, one can easily obtain the elements of $O_l$, the diagonalizing
matrix for the charged lepton sector by replacing $m_1$, -$m_2$,
$m_3$ with $m_e$, -$m_\mu$, $m_\tau$, e.g., the first element of
$O_l$ can be given as
\be
O_l(11)=\sqrt{\frac{(E_l -m_e)(D_l + E_l - m_e - m_\mu)(D_l + E_l
-m_e -m_\tau)}{(D_l +2 E_l -m_e -m_\mu -m_\tau)(m_e -m_\mu)
(m_e-m_\tau)}}. \ee In an analogous manner, diagonalizing matrix
$O_{\nu D}$ for Dirac neutrinos for the normal mass hierarchy (NH)
given by $m_{\nu 1} < m_{\nu 2} \ll m_{\nu 3}$ and for the
corresponding degenerate scenario given by $m_{\nu 1} \lesssim
m_{\nu 2} \sim m_{\nu 3}$ can be obtained from
equation (\ref{diat20}) by replacing $m_1$, -$m_2$, $m_3$ with $m_
{\nu 1}$, -$m_{\nu 2}$, $m_{\nu 3}$. For instance, first element
of $O_{\nu D}$ can be given as
\be
O_{\nu D}(11)=\sqrt{\frac{(E_\nu -m_{\nu 1})(D_\nu + E_\nu -
m_{\nu 1} - m_{\nu 2})(D_{\nu} + E_{\nu} -m_{\nu 1} -m_{\nu
3})}{(D_\nu +2 E_\nu -m_{\nu 1} -m_{\nu 2} -m_{\nu 3}) (m_{\nu 1}
-m_{\nu 2}) (m_{\nu 1}-m_{\nu 3})}}. \ee

Similarly, one can obtain the elements of diagonalizing
transformation for the inverted hierarchy (IH) case defined as
$m_{\nu 3} \ll  m_{\nu 1} < m_{\nu 2}$ and the corresponding
degenerate case given by $m_{\nu 3} \sim m_{\nu 1} \lesssim m_{\nu
3}$ by replacing $m_1$, -$m_2$, $m_3$ with $m_ {\nu 1}$, -$m_{\nu
2}$, -$m_{\nu 3}$ in eqn. (\ref{diat20}), e.g., the first element of
$O_{\nu D}$ in this scenario can be given as
\be
O_{\nu D}(11)=\sqrt{\frac{(E_\nu -m_{\nu 1})(D_\nu + E_\nu -
m_{\nu 1} - m_{\nu 2})(D_{\nu} + E_{\nu} -m_{\nu 1} +m_{\nu
3})}{(D_\nu +2 E_\nu -m_{\nu 1} -m_{\nu 2} + m_{\nu 3}) (m_{\nu 1}
-m_{\nu 2}) (m_{\nu 1} + m_{\nu 3})}}. \ee

The other elements of the diagonalizing transformations for the
charged lepton as well as neutrinos can be obtained in a similar
way. The lepton mixing matrix, the
`Pontecorvo-Maki-Nakagawa-Sakata (PMNS)' matrix \cite{pmns} can be
obtained from these orthogonal transformations using the relation
\be
U=O_l^\dagger Q_l P_{\nu D} O_{\nu D}, \ee where $Q_l P_{\nu D}$
can be taken as Diag($e^{-i \phi_1}$, 1, $e^{i \phi_2}$). The
parameters $\phi_1$ and $\phi_2$
 can be considered as free parameters and are related to the phases of mass matrices as $\phi_1~=~\alpha_{\nu D}-\alpha_l$,
$\phi_2~=~\beta_{\nu D}-\beta_l$.

\section{Inputs used for the analysis}
Before getting into the details of the analysis, we would like to
mention some of the essentials pertaining to various inputs. In the present analysis,
we have made use of the results of a latest global three neutrino oscillation analysis \cite{fogli2012}
, in table (\ref{data}) we present the $ 1 \sigma$ and $ 3 \sigma$ ranges of the neutrino oscillation parameters.

\begin{table}[ht]
\centering
\begin{tabular}{c c c }
\hline
Parameter & $1 \sigma$ range & $3 \sigma$ range  \\ [0.5ex]
\hline

$\Delta m_{sol}^2$ $[ 10^{-5} eV^2]$ & (7.32-7.80) & (6.99-8.18) \\
\hline
$\Delta m_{atm}^2$ $[ 10^{-3} eV^2]$ & (2.33-2.49)(NH); (2.31-2.49) (IH) & (2.19-2.62)(NH); (2.17-2.61)(IH)  \\
\hline
$sin^2 \theta_{13}$ $[10^{-2}]$ & (2.16-2.66)(NH); (2.19-2.67)(IH) & (1.69-3.13)(NH); (1.71-3.15) (IH)  \\
\hline
$sin^2 \theta_{12}$ $[10^{-1}]$ & (2.91-3.25) & (2.59-3.59)  \\
\hline
$sin^2 \theta_{23}$ $[10^{-1}]$ & (3.65-4.10)(NH);(3.70-4.31)(IH)  & (3.31-6.37)(NH);(3.35-6.63)(IH)  \\
\hline
\end{tabular}
\caption{The $1\sigma$ and $3\sigma$ ranges of neutrino oscillation parameters presented in \cite{fogli2012}}
\label{data}
\end{table}

While carrying out the analysis, the lightest neutrino mass, $m_1$ for the case of NH and $m_3$ for the case of IH,
is considered as a free
parameter.  For all the three possible mass
hierarchies of neutrinos i.e. normal, inverted and degenerate scenario, the explored range of the lightest neutrino mass is
taken to be $10^{-8}\,\rm{eV}-10^{-1}\,\rm{eV}$, our conclusions
remain unaffected even if the range is extended further.
In the absence of any constraint on the phases, $\phi_1$ and $\phi_2$
have been given full variation from 0 to $2\pi$. Although $D_{l,
\nu}$ and $C_{l, \nu}$ are free parameters, however, they have been constrained
such that diagonalizing transformations $O_l$ and $O_{\nu}$ always
remain real.

\section{Results and discussions}
Before discussing the results, we would like to briefly discuss the
possible non-minimal textures which we plan to analyse and
discuss. While considering such possibilities, we ignore WB
related mass matrix structures, in particular those related  through  permutations, as in
the case of parallel textures these are equivalent. In the case of
texture two zero matrices, therefore we have to consider one
possibility only, which we discuss for the case of three
hierarchies of neutrino masses. Beyond texture two zero, we have to
make specific assumptions of taking particular elements of $M_l$
and $M_{\nu D}$ being zero. In table  (\ref{3t4}) we have considered
four classes with their permutations covering all possibilities.
Texture five zero mass matrices can easily be derived from the corresponding texture four
zero ones. 
\subsection{Texture two zero lepton mass matrices}
\begin{figure}
\begin{tabular}{cc}
  \includegraphics[width=0.2\paperwidth,height=0.2\paperheight,angle=-90]{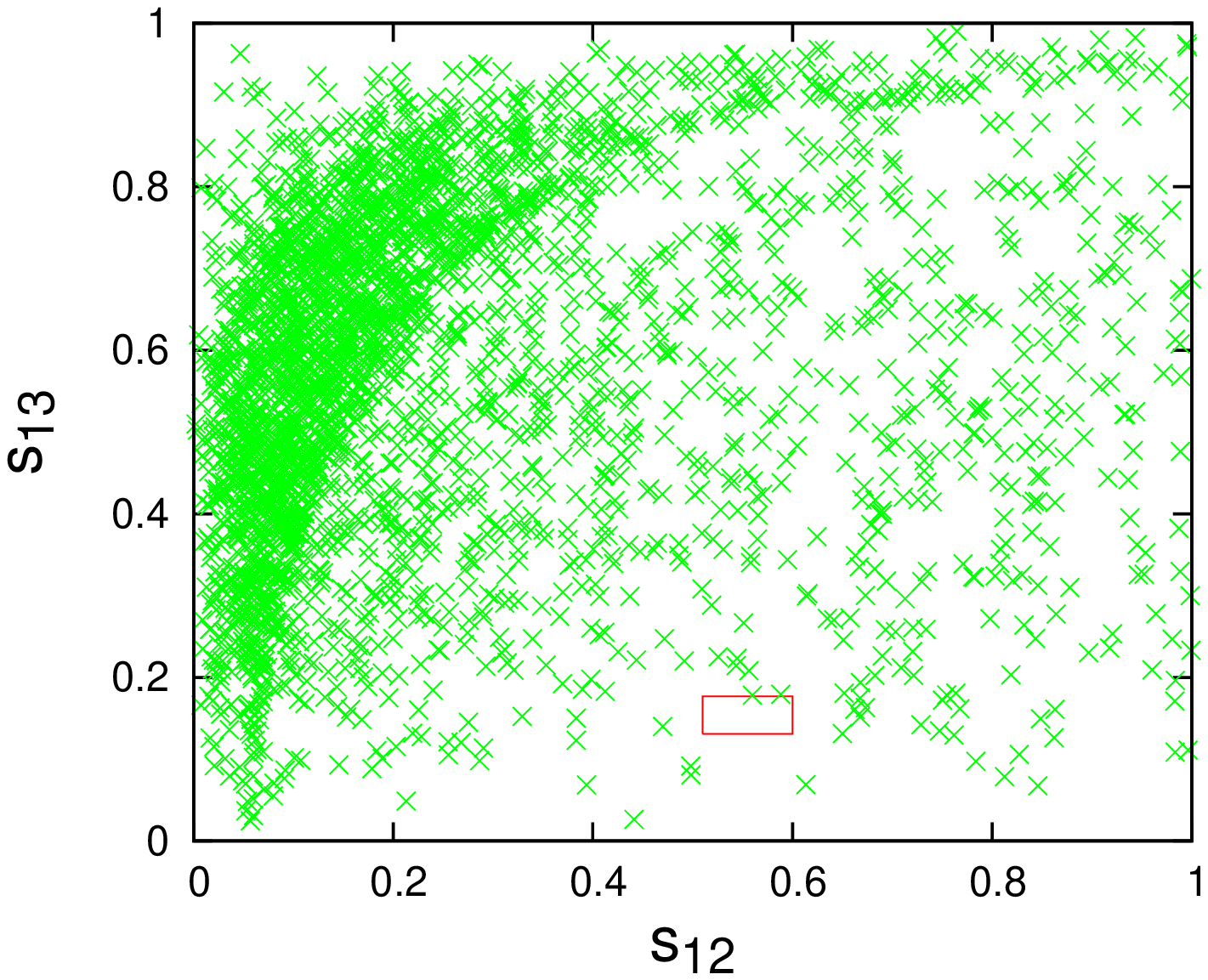}
  \includegraphics[width=0.2\paperwidth,height=0.2\paperheight,angle=-90]{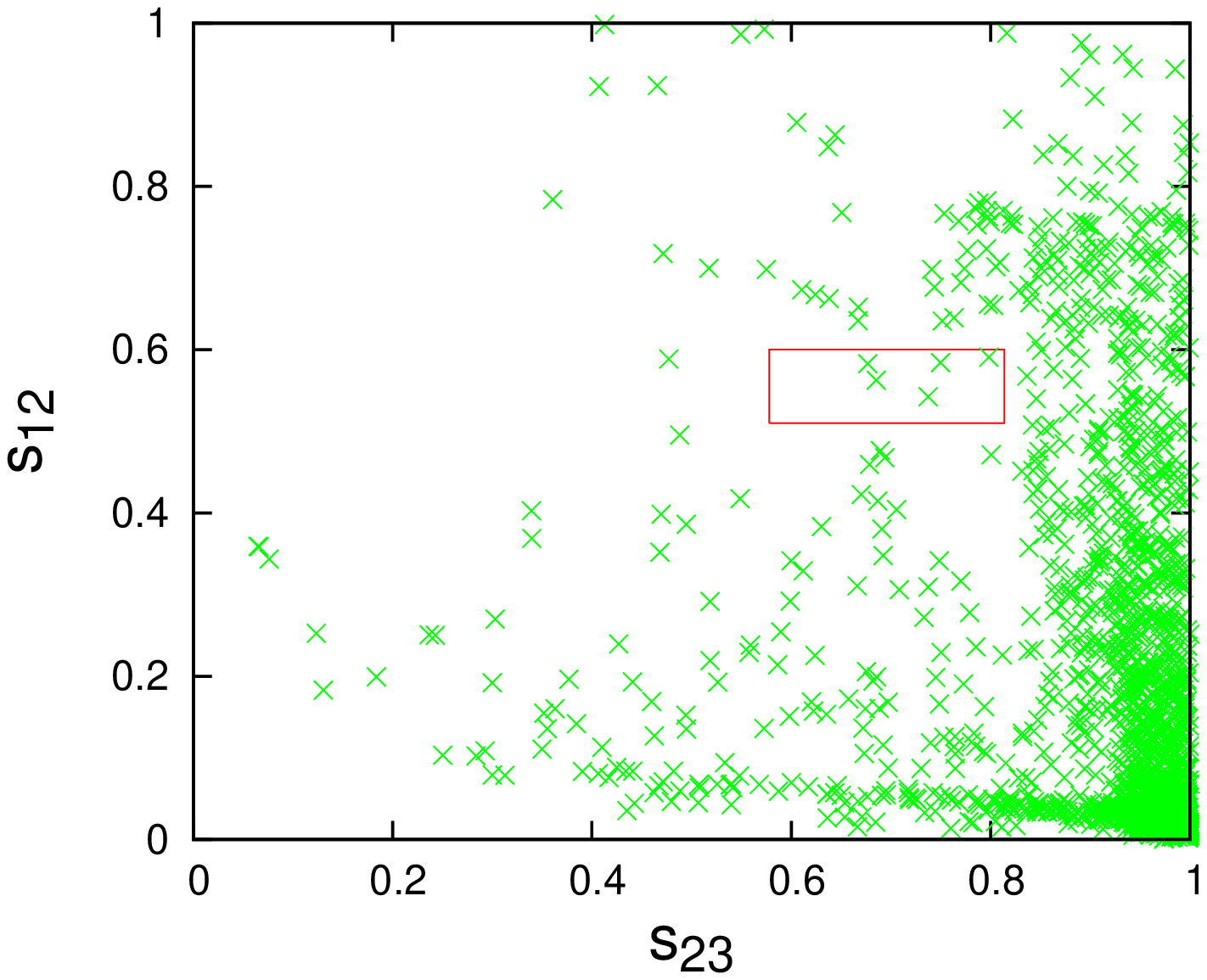}
  \includegraphics[width=0.2\paperwidth,height=0.2\paperheight,angle=-90]{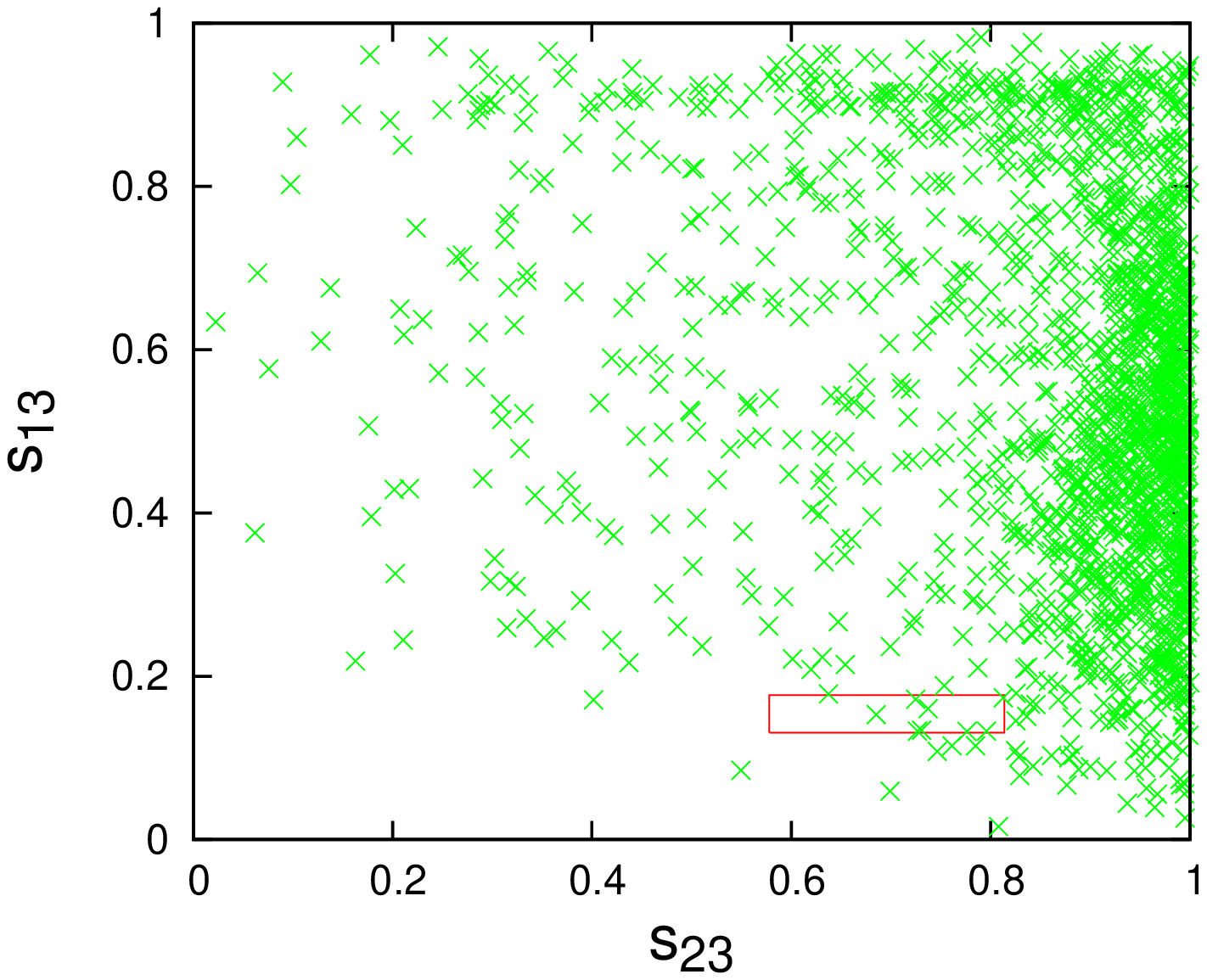}
\end{tabular}
\caption{Plots showing the parameter space for any two mixing
angles for texture two zero Dirac mass matrices ( inverted
hierarchy).} 
\label{t2ih1}
\end{figure}

To examine the compatibility of texture two zero lepton mass matrices given in equation (\ref{t20}) with the recent
mixing data, we carry out a detailed analysis pertaining to all three possible neutrino mass hierarchies.
To this end, in figure (\ref{t2ih1}) we present the plots showing the parameter space of two mixing angles,
with the third one being constrained by its $1\sigma$ experimental bounds for inverted hierarchy. The 
rectangular boxes in these figures show the $3 \sigma$ ranges for the two mixing angles being
considered in the figure. As is evident from
these figures, the parameter space for the mixing angles shows considerable overlap with the
experimentally allowed $3 \sigma$ region. Therefore, inverted hierarchy seems to be viable for
texture two zero lepton mass matrices given in equation (\ref{t20}).
\par
\begin{figure}
\begin{tabular}{cc}
  \includegraphics[width=0.2\paperwidth,height=0.2\paperheight,angle=-90]{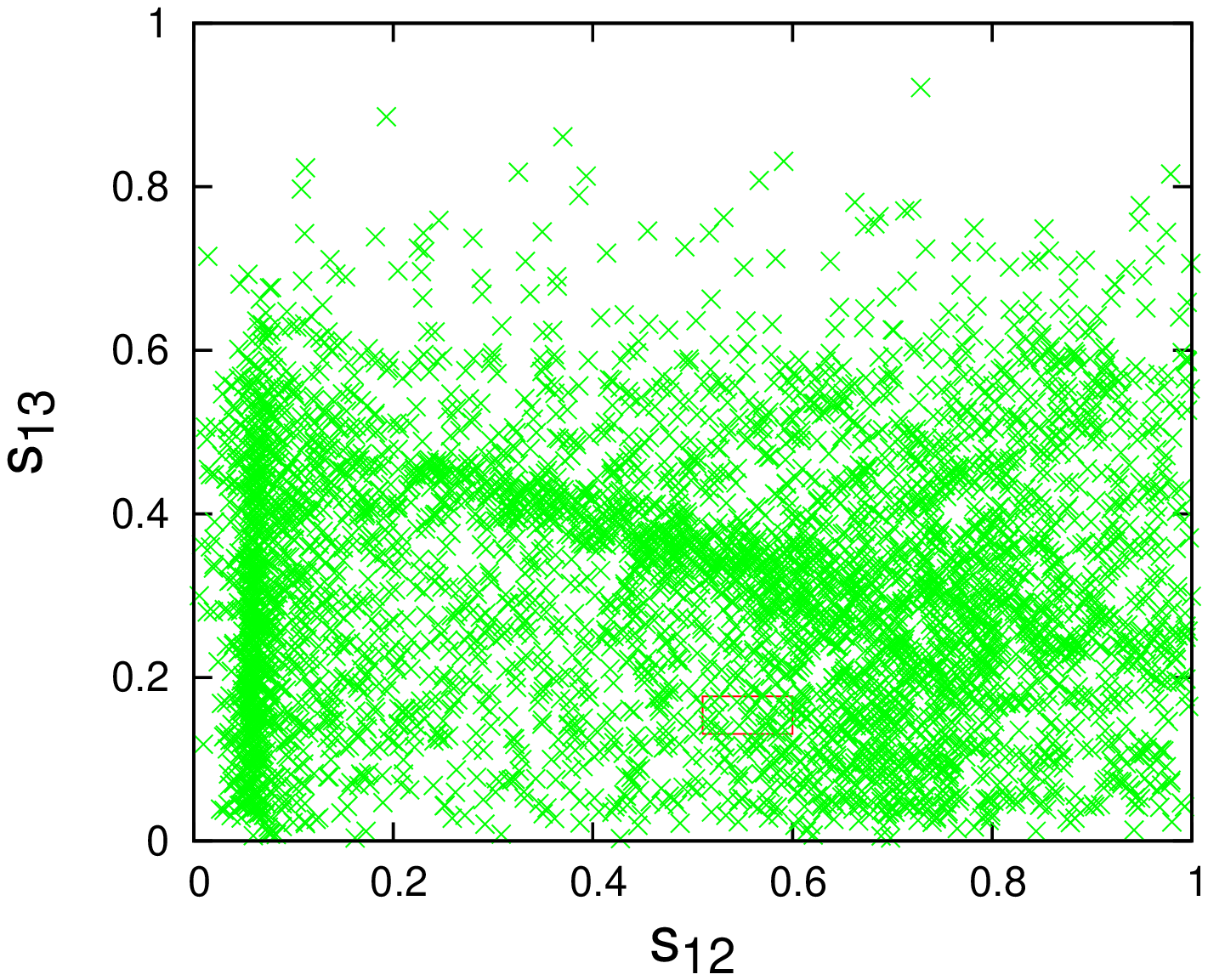}
  \includegraphics[width=0.2\paperwidth,height=0.2\paperheight,angle=-90]{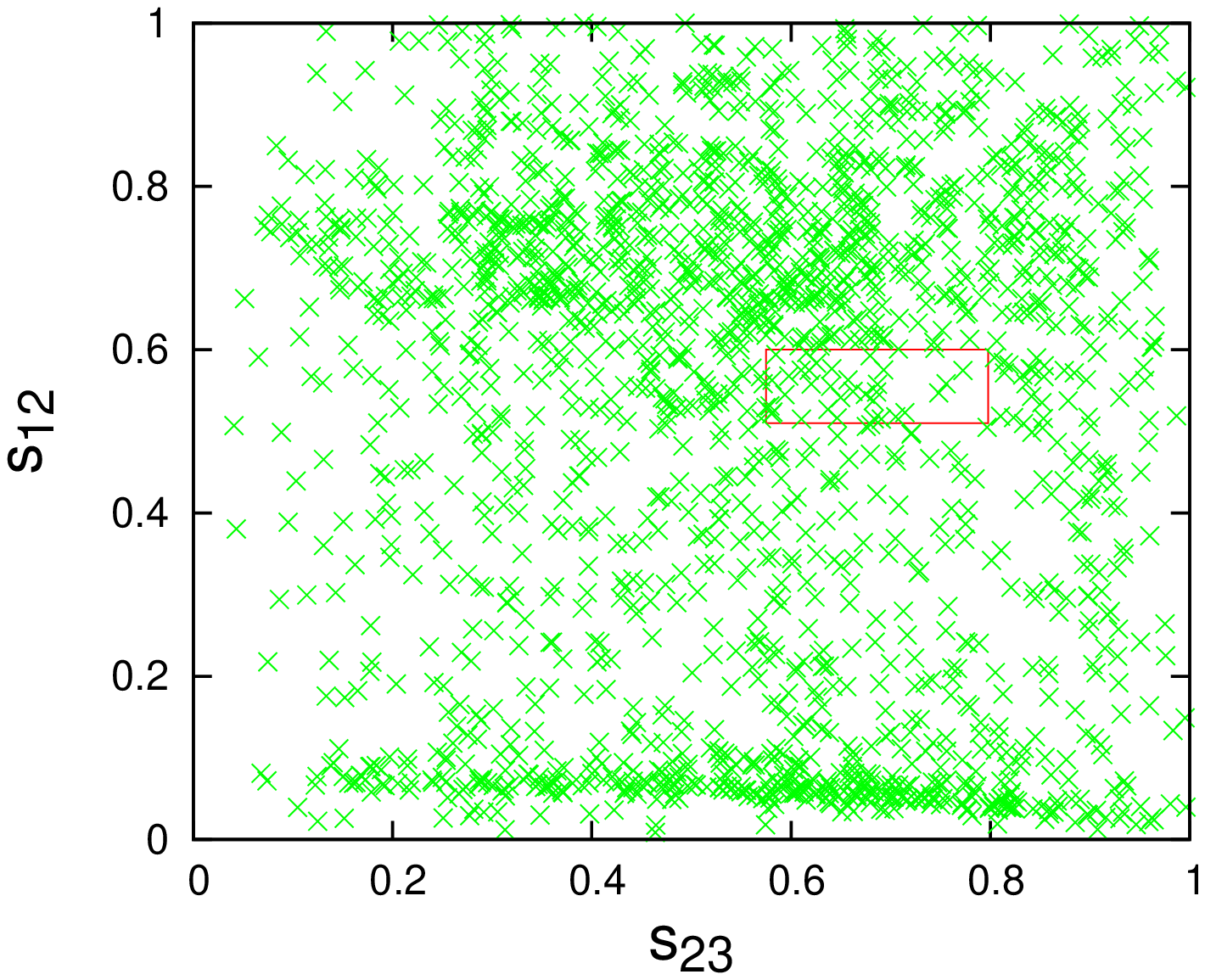}
  \includegraphics[width=0.2\paperwidth,height=0.2\paperheight,angle=-90]{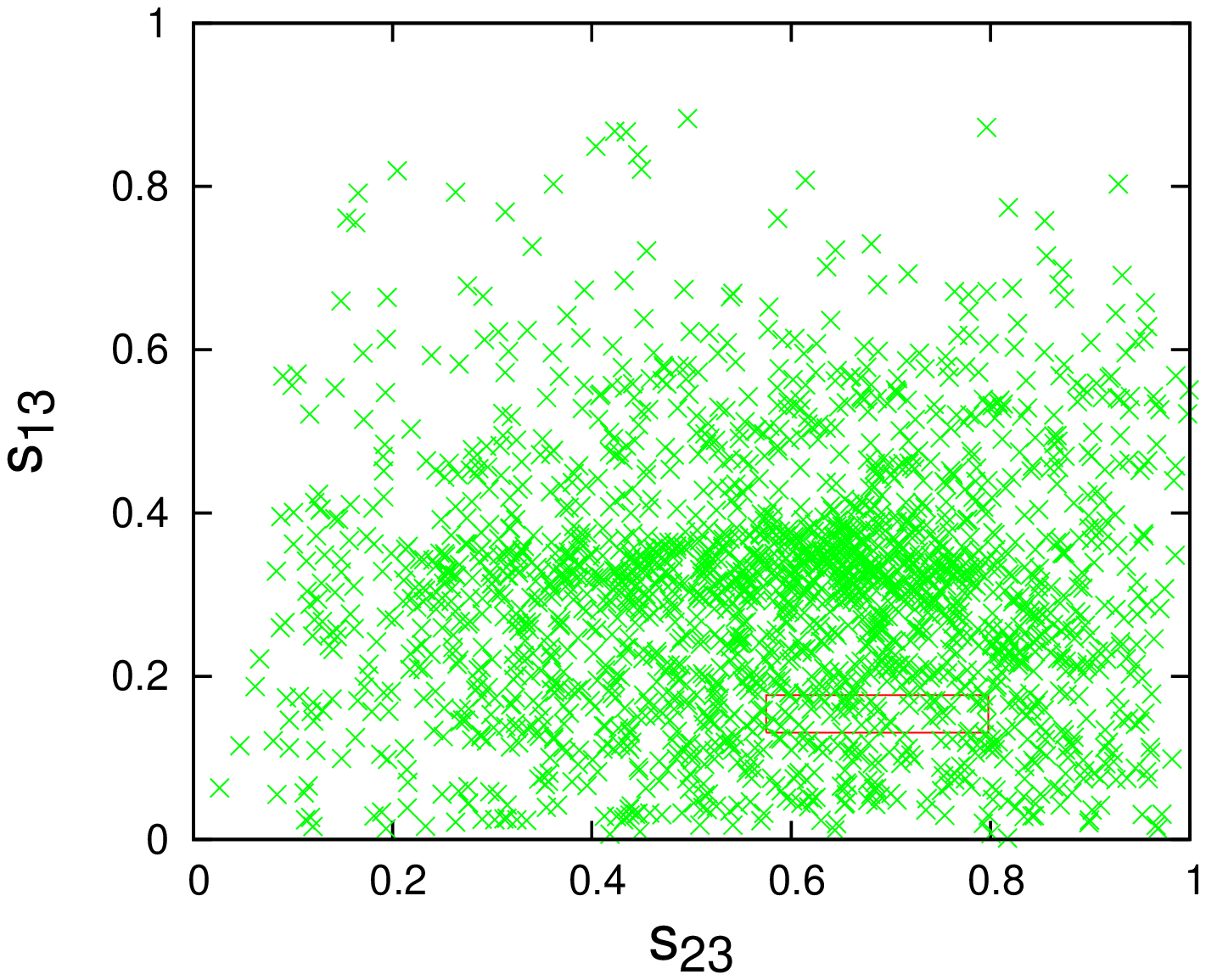}
\end{tabular}
\caption{Plots showing the parameter space for any two mixing
angles for  texture two zero Dirac mass matrices (normal hierarchy).}
\label{t2nh1}
\end{figure}

 After discussing the viability of inverted hierarchy for texture two zero mass matrices, we now examine the compatibility
of these matrices for the normal hierarchy case. To this end, in figure (\ref{t2nh1}) we present the plots showing the
parameter space allowed for two mixing angles when the third one is constrained by its $1\sigma$ experimental bound for normal
neutrino mass hierarchy. A general look at the figure (\ref{t2nh1}) reveals that the structure given in eqn.(\ref{t20}) is compatible
with the normal neutrino mass hierarchy.

\begin{figure}
\begin{tabular}{cc}
  \includegraphics[width=0.2\paperwidth,height=0.2\paperheight,angle=-90]{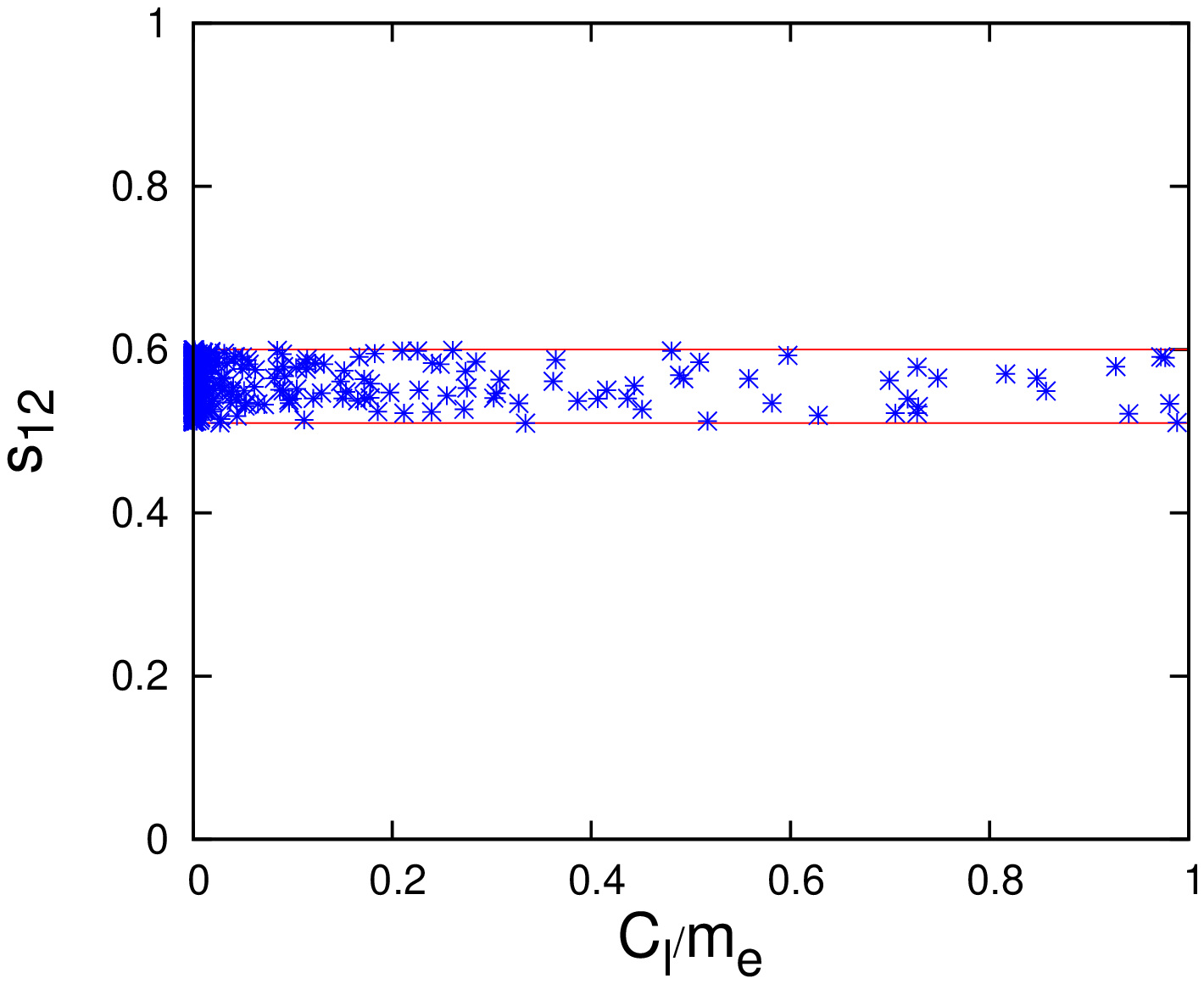}
  \includegraphics[width=0.2\paperwidth,height=0.2\paperheight,angle=-90]{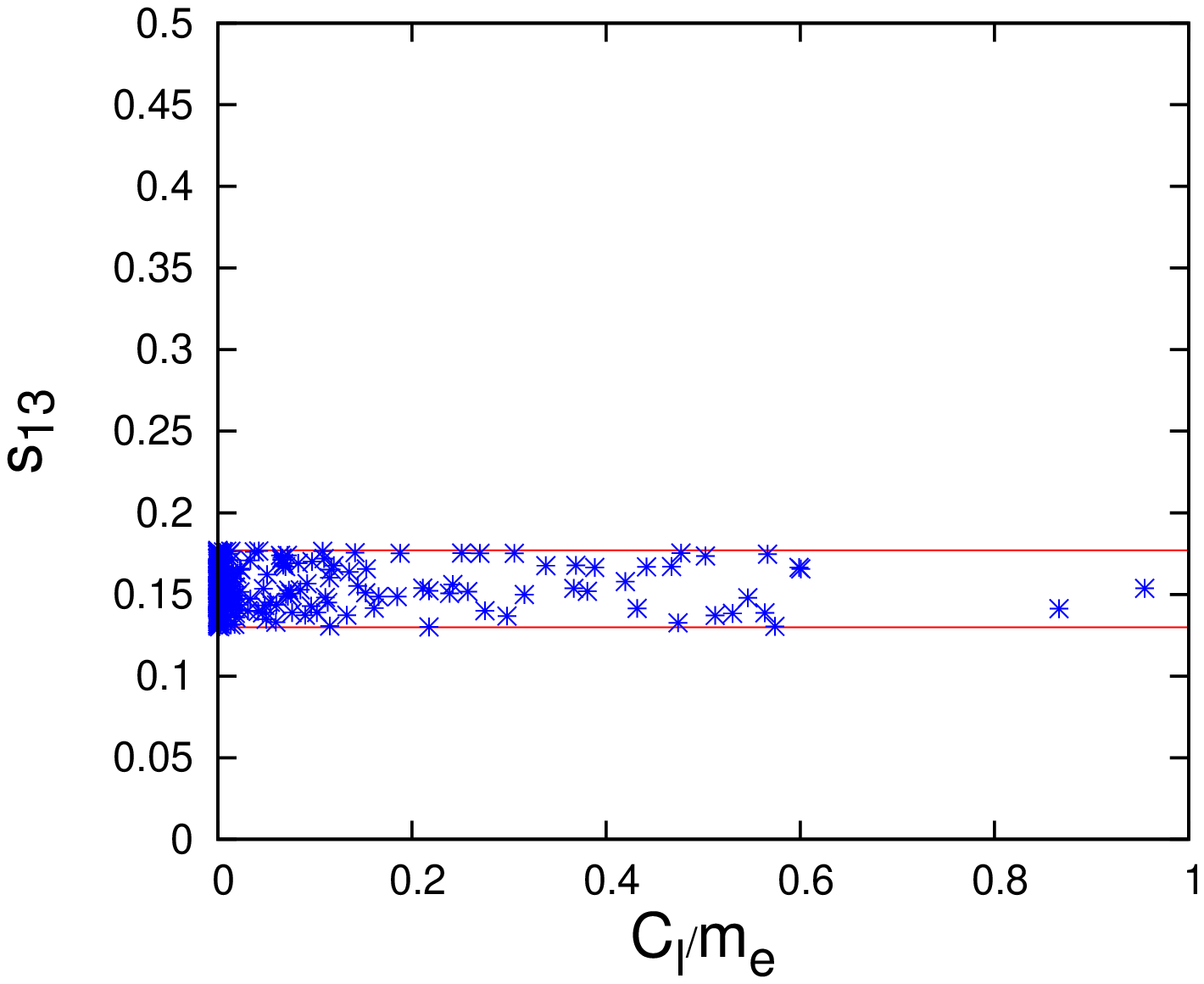}
  \includegraphics[width=0.2\paperwidth,height=0.2\paperheight,angle=-90]{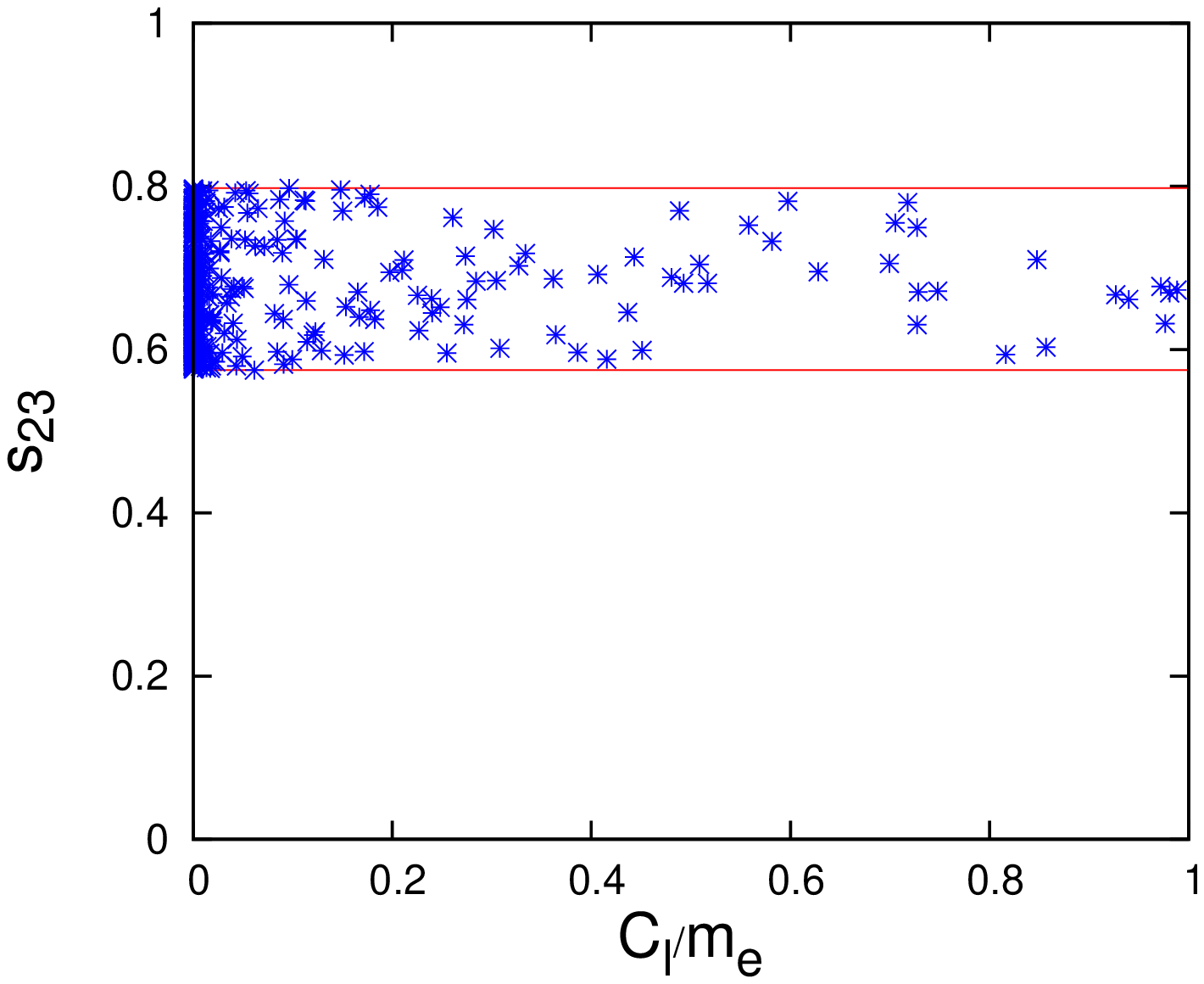}
\end{tabular}
\caption{Plots showing the variation of the mixing angles with the
(1,1) element of the charged lepton mass matrix for texture two
zero Dirac mass matrices (normal hierarchy).} 
\label{t2nh2}
\end{figure}

\begin{figure}
\begin{tabular}{cc}
  \includegraphics[width=0.2\paperwidth,height=0.2\paperheight,angle=-90]{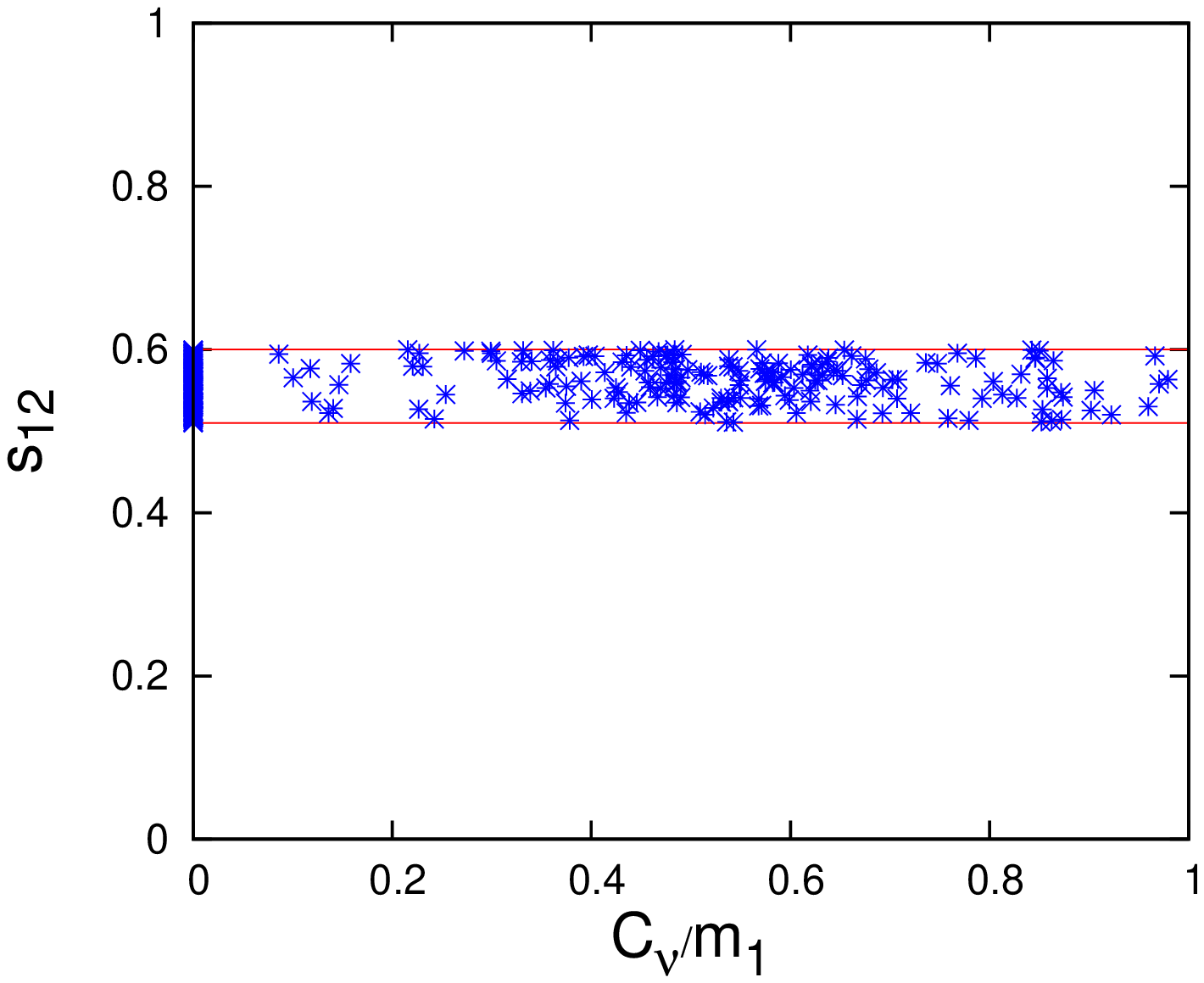}
  \includegraphics[width=0.2\paperwidth,height=0.2\paperheight,angle=-90]{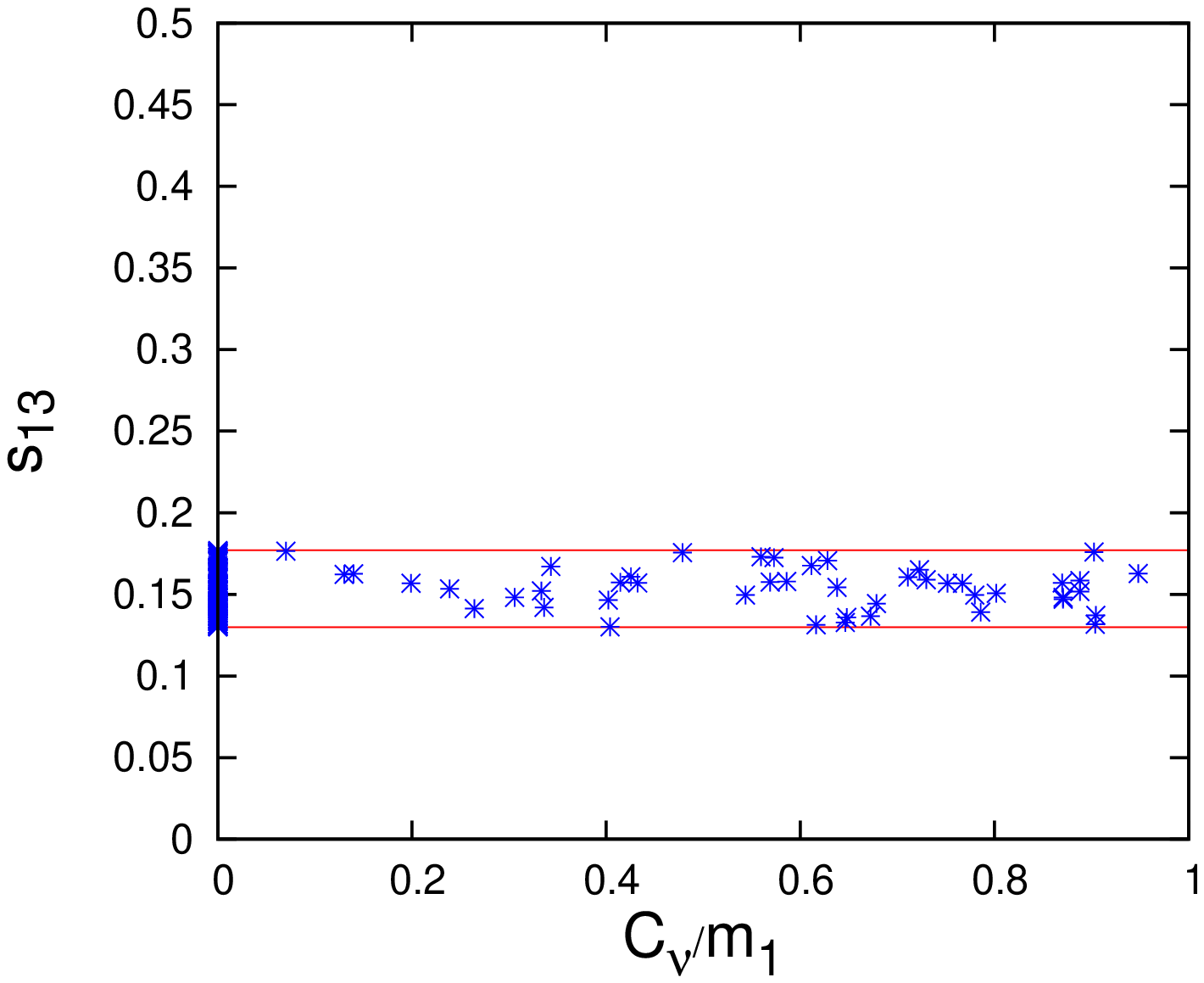}
  \includegraphics[width=0.2\paperwidth,height=0.2\paperheight,angle=-90]{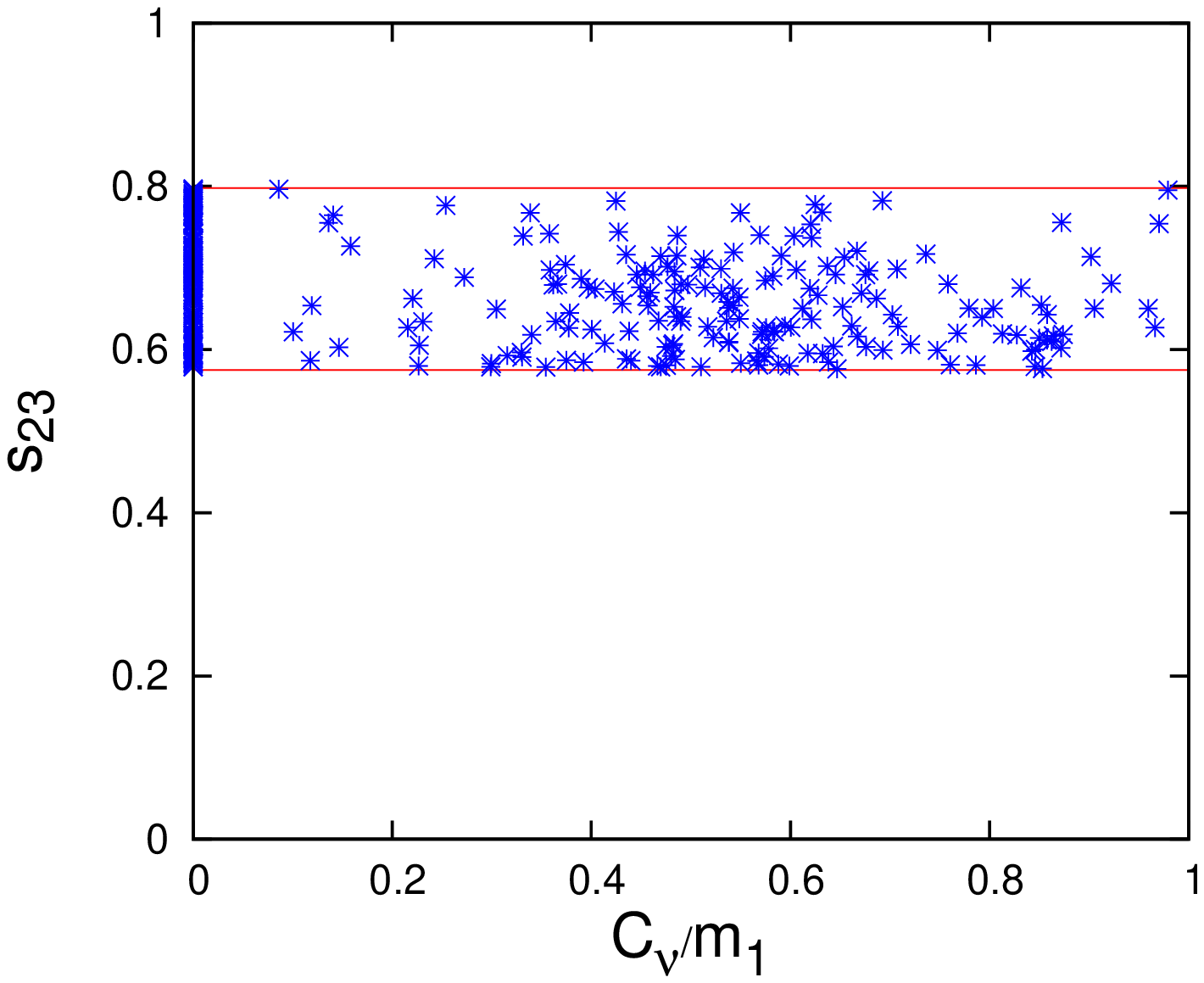}
\end{tabular}
\caption{Plots showing the variation of the mixing angles with the
(1,1) element of the neutrino mass matrix for  texture two zero
Dirac mass matrices (normal hierarchy).} 
\label{t2nh3}
\end{figure}

Further, in figures (\ref{t2nh2}) and (\ref{t2nh3}) we present the graphs showing variation of
the mixing angles with the parameters $C_l/m_e$ and $C_\nu/m_1$ respectively for structure (\ref{t20}) pertaining to
normal hierarchy of neutrino masses. While plotting these figures, all the three mixing angles have been constrained by their $3\sigma$ experimental
bounds, while all the free parameters have been given full variation. Taking a careful look at these
plots it is interesting to note that
the leptonic mixing angles donot have much dependence on the parameters $C_l$ and $C_\nu$.
Further one can see that a fit for all the three mixing angles can be obtained for the values of parameters being
$C_l \lesssim 0.9 m_e$ and $C_\nu \lesssim 0.9 m_1$.

\begin{figure}
\begin{minipage} {0.45\linewidth} \centering
\includegraphics[width=2.0in,angle=-90]{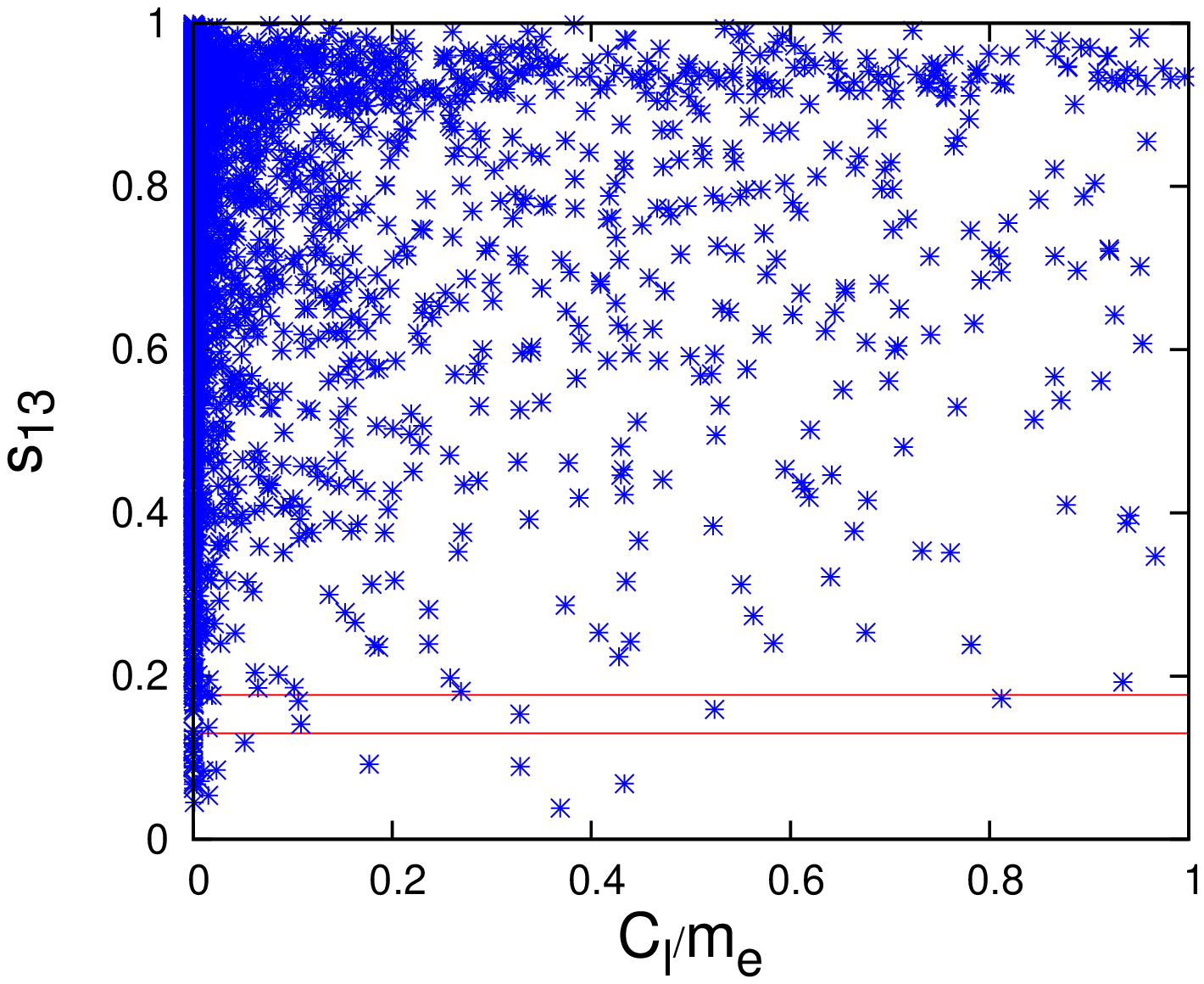}
 \end{minipage}
 \hspace{1.2cm}
 \begin{minipage} {0.45\linewidth} \centering
\includegraphics[width=2.0in,angle=-90]{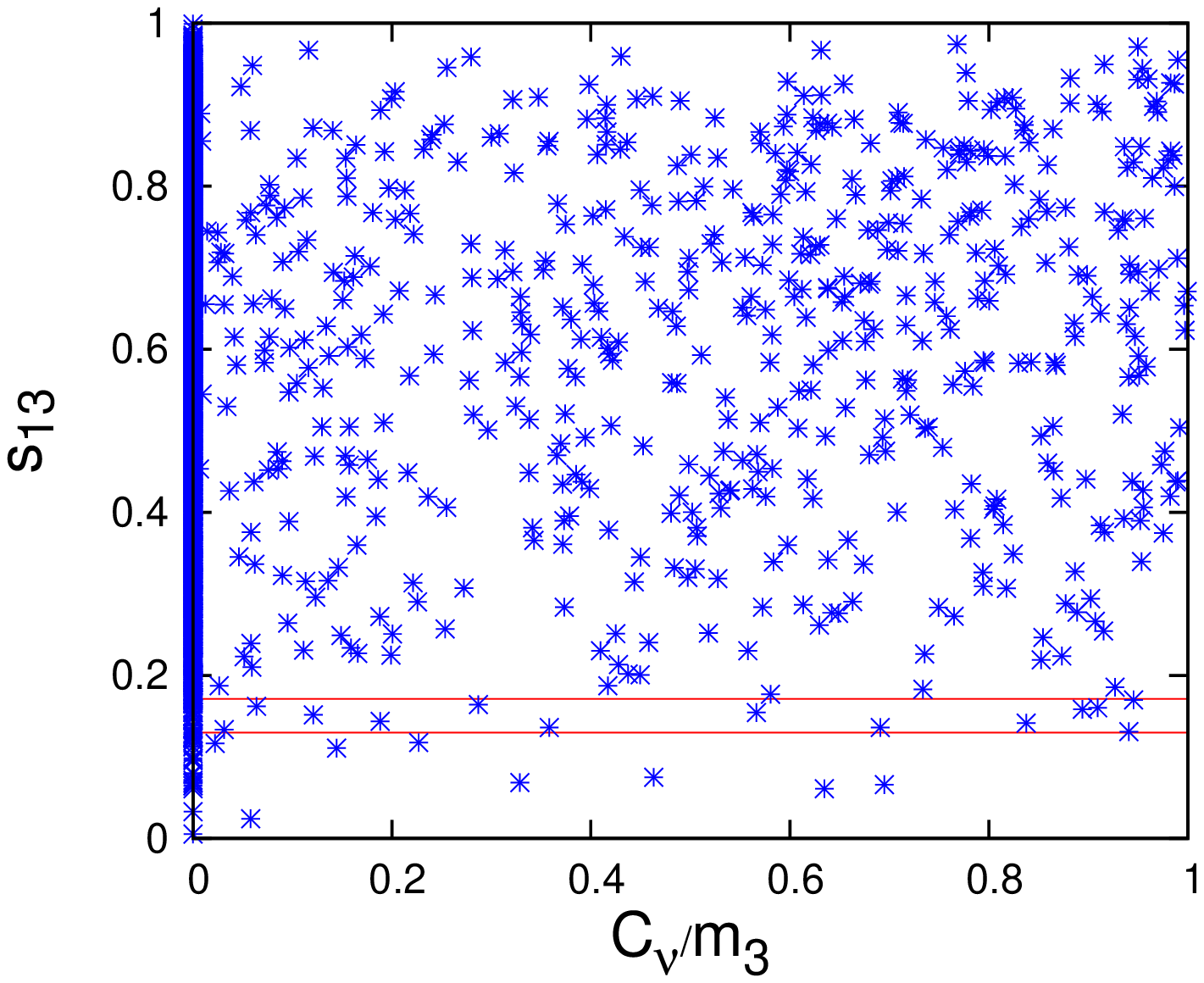}
 \end{minipage}
\caption{Plots showing the variation of the mixing angle $s_{13}$
with the (1,1) element of the charged lepton and neutrino mass
matrices for texture two zero Dirac mass matrices (inverted
hierarchy).} \label{t2ih2}
\end{figure}

 To further emphasise this conclusion, we present in figure (\ref{t2ih2})
the plots showing the dependence of leptonic mixing angle $s_{13}$ on $C_l/m_e$ and $C_\nu/m_3$, constraining the other two angles by their
$3 \sigma$ experimentally allowed ranges for inverted hierarchy of neutrino masses. The two parallel lines in these
figures show the $3 \sigma$ allowed range for the mixing angle $s_{13}$. Figure (\ref{t2ih2}) clearly shows that
in order to accomodate the latest $3 \sigma$ ranges for the mixing parameters, one requires $C_l/m_e$ and $C_\nu/m_3$ $\lesssim$ 1.
Therefore, one can conclude that very small values of the parameters $C_l$ and $C_\nu$ are required to fit the latest experimental
data for the structure (\ref{t20}) corresponding to both normal as well as inverted neutrino mass orderings and the structure
(\ref{t20}) thus essentially reduces to
\be
 M_{l}=\left( \ba{ccc}
0 & A _{l} & 0      \\
A_{l}^{*} & D_{l} &  B_{l}     \\
 0 &     B_{l}^{*}  &  E_{l} \ea \right), \qquad
M_{\nu D}=\left( \ba{ccc} 0 & A _{\nu } & 0\\
A_{\nu }^{*} & D_{\nu } & B_{\nu }     \\
 0 &     B_{\nu }^{*}  &  E_{\nu } \ea \right)
\label{flt40}\ee
The structure (\ref{flt40}) is referred to as Fritzsch-like texture four zero structure and is studied extensively in
literature \cite{t40lep}. However, no such attempt has been made after the recent measurement of the mixing angle $s_{13}$. Therefore, it becomes
interesting to analyse this structure in detail for its compatibility with the latest lepton mixing data.

\subsection{Texture four zero lepton mass matrices}
As discussed in the previous section, starting with the most general lepton mass matrices having `natural
hierarchy' \cite{nmm} among their elements, one essentially arrives at the Fritzsch-like texture four zero
structure as given in eqn.(\ref{flt40}). In this context, it is interesting to
note that different four-zero texture parallel matrices, with zeros
located in different positions, may have exactly the same physical content.
Indeed, they can be related by a WB transformation, performed by a permutation
 matrix P ,
\beqn
m_l^\prime = P^T m_l P\\ \nonumber
m_\nu ^\prime= P^T m_{\nu} P
\eeqn

which automatically preserves the parallel structure, but changes the position
of the zeros.
\begin{table}
{\renewcommand{\arraystretch}{1.0} \setlength\arraycolsep{0pt}
\bt{|c|c|c|c|c|} \hline
 & Class I & Class II  & Class III &
 Class IV \\ \hline &&&&\\
a & $\left ( \ba{ccc} {\bf 0} & Ae^{i\alpha} & {\bf 0} \\
Ae^{-i\alpha}  & D & Be^{i\beta} \\ {\bf 0} & Be^{-i\beta}  & E
\ea \right )$  &
 $\left ( \ba{ccc} D & Ae^{i\alpha} &
{\bf 0} \\ Ae^{-i\alpha}  & {\bf 0} &  Be^{i\beta} \\ {\bf 0} &
Be^{-i\beta}  & E \ea \right )$     &
 $\left ( \ba{ccc} {\bf 0} & Ae^{i\alpha} &
 Be^{i\gamma}\\ Ae^{-i\alpha}  &{\bf 0}  & De^{i\beta} \\ Be^{-i\gamma} & De^{-i\beta}  &
E \ea \right )$ & $\left ( \ba{ccc} A & {\bf 0} & {\bf 0} \\ {\bf
0}  & D & Be^{i\beta}\\ {\bf 0} & Be^{-i\beta} & E \ea \right )$\\
& & & &\\
\hline &&&&\\
b &  $\left ( \ba{ccc} {\bf 0} & {\bf 0}  & Ae^{i\alpha} \\ {\bf
0}  & E & Be^{i\beta} \\  Ae^{-i\alpha} & Be^{-i\beta}  & D \ea
\right )$  &
 $\left ( \ba{ccc} D & {\bf 0} & Ae^{i\alpha}
 \\ {\bf 0} & E & Be^{i\beta} \\  Ae^{-i\alpha} & Be^{-i\beta}  &
{\bf 0} \ea \right )$     & $\left ( \ba{ccc} {\bf 0} &
De^{i\gamma} & Ae^{i\alpha}
\\  De^{-i\gamma}  & E  & Be^{i\beta} \\  Ae^{-i\alpha} & Be^{-i\beta}  &
 {\bf 0}\ea \right )$ & $\left (\ba{ccc}E & {\bf 0} & Be^{i\alpha}
 \\ {\bf 0}  & A & {\bf 0} \\ Be^{-i\alpha}  & {\bf 0}  &
D \ea \right )$\\   & & && \\
\hline &&&&\\
c &  $\left ( \ba{ccc} D & Ae^{i\alpha} &
  Be^{i\beta}\\ Ae^{-i\alpha}  & {\bf 0}  & {\bf 0} \\  Be^{-i\beta} & {\bf 0}
& E \ea \right )$  & $\left ( \ba{ccc} {\bf 0} & Ae^{i\alpha} &
 Be^{i\beta} \\ Ae^{-i\alpha}  & D & {\bf 0}  \\  Be^{-i\beta}  & {\bf 0} &
E \ea \right )$     & $\left ( \ba{ccc} {\bf 0} & Ae^{i\alpha} &
Be^{i\beta}
\\ Ae^{-i\alpha} & {\bf 0} & De^{i\gamma} \\ Be^{-i\beta}  & De^{-i\gamma} &
E \ea \right )$ & $\left ( \ba{ccc} E & Be^{i\alpha} & {\bf 0} \\
Be^{-i\alpha}  & D & {\bf 0} \\ {\bf 0} & {\bf 0}  & A \ea \right
)$ \\  & & && \\
\hline &&&&\\
d &  $\left ( \ba{ccc} E & Be^{i\beta} & {\bf 0}
 \\ Be^{-i\beta} & D & Ae^{i\alpha}  \\ {\bf 0} & Ae^{-i\alpha} & {\bf 0}
 \ea \right )$   &  $\left ( \ba{ccc} E & Be^{i\alpha} &
{\bf 0} \\ Be^{-i\alpha}  & {\bf 0} &  Ae^{i\beta} \\ {\bf 0} &
Ae^{-i\beta}  & D \ea \right )$   &  $\left ( \ba{ccc} {\bf 0} &
Be^{i\alpha} &
 Ee^{i\gamma}\\ Be^{-i\alpha}  &{\bf 0}  & Ae^{i\beta} \\ Ee^{-i\gamma} & Ae^{-i\beta}  &
D \ea \right )$   & $\left ( \ba{ccc} A & {\bf 0} & {\bf 0} \\
{\bf 0}  & E & Be^{i\beta}\\ {\bf 0} & Be^{-i\beta} & D \ea \right
)$ \\  & & && \\
\hline &&&&\\

 e &  $\left ( \ba{ccc} D & Be^{i\beta} & Ae^{i\alpha}
  \\ Be^{-i\beta}  & E & {\bf 0} \\ Ae^{-i\alpha} & {\bf 0}
& {\bf 0} \ea \right )$  &  $\left ( \ba{ccc} E & {\bf 0} &
Be^{i\alpha}
 \\ {\bf 0} & D & Ae^{i\beta} \\  Be^{-i\alpha} & Ae^{-i\beta}  &
{\bf 0} \ea \right )$   & $\left ( \ba{ccc} {\bf 0} &
Ee^{i\gamma} & Be^{i\alpha}
\\  Ee^{-i\gamma}  & D & Ae^{i\beta} \\  Be^{-i\alpha} & Ae^{-i\beta}  &
 {\bf 0}\ea \right )$ & $\left (\ba{ccc}D & {\bf 0} & Be^{i\alpha}
 \\ {\bf 0}  & A & {\bf 0} \\ Be^{-i\alpha}  & {\bf 0}  &
C \ea \right )$ \\  & & && \\
\hline &&&&\\

f &  $\left ( \ba{ccc} E & {\bf 0} & Be^{i\beta}
 \\ {\bf 0} & {\bf 0} & Ae^{i\alpha}  \\ Be^{-i\beta} & Ae^{-i\alpha} & D
 \ea \right )$  & $\left ( \ba{ccc} {\bf 0} & Be^{i\alpha} &
 Ae^{i\beta} \\ Be^{-i\alpha}  & E & {\bf 0}  \\  Ae^{-i\beta}  & {\bf 0} &
D \ea \right )$  & $\left ( \ba{ccc} {\bf 0} & Be^{i\alpha} &
Ae^{i\beta}
\\ Be^{-i\alpha} & {\bf 0} & Ee^{i\gamma} \\ Ae^{-i\beta}  & Ee^{-i\gamma} &
D \ea \right )$ & $\left ( \ba{ccc} E & Be^{i\alpha} & {\bf 0} \\
Be^{-i\alpha}  & D & {\bf 0} \\ {\bf 0} & {\bf 0}  & A \ea \right
)$\\
  \hline
\et \vspace{0.7 cm}} \caption{Table showing various
phenomenologically allowed texture 2 zero possibilities,
categorized into four distinct classes.} \label{3t4}
\end{table}
The matrix P belongs to the group of six permutation matrices,
which are isomorphic to $S_3$ . The four-zero texture ansatz can then
be classified into four classes, as indicated in table (\ref{3t4}). It can easily be seen
that the matrices given in eqn.(\ref{flt40})
belong to class I of table (\ref{3t4}).
In this context, it becomes
interesting to confront  all the classes of lepton mass matrices
with the latest experimental data.  From the table, it is  clear that class IV is  not viable as all
the matrices in this class correspond to the scenario where one of
the generations gets decoupled from the other two. Such classification of all
possible parallel texture four zero structures was first presented
in the analysis by Branco {\it{et al.}} \cite{branco} wherein they carry out detailed
study of lepton mass matrices corresponding to all the four classes pertaining to Majorana
neutrinos. However, such analysis has not been carried out considering the neutrinos to be Dirac
like particles. Therefore, in this section, we carry out a detailed study of all classes of
texture four zero lepton mass matrices assuming the neutrinos to be Dirac type.
\subsubsection{Class I ansatz}
\begin{figure}
\begin{tabular}{cc}
  \includegraphics[width=0.2\paperwidth,height=0.2\paperheight,angle=-90]{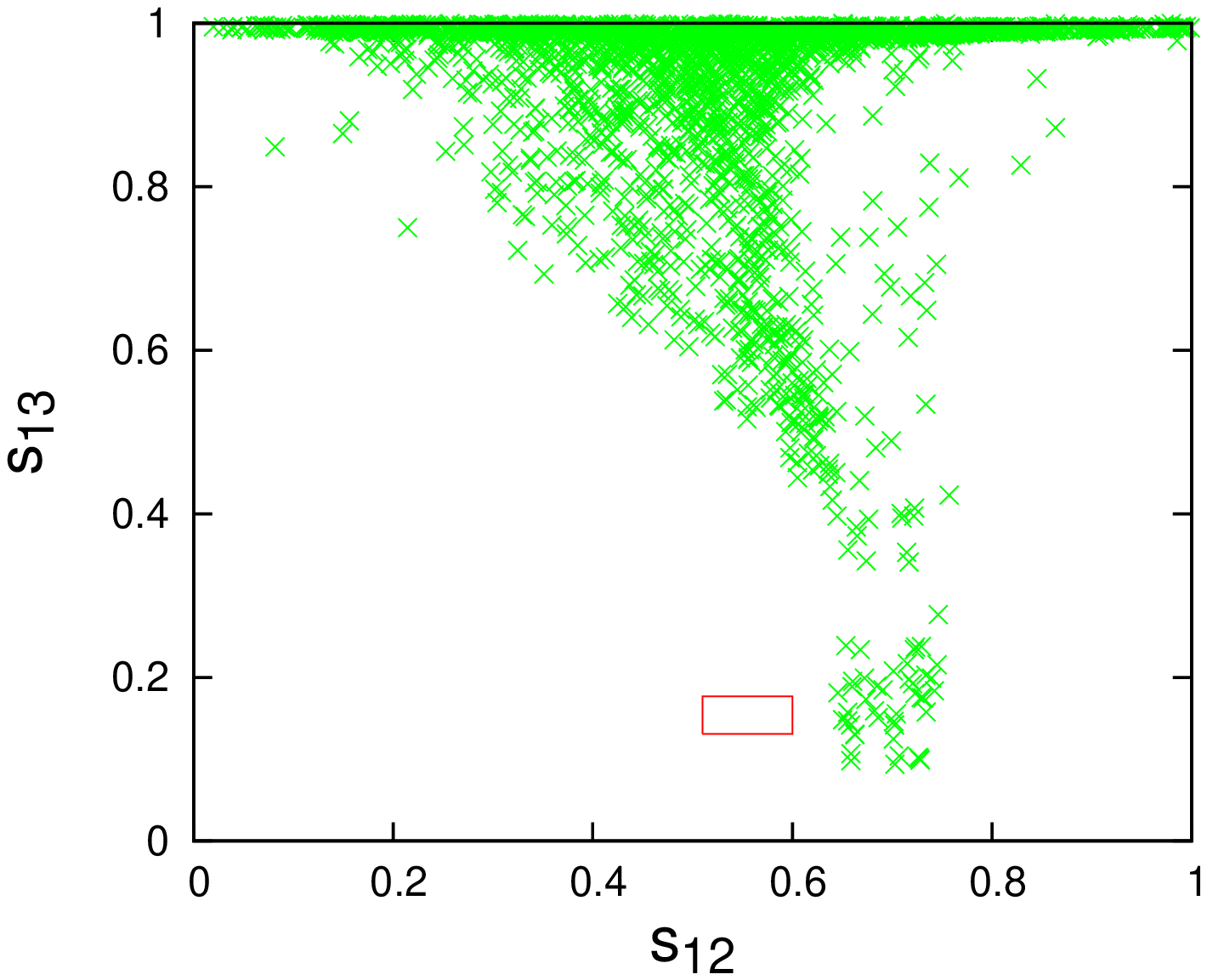}
  \includegraphics[width=0.2\paperwidth,height=0.2\paperheight,angle=-90]{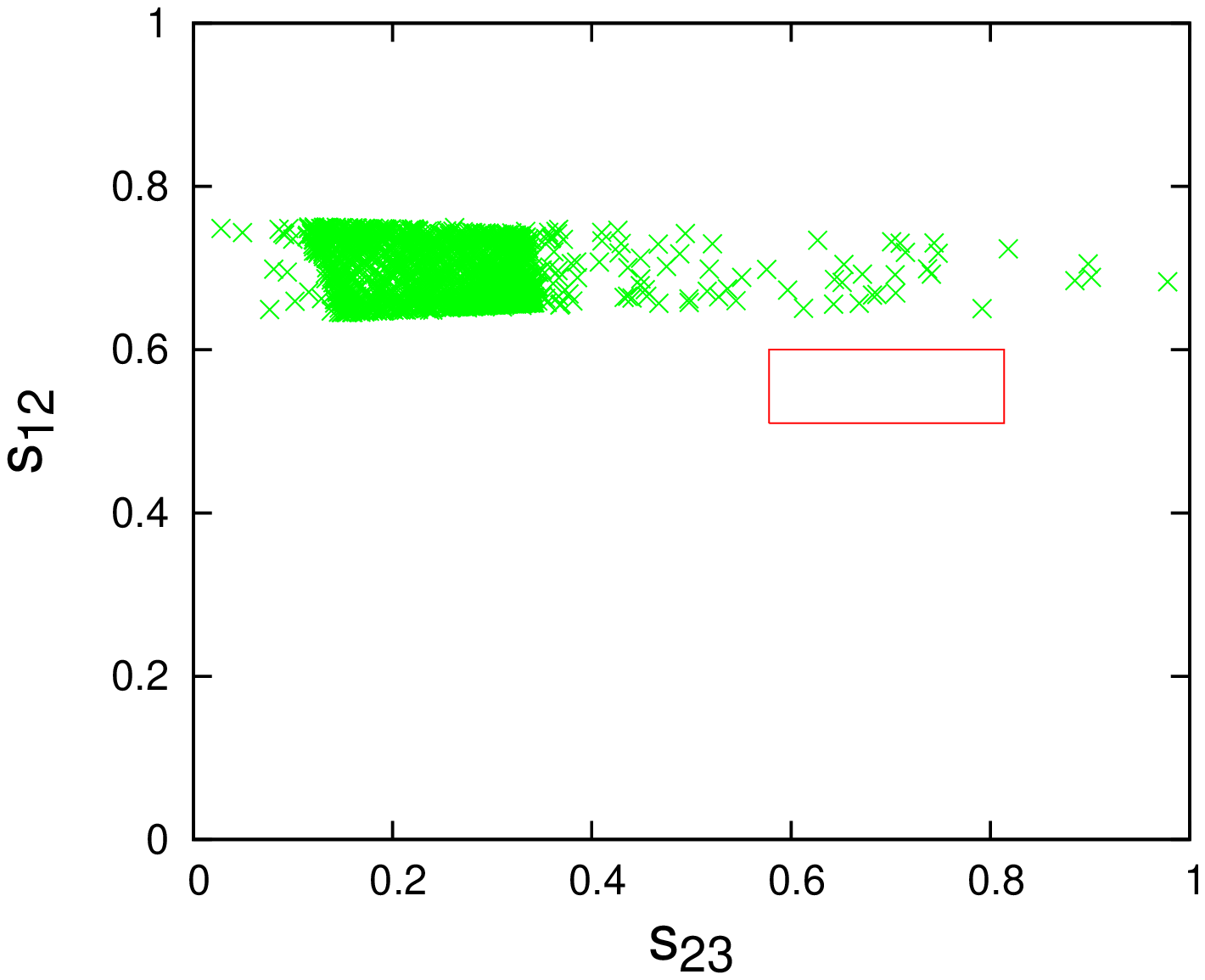}
  \includegraphics[width=0.2\paperwidth,height=0.2\paperheight,angle=-90]{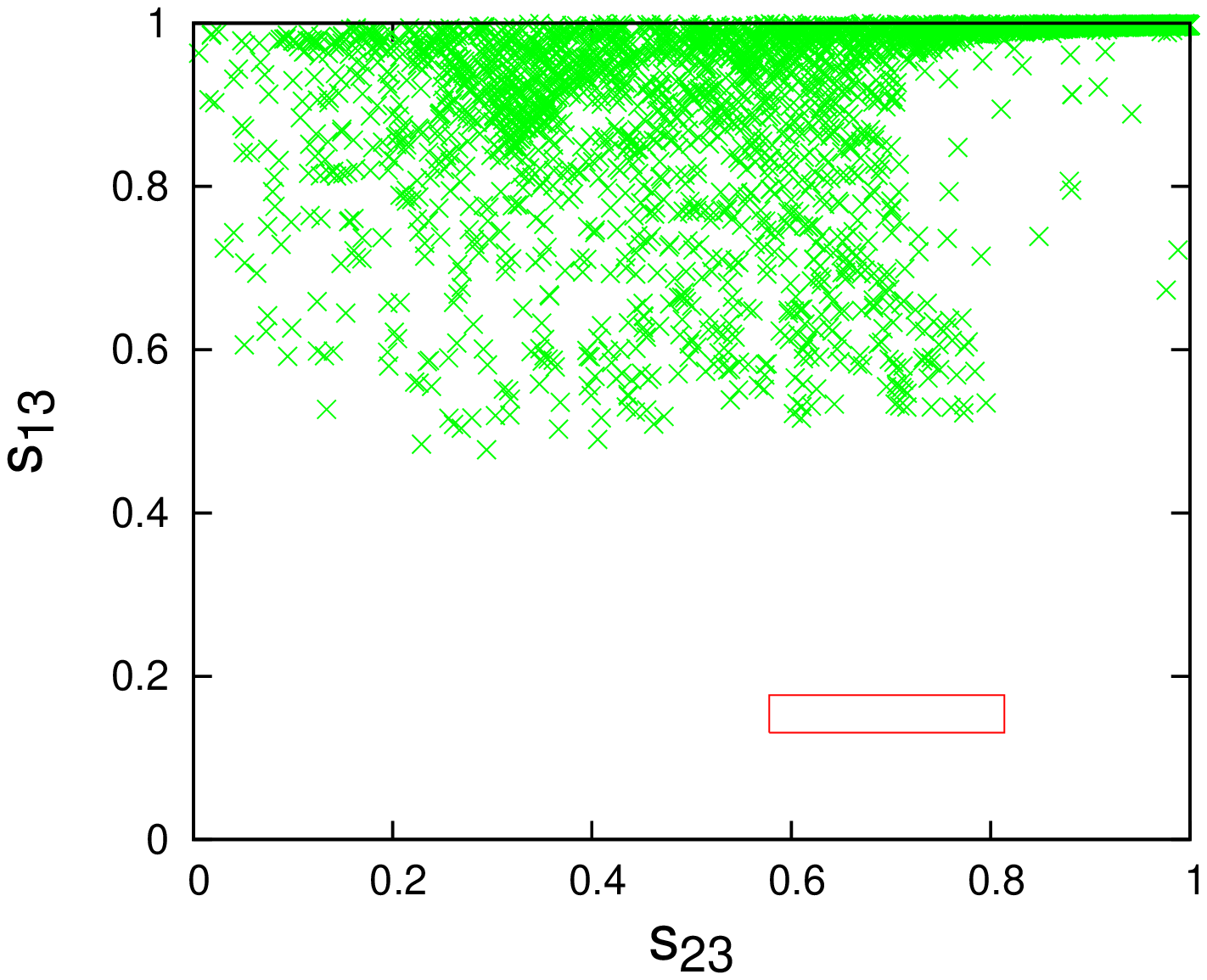}
\end{tabular}
\caption{Plots showing the parameter space for any two mixing
angles when the third angle is constrained by its  $1 \sigma$
range for Class I ansatz of texture four zero  Dirac mass matrices
(inverted hierarchy) .} \label{t4ih1}
\end{figure}

\begin{figure}
\begin{tabular}{cc}
  \includegraphics[width=0.2\paperwidth,height=0.2\paperheight,angle=-90]{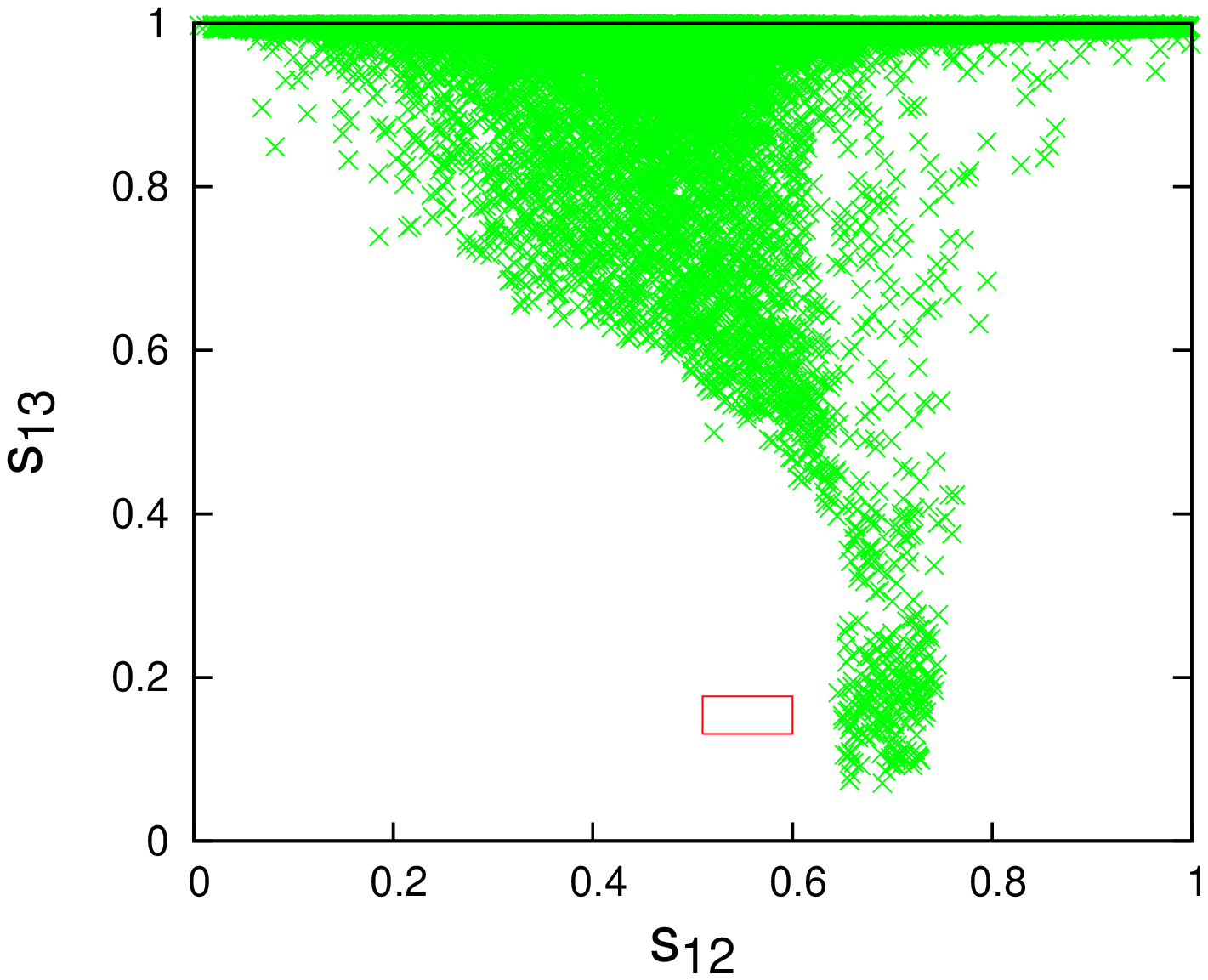}
  \includegraphics[width=0.2\paperwidth,height=0.2\paperheight,angle=-90]{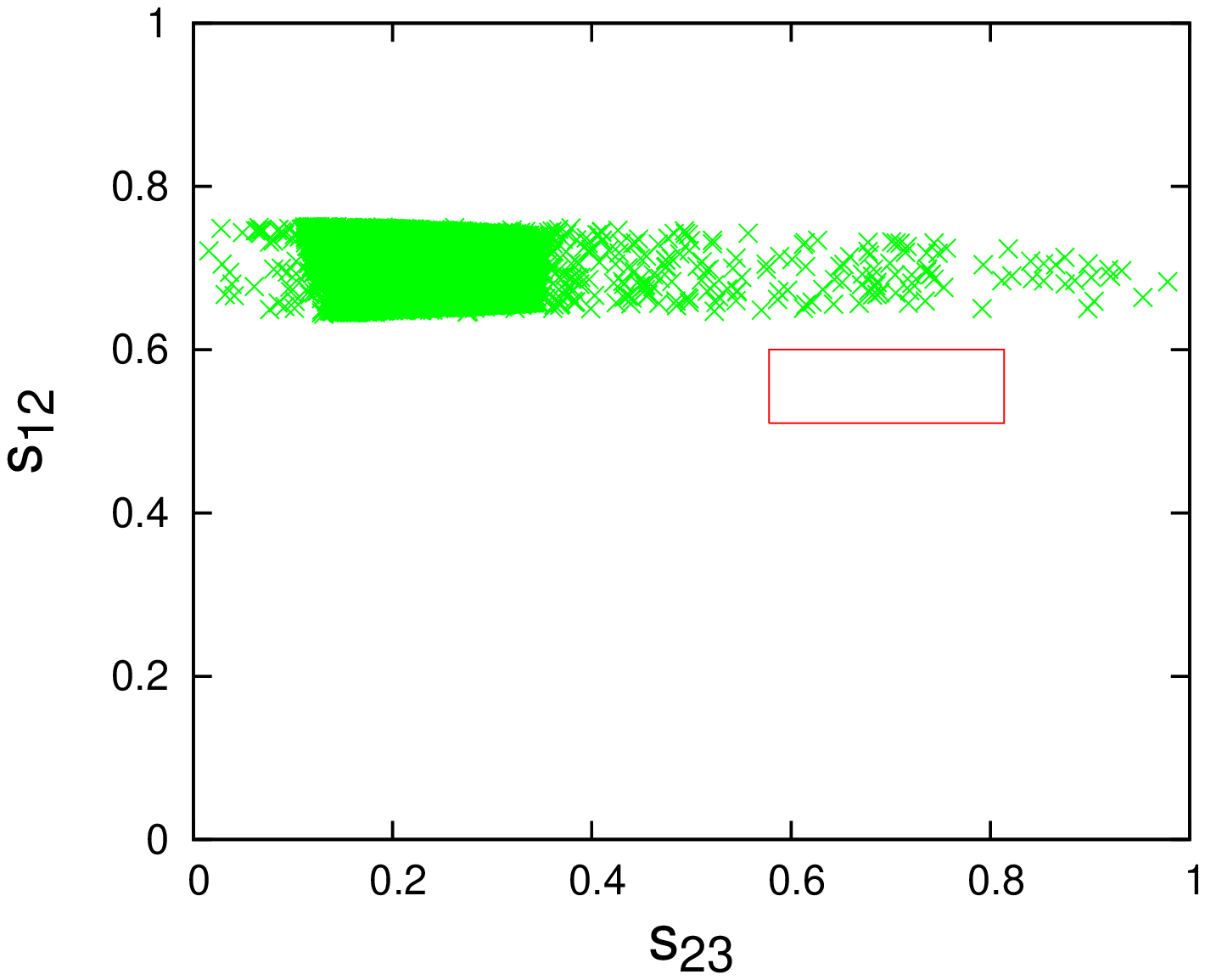}
  \includegraphics[width=0.2\paperwidth,height=0.2\paperheight,angle=-90]{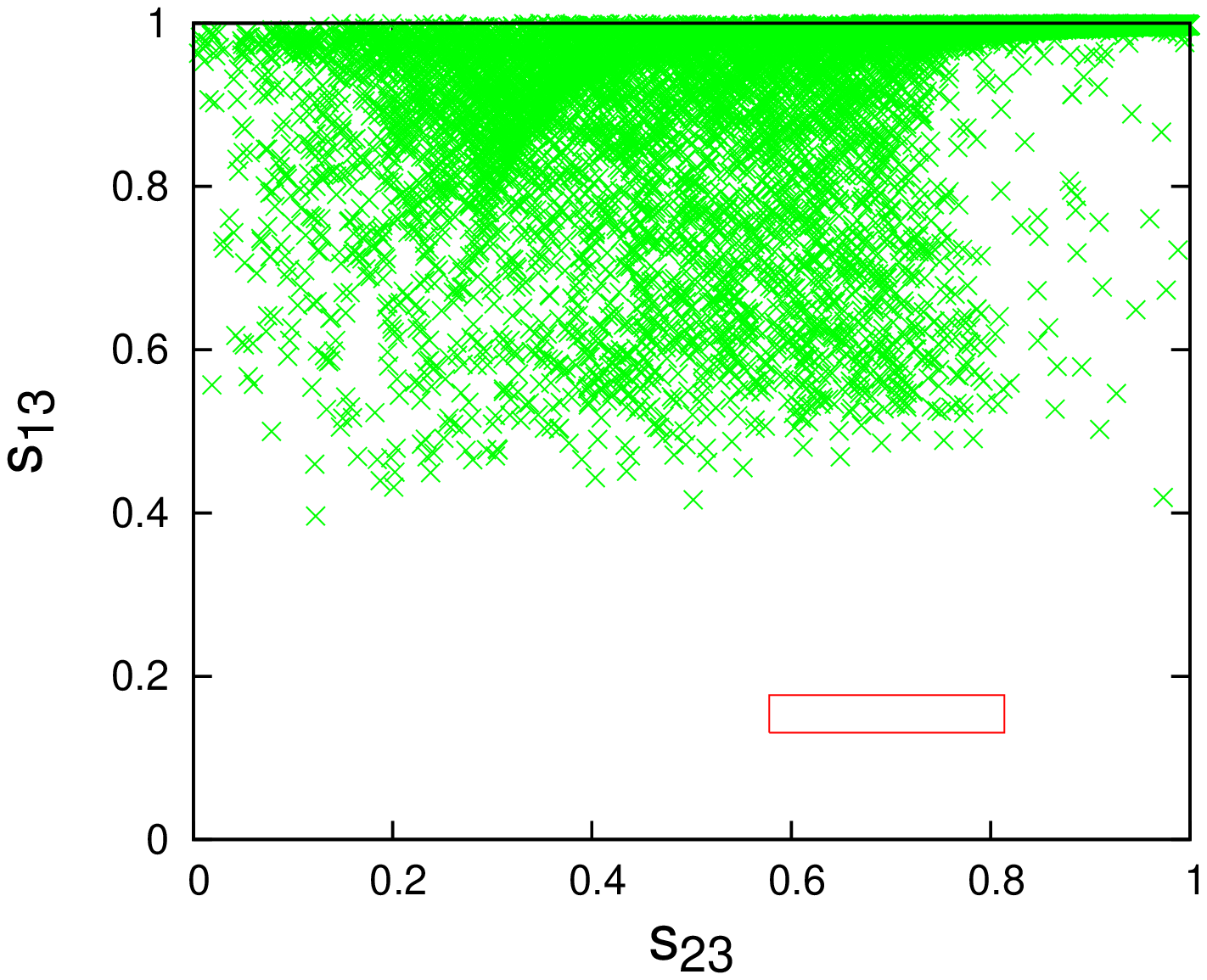}
\end{tabular}
\caption{Plots showing the parameter space for any two mixing angles when the third angle is constrained by
its  $3 \sigma$ range for Class I ansatz of texture four zero  Dirac mass matrices (inverted
hierarchy).}
\label{t4ih2}
\end{figure}

To begin with, we carry out a detailed analysis for texture four zero mass matrices belonging to class I
of table (\ref{3t4}), i.e.,
\be
 M_{i}=\left( \ba{ccc}
0 & A _{i}e^{i\alpha_i} & 0      \\
A_{i}e^{-i\alpha_i} & D_{i} &  B_{i}e^{i\beta_i}     \\
 0 &     B_{i}e^{-i\beta_i} &  E_{i} \ea \right),
\label{cl1}
\ee
where $i=l,~\nu_D$ corresponds to the charged lepton and Dirac neutrino mass matrices
respectively. The diagonalizing transformations for this class can be obtained by substituting
$C_l=C_\nu=0$ in eqn. (\ref{diat20}). For the purpose of calculations, the elements
$D_l$, $D_\nu$ as well as the phases $\phi_1$ and $\phi_2$ have been considered as free parameters.
Following the methodology as discussed in section (\ref{metho}), we attempt to carry out a detailed
study pertaining to normal, inverted as well as degenerate neutrino mass orderings. Firstly, we examine
the compatibility of mass matrices given in eqn. (\ref{flt40}) 
with the inverted hierarchy of neutrino masses. For this purpose in figures
(\ref{t4ih1}) and (\ref{t4ih2}), we present the parameter space corresponding to any two mixing angles while
the third one being constrained by its $1 \sigma$ and $3\sigma$ range respectively. The blank rectangular regions
in these figures represent the $3\sigma$ allowed ranges of the two mixing angles being considered. As can be
seen from all these plots, the parameter space of two angles does not show any overlap with the experimentally
allowed region. Therefore, we find that the inverted hierarchy 
seems to be ruled out at both the $ 1 \sigma$ as well as the $ 3\sigma$ level for
class I ansatz of texture four zero Dirac neutrino mass matrices.

\begin{figure}
\begin{tabular}{cc}
  \includegraphics[width=0.2\paperwidth,height=0.2\paperheight,angle=-90]{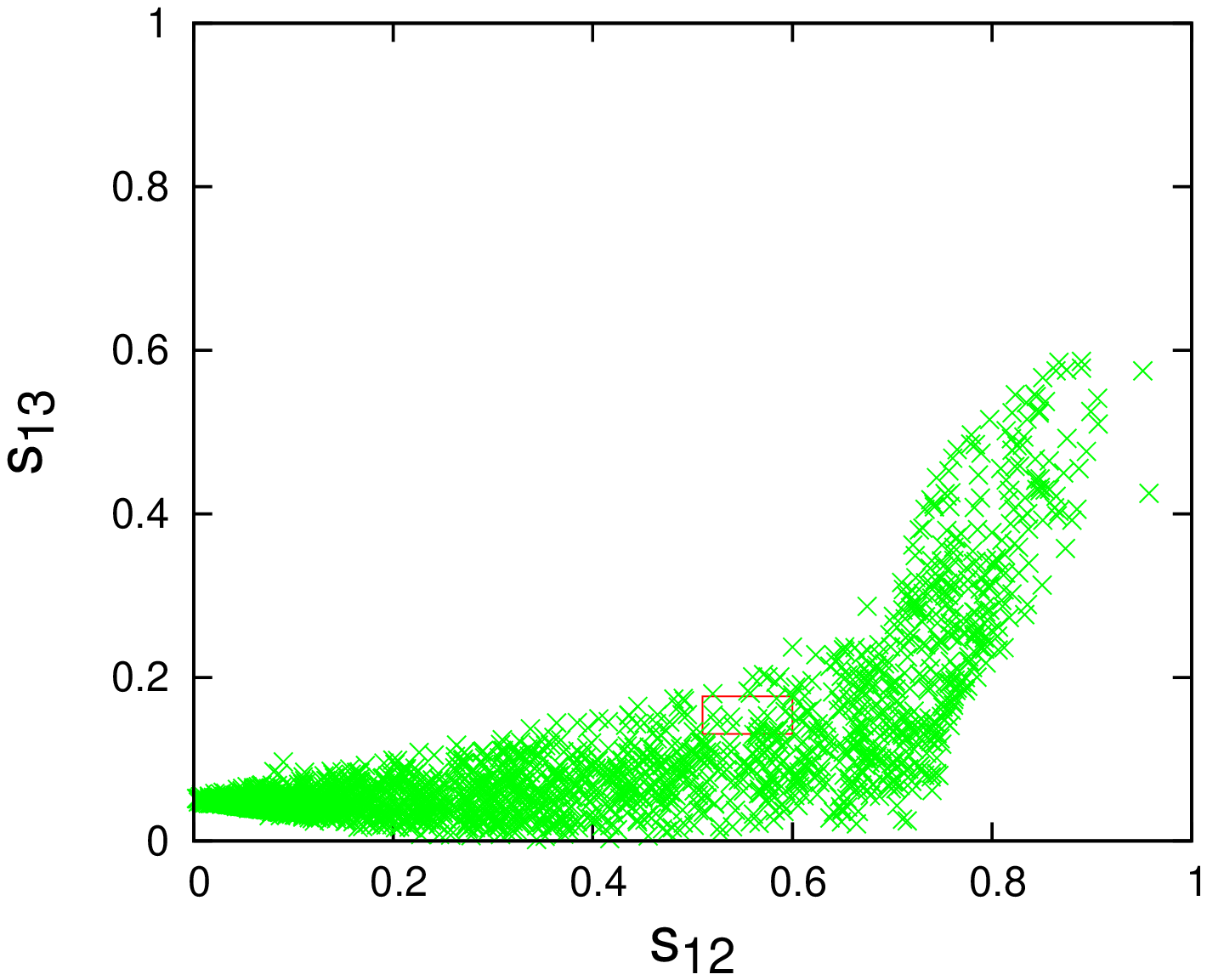}
  \includegraphics[width=0.2\paperwidth,height=0.2\paperheight,angle=-90]{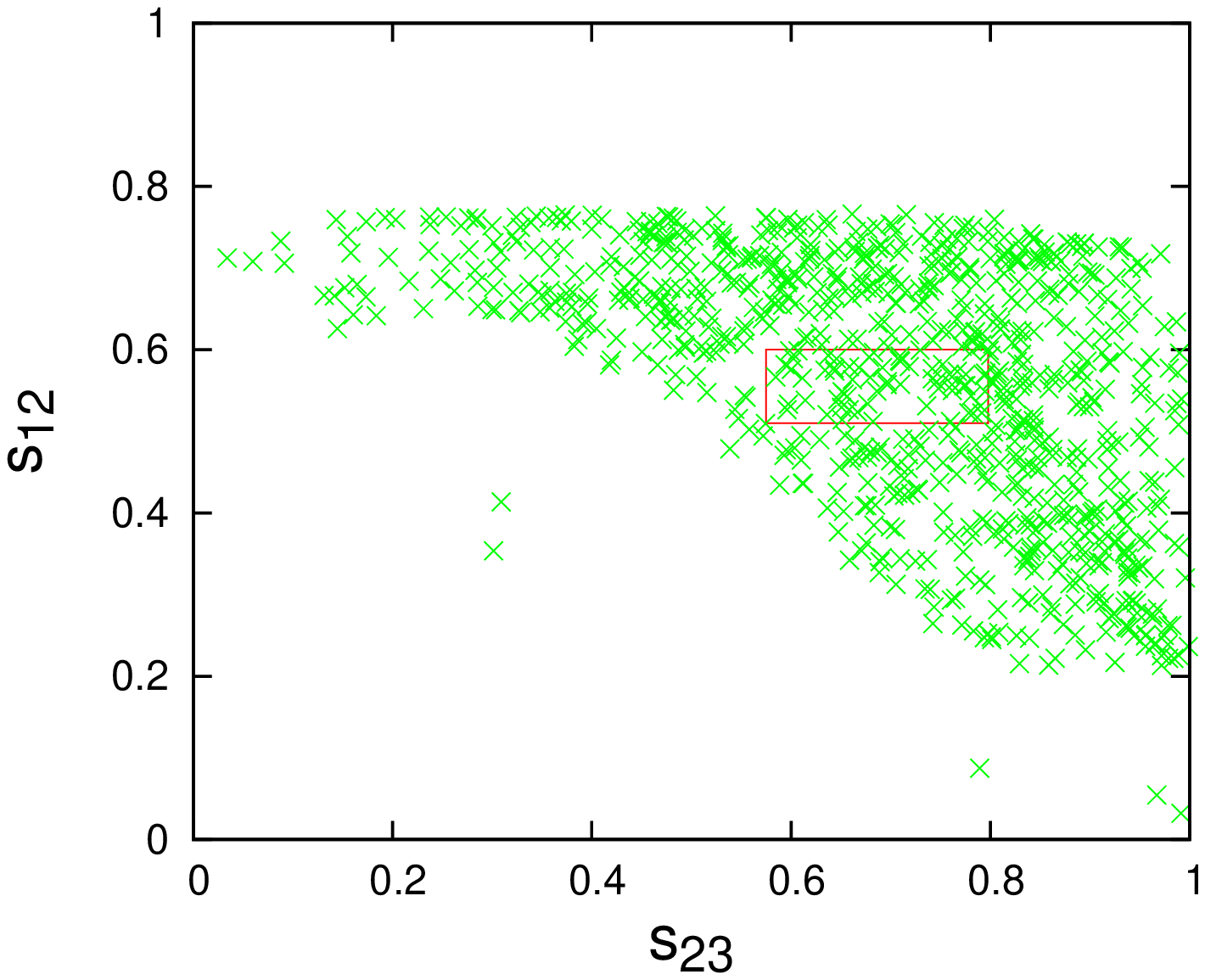}
  \includegraphics[width=0.2\paperwidth,height=0.2\paperheight,angle=-90]{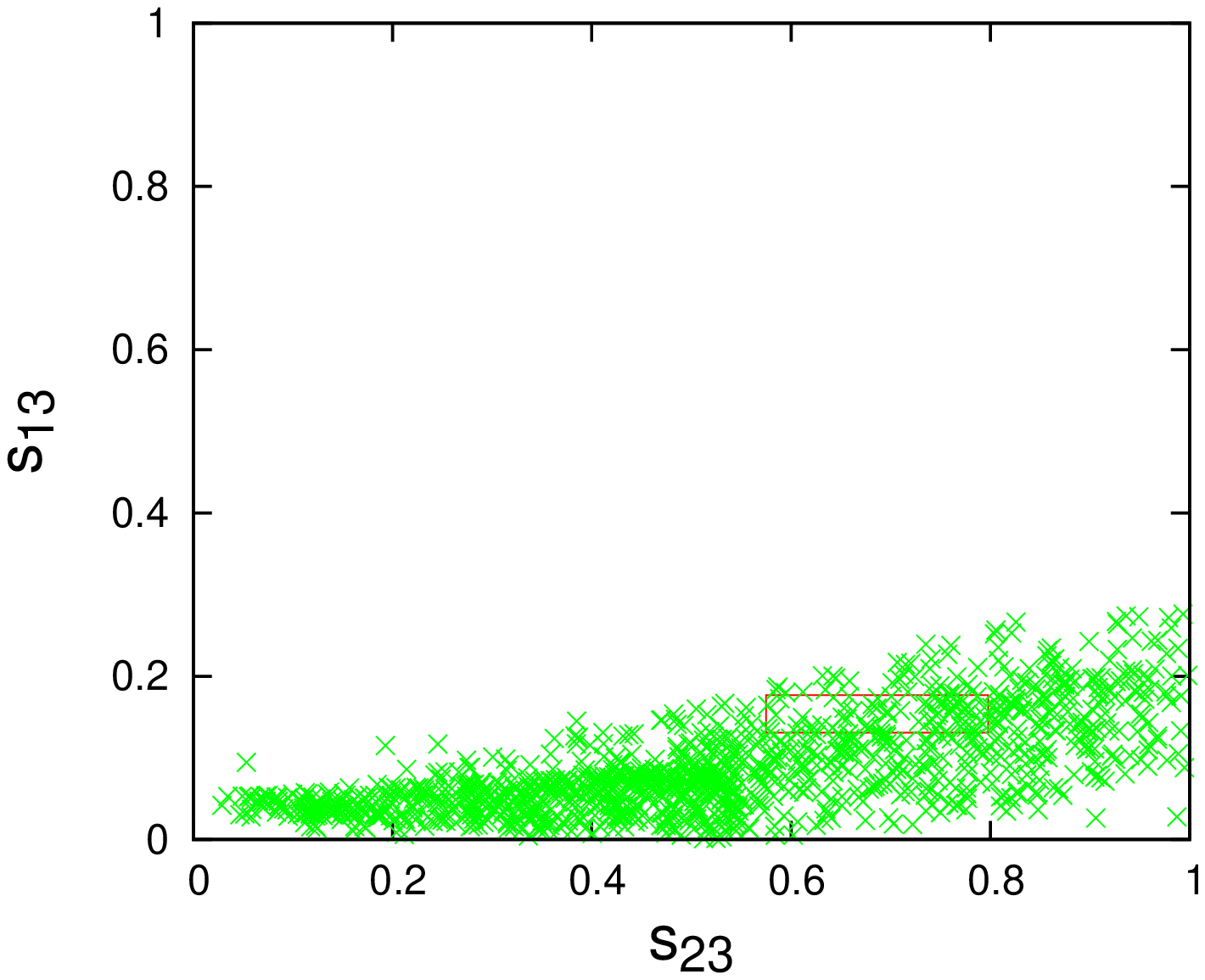}
\end{tabular}
\caption{Plots showing the parameter space for any two mixing
angles when the third angle is constrained by its  $1 \sigma$
range for Class I ansatz of texture four zero  Dirac mass matrices
(normal hierarchy) .} \label{t4nh1}
\end{figure}

\begin{figure}
\begin{tabular}{cc}
  \includegraphics[width=0.2\paperwidth,height=0.2\paperheight,angle=-90]{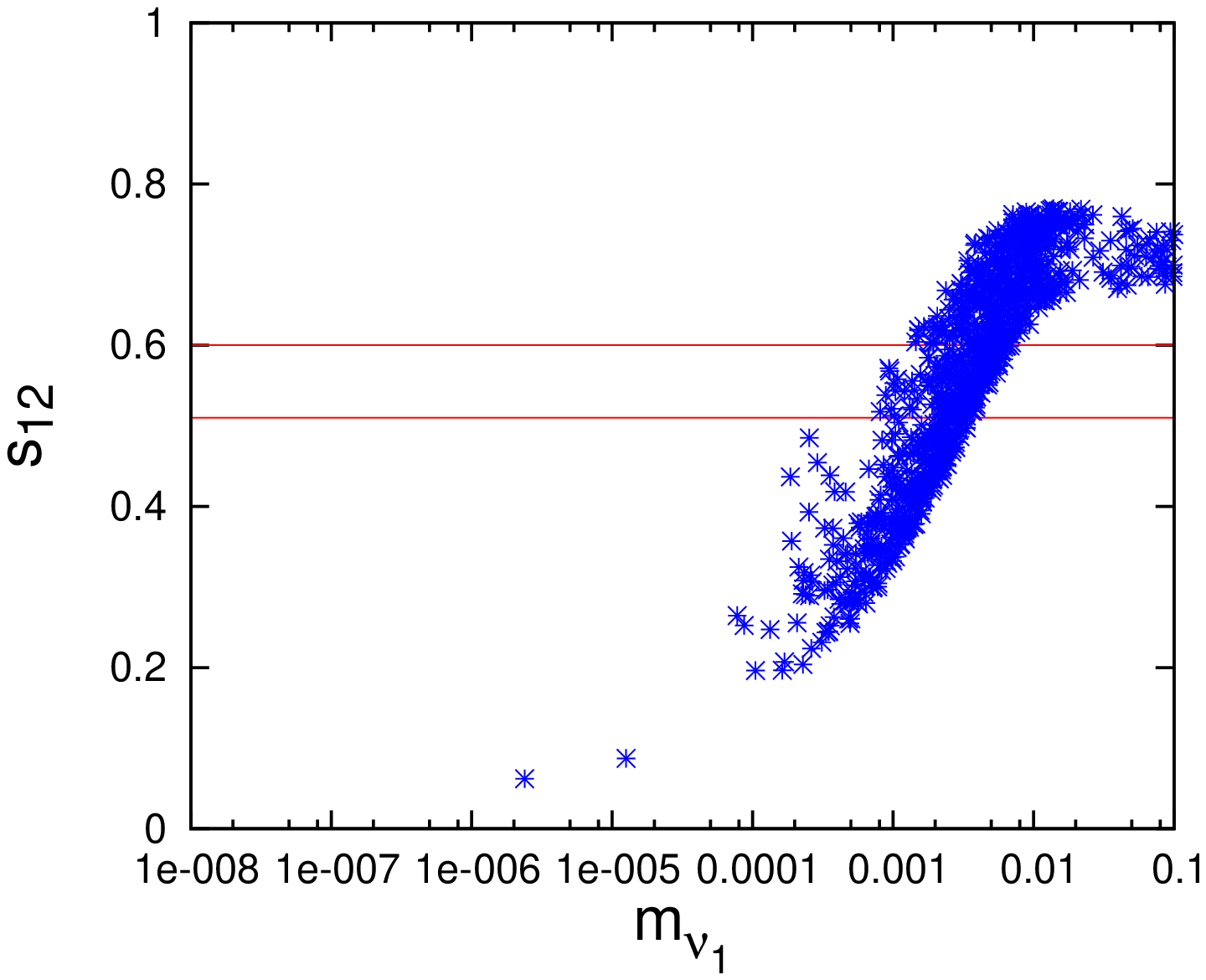}
  \includegraphics[width=0.2\paperwidth,height=0.2\paperheight,angle=-90]{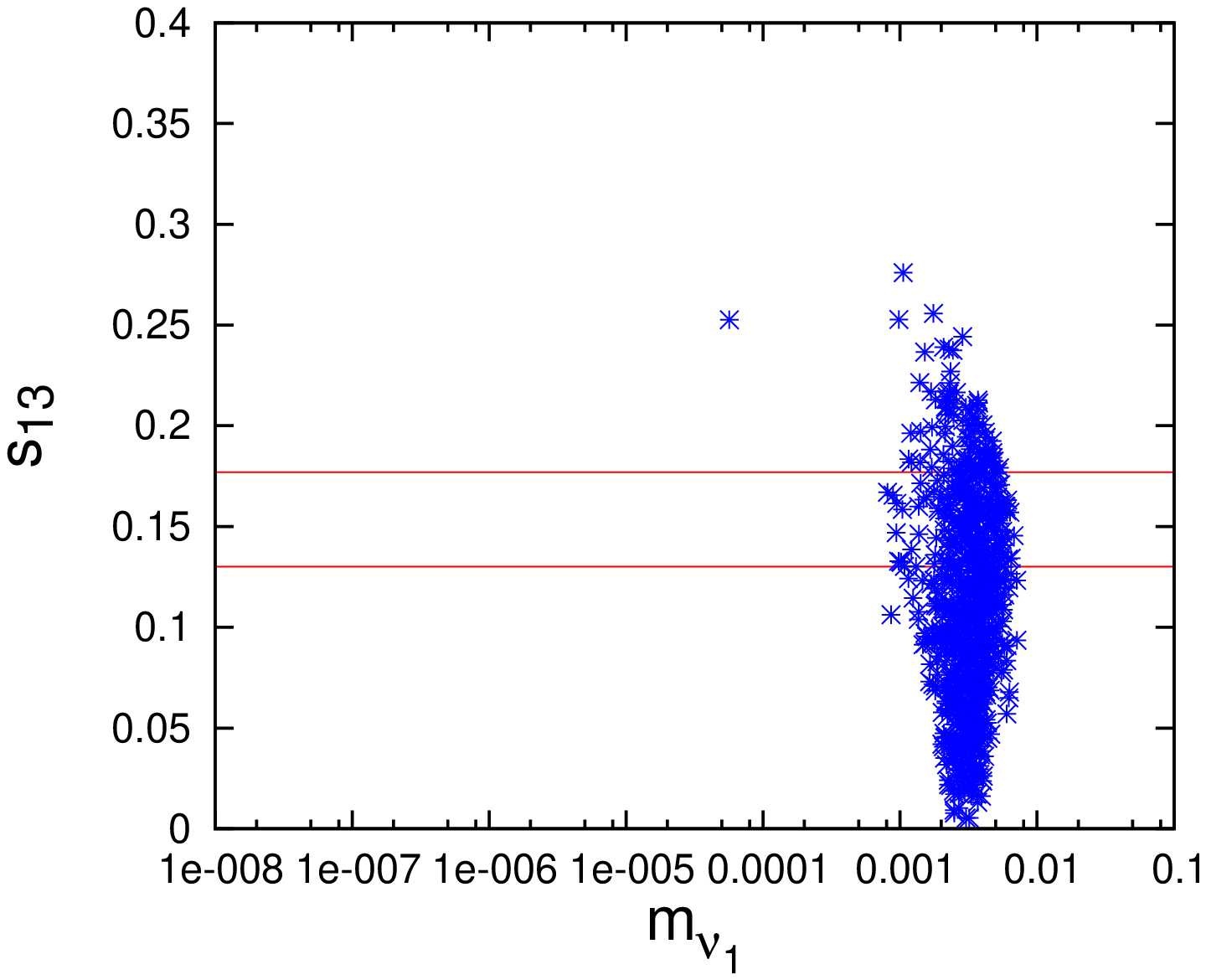}
  \includegraphics[width=0.2\paperwidth,height=0.2\paperheight,angle=-90]{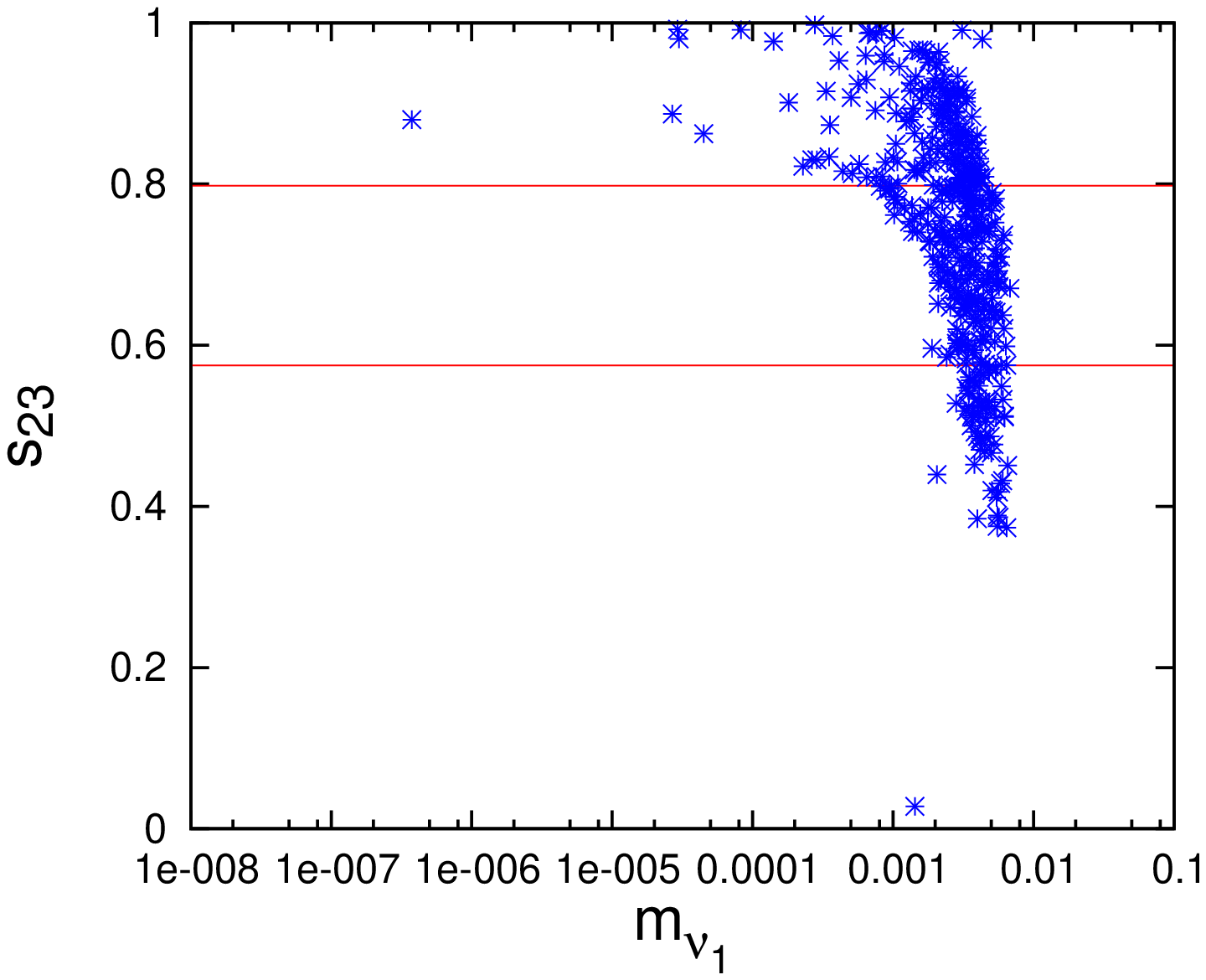}
\end{tabular}
\caption{Plots showing the dependence of mixing angles on the
lightest neutrino mass when the other two angles are constrained
by their $3 \sigma$ ranges  for  Class I ansatz of texture four
zero  Dirac mass matrices (normal hierarchy ).} \label{t4nh2}
\end{figure}

After ruling out the structure given in eqn.(\ref{cl1}) for the inverted hierarchy, we now proceed to examine the compatibility
of these matrices for the normal hierarchy case. To this end, in figure (\ref{t4nh1}) we present the plots showing the
parameter space corresponding to any two mixing angles wherein the third one is constrained by its $1 \sigma$ range.
Interestingly, normal hierarchy seems to be viable in this case as can be seen from these plots,
wherein the parameter space shows significant overlap with the experimentally allowed
$3 \sigma$ region shown by the rectangular boxes in each plot.
Further, in figure (\ref{t4nh2}) we present the graphs showing the variation
of the lightest neutrino mass with the mixing angles for normal hierarchy, keeping the
other two mixing angles constrained by their $3\sigma$ bounds. The parallel lines in each plot
show the $3\sigma$ range of the mixing angle being considered. Taking a careful look at these graphs,
one can find the range the lightest neutrino mass to be $0.001eV \lesssim m_{\nu 1} \lesssim 0.01 eV$ approximately.

\begin{figure}
\begin{tabular}{cc}
  \includegraphics[width=0.2\paperwidth,height=0.2\paperheight,angle=-90]{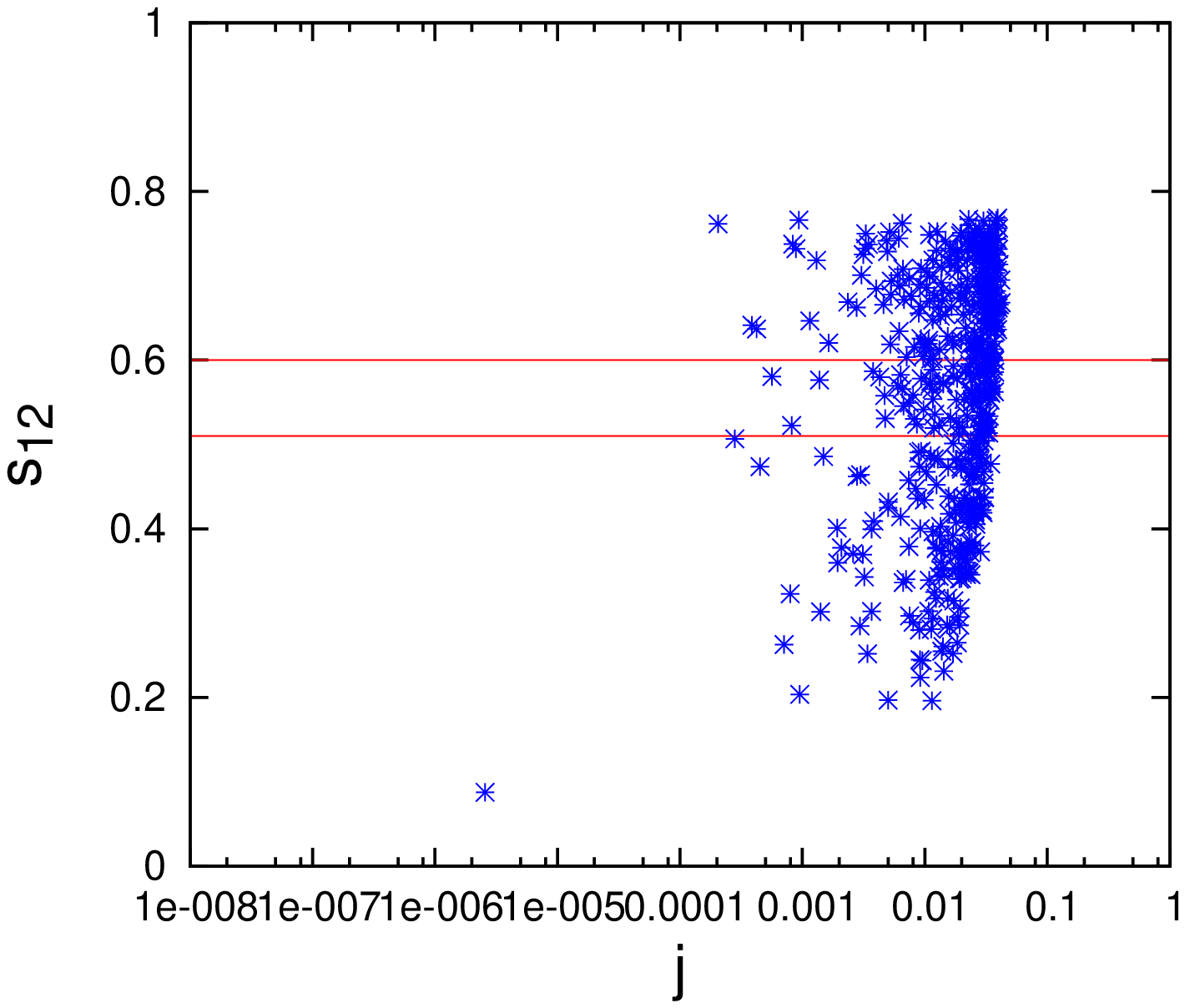}
  \includegraphics[width=0.2\paperwidth,height=0.2\paperheight,angle=-90]{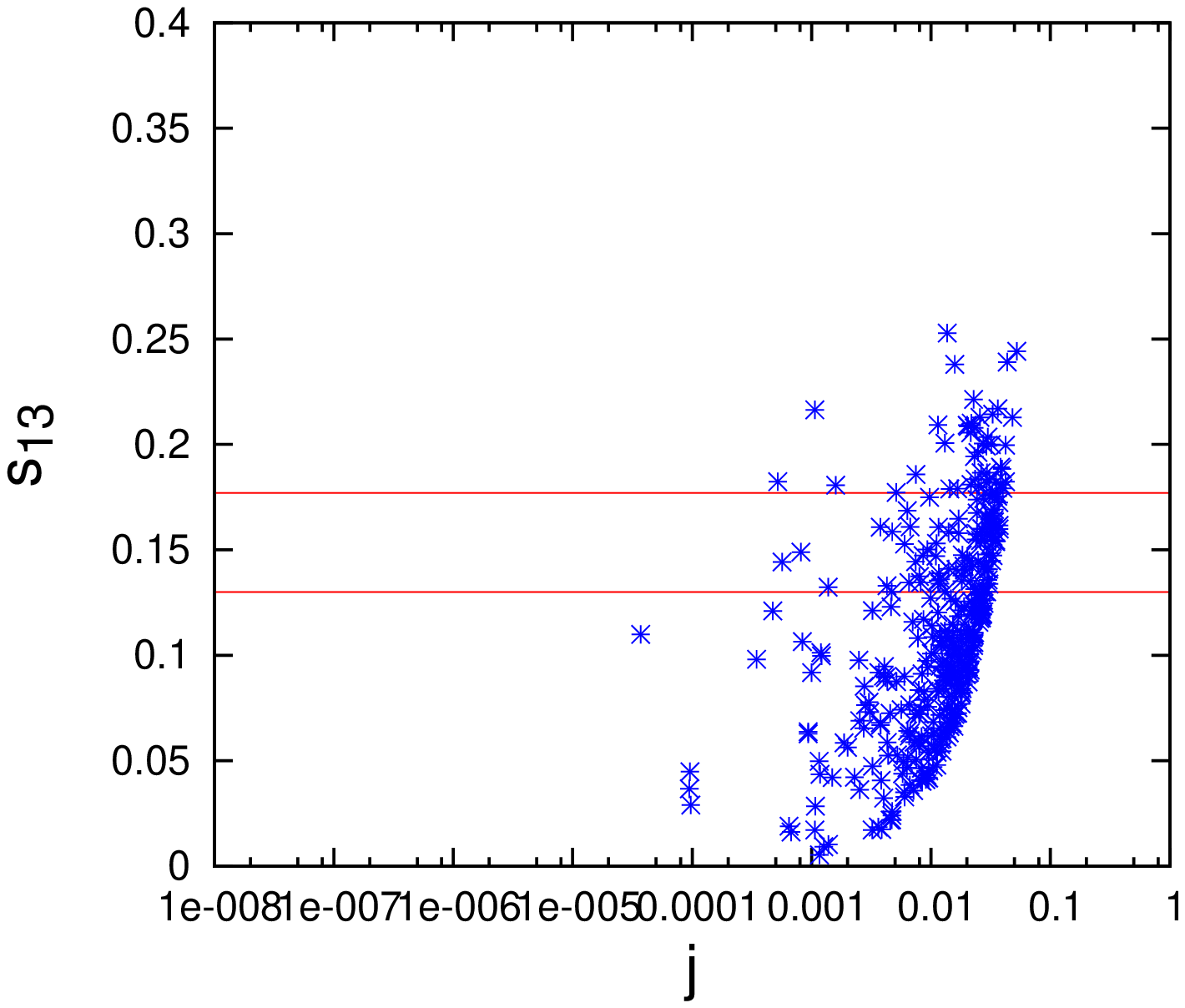}
  \includegraphics[width=0.2\paperwidth,height=0.2\paperheight,angle=-90]{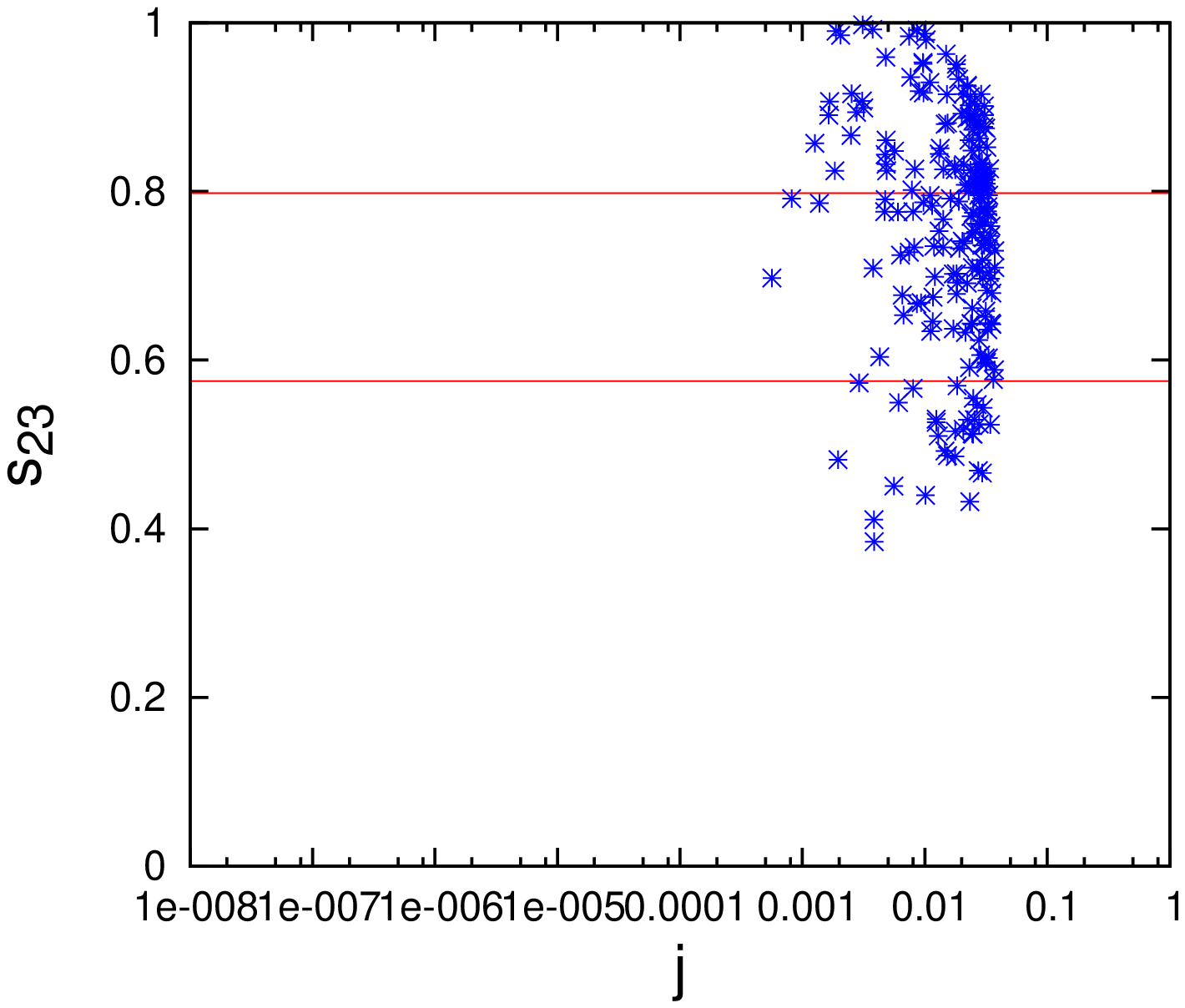}
\end{tabular}
\caption{Plots showing the variation of Jarlskog CP violating
parameter with mixing angles when the other two angles are
constrained by their $3 \sigma$ ranges  for  Class I ansatz of
texture four zero  Dirac mass matrices (normal hierarchy).}
\label{t4nh3}
\end{figure}

Next, in order to explore the possibilty of CP violation in the leptonic sector, in figure (\ref{t4nh3}) we study the dependence
of Jarlskog's rephasing invariant parameter ${\it {J}}$ on each of the three mixing angles, while keeping
the other two mixing angles constrained by their $3\sigma$ ranges. From these graphs, there appears to be
significant possibility of CP violation in the leptonic sector and the range for the magnitude of ${\it {J}}$ seems to be
$0.0005 - 0.03$ approximately.


\par The degenerate scenario of the neutrino masses can be characterized by either $m_{\nu 1} \lesssim m_{\nu 2} \sim m_{\nu3} \sim 0.1 eV$
or $m_{\nu 3} \sim m_{\nu 1} \lesssim m_{\nu2} \sim~0.1eV$, corresponding to normal hierarchy and inverted hierarchy respectively. Since
while carrying out the calculations pertaining to both the normal as well as inverted hierarchy cases, the range of lightest neutrino
mass is taken to be $10^{-8} -10^{-1}$ eV, which includes the neutrino masses corresponding to the degenerate scenario; therefore, by
discussion similar to the one given for ruling out inverted hierarchy, class I ansatz seems to be ruled out for degenerate scenario of neutrino
masses as well. Similarly, from figure (\ref{t4nh2}) the value of the lightest neutrino mass pertaining to the degenerate scenario,
$m_{\nu 1} \sim 0.1 eV$, seems to be outside the experimentally allowed region, thereby ruling out Class I ansatz for both the 
cases of degenerate scenario of neutrino masses.

\subsubsection{Class II ansatz}
\begin{figure}
\begin{tabular}{cc}
  \includegraphics[width=0.2\paperwidth,height=0.2\paperheight,angle=-90]{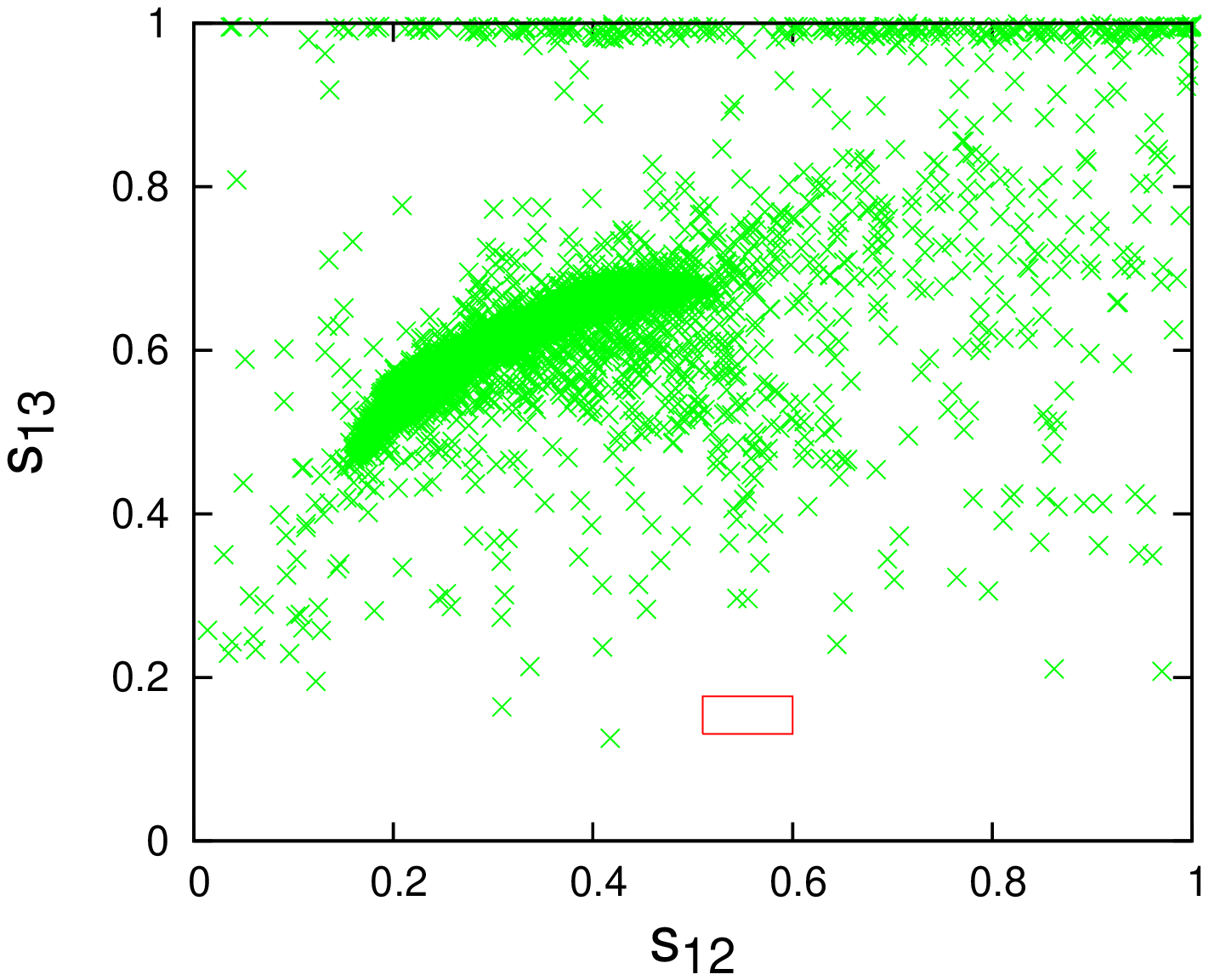}
  \includegraphics[width=0.2\paperwidth,height=0.2\paperheight,angle=-90]{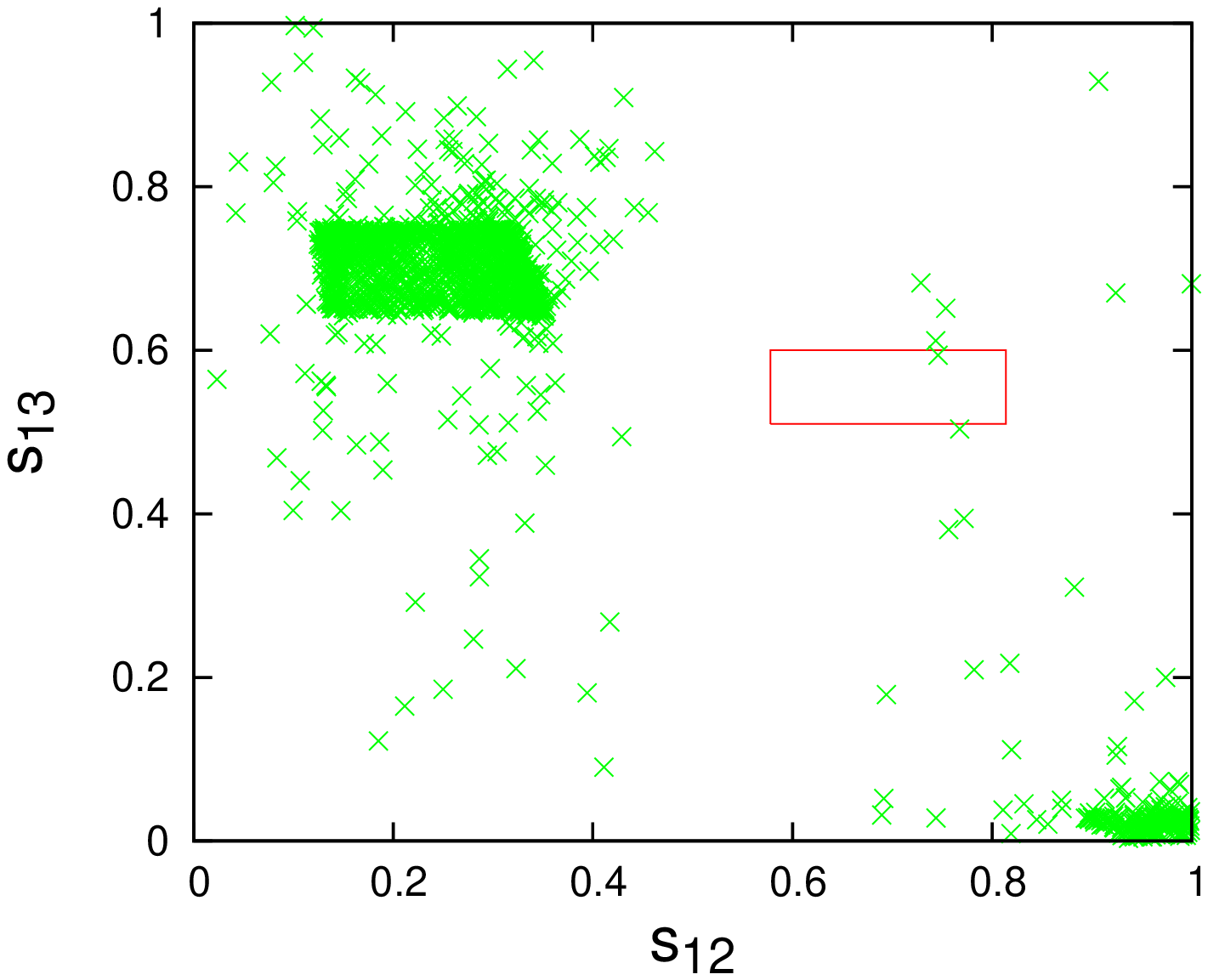}
  \includegraphics[width=0.2\paperwidth,height=0.2\paperheight,angle=-90]{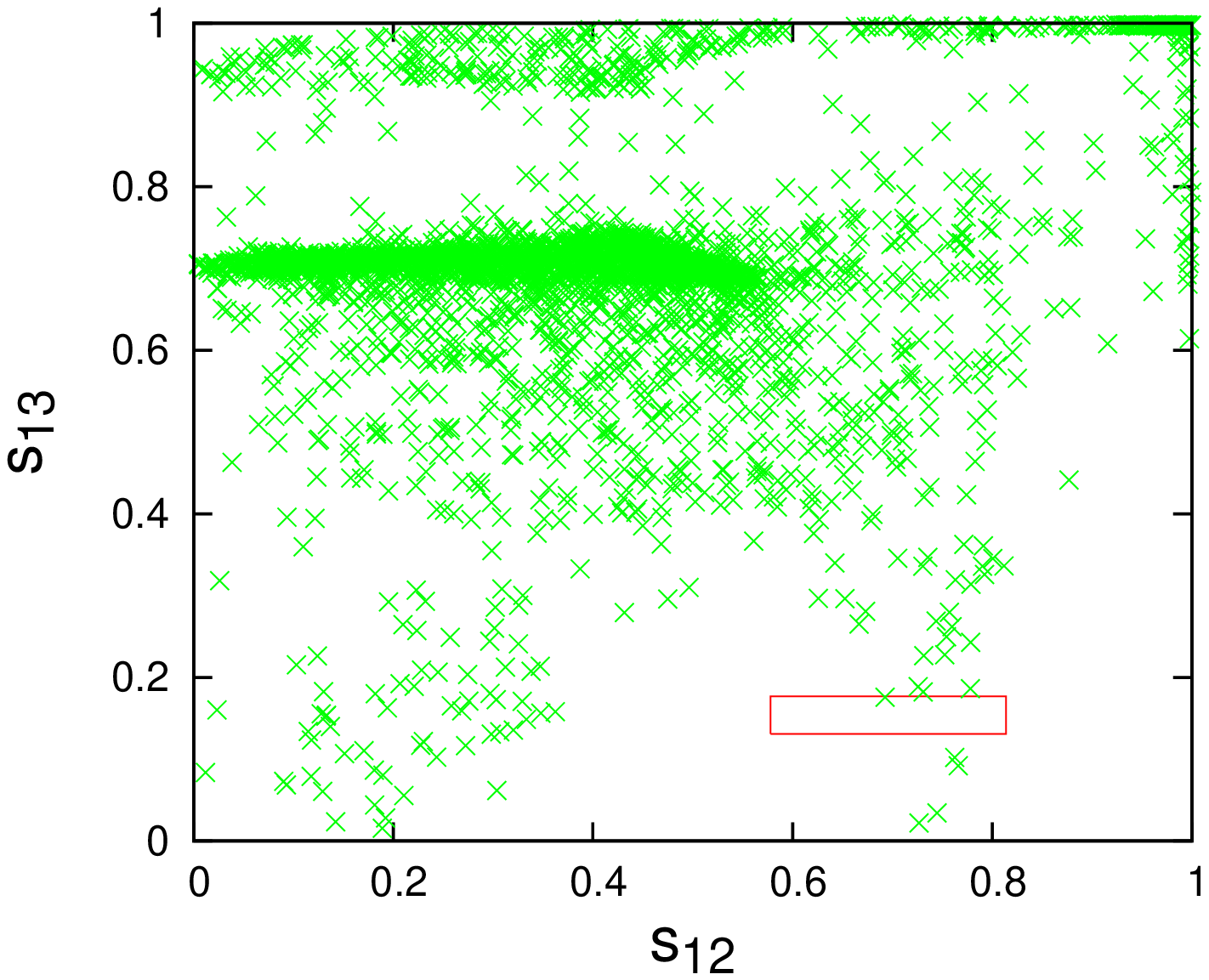}
\end{tabular}
\caption{Plots showing the parameter space for any two mixing
angles when the third angle is constrained by its  $1 \sigma$
range for  Class II ansatz of texture four zero  Dirac mass
matrices (inverted hierarchy).} \label{4aih1}
\end{figure}

\begin{figure}
\begin{tabular}{cc}
  \includegraphics[width=0.2\paperwidth,height=0.2\paperheight,angle=-90]{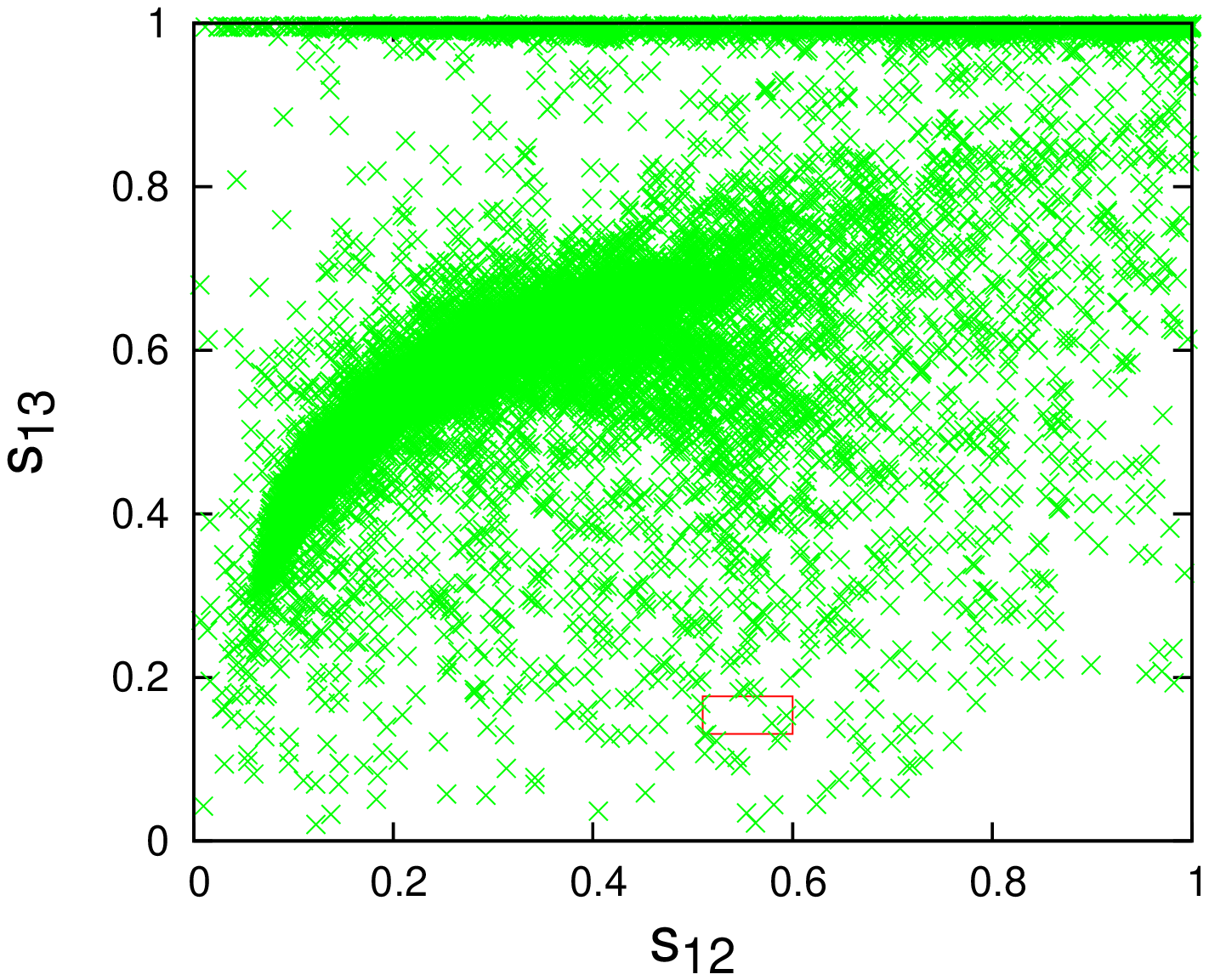}
  \includegraphics[width=0.2\paperwidth,height=0.2\paperheight,angle=-90]{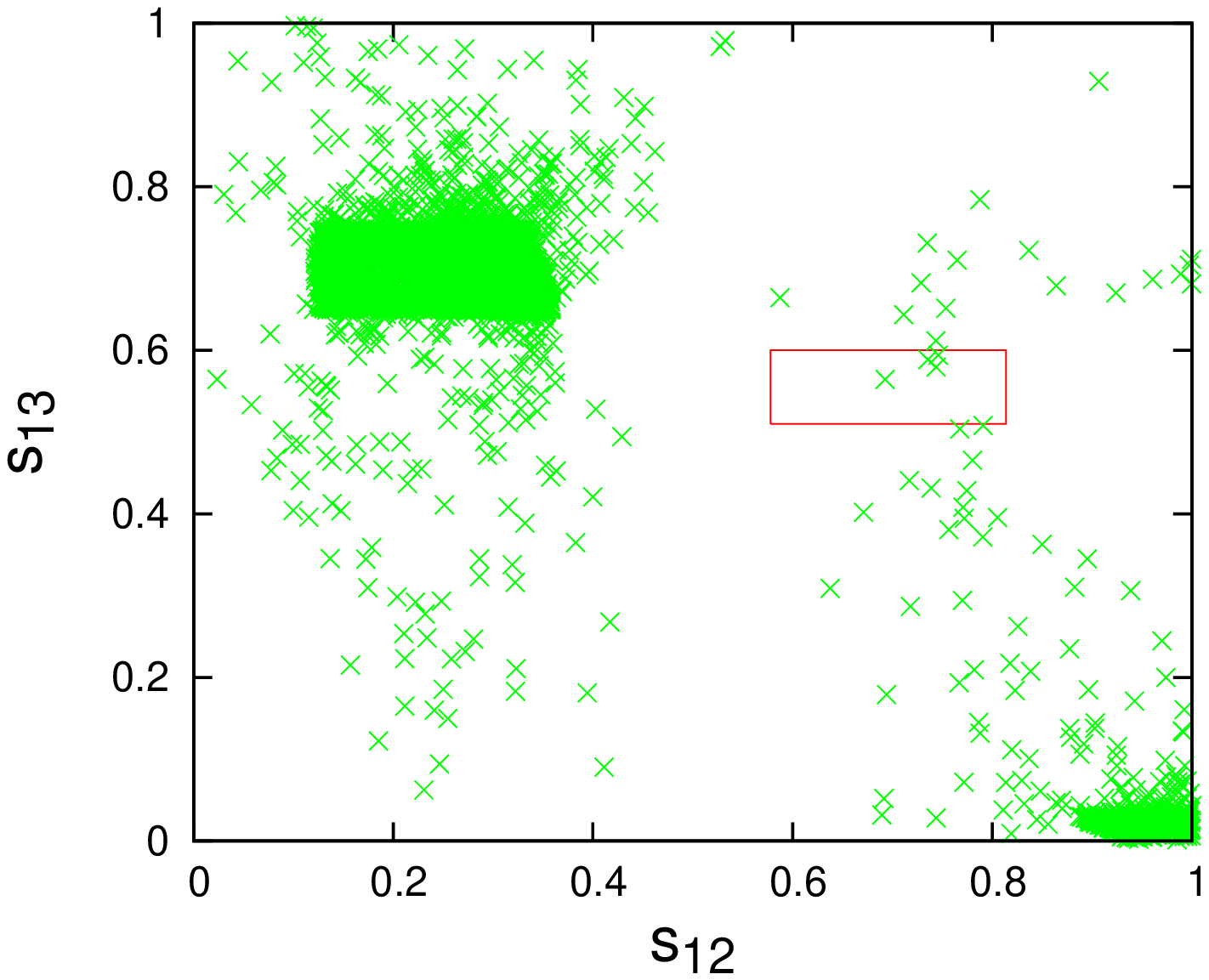}
  \includegraphics[width=0.2\paperwidth,height=0.2\paperheight,angle=-90]{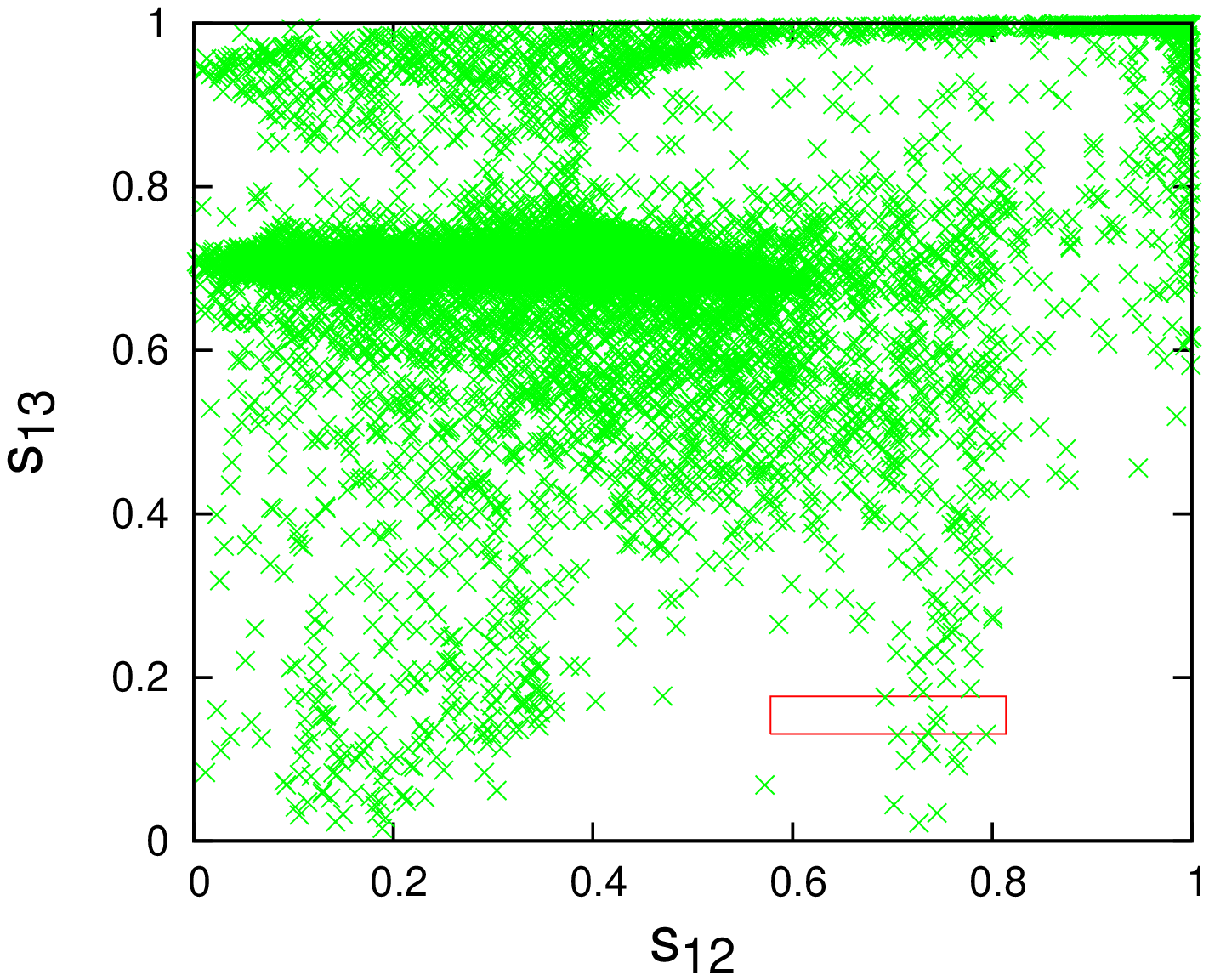}
\end{tabular}
\caption{Plots showing the parameter space for any two mixing
angles when the third angle is constrained by its  $3 \sigma$
range for  Class II ansatz of texture four zero  Dirac mass
matrices (inverted hierarchy).} \label{4aih2}
\end{figure}

To analyse this class we follow the same procedure as for class I
ansatz. The charged lepton and neutrino mass matrices which we
choose to analyse for this class can be given as,
\be
 M_{i}=\left( \ba{ccc}
D_i & A _{i}e^{i\alpha_i} & 0     \\
A_{i}e^{-i\alpha_i} & 0 &  B_{i}e^{i\beta_i}     \\
 0 & B_{i}e^{-i\beta_i} &  E_{i} \ea \right),
\label{cl2}
\ee
where $i=l,~\nu_D$ corresponds to the charged lepton and Dirac neutrino mass matrices
respectively. As for class I, the elements $D_l$ and $D_\nu$ as well
as the phases $\phi_1$ and $\phi_2$ are considered to be the free parameters. Firstly, we examine
the viability of inverted hierarchy for the structure given in eqn.(\ref{cl2}). To this end in figures (\ref{4aih1}) and (\ref{4aih2}),
we present the plots showing the parameter space for two mixing angles wherein the third angle is constrained by its $1\sigma$ range
and $3 \sigma$ range respectively. The rectangular boxes in these plots show the $3\sigma$ ranges for the
two mixing angles being considered. It is interesting to note that the inverted hierarchy seems to be ruled at by the $1\sigma$
ranges for the present mixing data, while it seems to be compatible with the $3\sigma$ experimental bounds.

\begin{figure}
\begin{tabular}{cc}
  \includegraphics[width=0.2\paperwidth,height=0.2\paperheight,angle=-90]{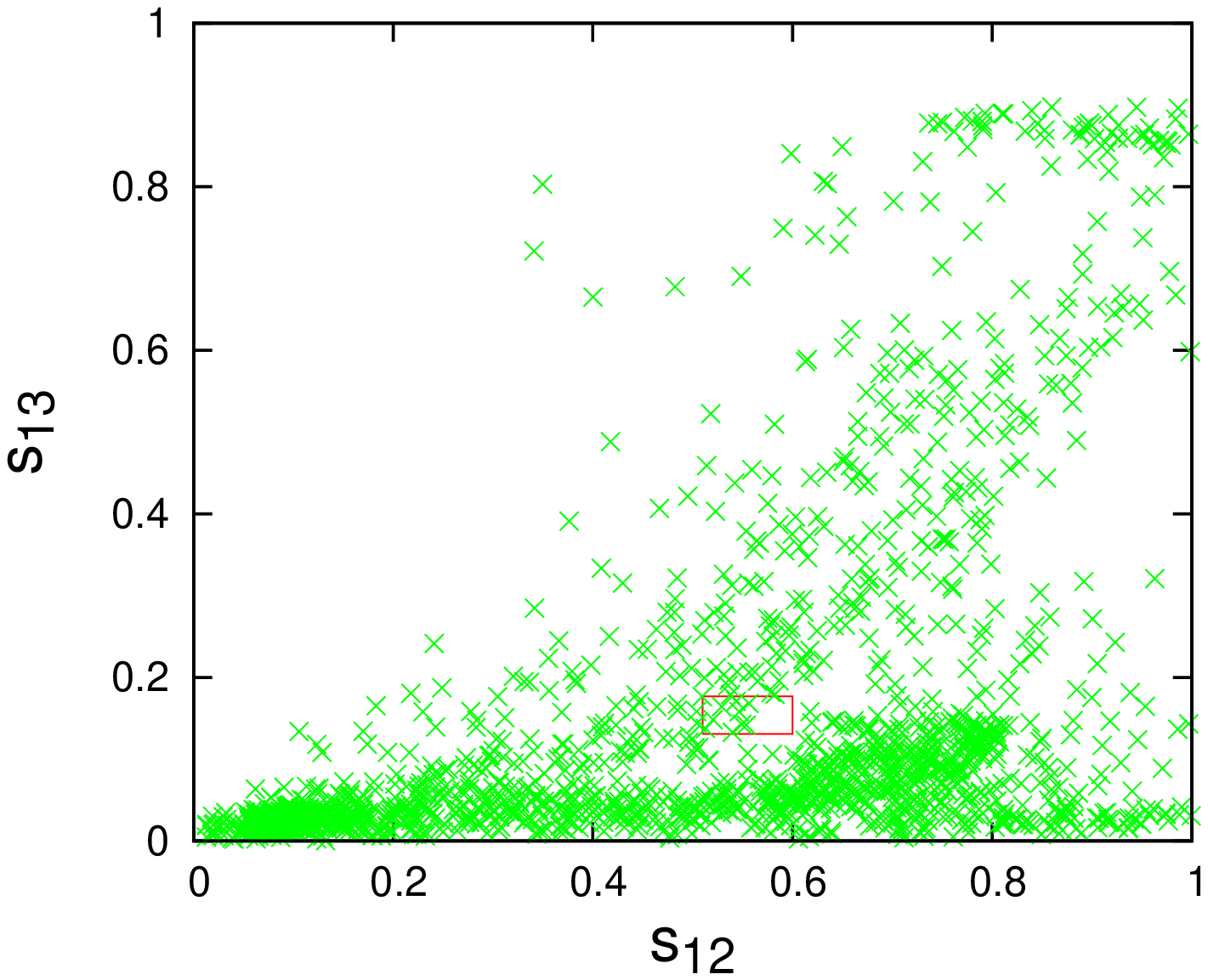}
  \includegraphics[width=0.2\paperwidth,height=0.2\paperheight,angle=-90]{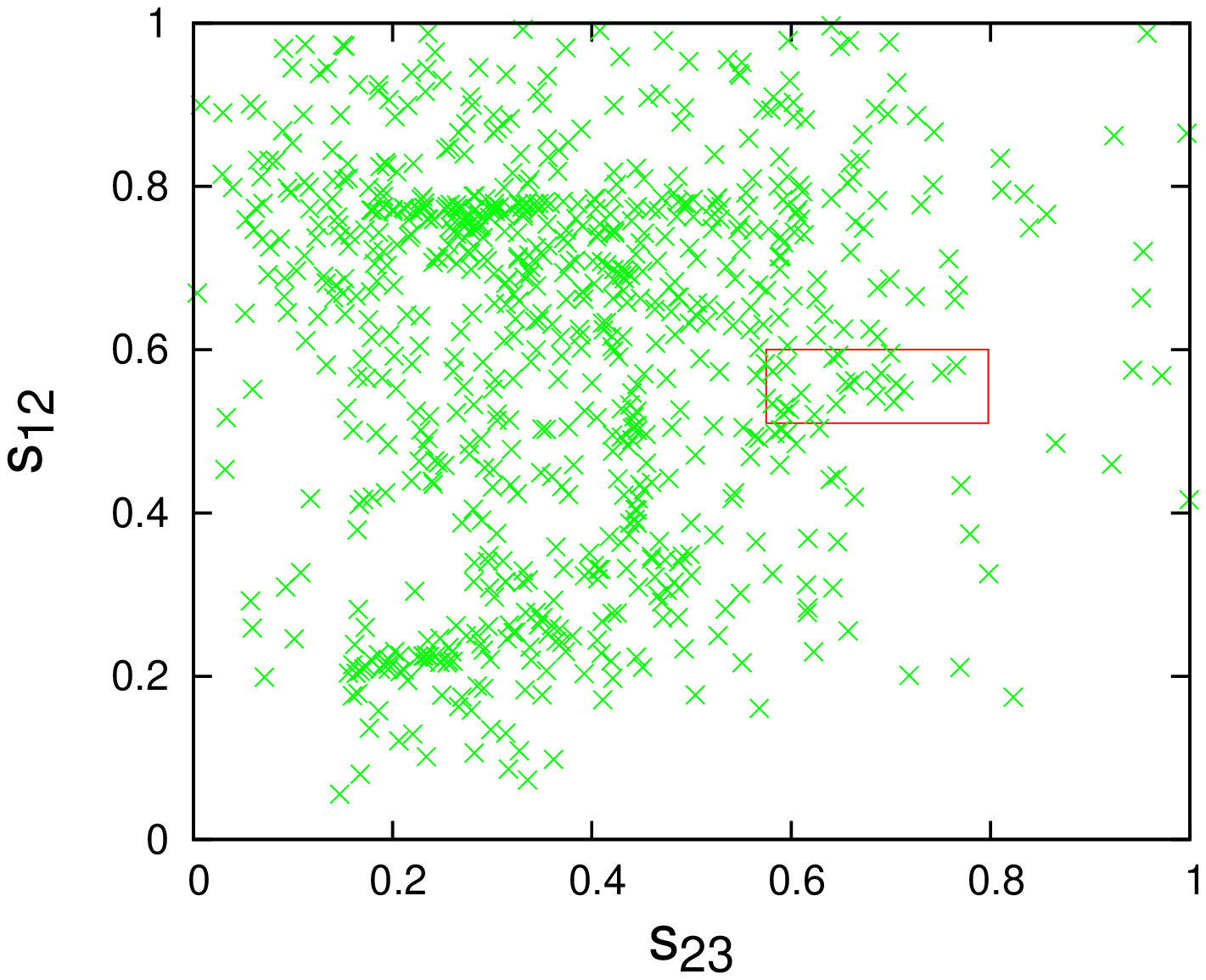}
  \includegraphics[width=0.2\paperwidth,height=0.2\paperheight,angle=-90]{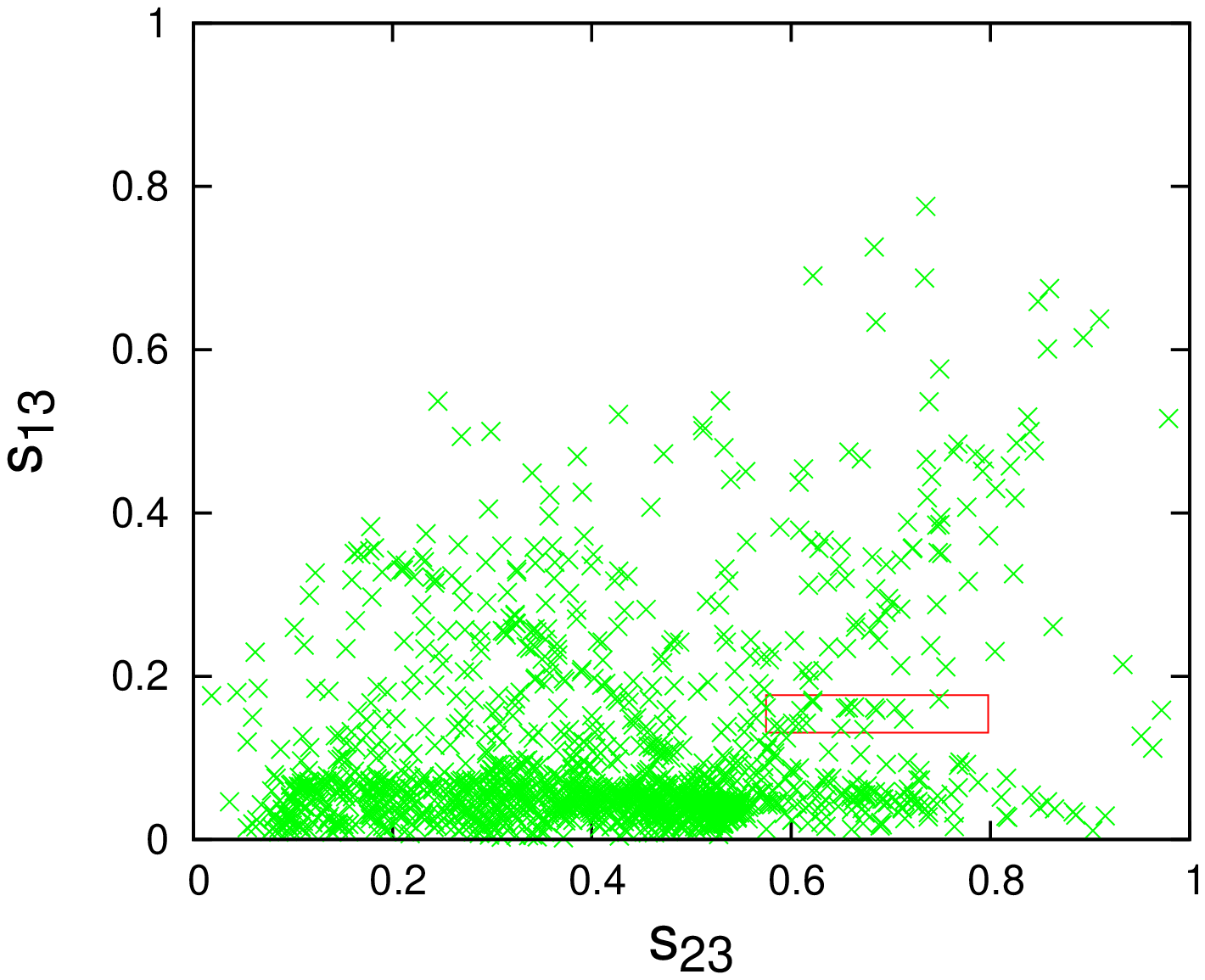}
\end{tabular}
\caption{Plots showing the parameter space for any two mixing
angles when the third angle is constrained by its  $1 \sigma$
range for Class II ansatz of texture four zero  Dirac mass
matrices (normal hierarchy).} \label{4anh1}
\end{figure}
\begin{figure}
\begin{tabular}{cc}
  \includegraphics[width=0.2\paperwidth,height=0.2\paperheight,angle=-90]{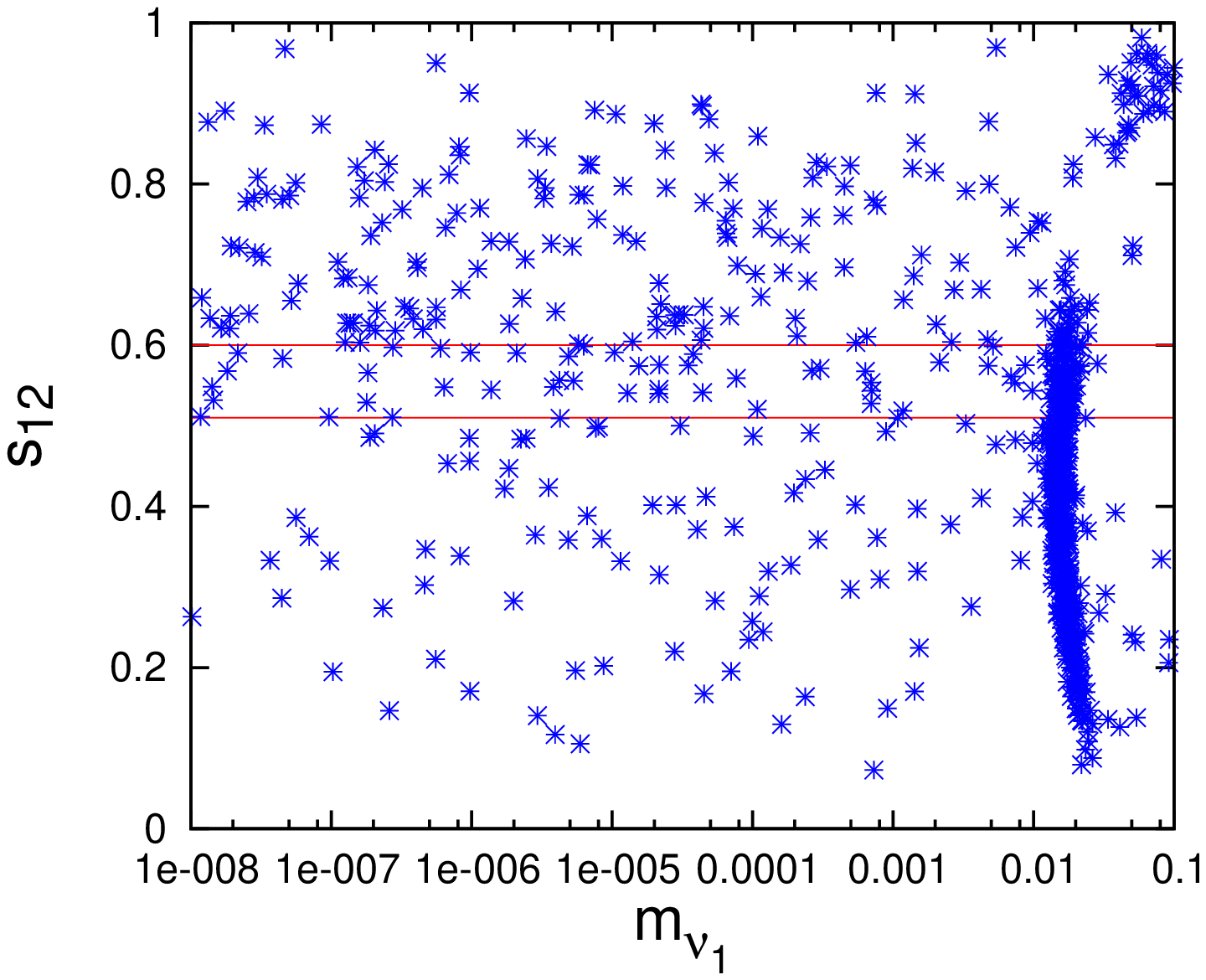}
  \includegraphics[width=0.2\paperwidth,height=0.2\paperheight,angle=-90]{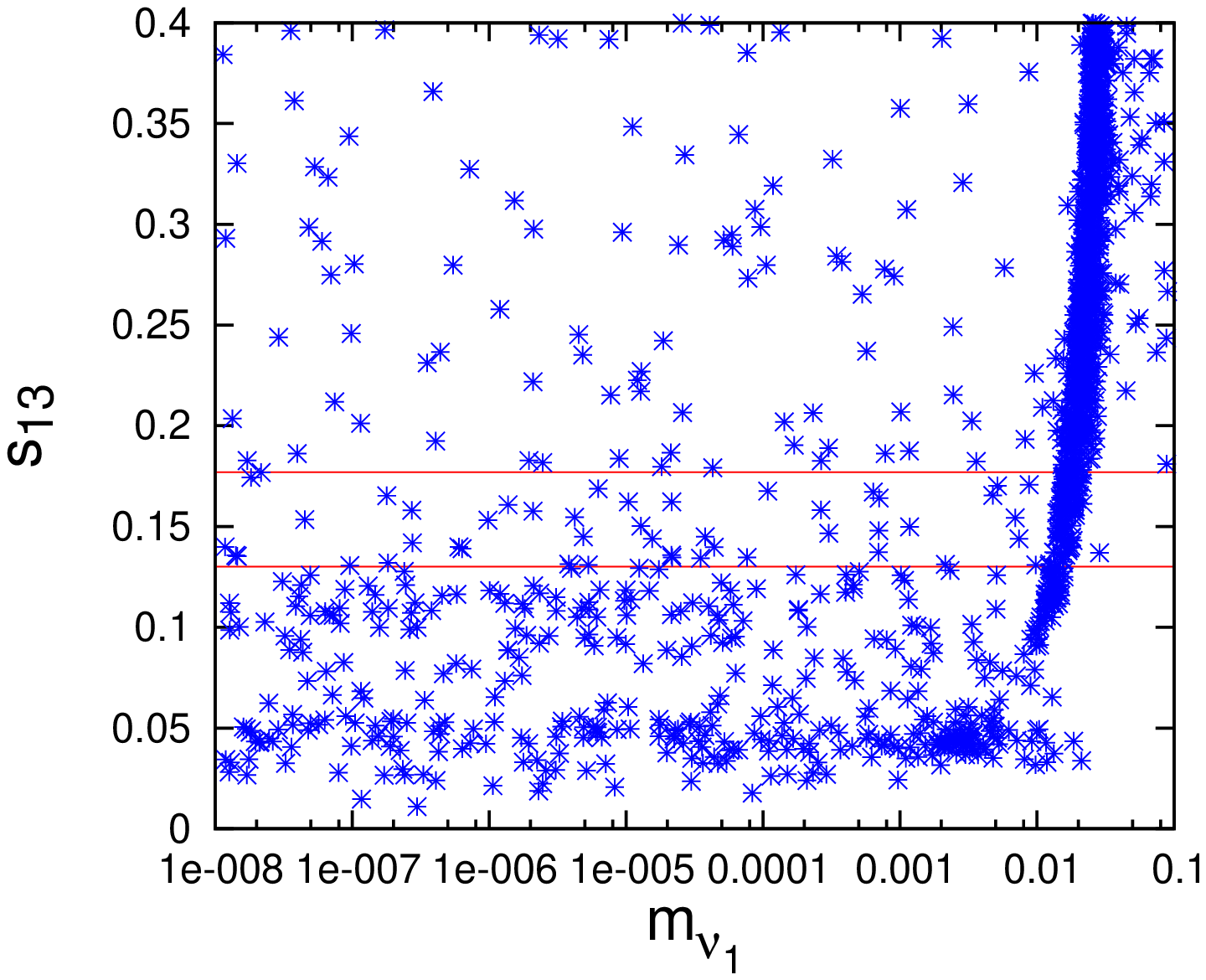}
  \includegraphics[width=0.2\paperwidth,height=0.2\paperheight,angle=-90]{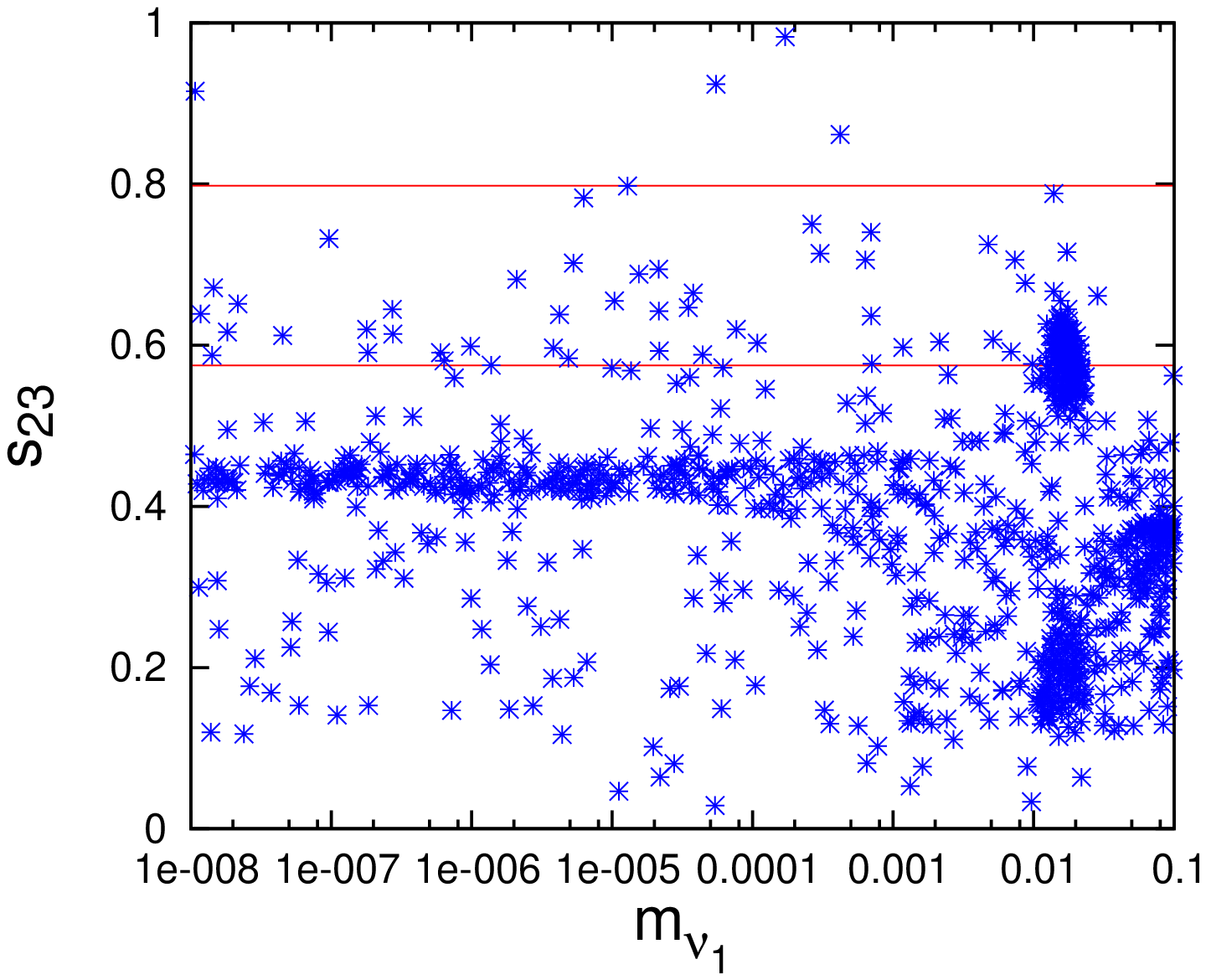}
\end{tabular}
\caption{Plots showing the dependence of mixing angles on the
lightest neutrino mass when the other two angles are constrained
by their $3 \sigma$ ranges  for Class II ansatz of texture four
zero  Dirac mass matrices ( normal hierarchy).} \label{4anh2}
\end{figure}
\begin{figure}
\begin{tabular}{cc}
  \includegraphics[width=0.2\paperwidth,height=0.2\paperheight,angle=-90]{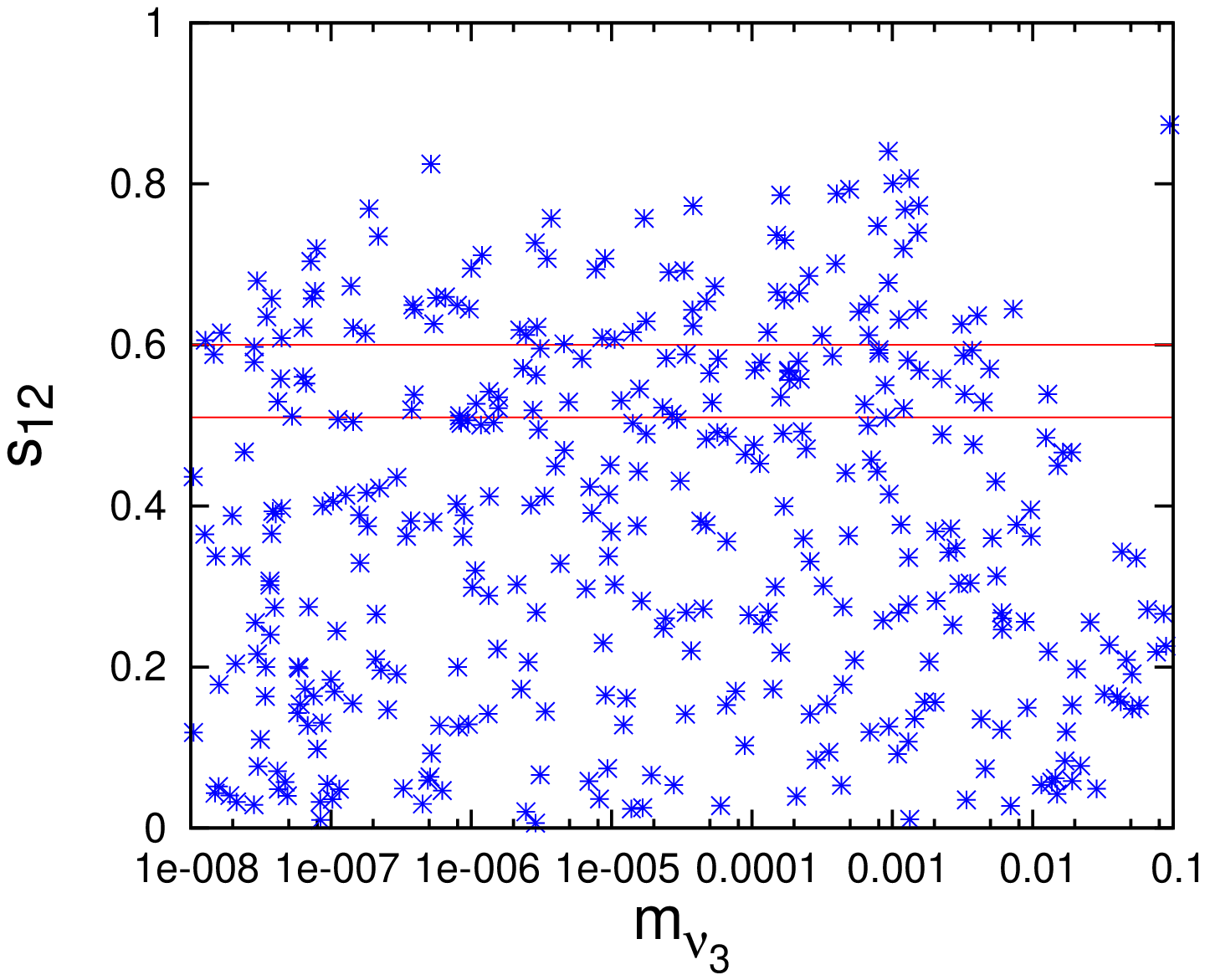}
  \includegraphics[width=0.2\paperwidth,height=0.2\paperheight,angle=-90]{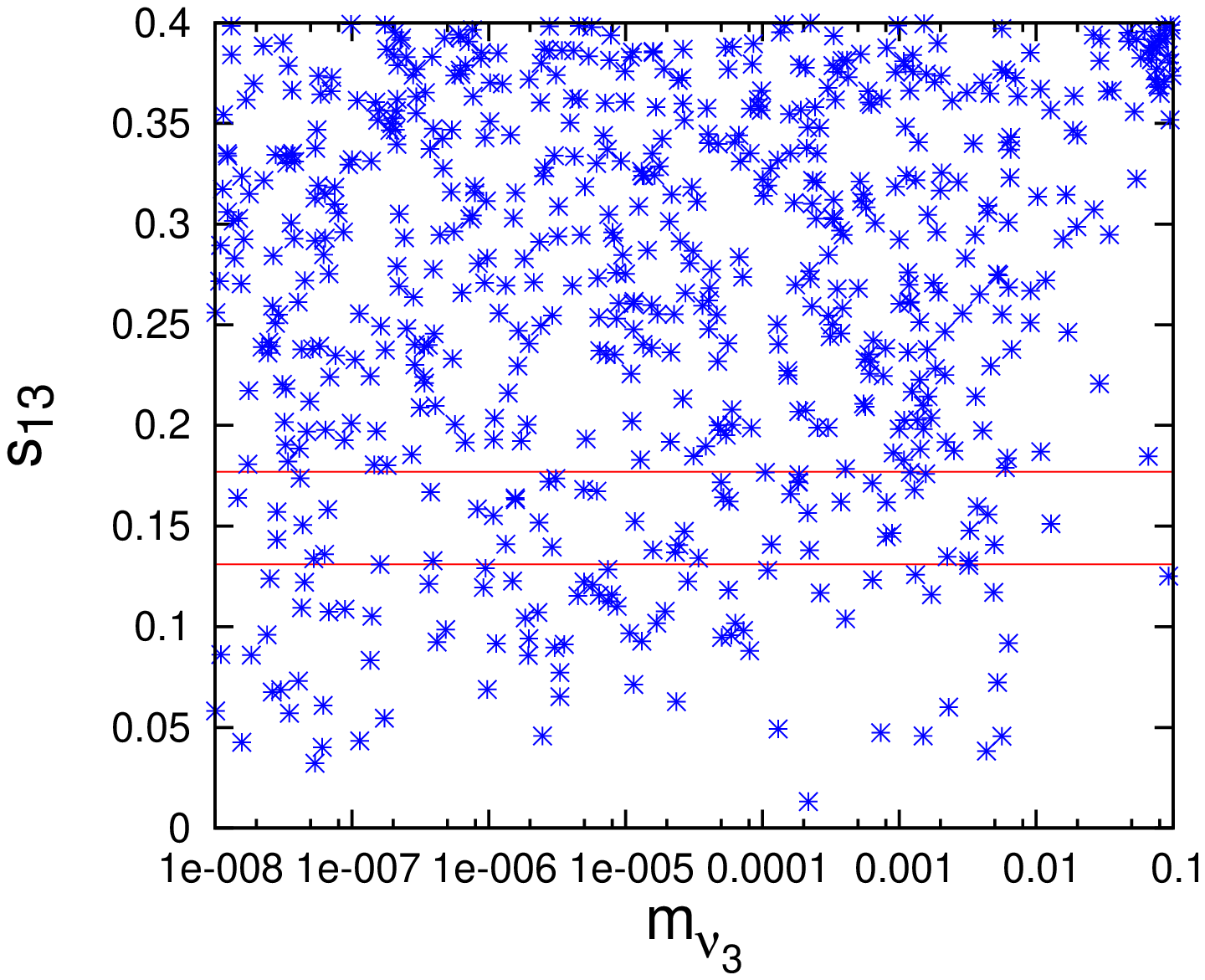}
  \includegraphics[width=0.2\paperwidth,height=0.2\paperheight,angle=-90]{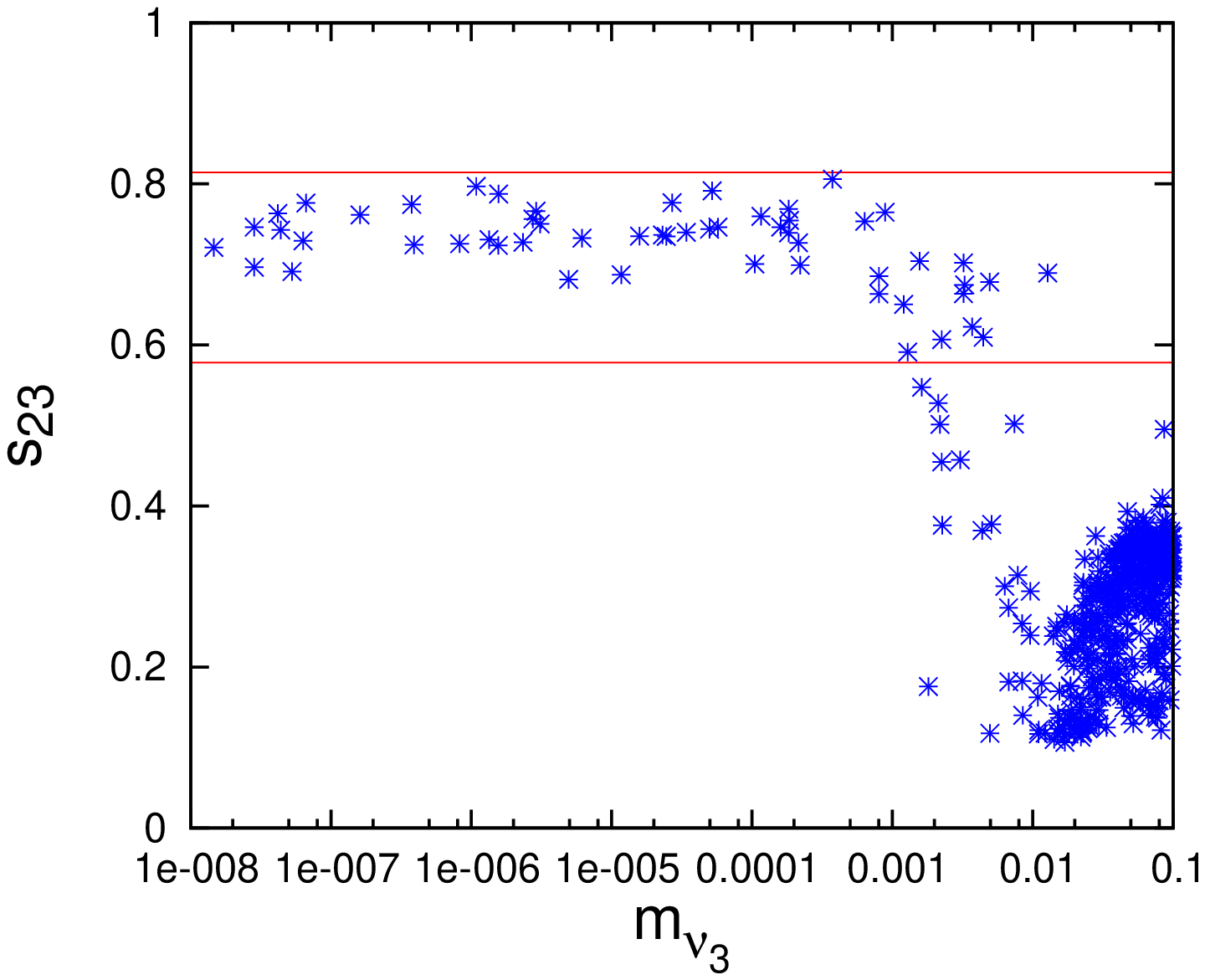}
\end{tabular}
\caption{Plots showing the dependence of mixing angles on the
lightest neutrino mass when the other two angles are constrained
by their $3 \sigma$ ranges  for  Class II ansatz of texture four
zero  Dirac mass matrices (inverted hierarchy).} \label{4aih3}
\end{figure}

\begin{figure}
\begin{tabular}{cc}
  \includegraphics[width=0.2\paperwidth,height=0.2\paperheight,angle=-90]{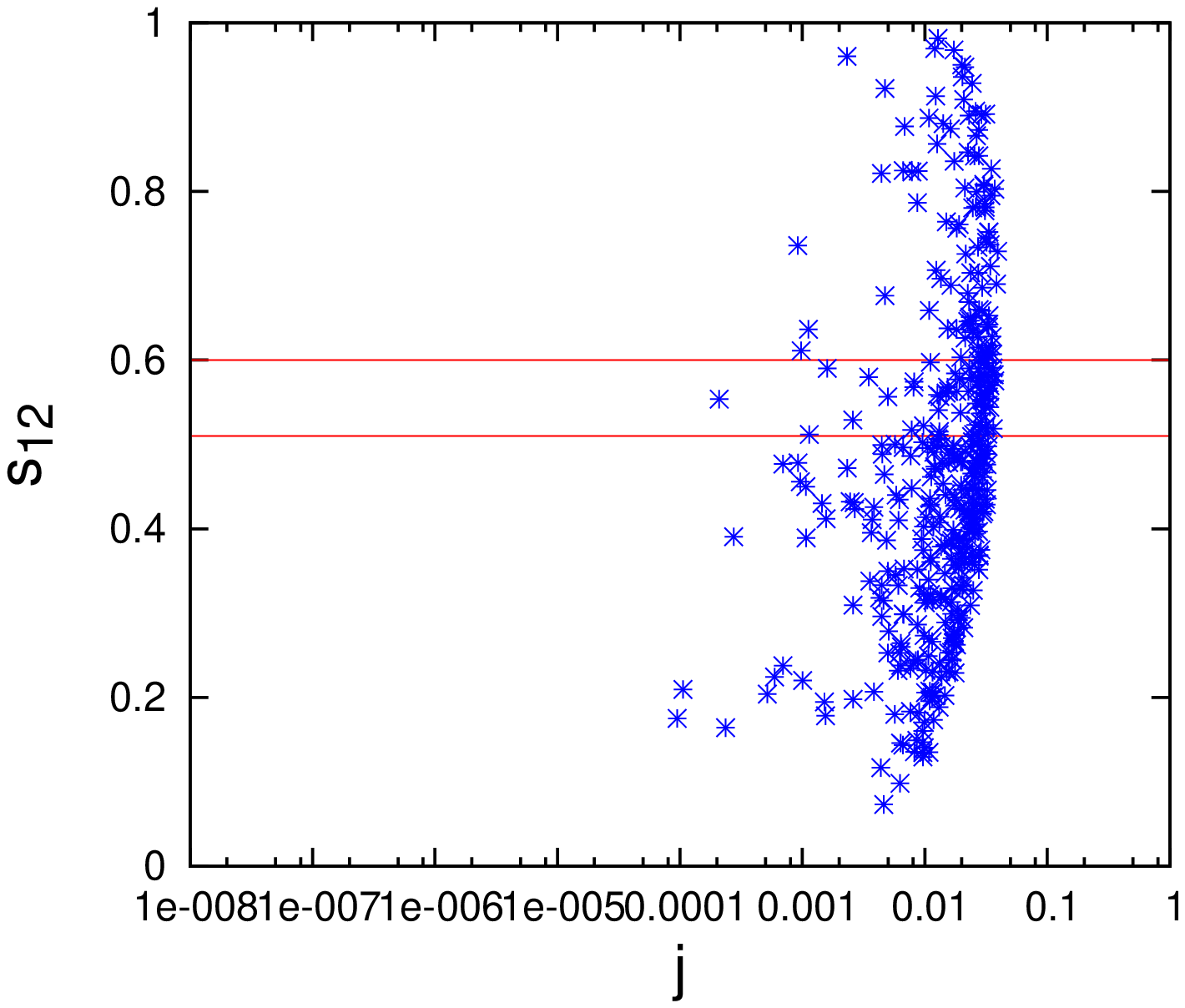}
  \includegraphics[width=0.2\paperwidth,height=0.2\paperheight,angle=-90]{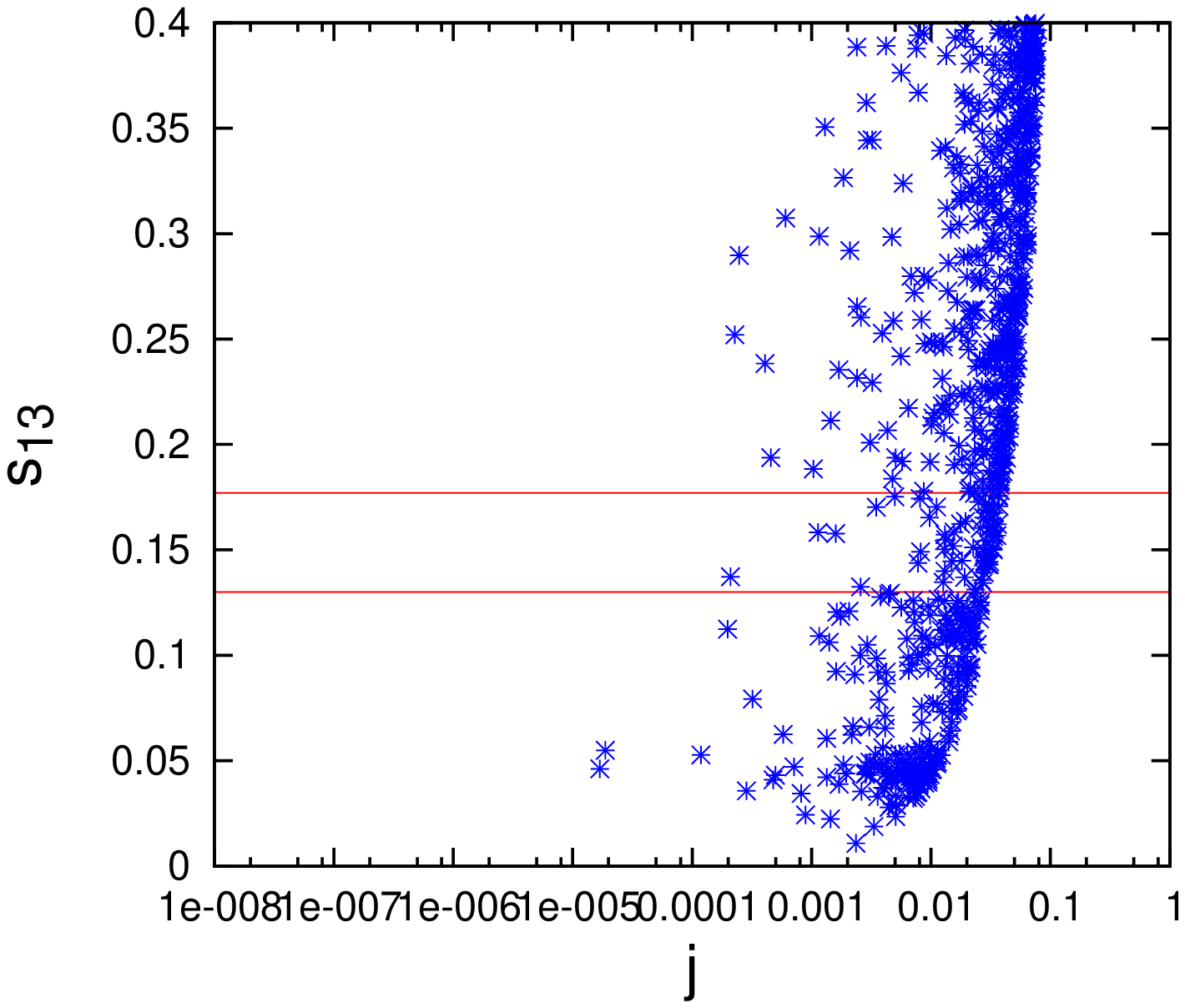}
  \includegraphics[width=0.2\paperwidth,height=0.2\paperheight,angle=-90]{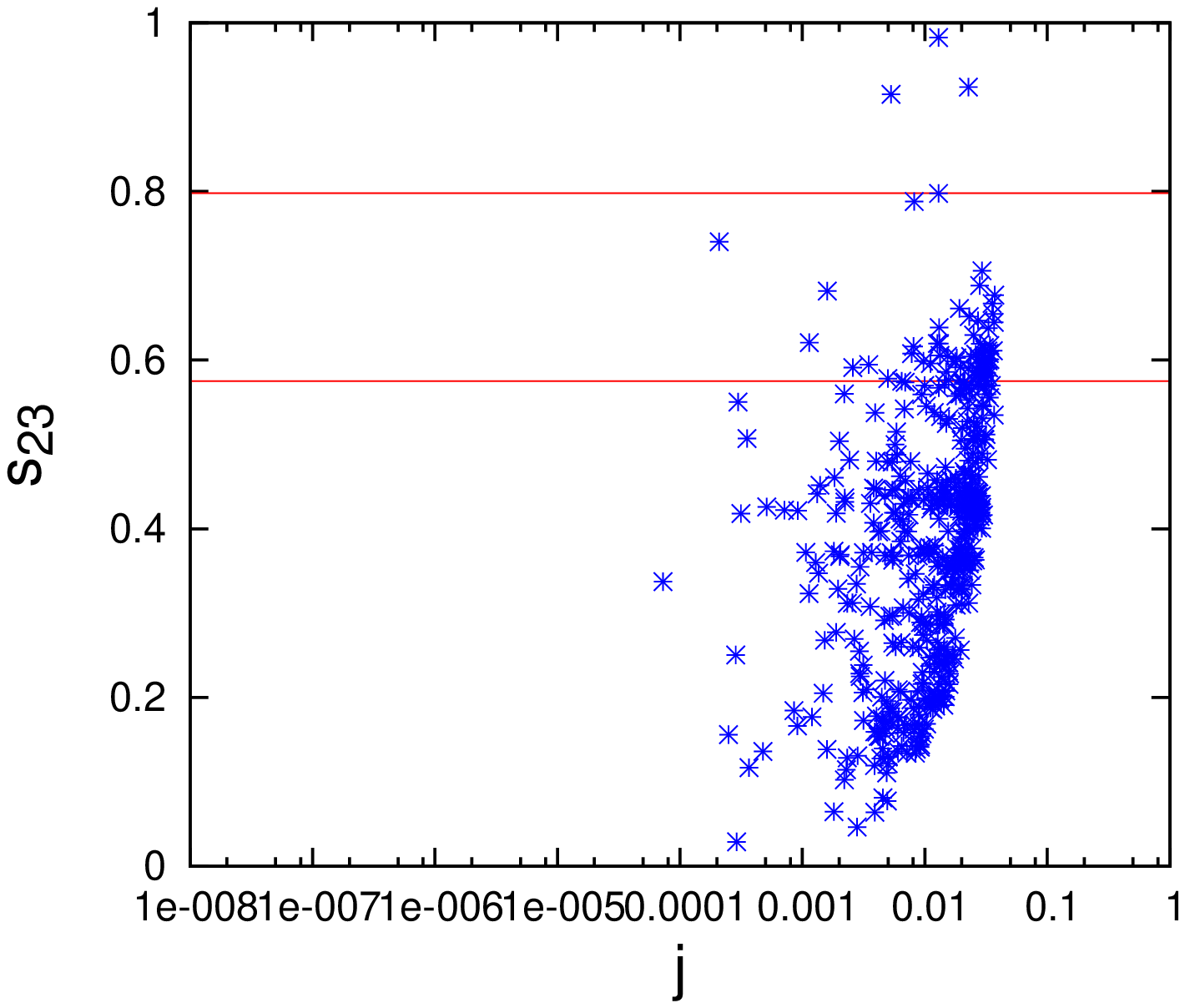}
\end{tabular}
\caption{Plots showing the variation of Jarlskog CP violating
parameter with mixing angles when the other two angles are
constrained by their $3 \sigma$ ranges  for  Class II ansatz of
texture four zero  Dirac mass matrices (normal hierarchy).}
\label{4anh3}
\end{figure}

\begin{figure}
\begin{tabular}{cc}
  \includegraphics[width=0.2\paperwidth,height=0.2\paperheight,angle=-90]{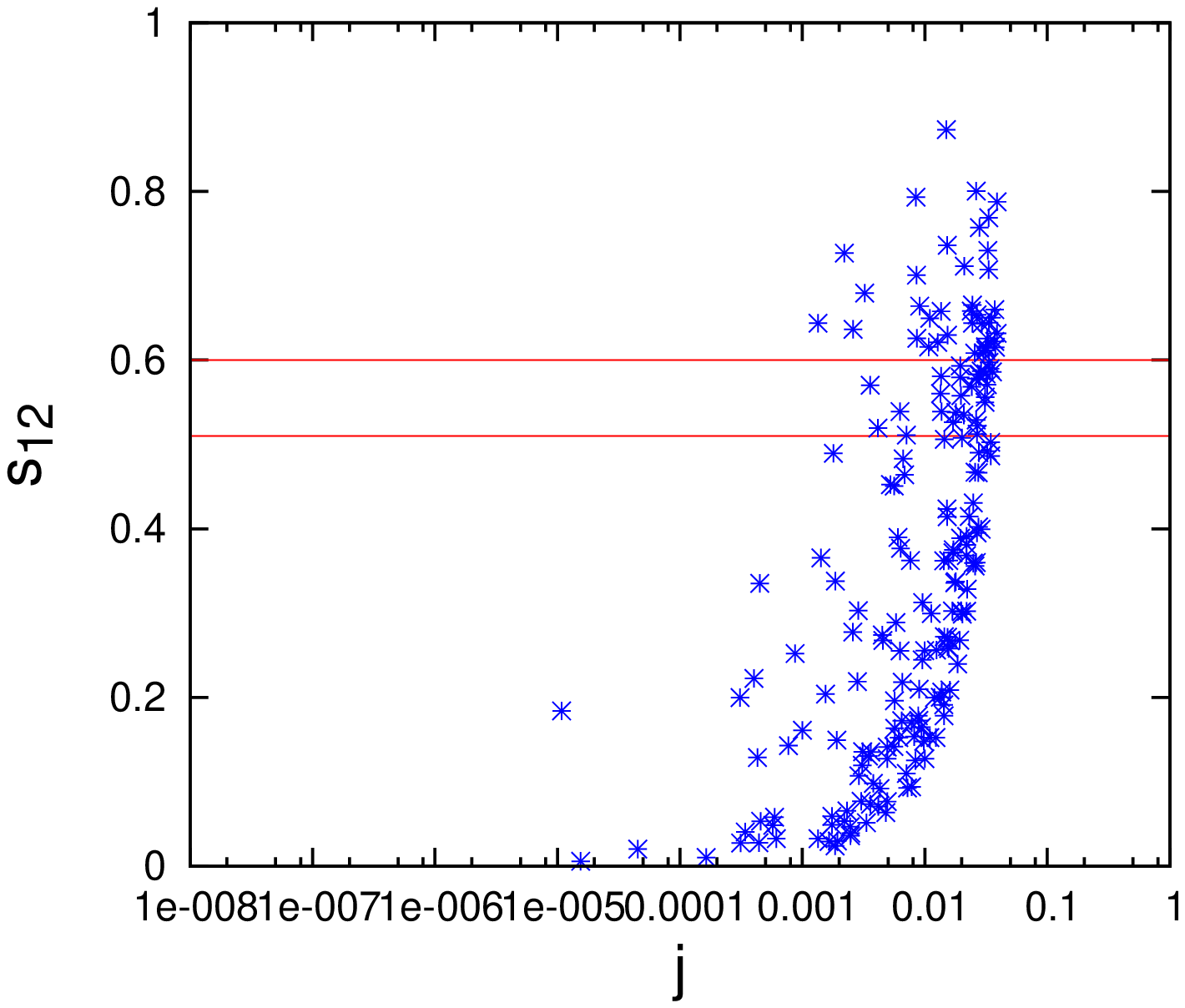}
  \includegraphics[width=0.2\paperwidth,height=0.2\paperheight,angle=-90]{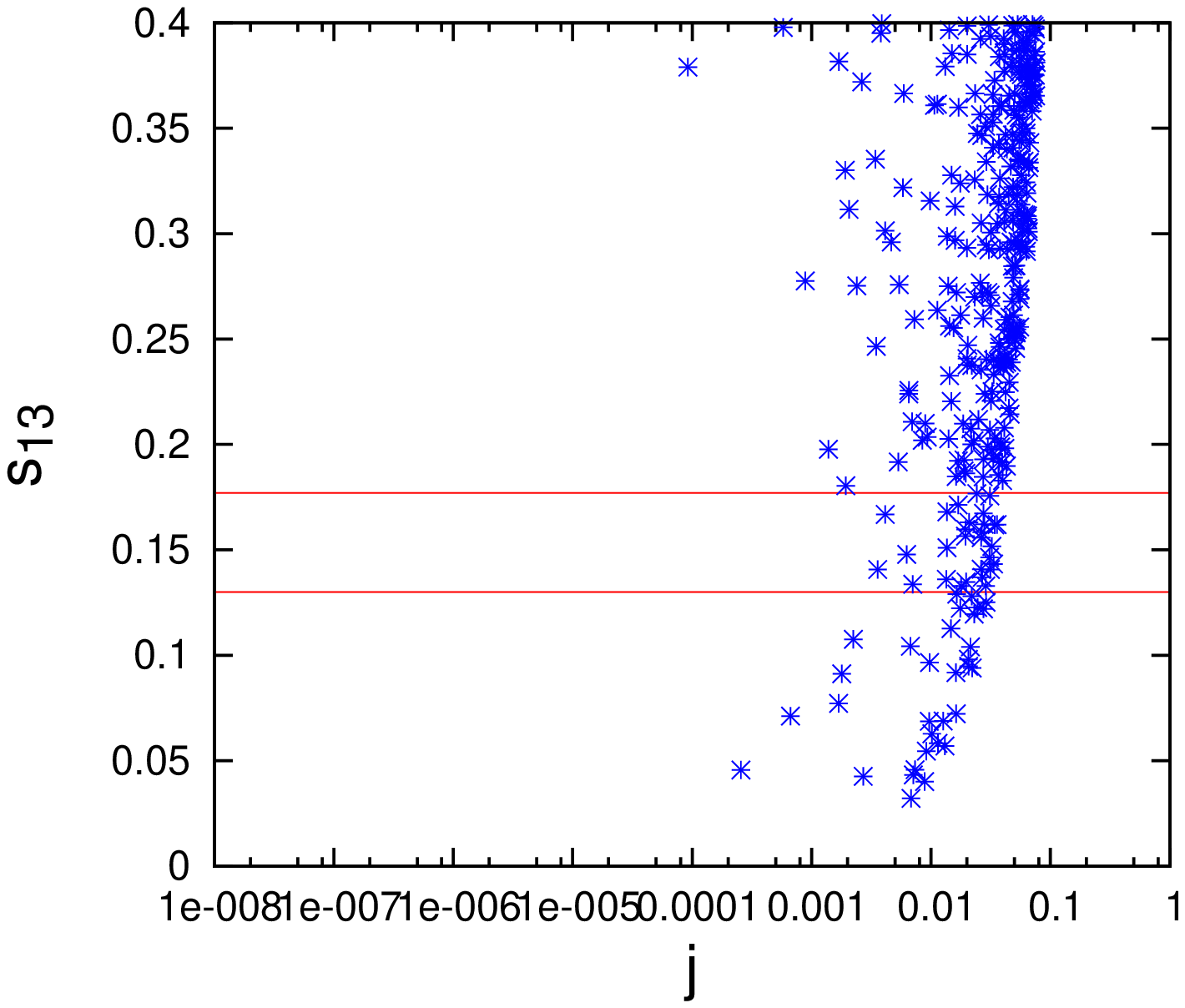}
  \includegraphics[width=0.2\paperwidth,height=0.2\paperheight,angle=-90]{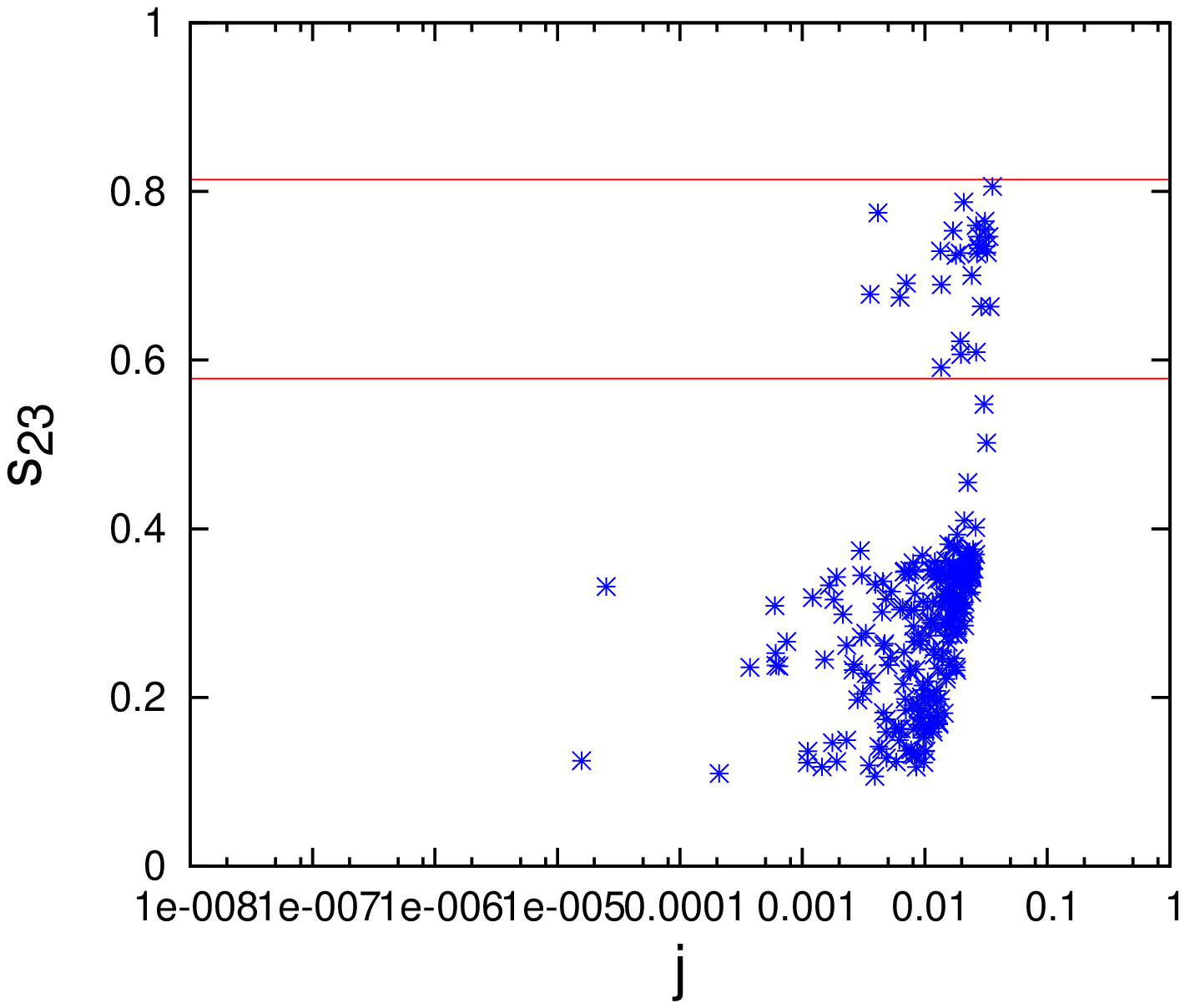}
\end{tabular}
\caption{Plots showing the variation of Jarlskog CP violating
parameter with mixing angles when the other two angles are
constrained by their $3 \sigma$ ranges  for Class II ansatz of
texture four zero  Dirac mass matrices (inverted hierarchy).}
\label{4aih4}
\end{figure}

Next, in order to examine the compatibility of structure given in eqn.(\ref{cl2}) with the normal hierarchy, in figure
(\ref{4anh1}) we present the plots showing the parameter space for
two mixing angles wherein the third angle is constrained by its $1\sigma$ experimental range. A general look at figure
(\ref{4anh1}) reveals that normal hierarchy is compatible with the structure (\ref{cl2}) at $1\sigma$ level.

Thus, we find that both the normal as well as inverted neutrino mass hierarchies are compatible
with the $3\sigma$ ranges for the lepton mixing data. As a next step, in figures
(\ref{4anh2}) and (\ref{4aih3}) we attempt to study the
implications of the $3\sigma$ ranges of the leptonic mixing angles on the lightest
neutrino mass for this class
for the normal and inverted neutrino mass hierarchy respectively. As can be seen from
these figures, for both the neutrino mass hierarchies the $3\sigma$ ranges of the mixing
angles provide only an upper bound on the lightest neutrino mass , viz. $m_{\nu 1} \lesssim 0.01 eV$
and $m_{\nu 3} \lesssim 0.01 eV$ for the normal and inverted hierarchy respectively. This bound seems to rule out
both the degenerate neutrino mass ordering scenarios as the bound on the lightest neutrino mass does not include the
value of the lightest neutrino mass pertaining to the degenerate scenario $(\sim 0.1 eV)$.

\par Further, we study the possiblity of CP violation in the leptonic sector for this class by studying the
variation of Jarlskog's rephasing invariant ${\it{J}}$ with all the mixing angles for normal as well as inverted
neutrino mass hierarchy in figures (\ref{4anh3}) and (\ref{4aih4}) respectively. While plotting these graphs the two mixing angles,
other than the one being considered, are constrained by their $3\sigma$ ranges. The parallel lines in these plots show the
$3\sigma$ experimental ranges for the mixing angle being considered. It is interesting to note that for the case
of normal hierarchy one gets a broader range
for ${\it{J}}$ as compared to the one for the case of inverted hierarchy. To be more explicit, one finds
$0.0001\lesssim {\it{J}} \lesssim 0.05$ and $0.003\lesssim {\it{J}} \lesssim 0.02$ corresponding to 
normal and inverted hierarchy respectively.

\subsubsection{Class III ansatz}
\begin{figure}
\begin{tabular}{cc}
  \includegraphics[width=0.2\paperwidth,height=0.2\paperheight,angle=-90]{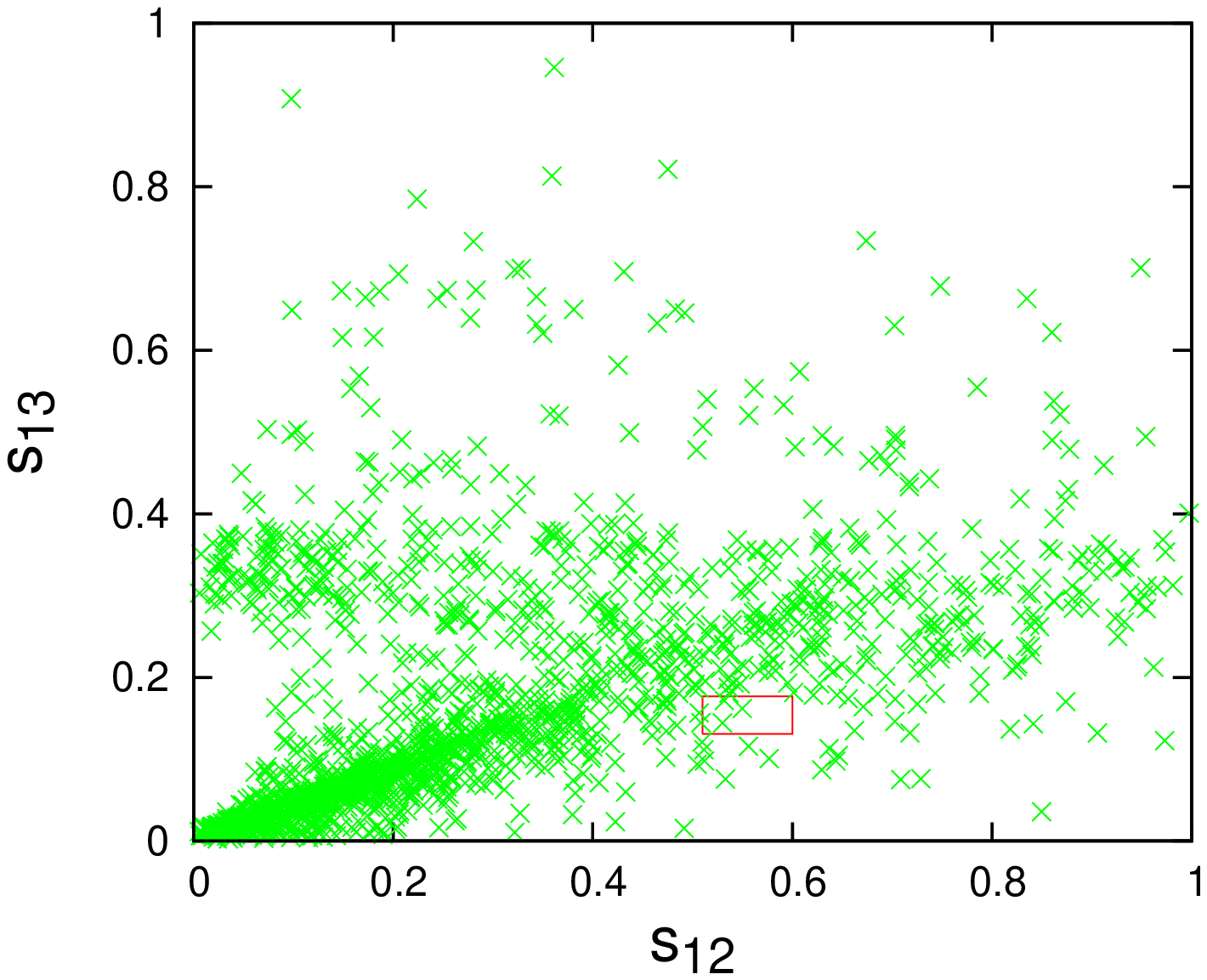}
  \includegraphics[width=0.2\paperwidth,height=0.2\paperheight,angle=-90]{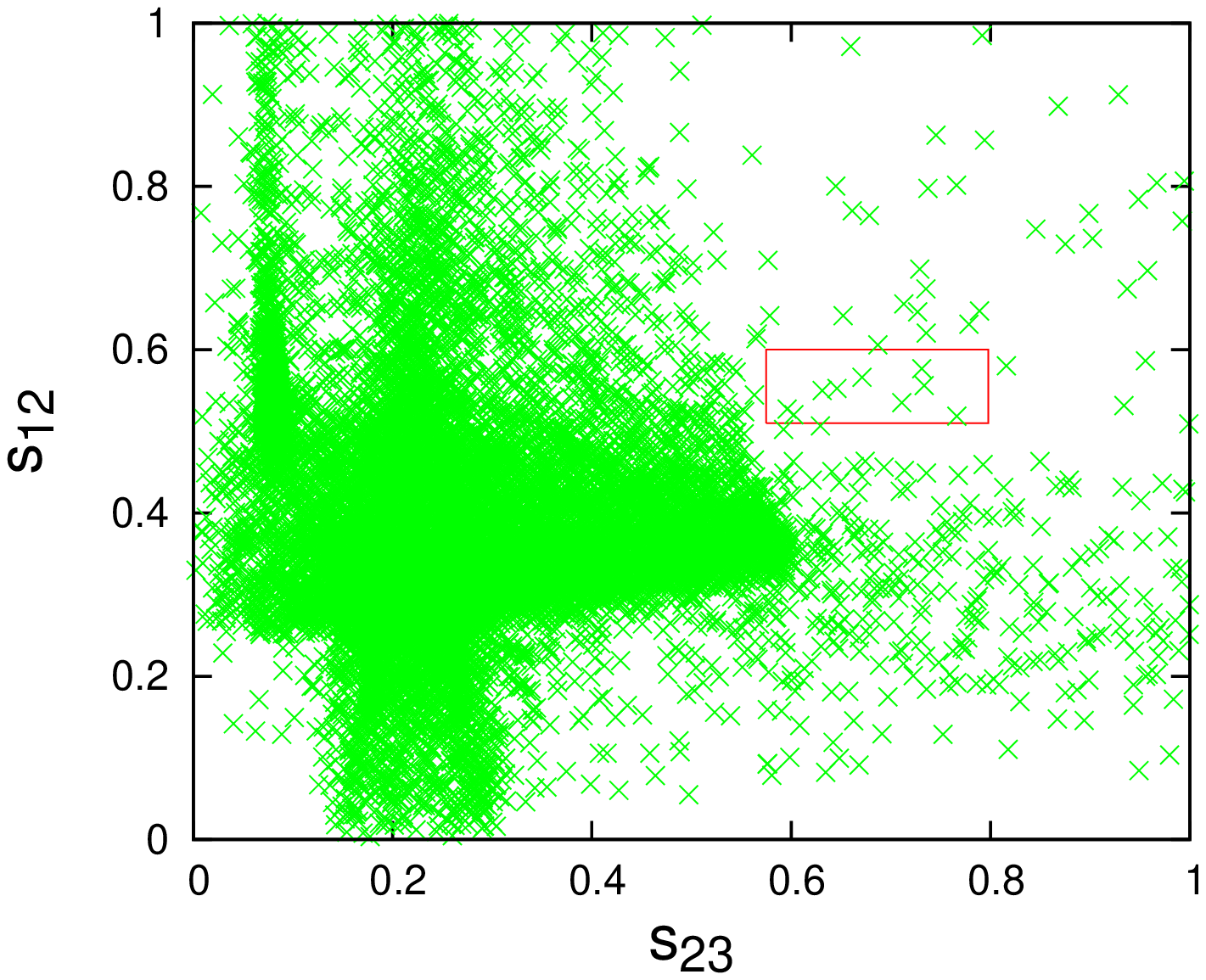}
  \includegraphics[width=0.2\paperwidth,height=0.2\paperheight,angle=-90]{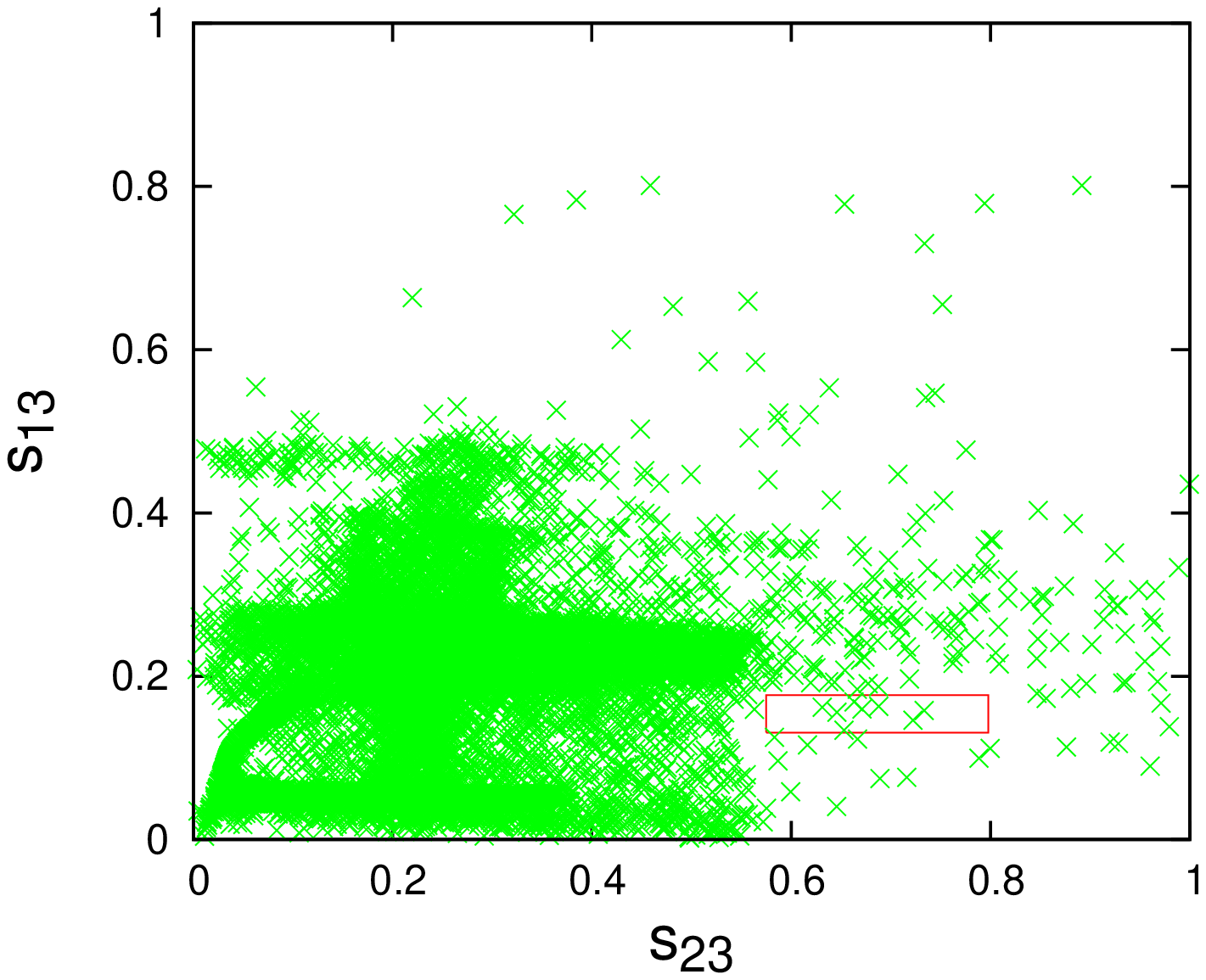}
\end{tabular}
\caption{Plots showing the parameter space for any two mixing
angles when the third angle is constrained by its  $1 \sigma$
range for Class III ansatz of texture four zero  Dirac mass
matrices (normal hierarchy).} \label{5anh1}
\end{figure}

\begin{figure}
\begin{tabular}{cc}
  \includegraphics[width=0.2\paperwidth,height=0.2\paperheight,angle=-90]{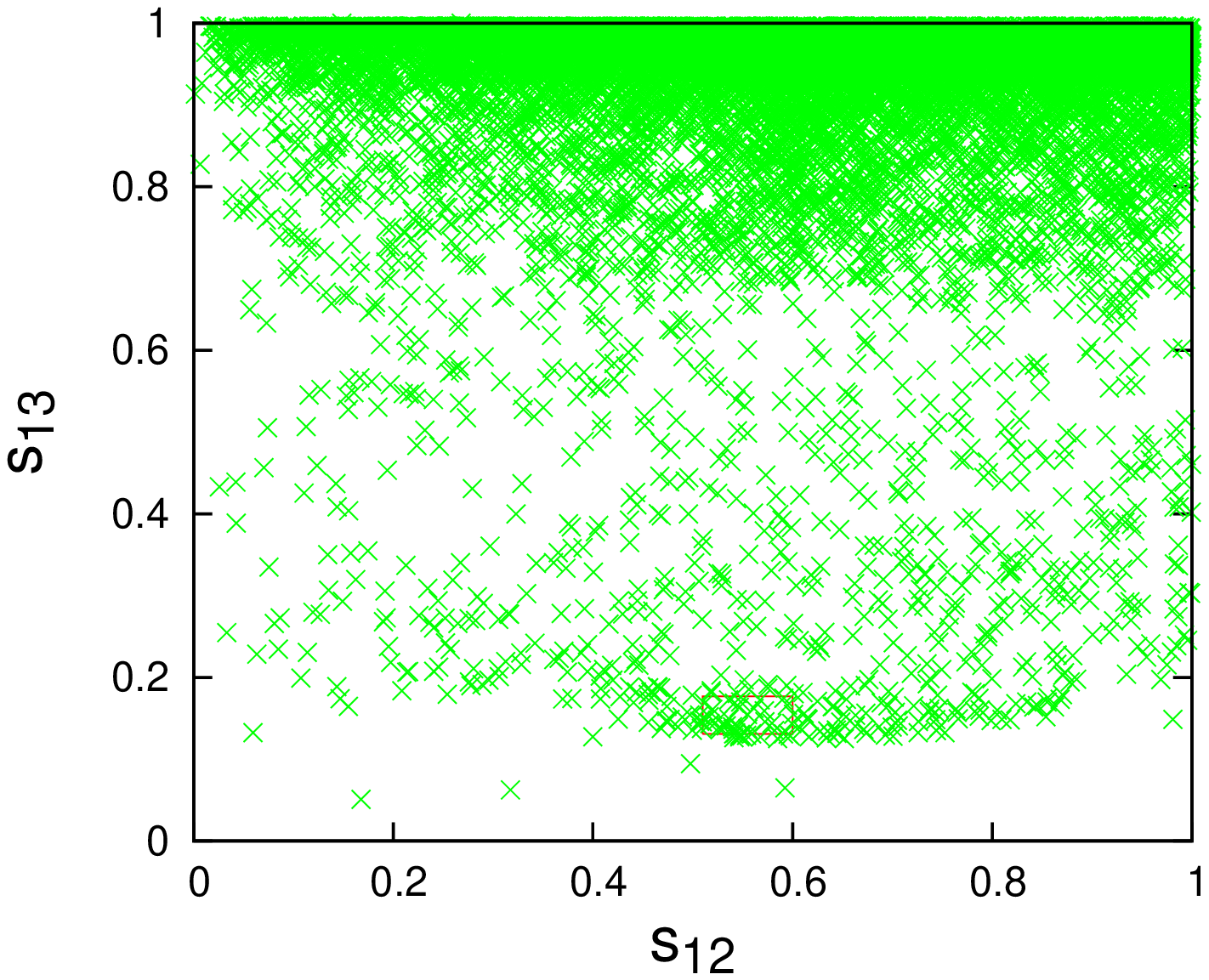}
  \includegraphics[width=0.2\paperwidth,height=0.2\paperheight,angle=-90]{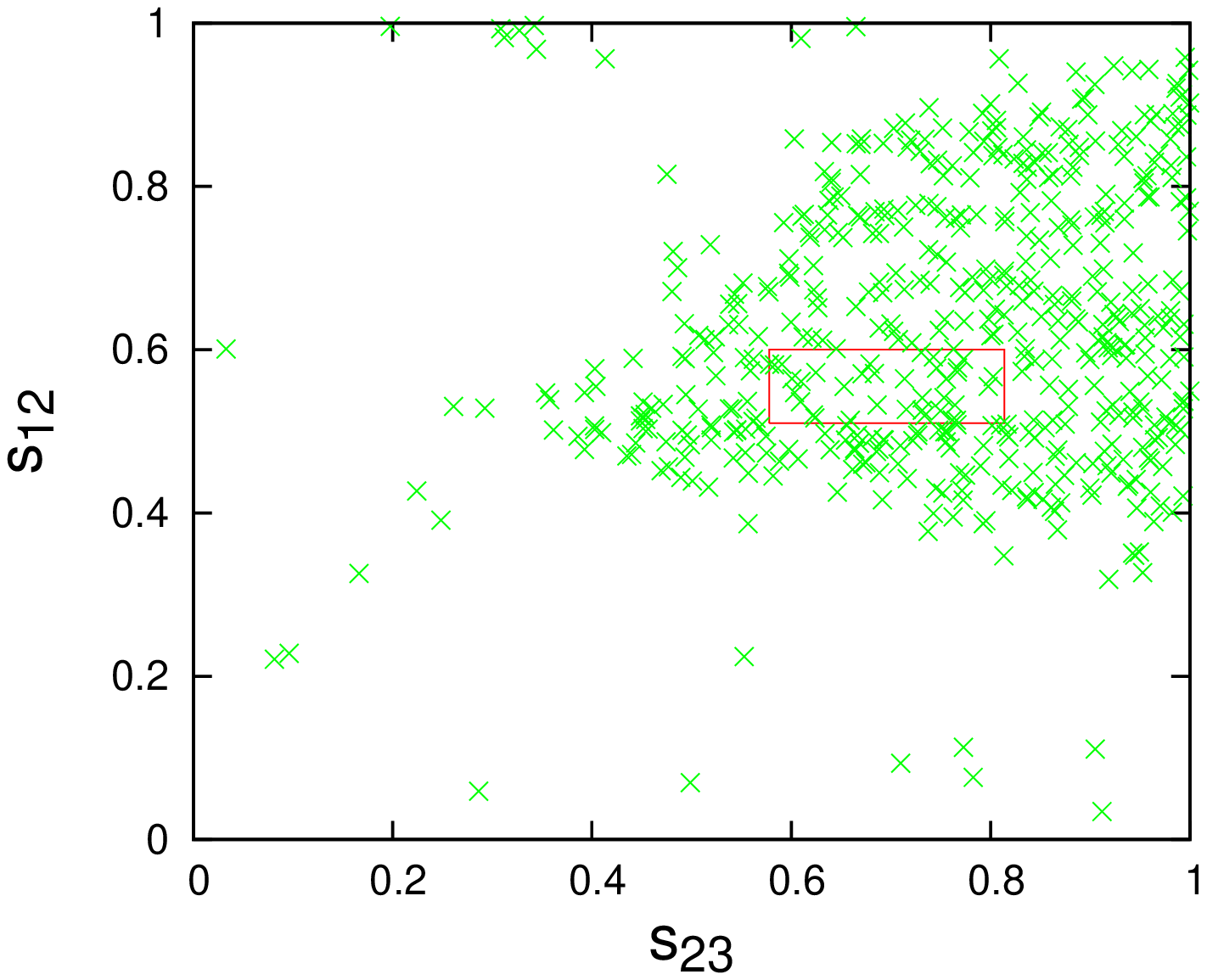}
  \includegraphics[width=0.2\paperwidth,height=0.2\paperheight,angle=-90]{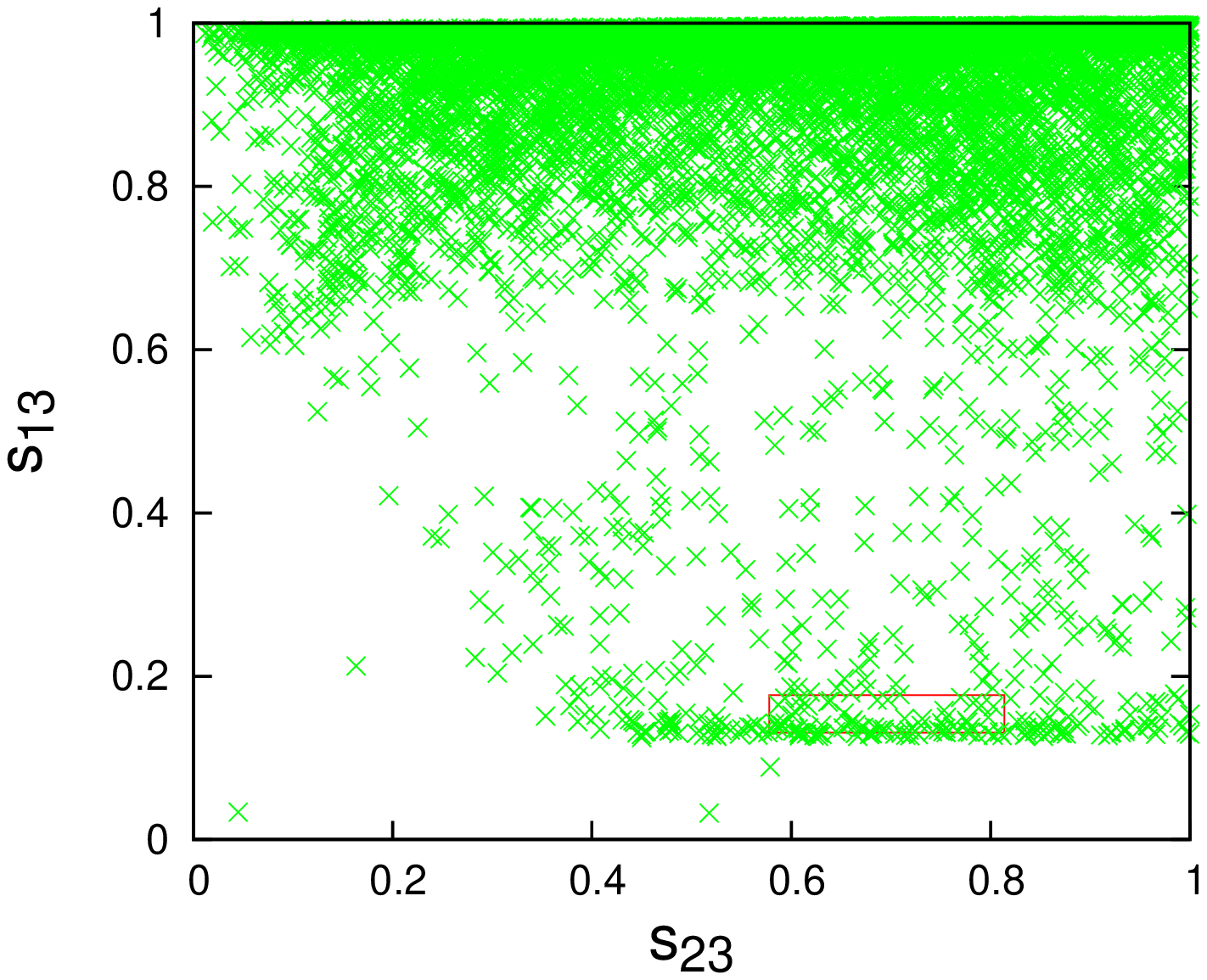}
\end{tabular}
\caption{Plots showing the parameter space for any two mixing
angles when the third angle is constrained by its  $1 \sigma$
range for  Class III ansatz of texture four zero  Dirac mass
matrices (inverted hierarchy).} \label{5aih1}
\end{figure}

To study the lepton mass matrices for this class, we choose to analyse the following
structure,

\be
 M_{i}=\left( \ba{ccc}
0 & A _{i}e^{i\alpha_i} & B_{i}e^{i\gamma_i}     \\
A_{i}e^{-i\alpha_i} & 0 &   D_{i}e^{i\beta_i}      \\
 B_{i}e^{-i\gamma_i} & D_{i}e^{-i\beta_i}  &  E_{i} \ea \right),
\label{cl3}
\ee
where $i=l,~\nu_D$ corresponds to the charged lepton and Dirac neutrino mass matrices
respectively. In the case of factorizable phases, the lepton mass matrices belonging to this
class can be analysed following a methodology similar to that for class I and class II ansatz.
For the purpose of calculations, the (2,3) element in each sector, $D_l$ and $D_\nu$, as well
as the phases $\phi_1$ and $\phi_2$ are considered to be the free parameters. It is interesting to note that
in this class as well both the normal as well as inverted mass neutrino mass orderings seem to be viable as can be
seen from figures (\ref{5anh1}) and (\ref{5aih1}).

\begin{figure}
\begin{tabular}{cc}
  \includegraphics[width=0.2\paperwidth,height=0.2\paperheight,angle=-90]{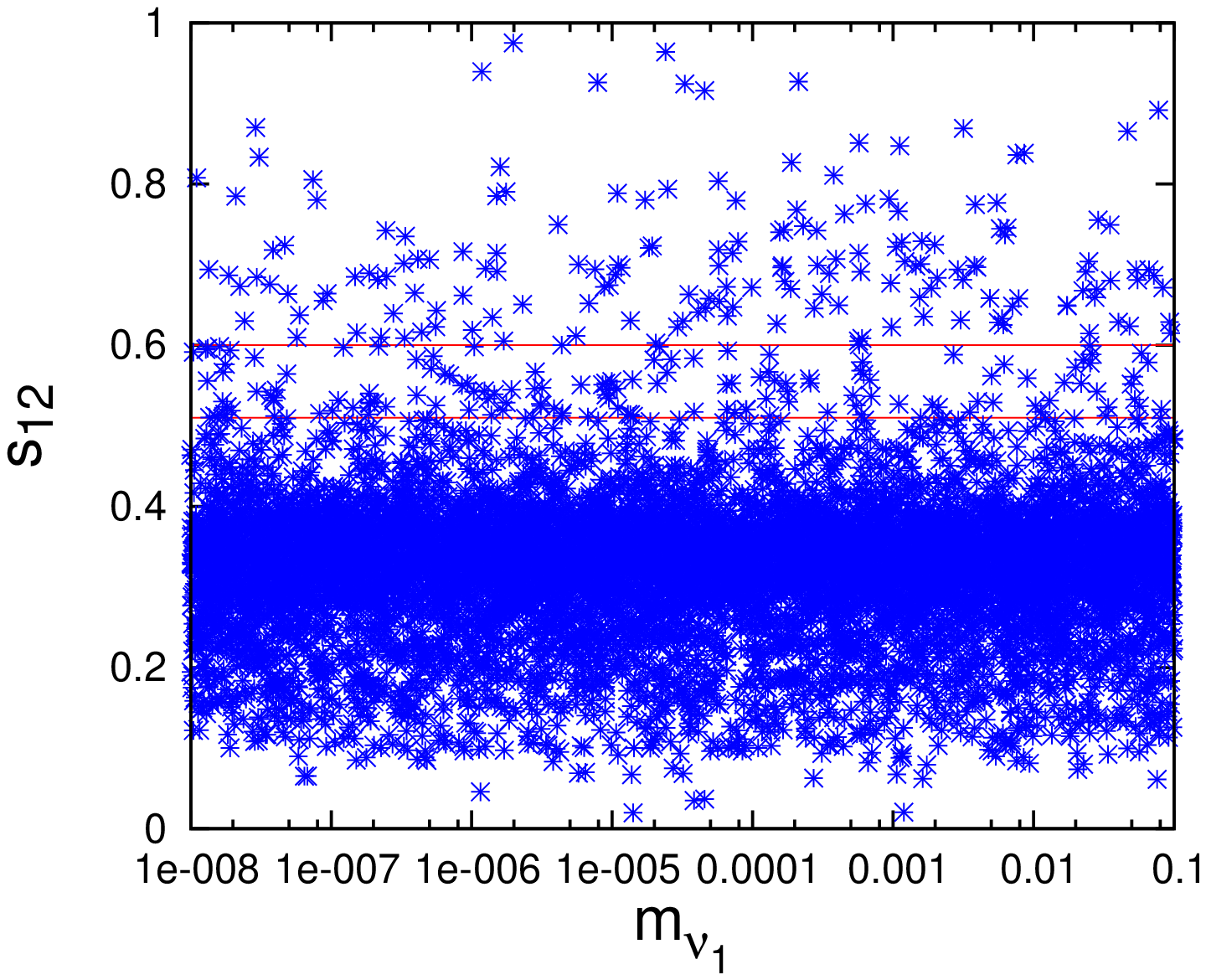}
  \includegraphics[width=0.2\paperwidth,height=0.2\paperheight,angle=-90]{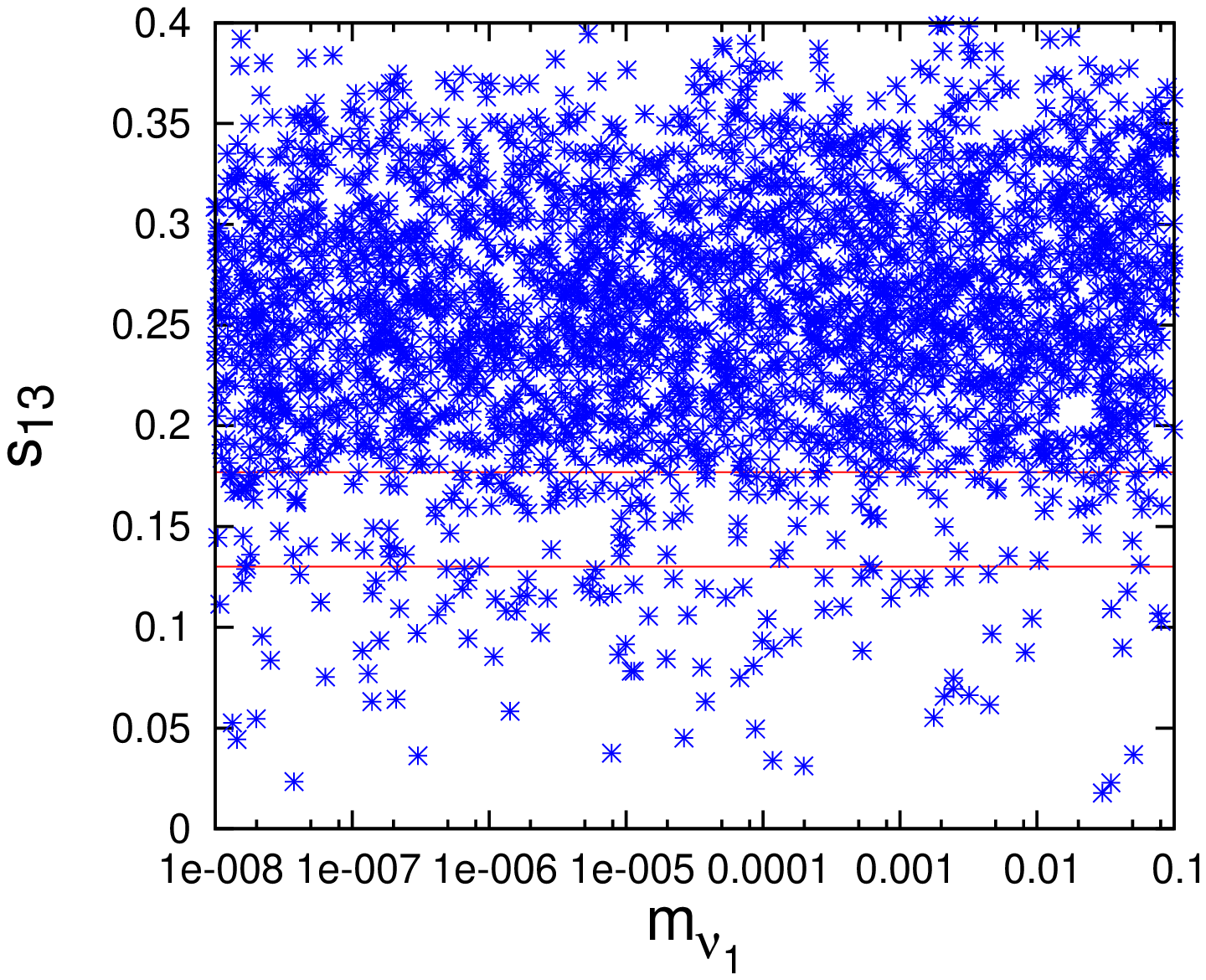}
  \includegraphics[width=0.2\paperwidth,height=0.2\paperheight,angle=-90]{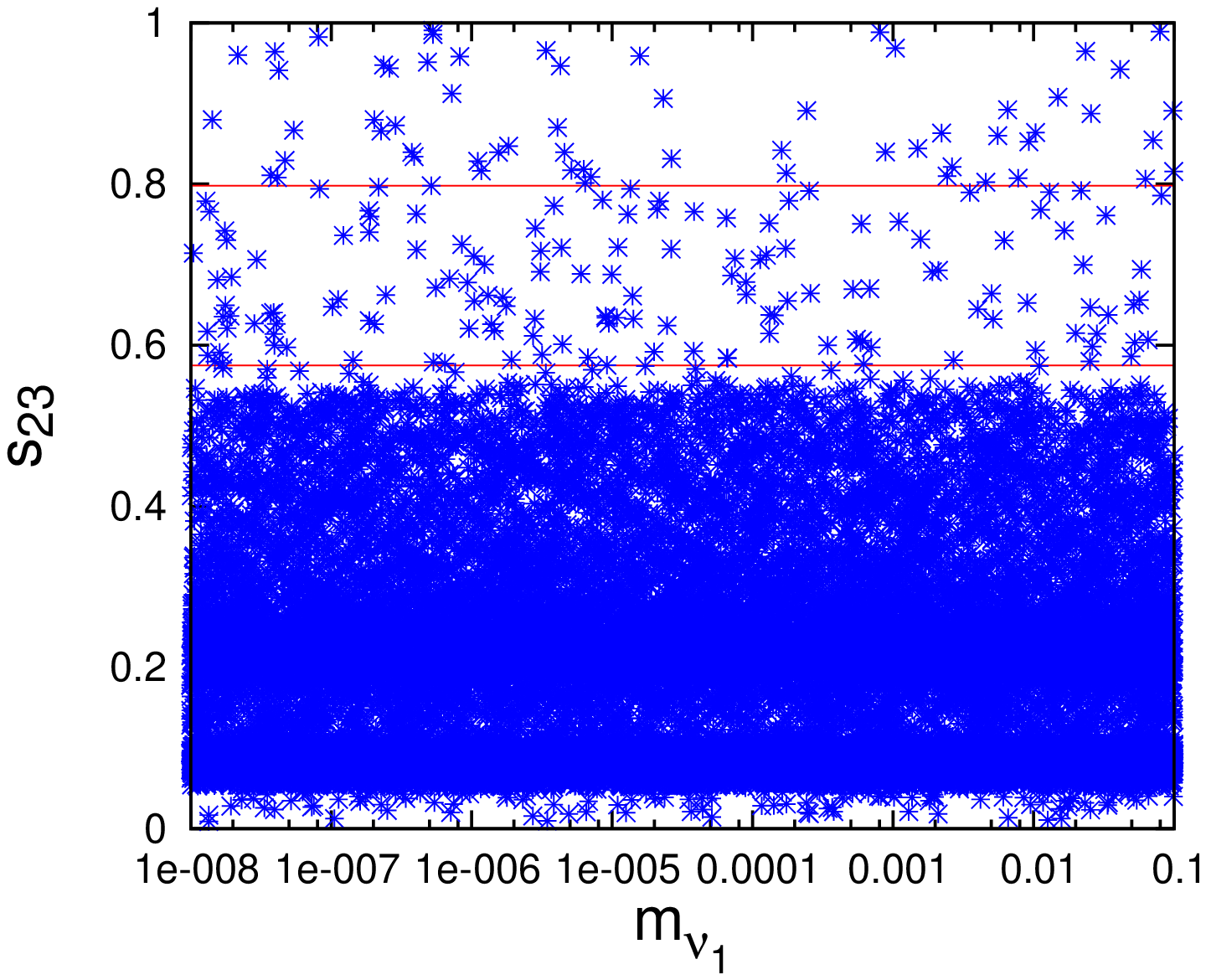}
\end{tabular}
\caption{Plots showing the dependence of mixing angles on the
lightest neutrino mass when the other two angles are constrained
by their $3 \sigma$ ranges  for  Class III ansatz of texture four
zero  Dirac mass matrices (normal hierarchy).} \label{5anh2}
\end{figure}

\begin{figure}
\begin{tabular}{cc}
  \includegraphics[width=0.2\paperwidth,height=0.2\paperheight,angle=-90]{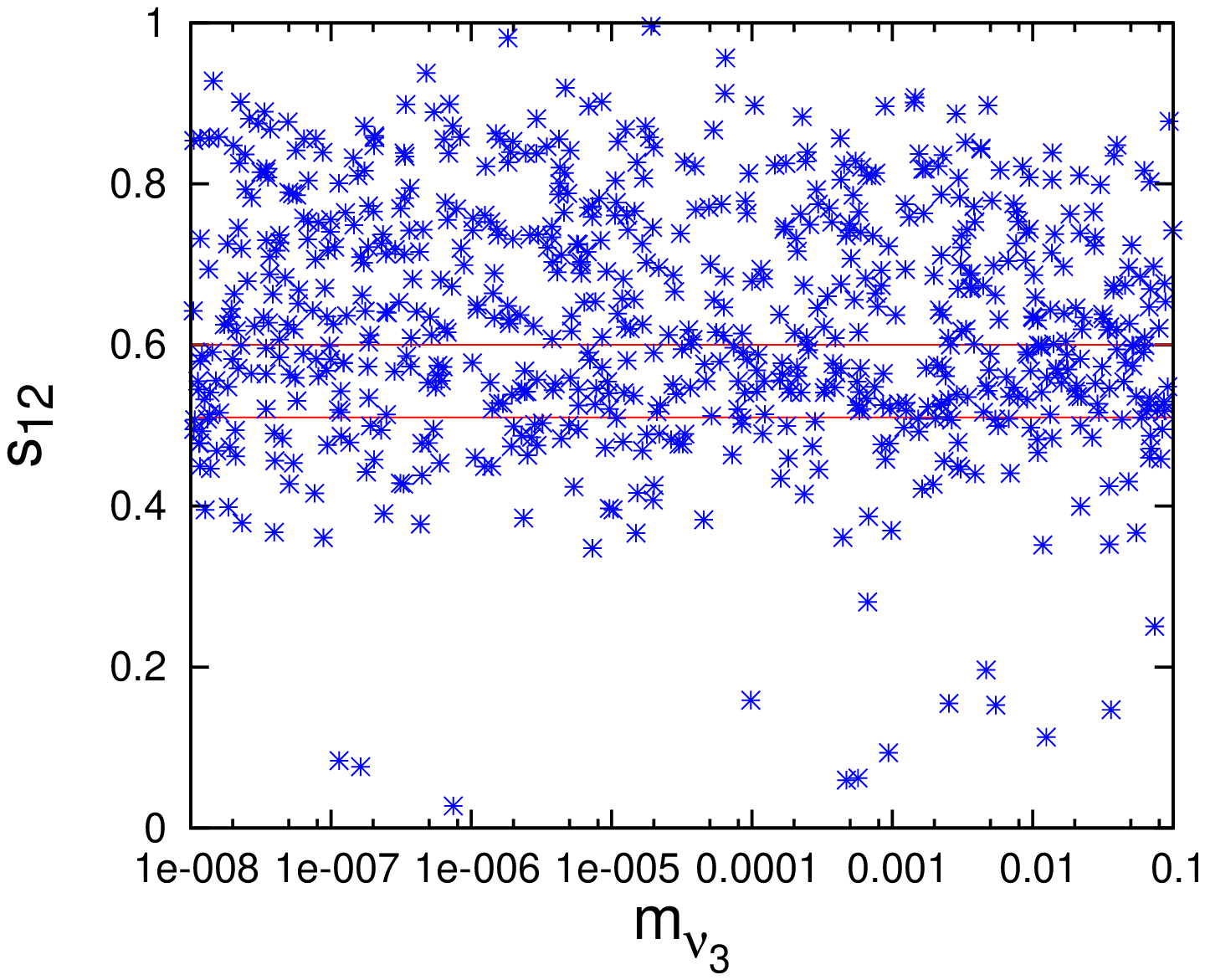}
  \includegraphics[width=0.2\paperwidth,height=0.2\paperheight,angle=-90]{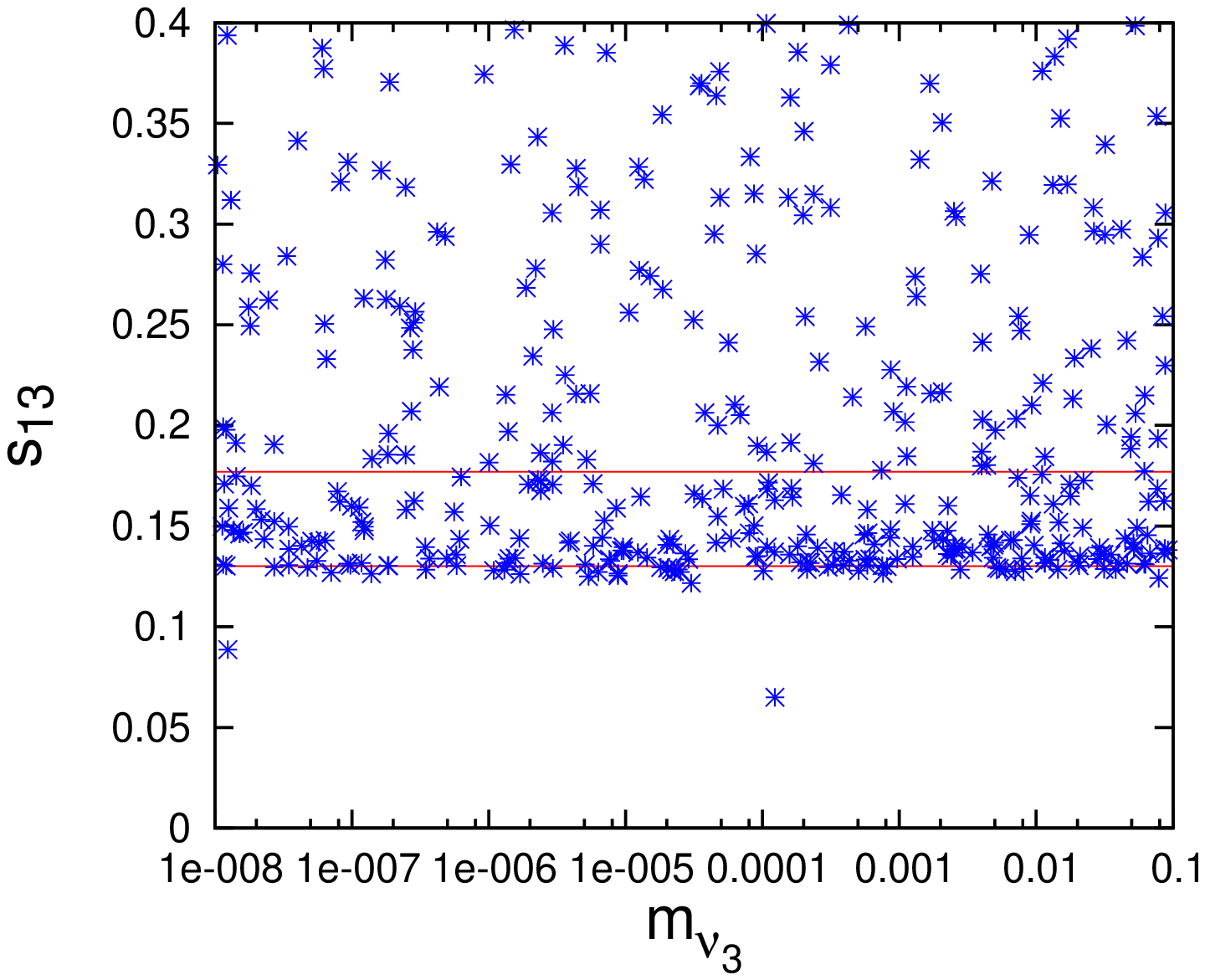}
  \includegraphics[width=0.2\paperwidth,height=0.2\paperheight,angle=-90]{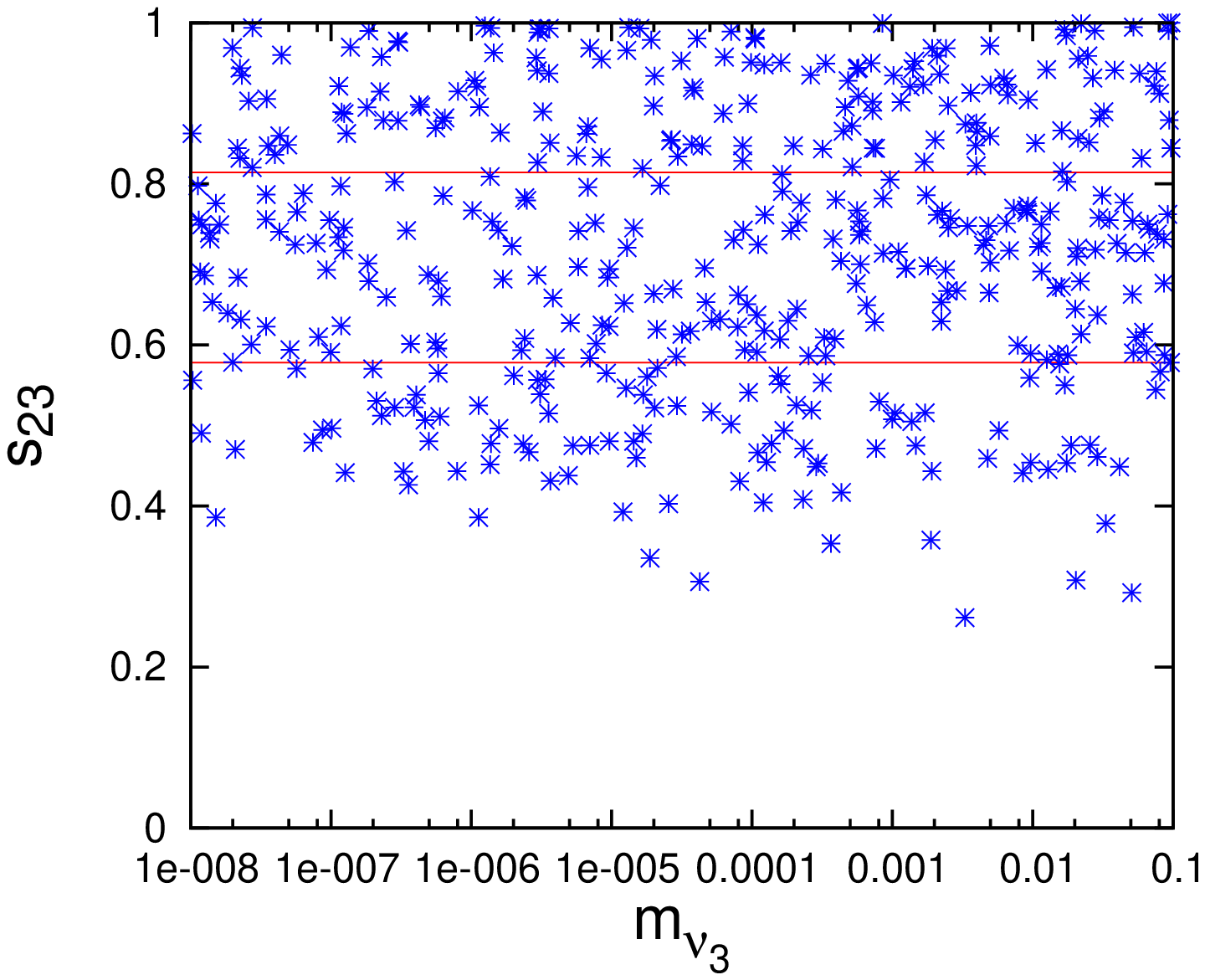}
\end{tabular}
\caption{Plots showing the dependence of mixing angles on the
lightest neutrino mass when the other two angles are constrained
by their $3 \sigma$ ranges  for  Class III ansatz of texture four
zero  Dirac mass matrices (inverted hierarchy).} \label{5aih2}
\end{figure}

Further, in figures (\ref{5anh2}) and (\ref{5aih2}) we attempt to study the
dependence of the leptonic mixing angles on the lightest
neutrino mass for this class
for the normal and inverted neutrino mass hierarchy respectively. As can be seen from
these figures, the lightest neutrino mass seems to be unrestricted by the $3\sigma$ ranges of the mixing angles
for both the the neutrino mass hierarchies. Since the value of the
lightest neutrino mass pertaining to the degenerate scenario $(\sim 0.1 eV)$ is included in the
range allowed by the $3\sigma$ ranges of the mixing angles, therefore both the degenerate scenarios may be viable
for this class of lepton mass matrices.

\par
\begin{figure}
\begin{tabular}{cc}
  \includegraphics[width=0.2\paperwidth,height=0.2\paperheight,angle=-90]{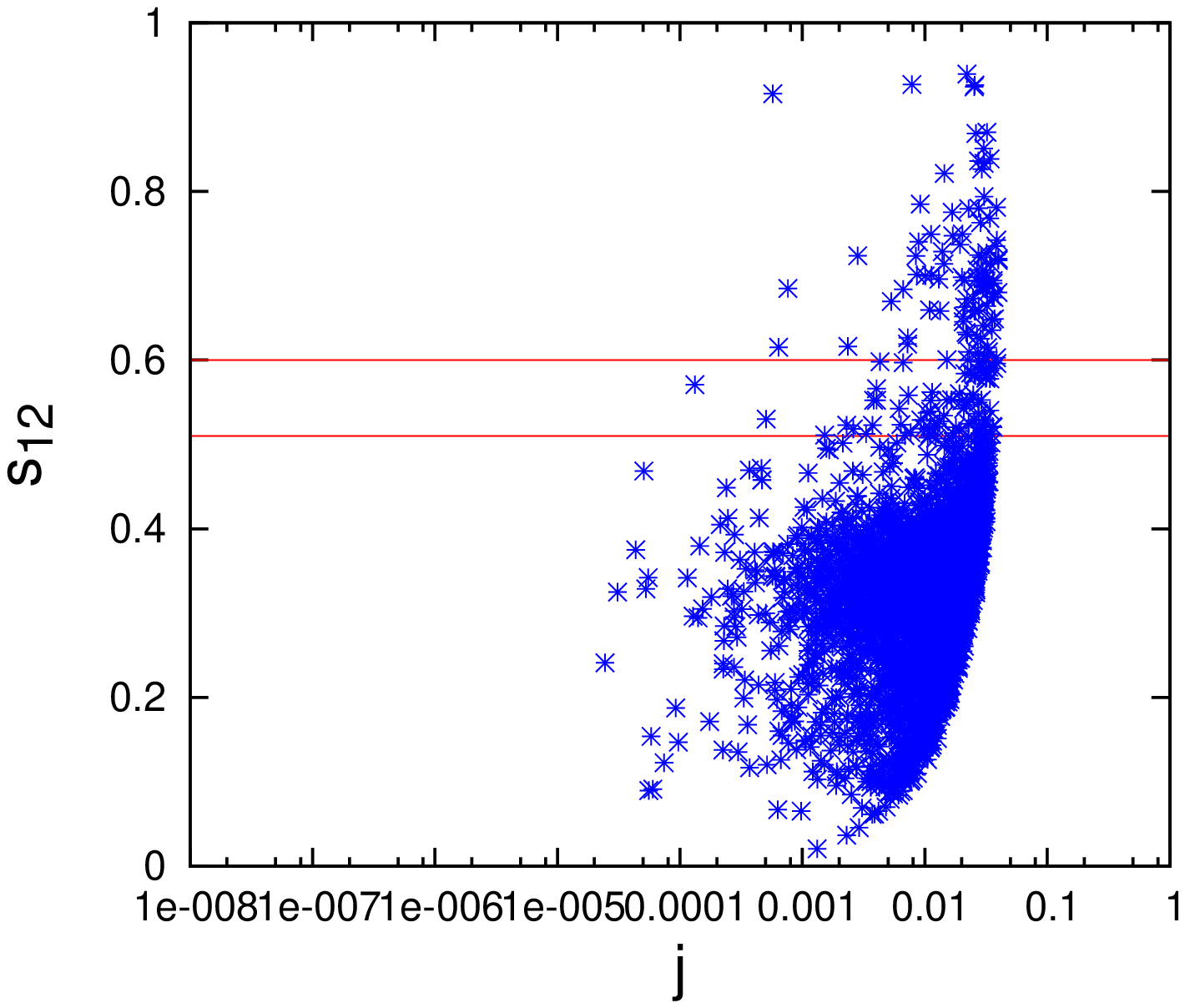}
  \includegraphics[width=0.2\paperwidth,height=0.2\paperheight,angle=-90]{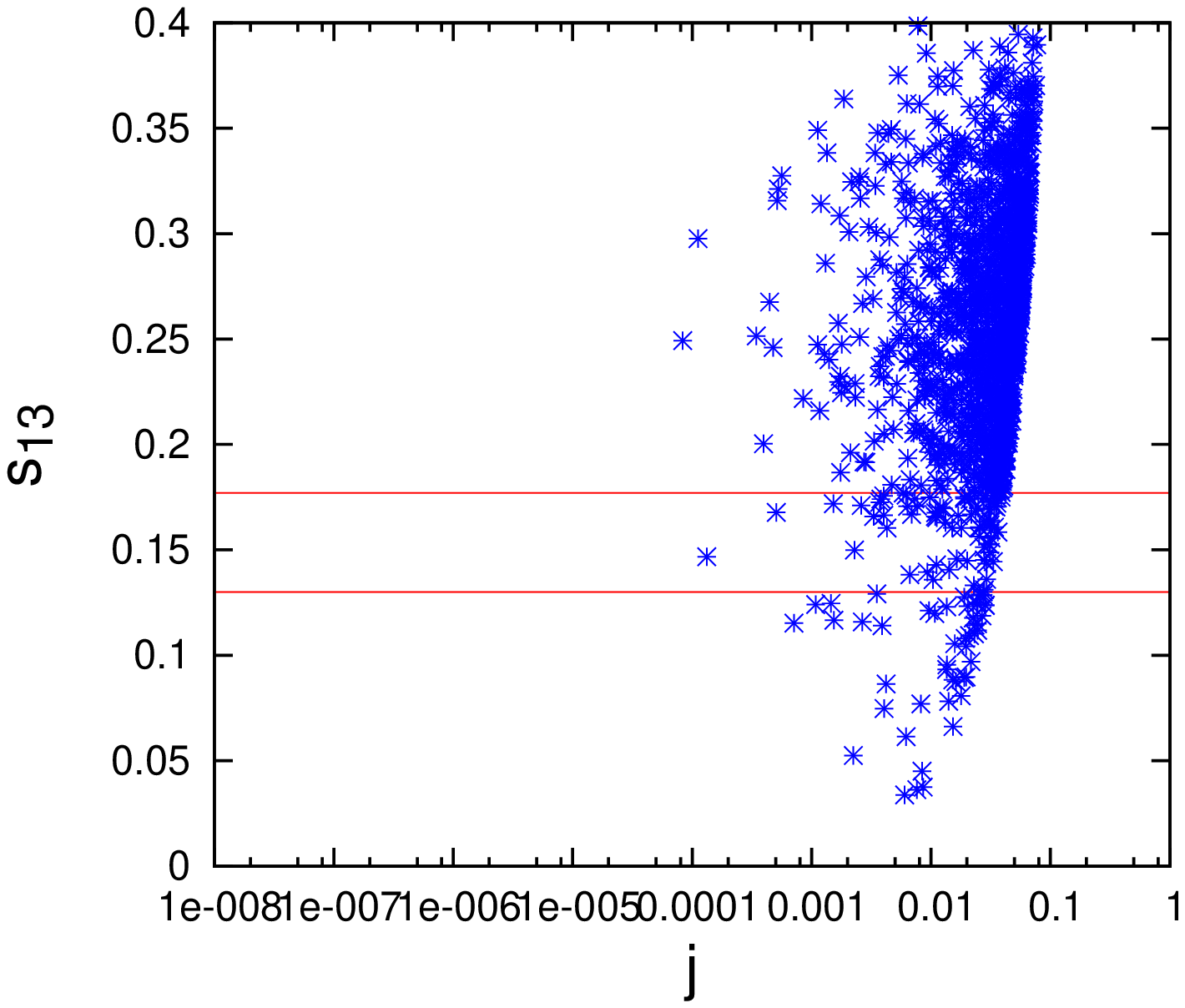}
  \includegraphics[width=0.2\paperwidth,height=0.2\paperheight,angle=-90]{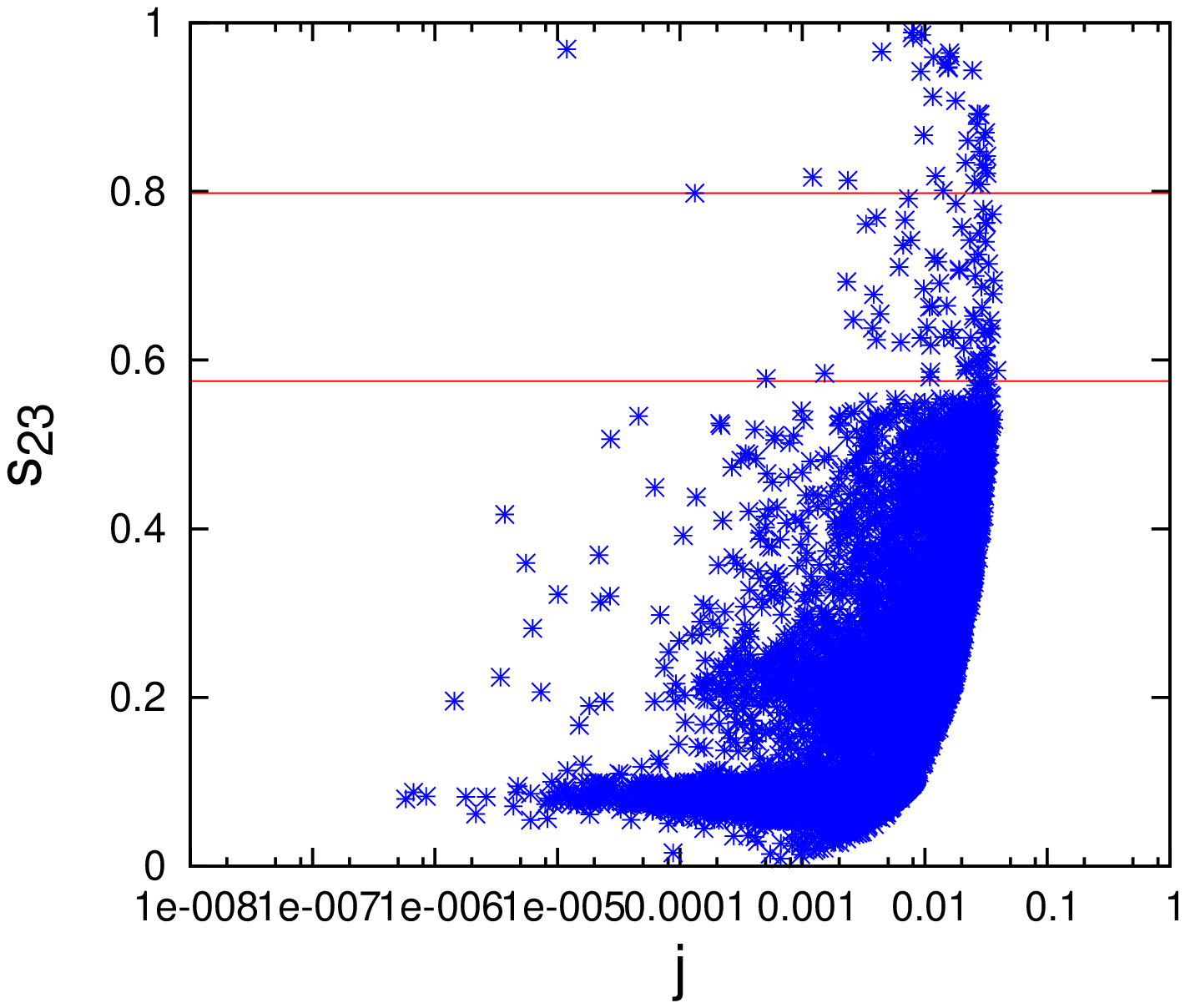}
\end{tabular}
\caption{Plots showing the variation of Jarlskog CP violating
parameter with mixing angles when the other two angles are
constrained by their $3 \sigma$ ranges  for  Class III ansatz of
texture four zero  Dirac mass matrices (normal hierarchy).}
\label{5anh3}
\end{figure}

\begin{figure}
\begin{tabular}{cc}
  \includegraphics[width=0.2\paperwidth,height=0.2\paperheight,angle=-90]{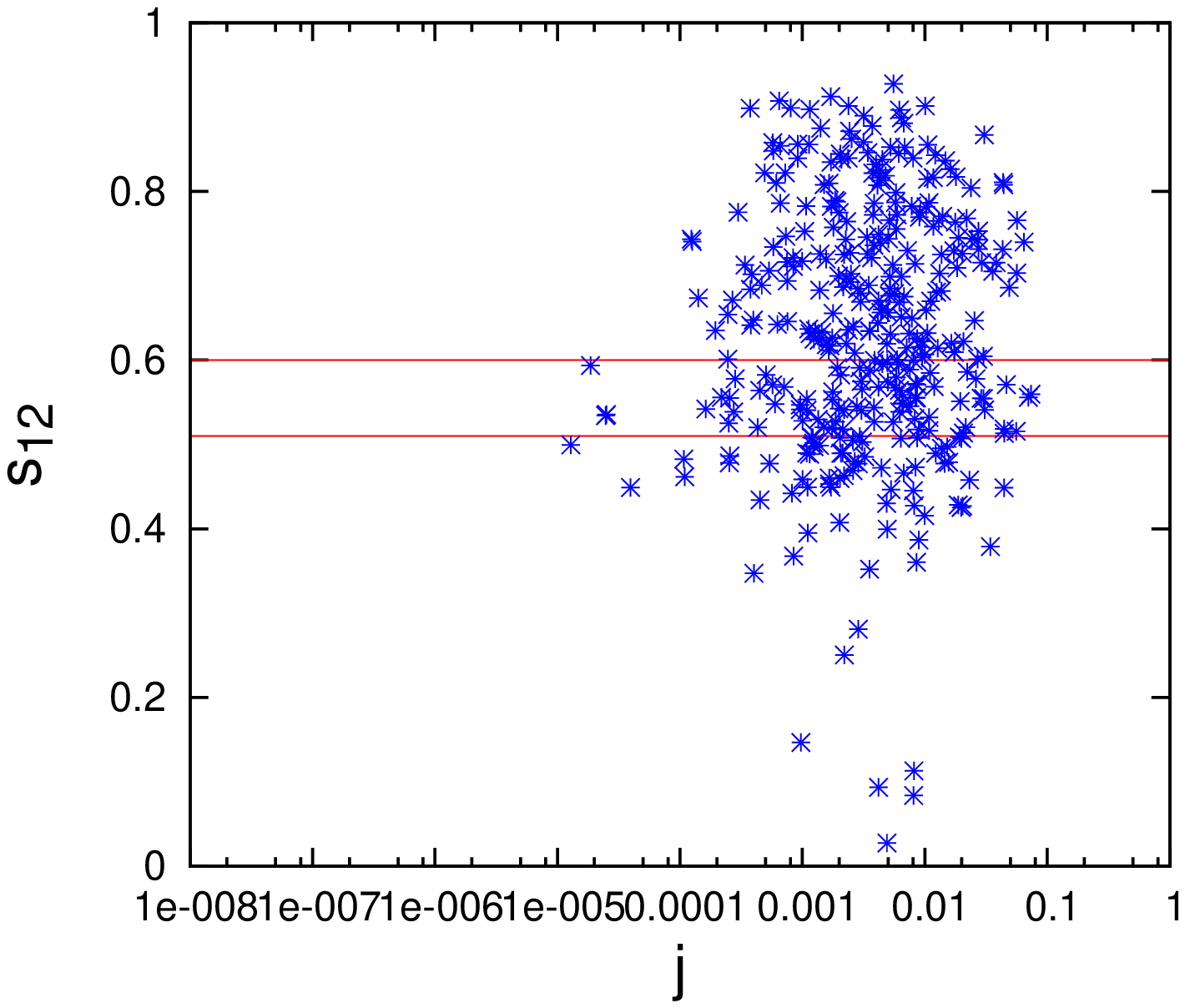}
  \includegraphics[width=0.2\paperwidth,height=0.2\paperheight,angle=-90]{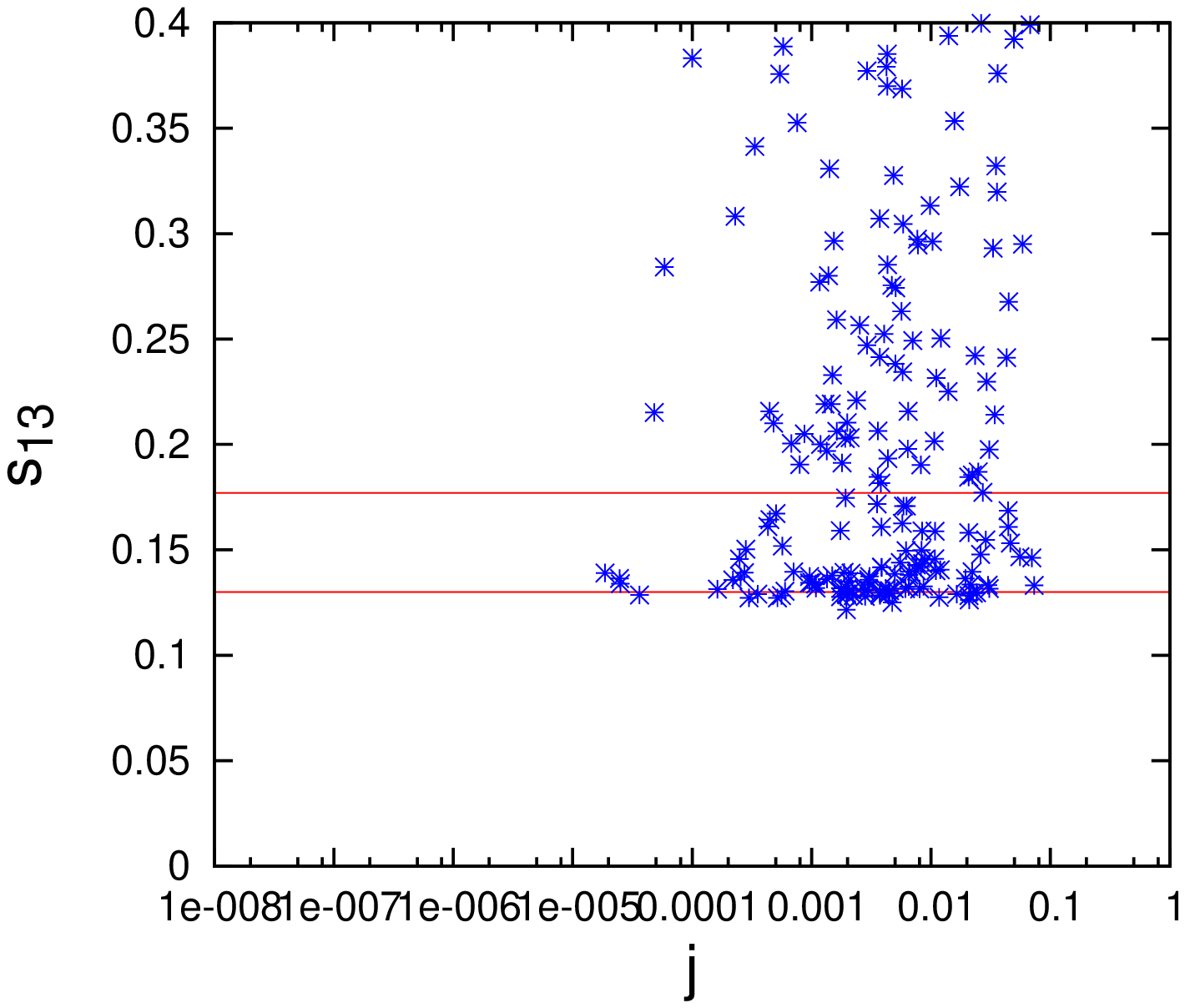}
  \includegraphics[width=0.2\paperwidth,height=0.2\paperheight,angle=-90]{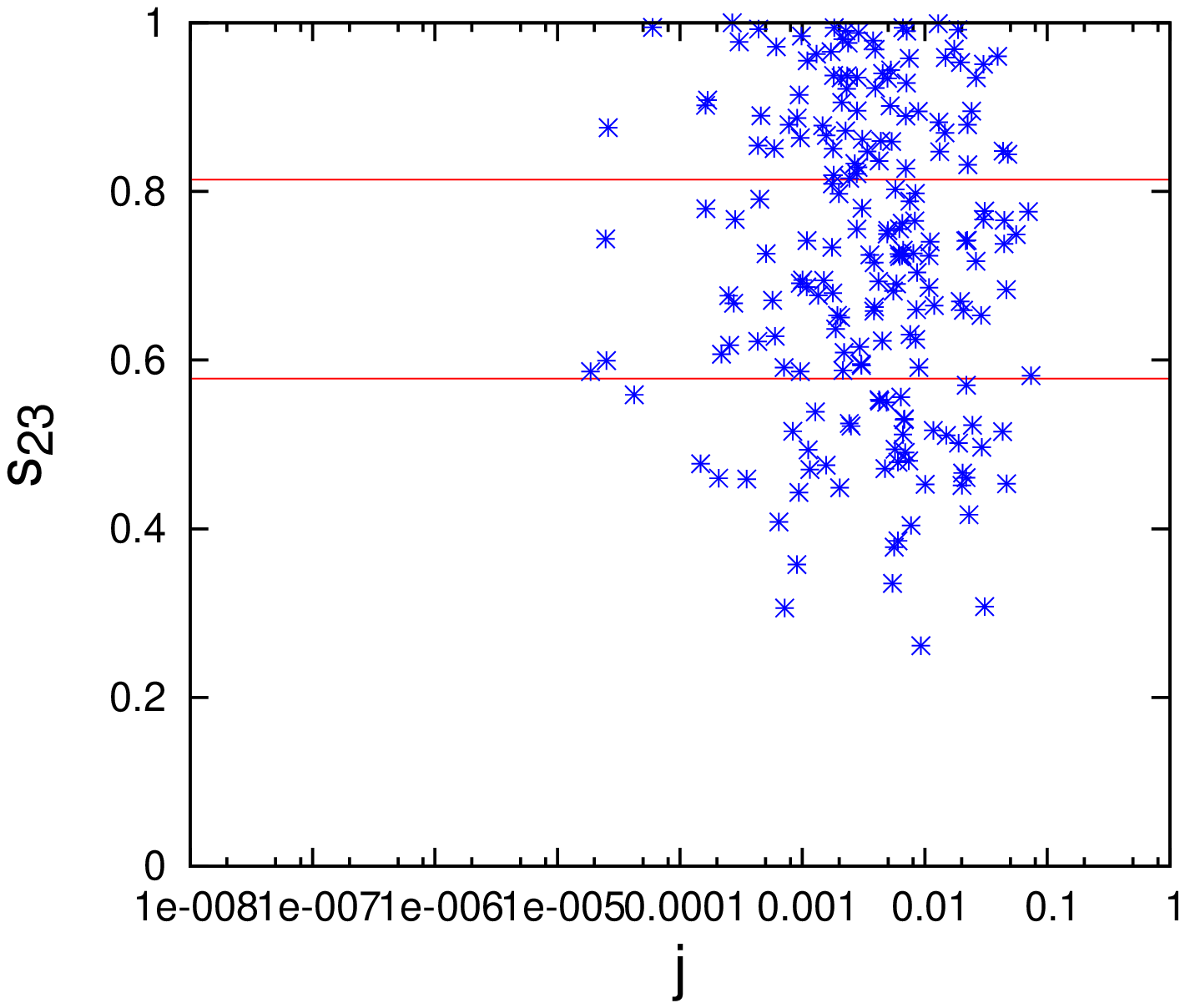}
\end{tabular}
\caption{Plots showing the variation of Jarlskog CP violating
parameter with mixing angles when the other two angles are
constrained by their $3 \sigma$ ranges  for  Class III ansatz of
texture four zero  Dirac mass matrices (inverted hierarchy).}
\label{5aih3}
\end{figure}

 Further, we study the possiblity of CP violation in the leptonic sector for this class by studying the
variation of Jarlskog's rephasing invariant ${\it{J}}$ with all the mixing angles for normal as well as inverted
neutrino mass hierarchy in figures (\ref{5anh3}) and (\ref{5aih3}) respectively. While plotting these graphs the two mixing angles,
other than the one being considered, are constrained by their $3\sigma$ ranges. The parallel lines in these plots show the
$3\sigma$ experimental ranges for the mixing angle being considered. It is interesting to note that for the case
of inverted hierarchy one gets a broader range
for ${\it{J}}$ as compared to the one for the case of normal hierarchy, contrary to the
observation in class II ansatz. To be more explicit, one finds
$0.0001\lesssim {\it{J}} \lesssim 0.05$ and $0.00001\lesssim {\it{J}} \lesssim 0.1$ corresponding to 
normal and inverted hierarchy scenario respectively.

\subsection{Texture five zero lepton mass matrices}
After studying all possible texture four zero lepton mass matrices, it becomes
interesting to explore the parallel texture five zero structures for each class which can be derived by
substituting  either $D_l=0$, $D_\nu \neq 0$ or $D_l\neq 0$, $D_l = 0$ in the corresponding
texture four zero mass matrices. In this subsection, we carry out a detailed study of all classes of
texture five zero lepton mass matrices for Dirac neutrinos pertaining to both the
possibilities leading to texture five zero structures.
\subsubsection{Class I ansatz}
The two possibilities for texture five zero lepton mass matrices for this class can be given as,
\be
 M_{l}=\left( \ba{ccc}
0 & A _{l} & 0      \\
A_{l}^{*} & 0 &  B_{l}     \\
 0 &     B_{l}^{*}  &  E_{l} \ea \right), \qquad
 M_{\nu}=\left( \ba{ccc}
0 & A _{\nu} & 0      \\
A_{\nu}^{*} & D_{\nu} &  B_{\nu}     \\
 0 &     B_{\nu}^{*}  &  E_{\nu} \ea \right),
\label{cl1t51}\ee
or
\be
 M_{l}=\left( \ba{ccc}
0 & A _{l} & 0      \\
A_{l}^{*} & D_l &  B_{l}     \\
 0 &     B_{l}^{*}  &  E_{l} \ea \right), \qquad
 M_{\nu}=\left( \ba{ccc}
0 & A _{\nu} & 0      \\
A_{\nu}^{*} & 0 &  B_{\nu}     \\
 0 &     B_{\nu}^{*}  &  E_{\nu} \ea \right).
\label{cl1t52}\ee

\begin{figure}
\begin{tabular}{cc}
  \includegraphics[width=0.2\paperwidth,height=0.2\paperheight,angle=-90]{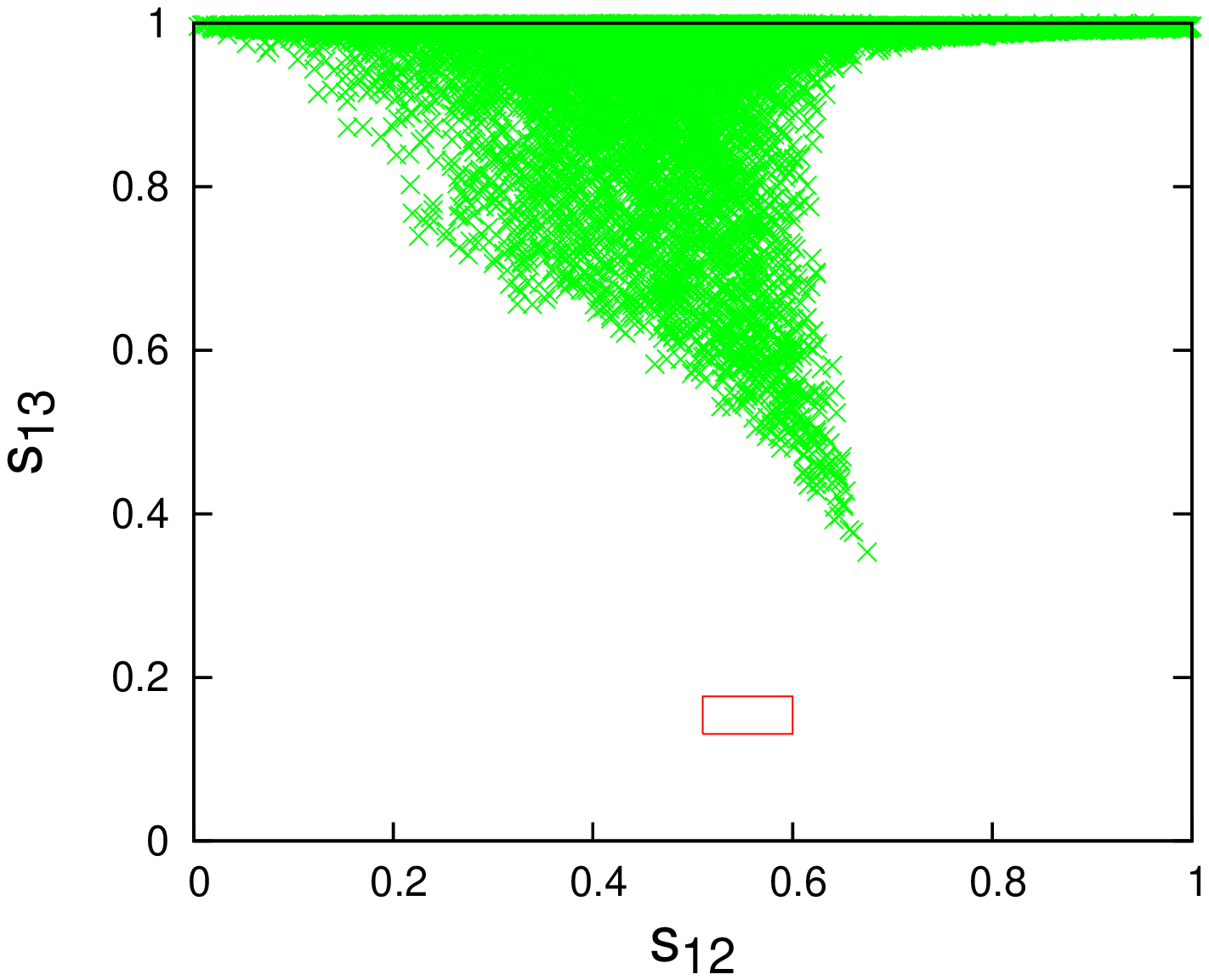}
  \includegraphics[width=0.2\paperwidth,height=0.2\paperheight,angle=-90]{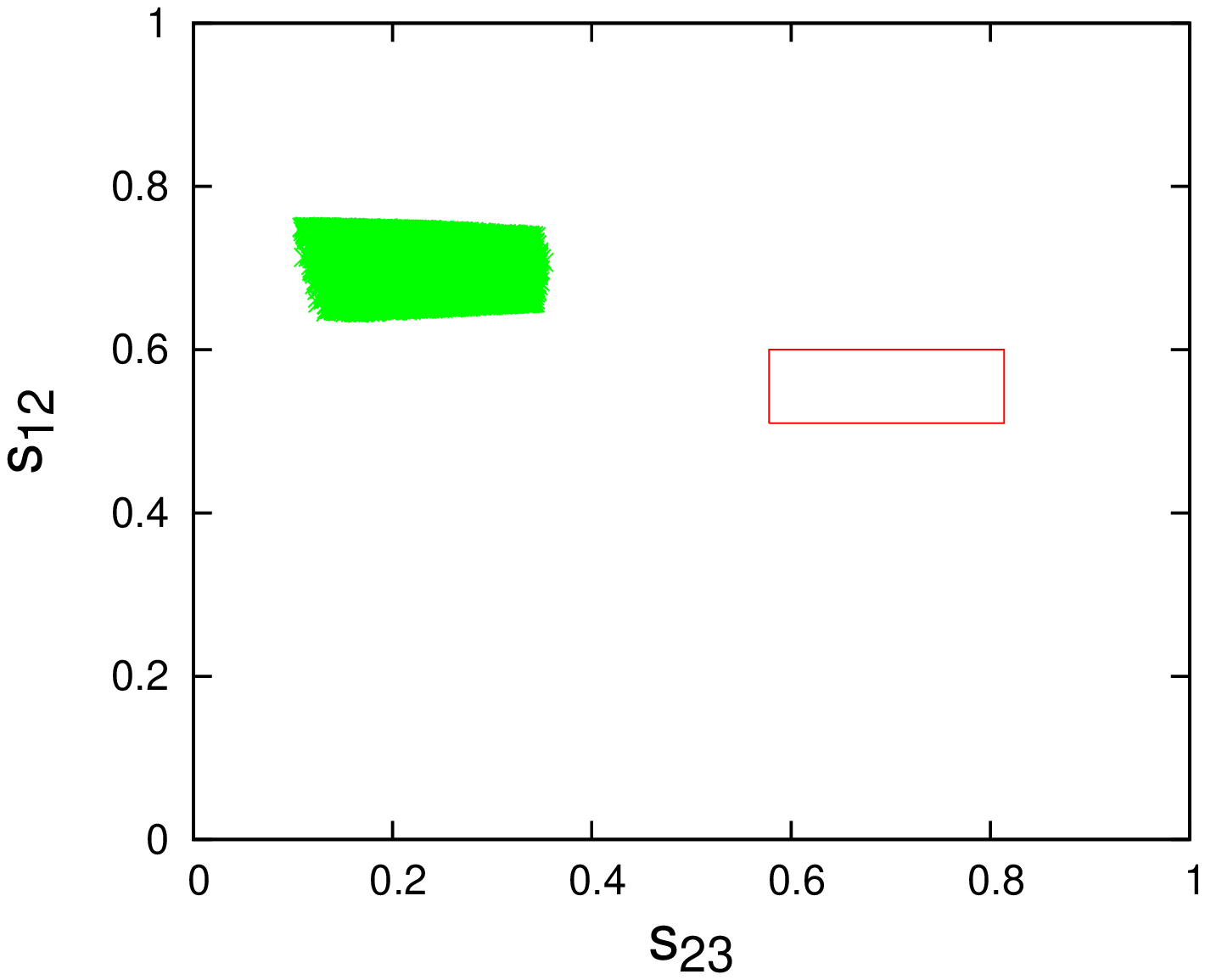}
  \includegraphics[width=0.2\paperwidth,height=0.2\paperheight,angle=-90]{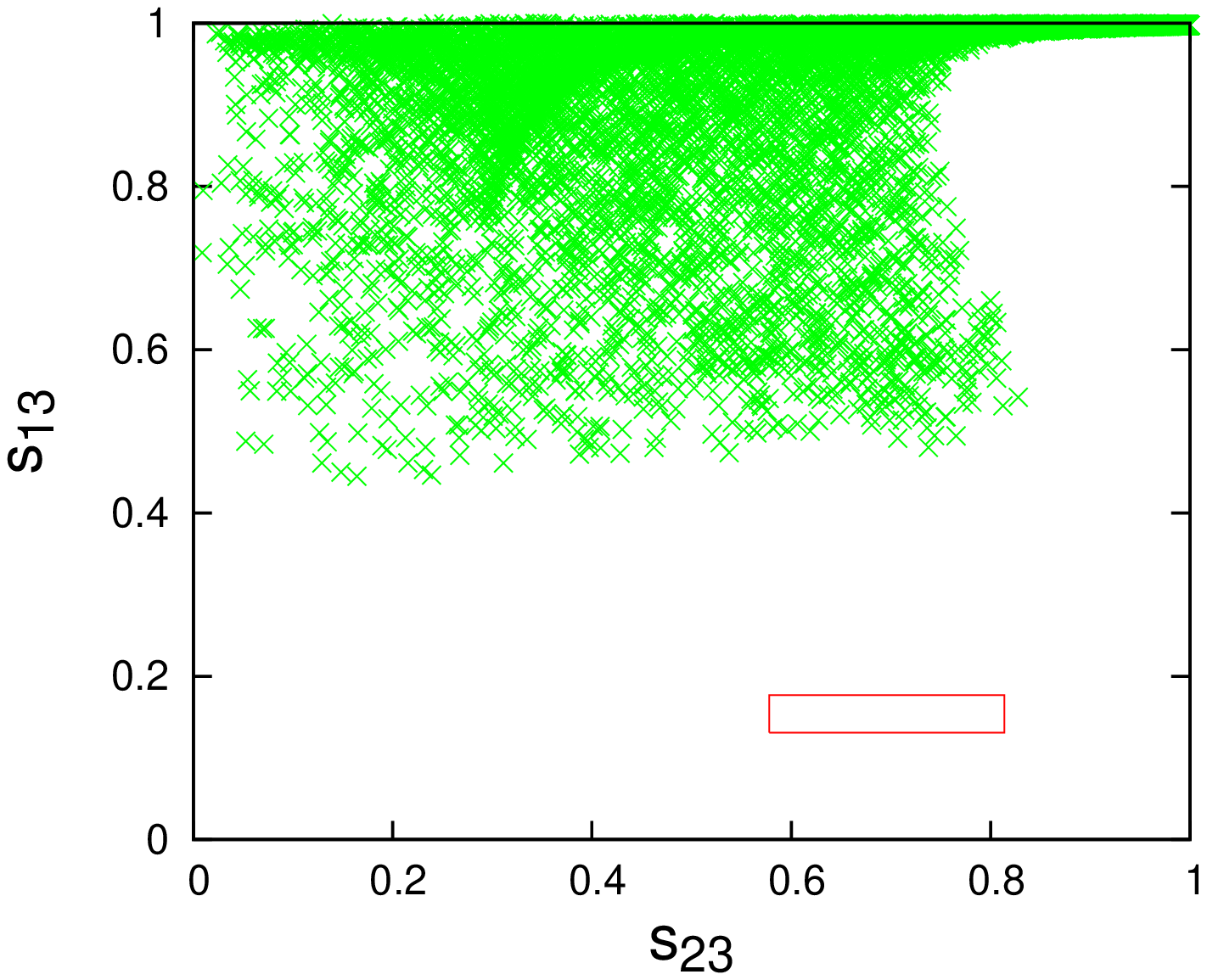}
\end{tabular}
\caption{Plots showing the parameter space for any two mixing
angles when the third angle is constrained by its  $3 \sigma$
range in the $D_l =0$ and $D_\nu\neq 0$ scenario for Class I
ansatz of texture five zero  Dirac mass matrices (inverted
hierarchy).} \label{t5cl1ih1}
\end{figure}

\begin{figure}
\begin{tabular}{cc}
  \includegraphics[width=0.2\paperwidth,height=0.2\paperheight,angle=-90]{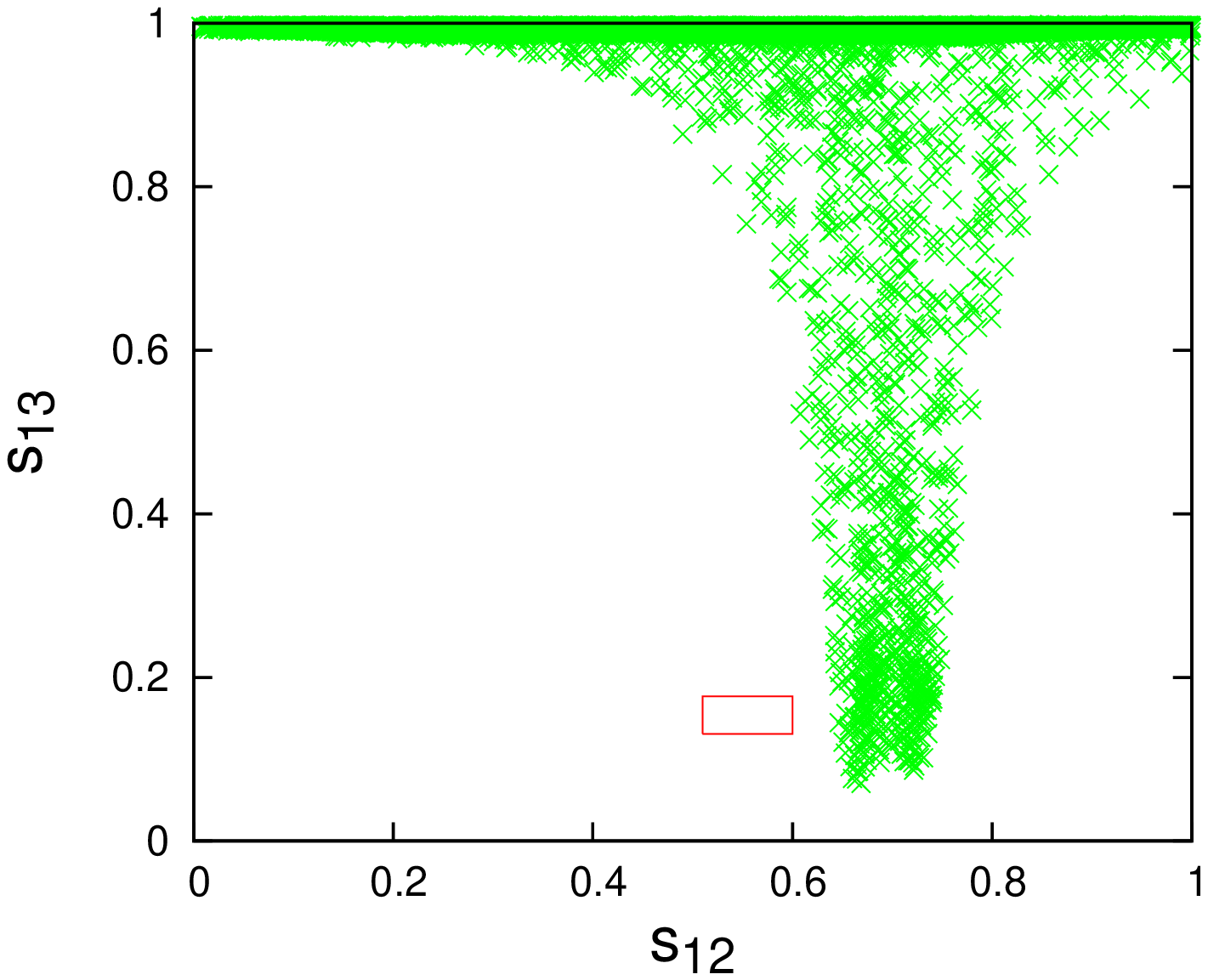}
  \includegraphics[width=0.2\paperwidth,height=0.2\paperheight,angle=-90]{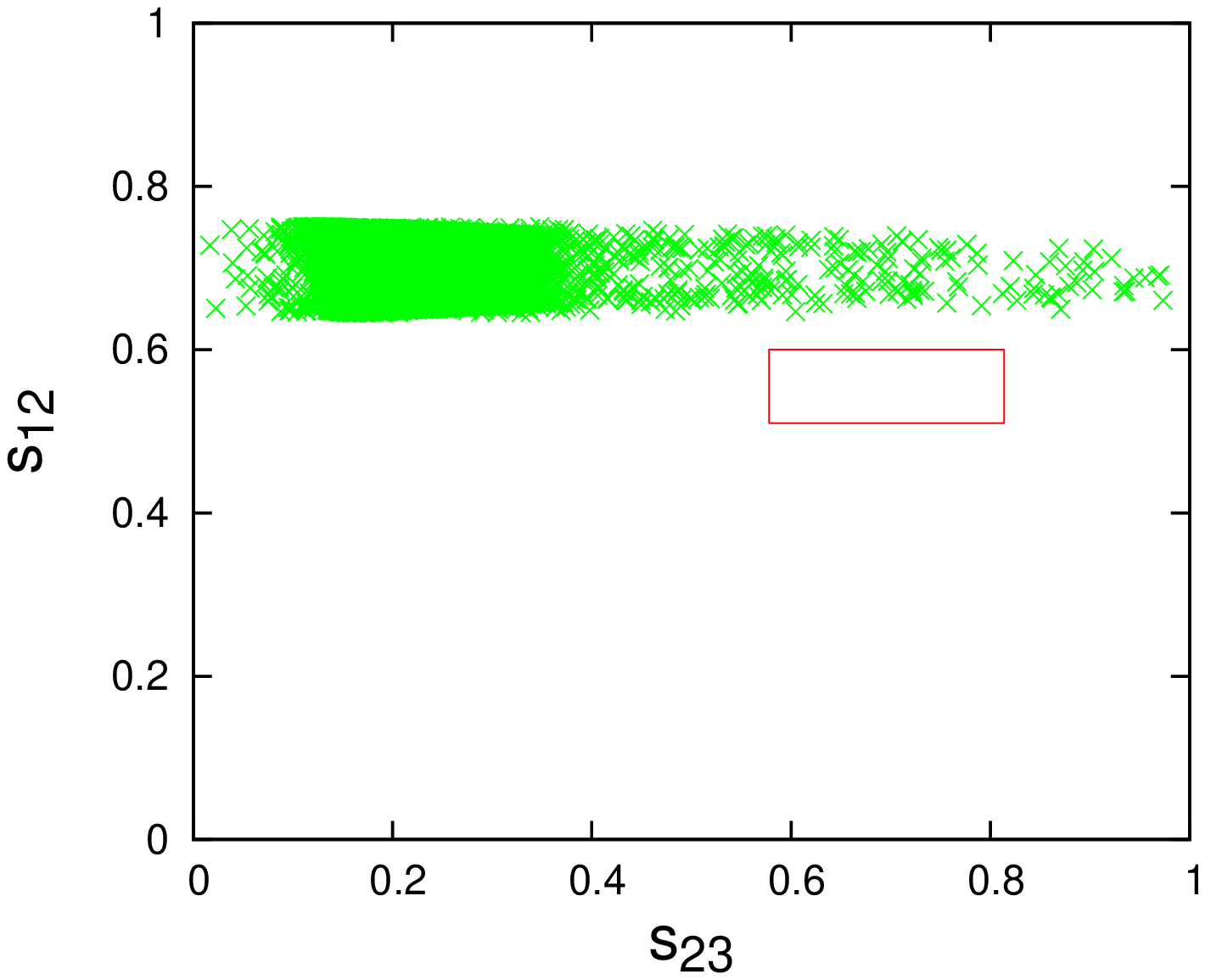}
  \includegraphics[width=0.2\paperwidth,height=0.2\paperheight,angle=-90]{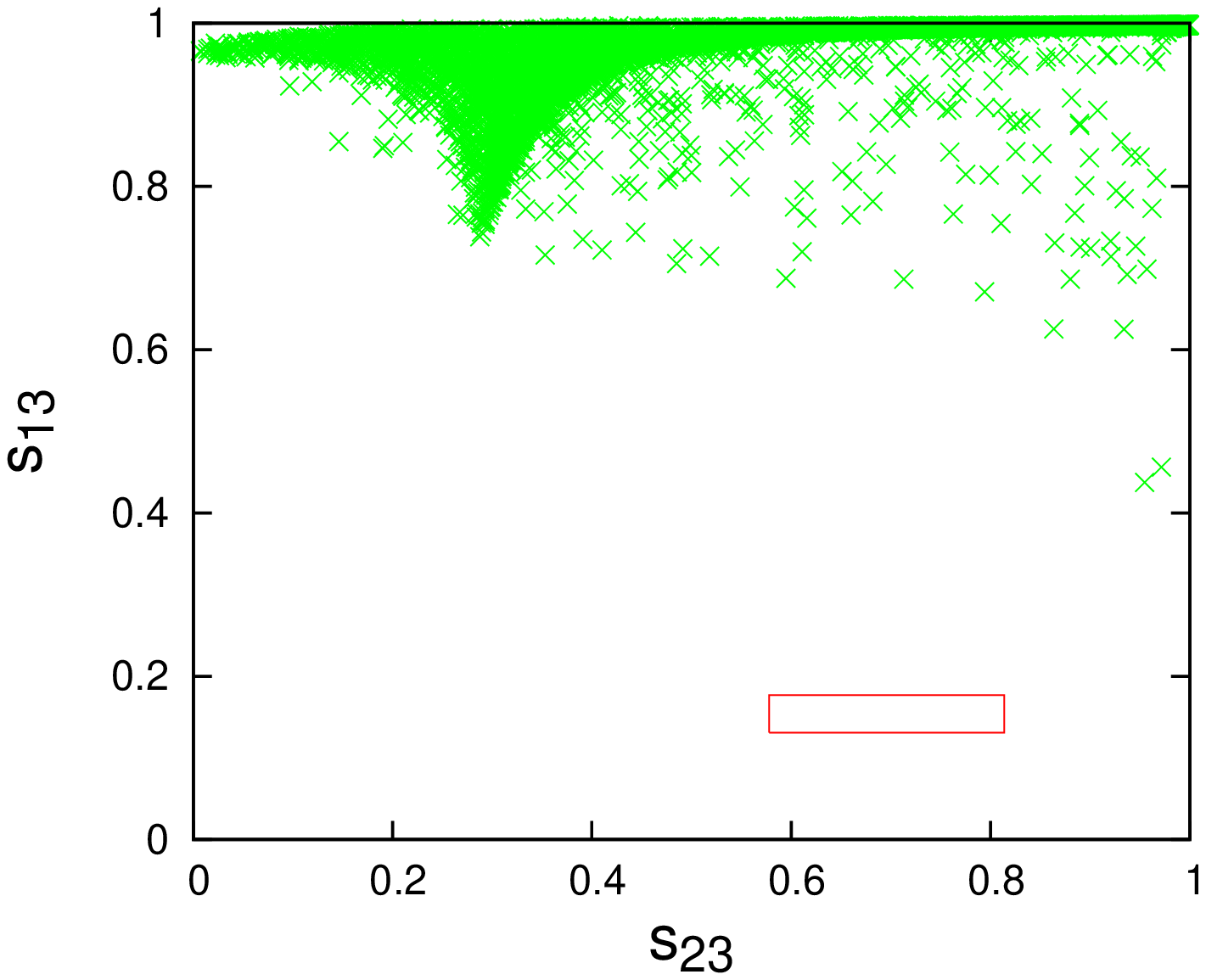}
\end{tabular}
\caption{Plots showing the parameter space for any two mixing
angles when the third angle is constrained by its  $3 \sigma$
range in the $D_l \neq 0$ and $D_\nu = 0$ scenario for Class I
ansatz of texture five zero  Dirac mass matrices (inverted
hierarchy).} \label{t5cl1ih2}
\end{figure}

 We study both these possibilities in detail for all the neutrino mass orderings. Firstly, we examine
 the compatibility of matrices (\ref{cl1t51}) and (\ref{cl1t52}) with the inverted hierarchy
of neutrino masses.
For this purpose, in figures (\ref{t5cl1ih1}) and
 (\ref{t5cl1ih2}), we present the plots showing the parameter space allowed by this ansatz for any two mixing angles wherein
the third one  is constrained by its $3\sigma$ experimental bound for inverted hierarchy of neutrino masses.
The rectangular regions in these plots represent the $3\sigma$ ranges for the two mixing angles being considered.
A general look at these plots shows that inverted hierarchy is ruled out for both the texture five zero possibilities
for this class.

\begin{figure}
\begin{tabular}{cc}
  \includegraphics[width=0.2\paperwidth,height=0.2\paperheight,angle=-90]{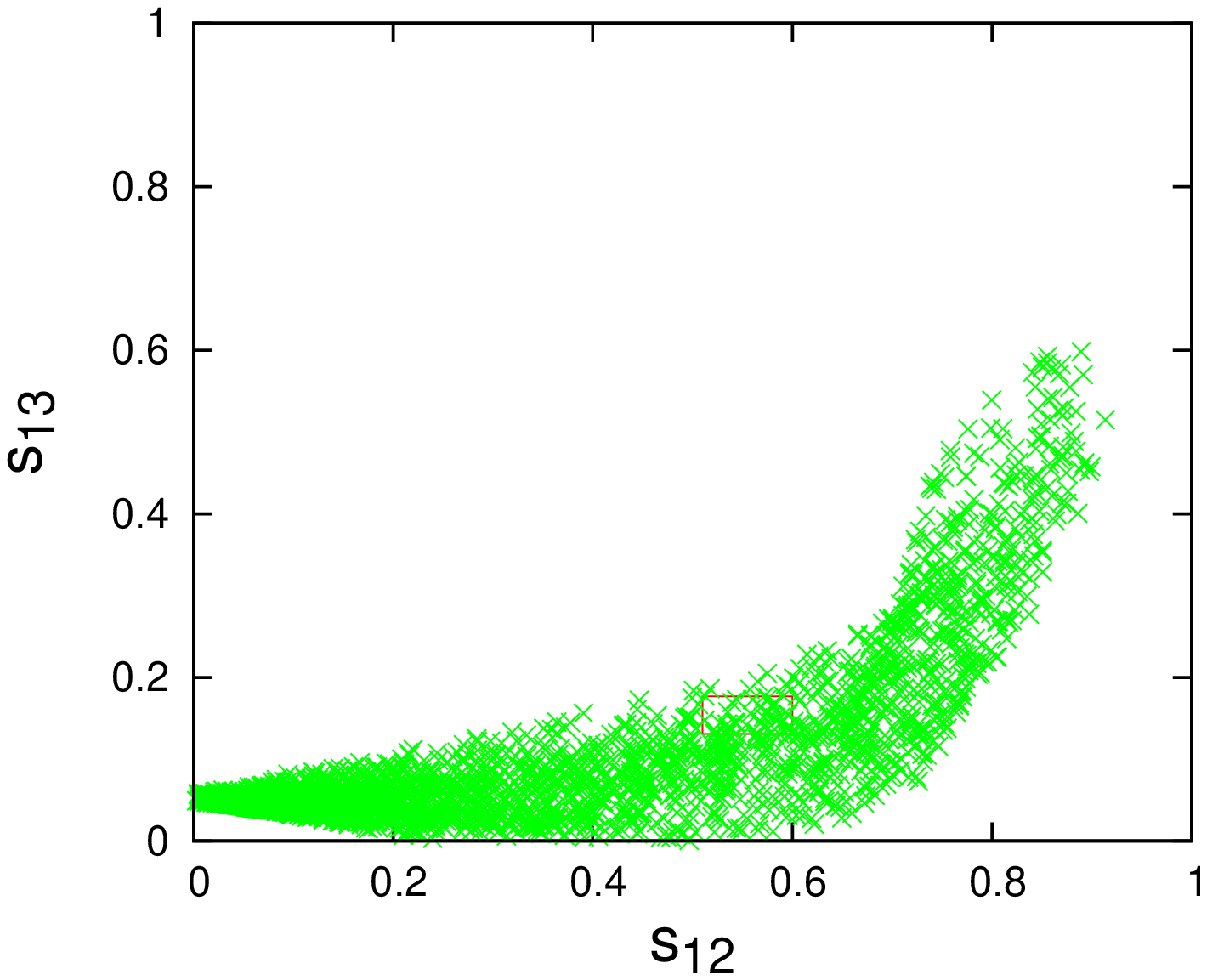}
  \includegraphics[width=0.2\paperwidth,height=0.2\paperheight,angle=-90]{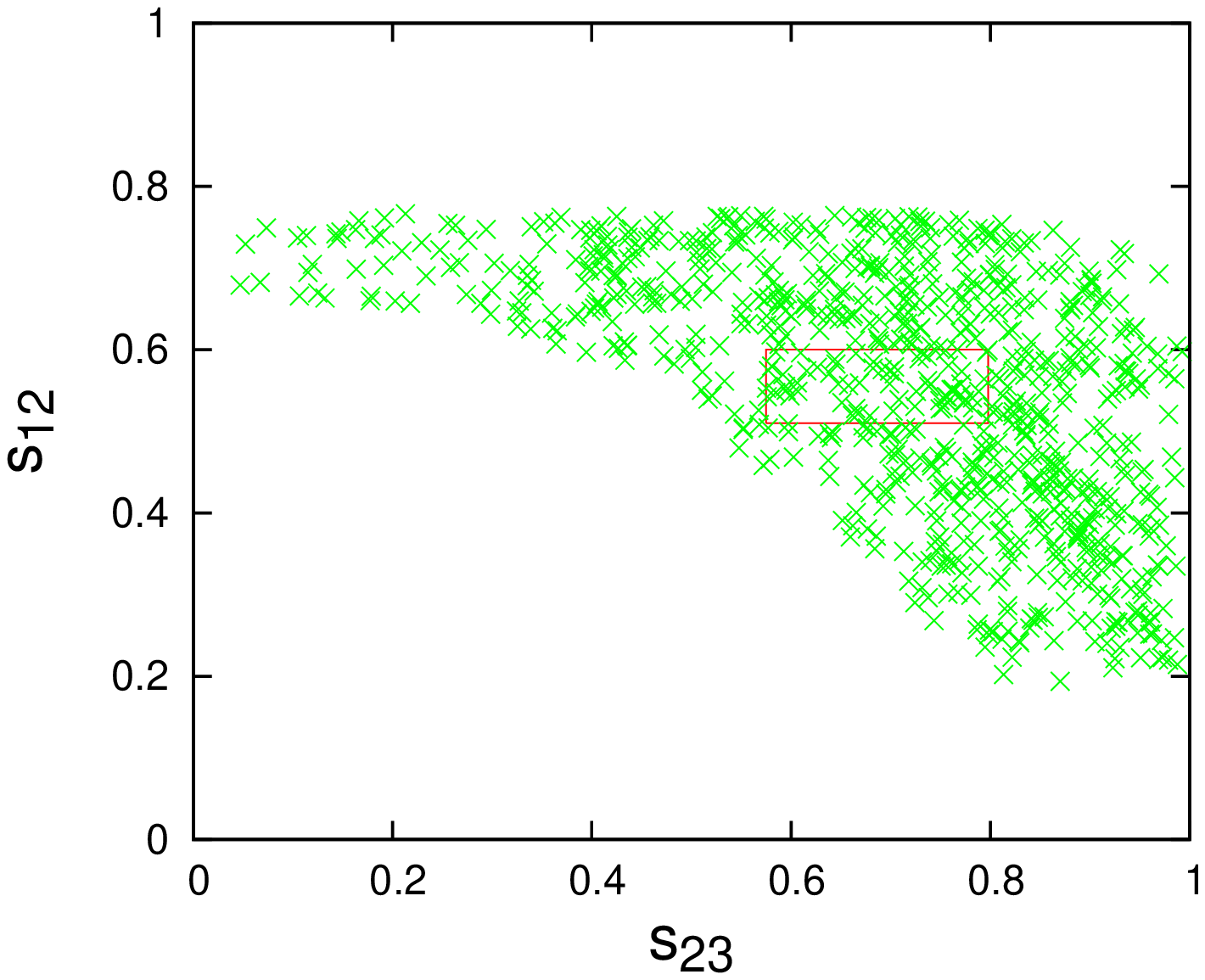}
  \includegraphics[width=0.2\paperwidth,height=0.2\paperheight,angle=-90]{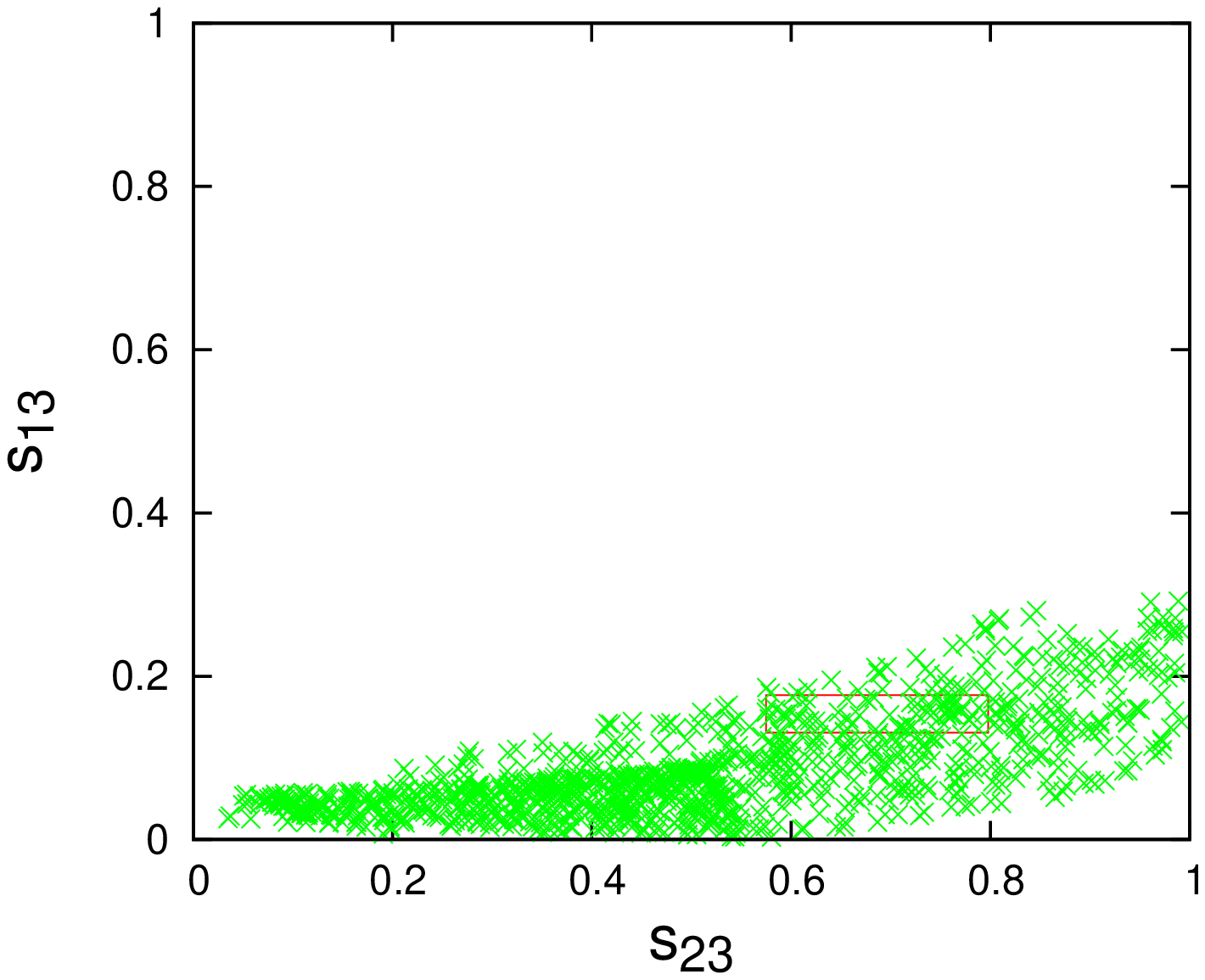}
\end{tabular}
\caption{Plots showing the parameter space for any two mixing
angles when the third angle is constrained by its  $3 \sigma$
range in the $D_l = 0$ and $D_\nu \neq 0$ scenario for Class I
ansatz of texture five zero  Dirac mass matrices (normal
hierarchy).} \label{t5cl1nh1}
\end{figure}
\begin{figure}
\begin{tabular}{cc}
  \includegraphics[width=0.2\paperwidth,height=0.2\paperheight,angle=-90]{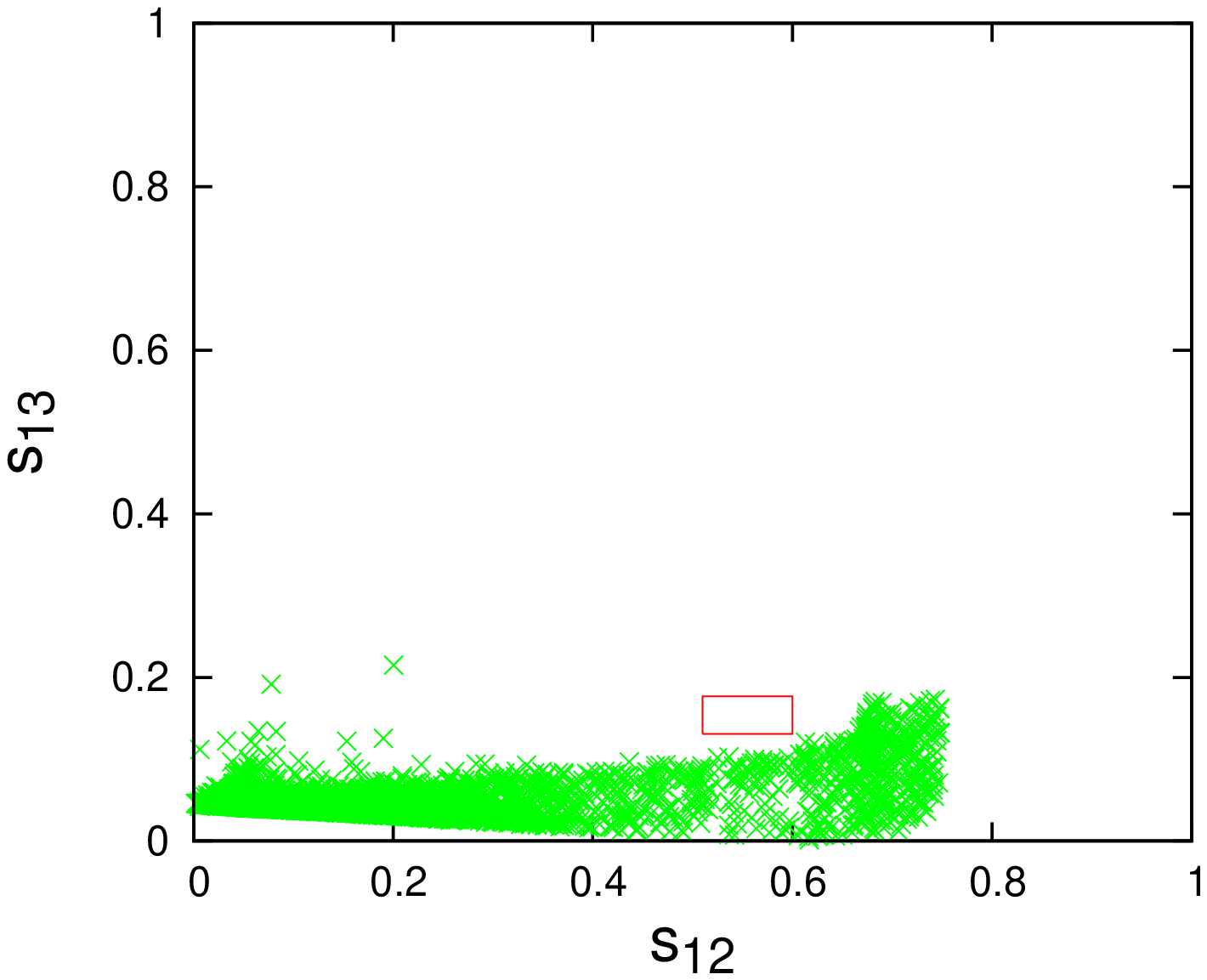}
  \includegraphics[width=0.2\paperwidth,height=0.2\paperheight,angle=-90]{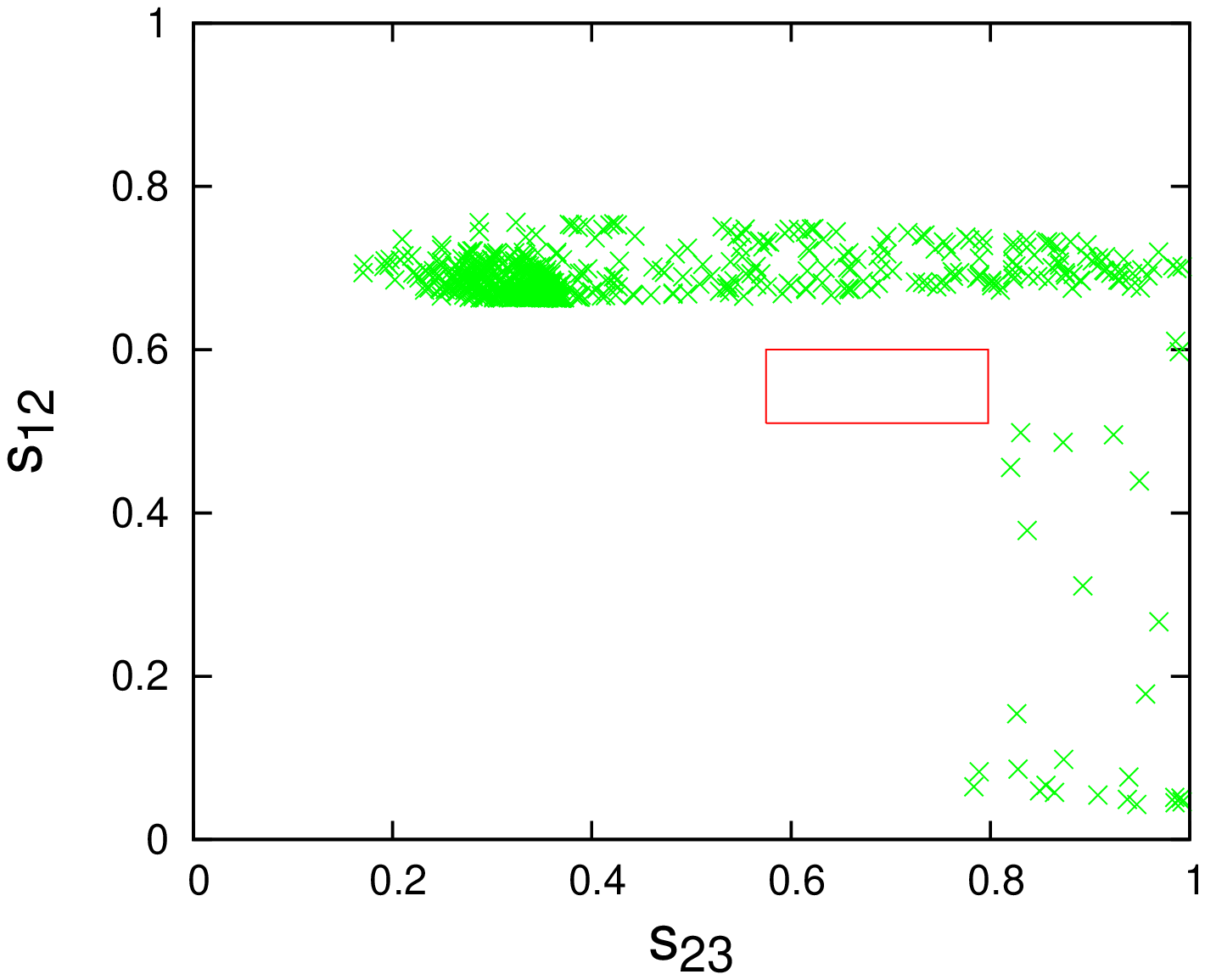}
  \includegraphics[width=0.2\paperwidth,height=0.2\paperheight,angle=-90]{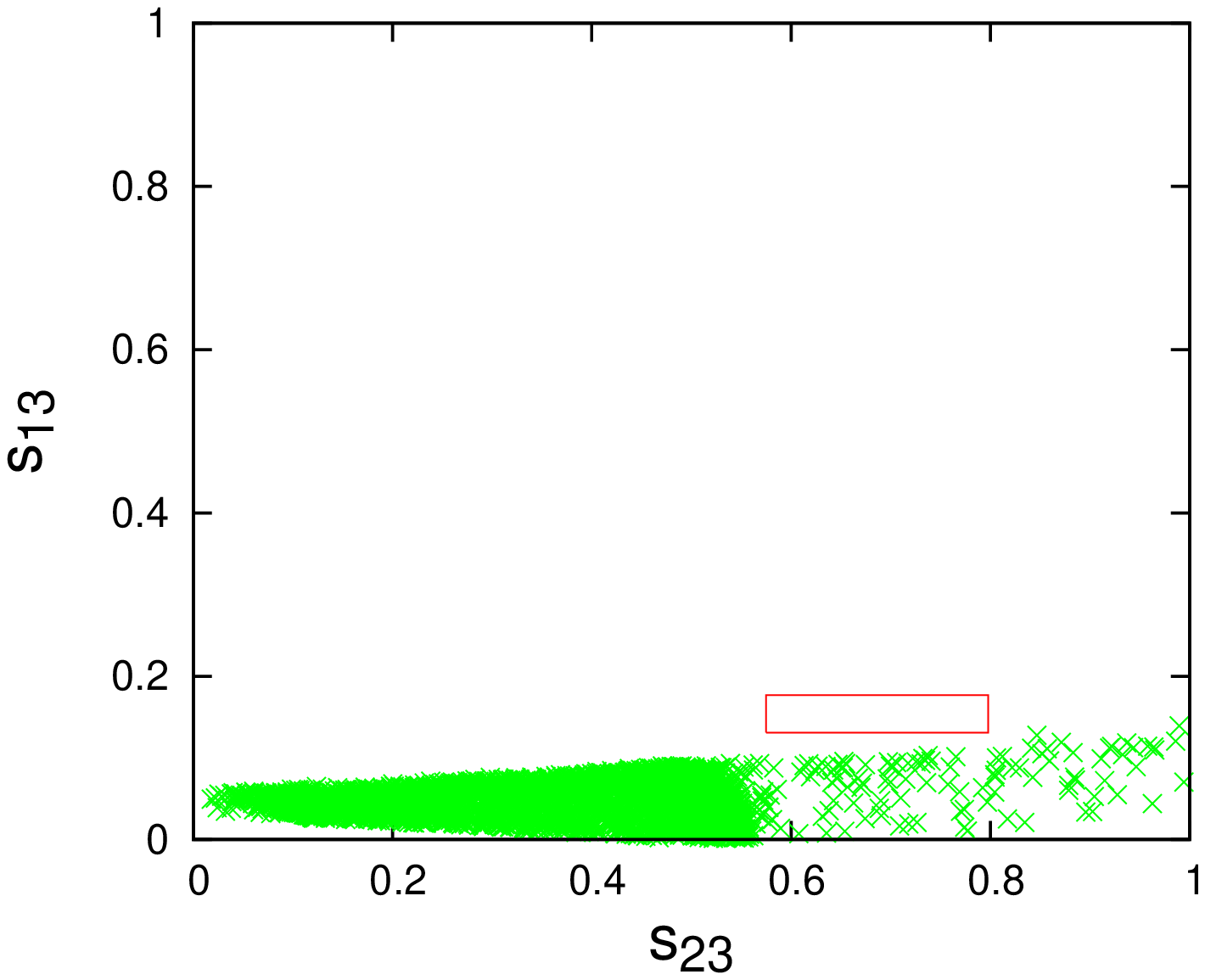}
\end{tabular}
\caption{Plots showing the parameter space for any two mixing
angles when the third angle is constrained by its  $1 \sigma$
range  in the $D_l\neq 0$ and $D_\nu = 0$ scenario for Class I
ansatz of texture five zero  Dirac mass matrices (normal
hierarchy).} \label{t5cl1nh2}
\end{figure}

\begin{figure}
\begin{tabular}{cc}
  \includegraphics[width=0.2\paperwidth,height=0.2\paperheight,angle=-90]{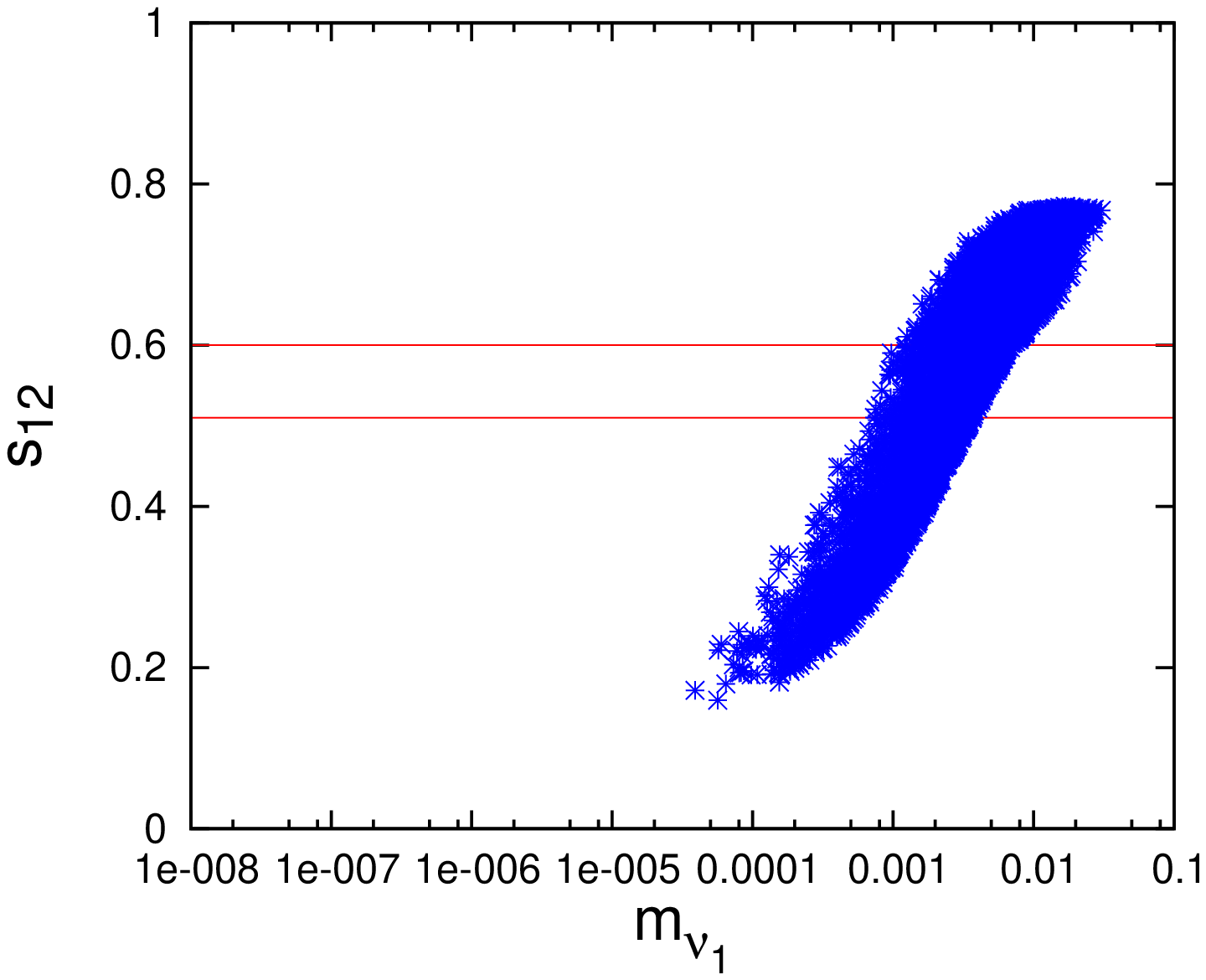}
  \includegraphics[width=0.2\paperwidth,height=0.2\paperheight,angle=-90]{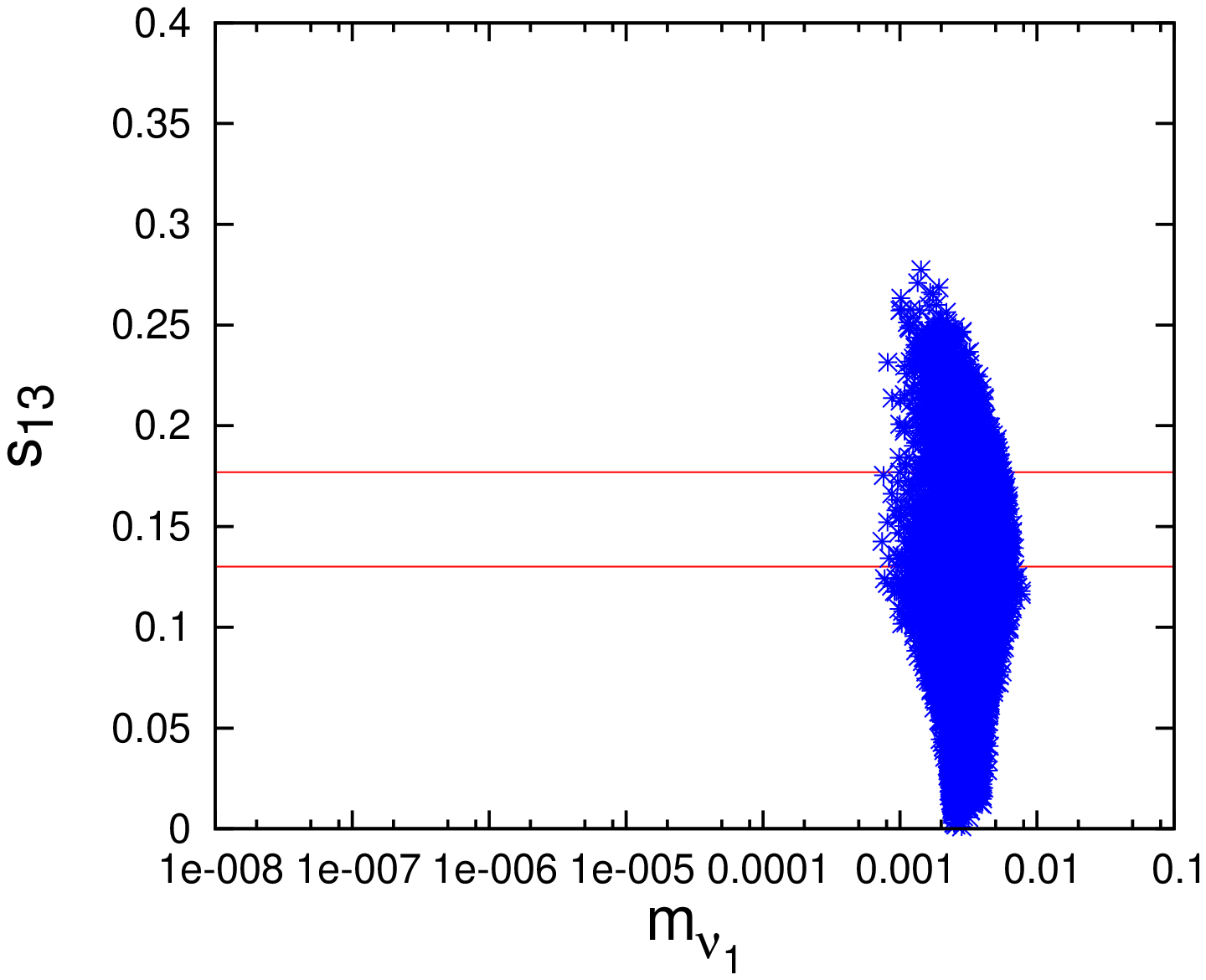}
  \includegraphics[width=0.2\paperwidth,height=0.2\paperheight,angle=-90]{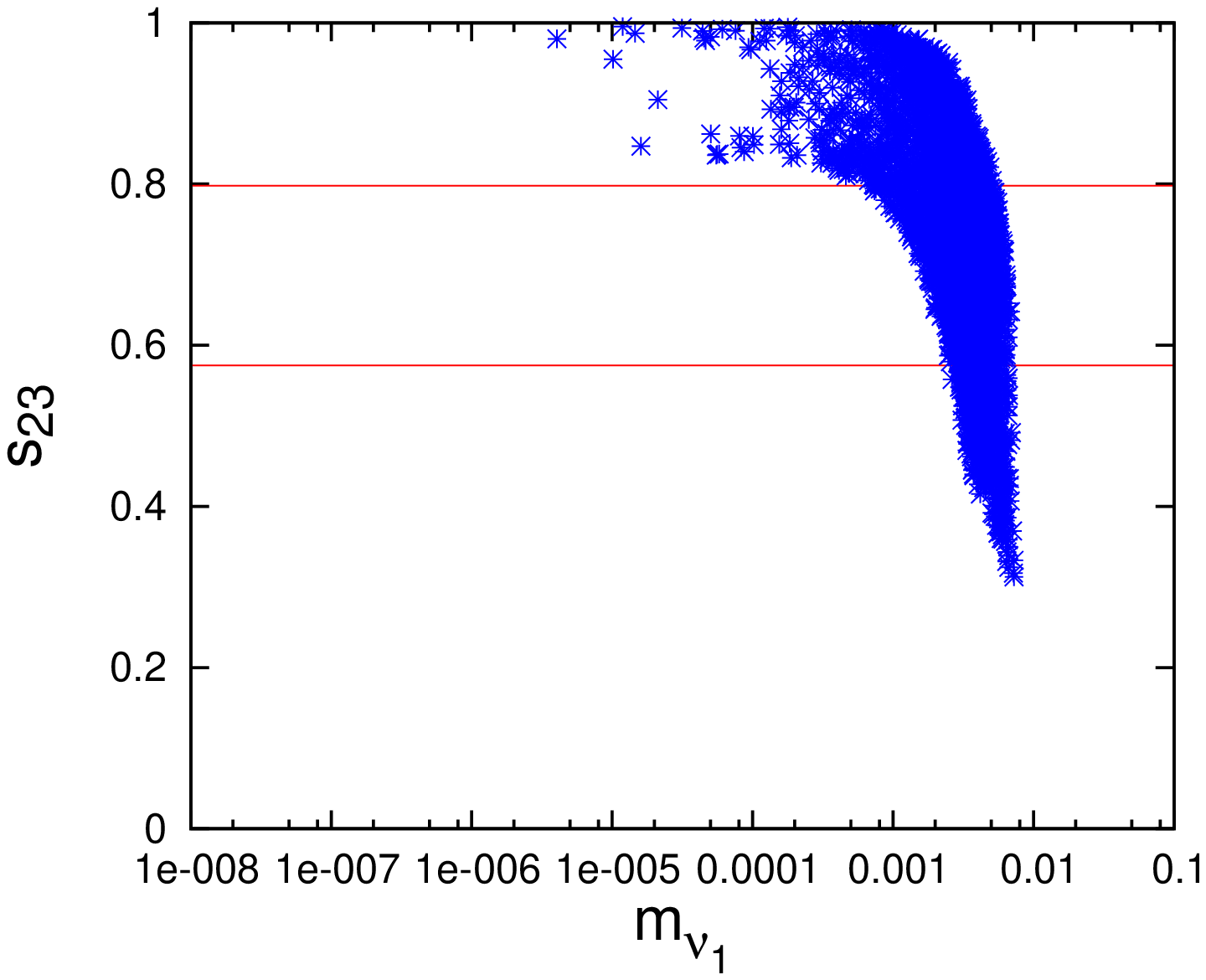}
\end{tabular}
\caption{Plots showing the lightest neutrino mass with mixing
angles when the other two angles are constrained by their $3
\sigma$ ranges   $D_l= 0$ and $D_\nu \neq 0$ scenario for Class I
ansatz of texture five zero  Dirac mass matrices (normal
hierarchy).} \label{cl1t5nh2}
\end{figure}
\begin{figure}
\begin{tabular}{cc}
  \includegraphics[width=0.2\paperwidth,height=0.2\paperheight,angle=-90]{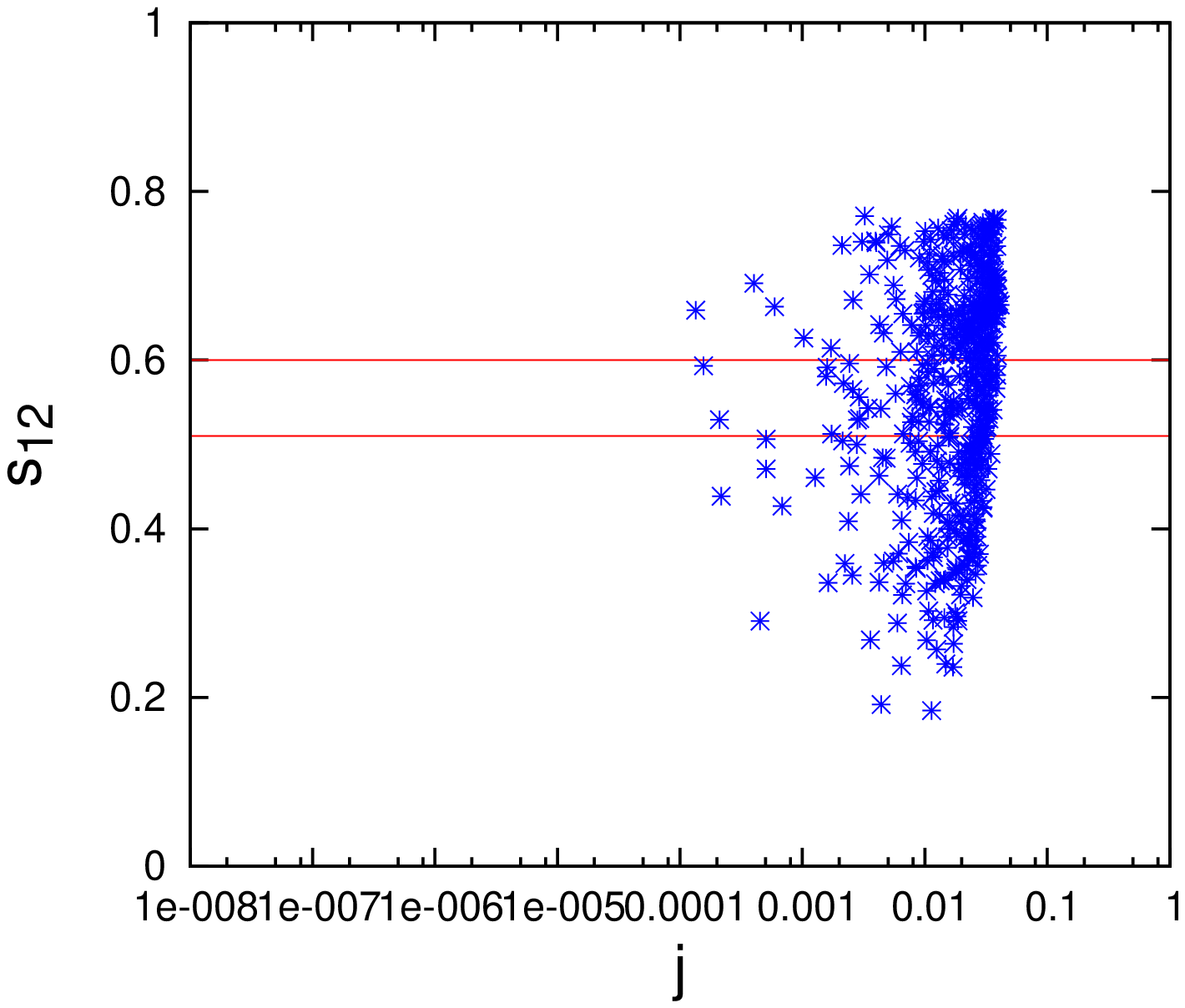}
  \includegraphics[width=0.2\paperwidth,height=0.2\paperheight,angle=-90]{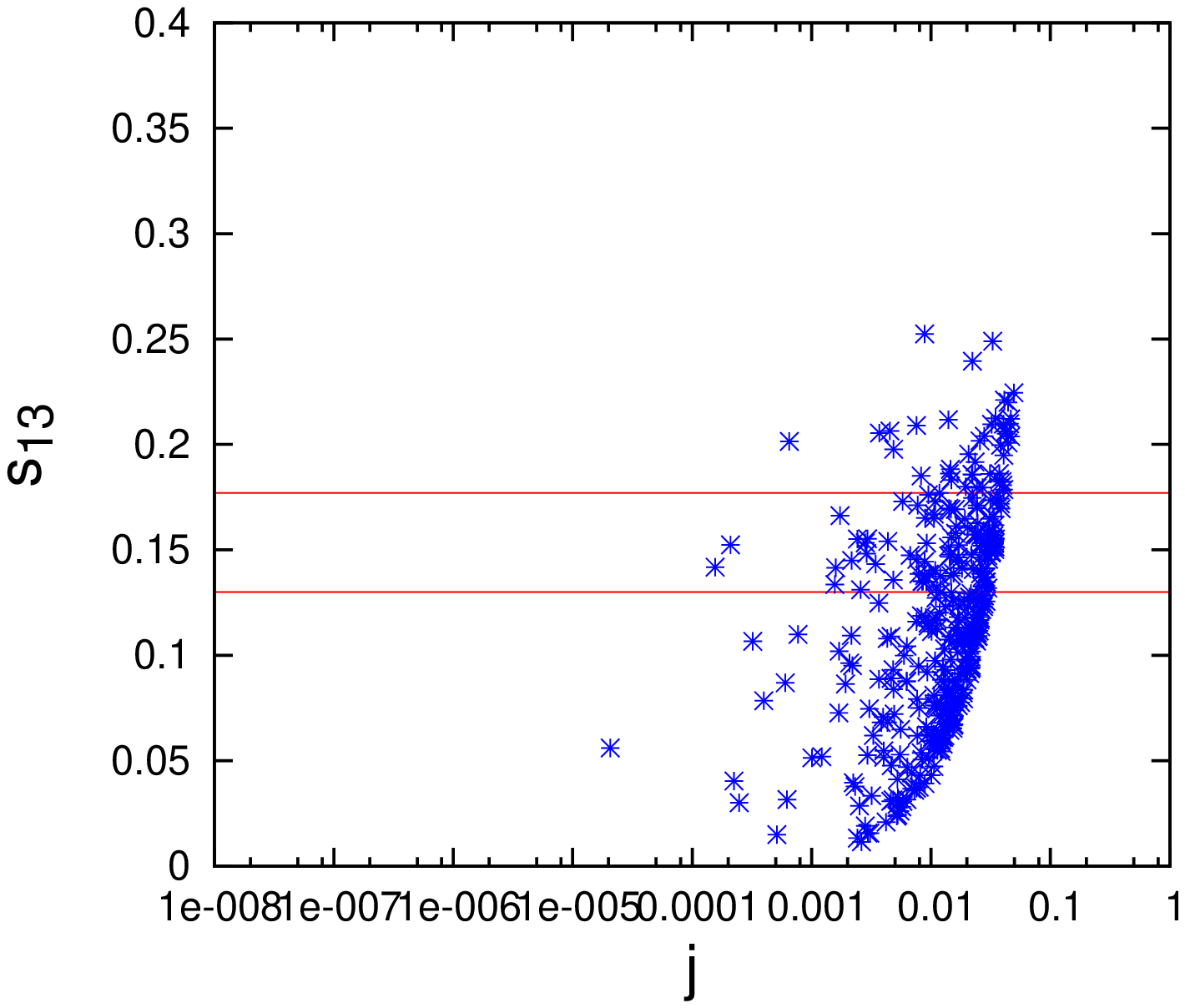}
  \includegraphics[width=0.2\paperwidth,height=0.2\paperheight,angle=-90]{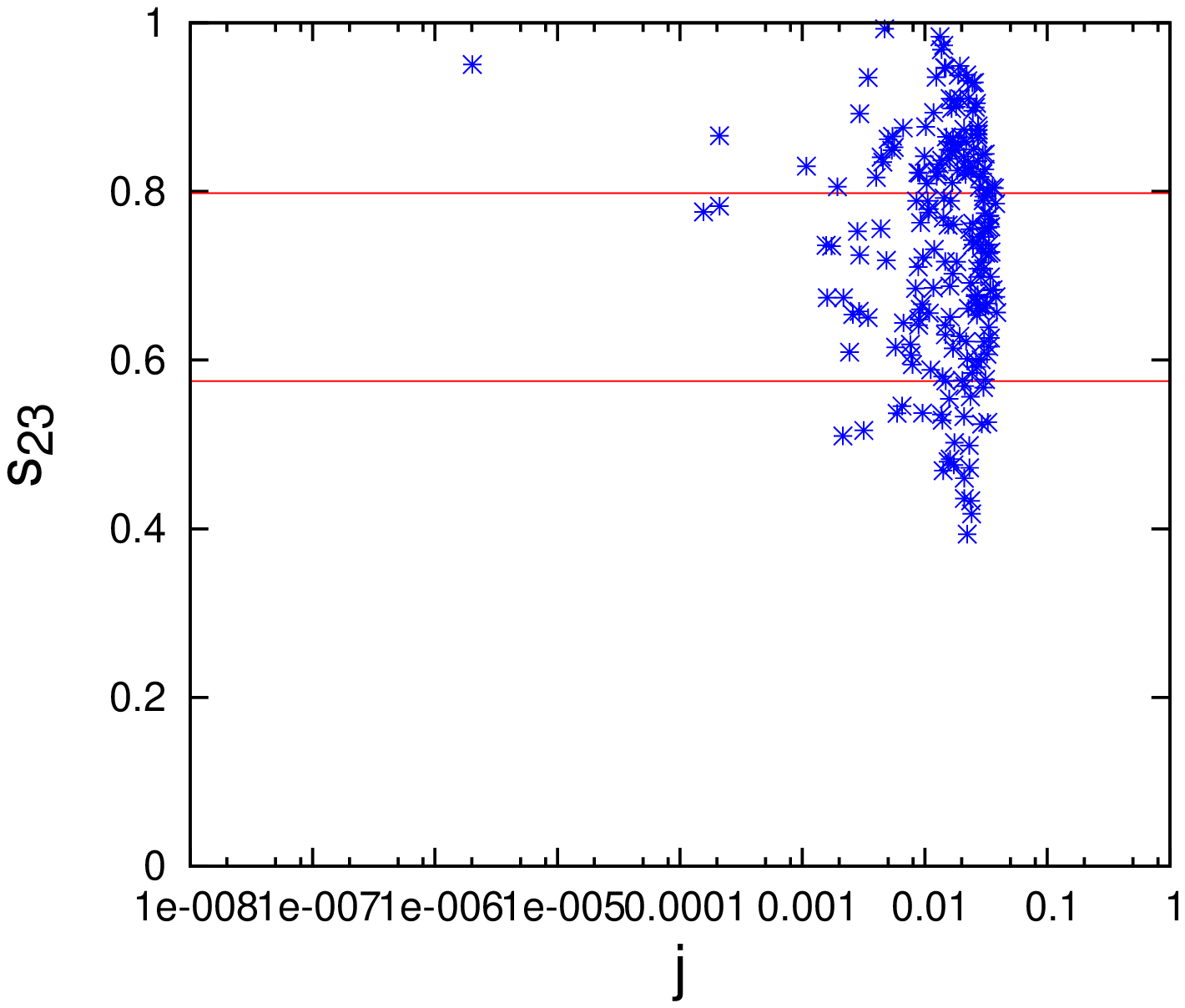}
\end{tabular}
\caption{Plots showing the variation of Jarlskog CP violating
parameter with mixing angles when the other two angles are
constrained by their $3 \sigma$ ranges  $D_l= 0$ and $D_\nu \neq
0$ scenario for Class I ansatz of texture five zero  Dirac mass
matrices (normal hierarchy).} \label{cl1t5nh3}
\end{figure}
After ruling out the structures (\ref{cl1t51}) and (\ref{cl1t52}) for the inverted hierarchy, we now proceed to
examine the compatibility of these matrices for the normal hierarchy case. To this
end, in figures (\ref{t5cl1nh1}) and (\ref{t5cl1nh2}), we present the plots showing the parameter space corresponding
to any two mixing angles wherein the third one is constrained by its $3\sigma$ range. Interestingly, normal
hierarchy seems to be ruled out for the case $D_\nu=0$ and $D_l \neq 0$, whereas
for the case $D_\nu \neq 0$ and $D_l=0$ it seems to be viable as can be seen from the
figure (\ref{t5cl1nh1}), wherein the parameter space shows significant overlap with the experimentally
allowed $3\sigma$ region shown by the rectangular boxes in each plot.
For the $D_l =0$ and $D_\nu\neq 0$ case for texture five zero mass matrices, wherein normal hierarchy has been
shown to be viable, we proceed to study the dependence of the lightest neutrino mass and Jarlskog's parameter on the the
leptonic mixing angles. To this end,  we present the plots showing variation of
the lightest neutrino mass and Jarlskog's parameter with the mixing angles in figures (\ref{cl1t5nh2}) and (\ref{cl1t5nh3})
respectively. While plotting these graphs, the other two mixing angles have been constrained by their $3\sigma$ ranges.
The parallel lines in these plots show the
$3\sigma$ experimental ranges for the mixing angle being considered.
Having a careful look at these plots, one can find the ranges for the lightest neutrino mass and the Jarlskog's
parameter for this case of texture five zero matrices, viz.,
$0.001 eV\lesssim m_{\nu1} \lesssim 0.01eV$, ~$0.00001 \lesssim j \lesssim 0.05 $.
Further, since inverted hierarchy is ruled out for both the cases for texture five zero and
for the normal hierarchy the range of the lightest neutrino mass does not include that for the degenerate
neutrino mass ordering, therefore both the cases for degenerate scenario seems to be ruled out for this class.

\subsubsection{Class II ansatz}
The two possibilities for texture five zero lepton mass matrices for this class can be given as,
\be
 M_{l}=\left( \ba{ccc}
0 & A _{l} & 0    \\
A_{l}^{*} & 0 &  B_l    \\
 0 &   B_{l}^{*}     &  E_{l} \ea \right), \qquad
 M_{\nu}=\left( \ba{ccc}
D_\nu & A _{\nu} & 0    \\
A_{\nu}^{*} & 0 &  B_{\nu}    \\
 0  &  B_{\nu}^{*}    &  E_{\nu} \ea \right),
\label{cl2t51}\ee
or
\be
  M_{l}=\left( \ba{ccc}
D_l & A _{l} & 0   \\
A_{l}^{*} & 0 &   B_{l}     \\
 0 &  B_{l}^{*}     &  E_{l} \ea \right), \qquad
 M_{\nu}=\left( \ba{ccc}
0 & A _{\nu} & 0   \\
A_{\nu}^{*} & 0 &  B_{\nu}      \\
 0  &  B_{\nu}^{*}    &  E_{\nu} \ea \right),
\label{cl2t52}\ee

\begin{figure}
\begin{tabular}{cc}
  \includegraphics[width=0.2\paperwidth,height=0.2\paperheight,angle=-90]{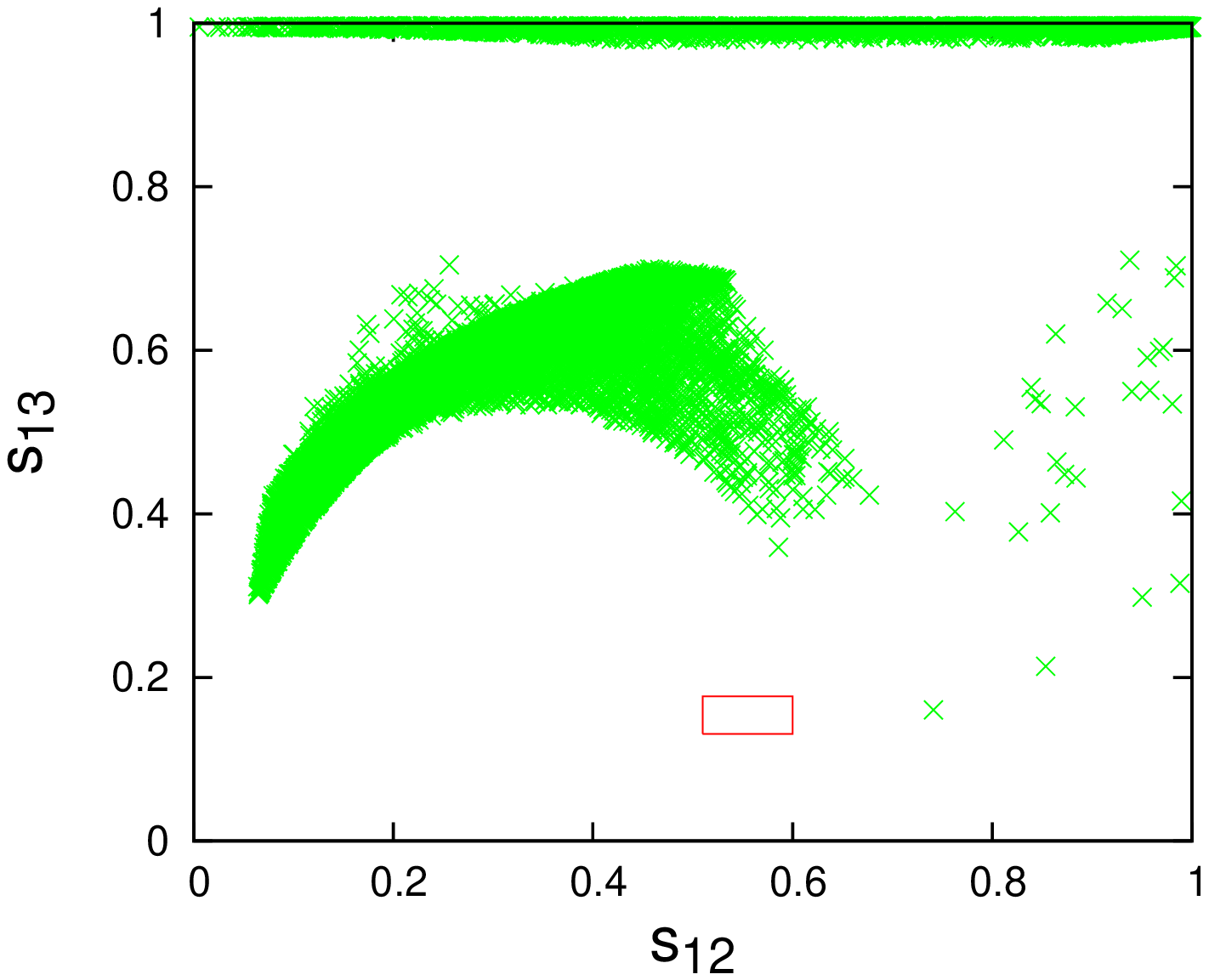}
  \includegraphics[width=0.2\paperwidth,height=0.2\paperheight,angle=-90]{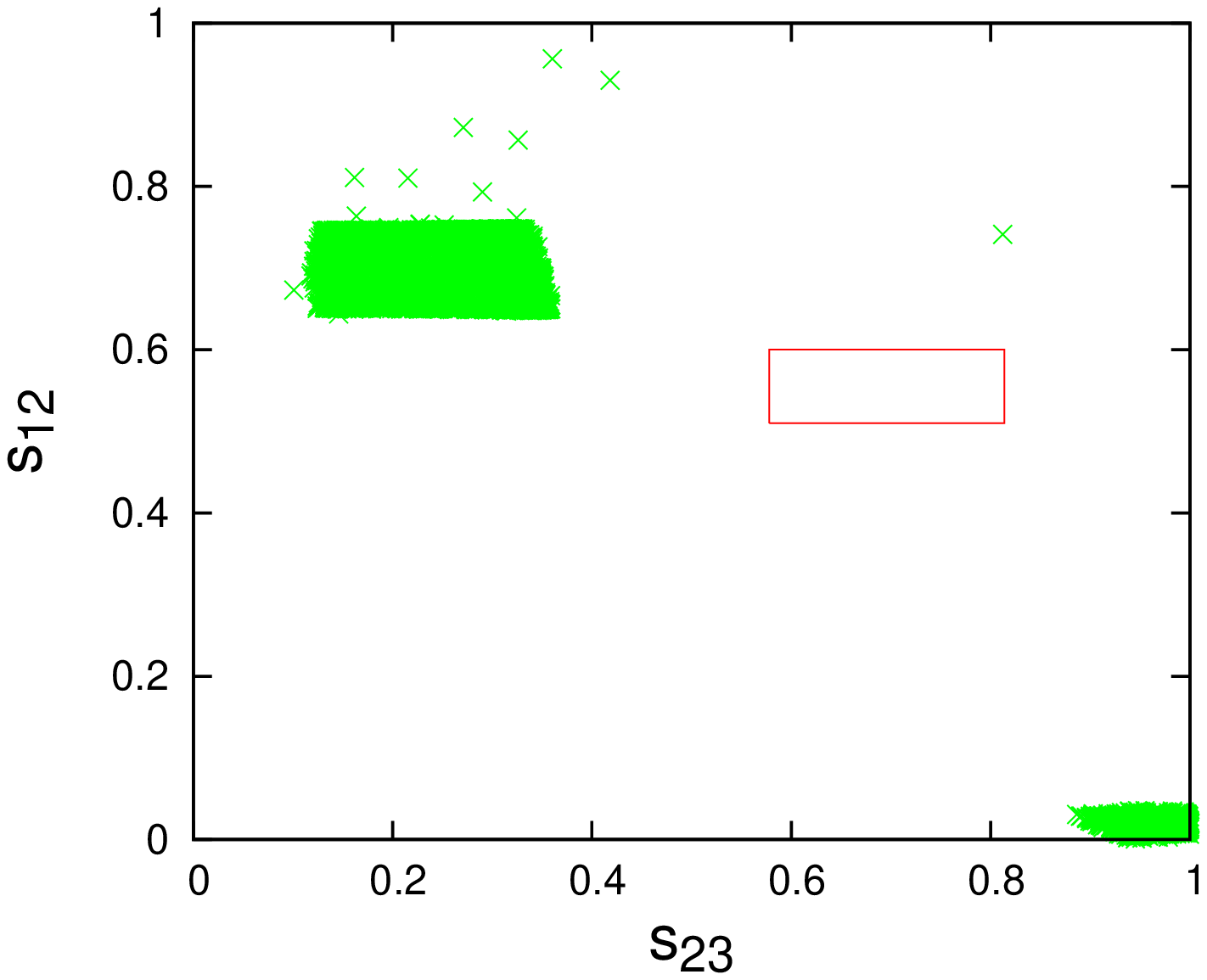}
  \includegraphics[width=0.2\paperwidth,height=0.2\paperheight,angle=-90]{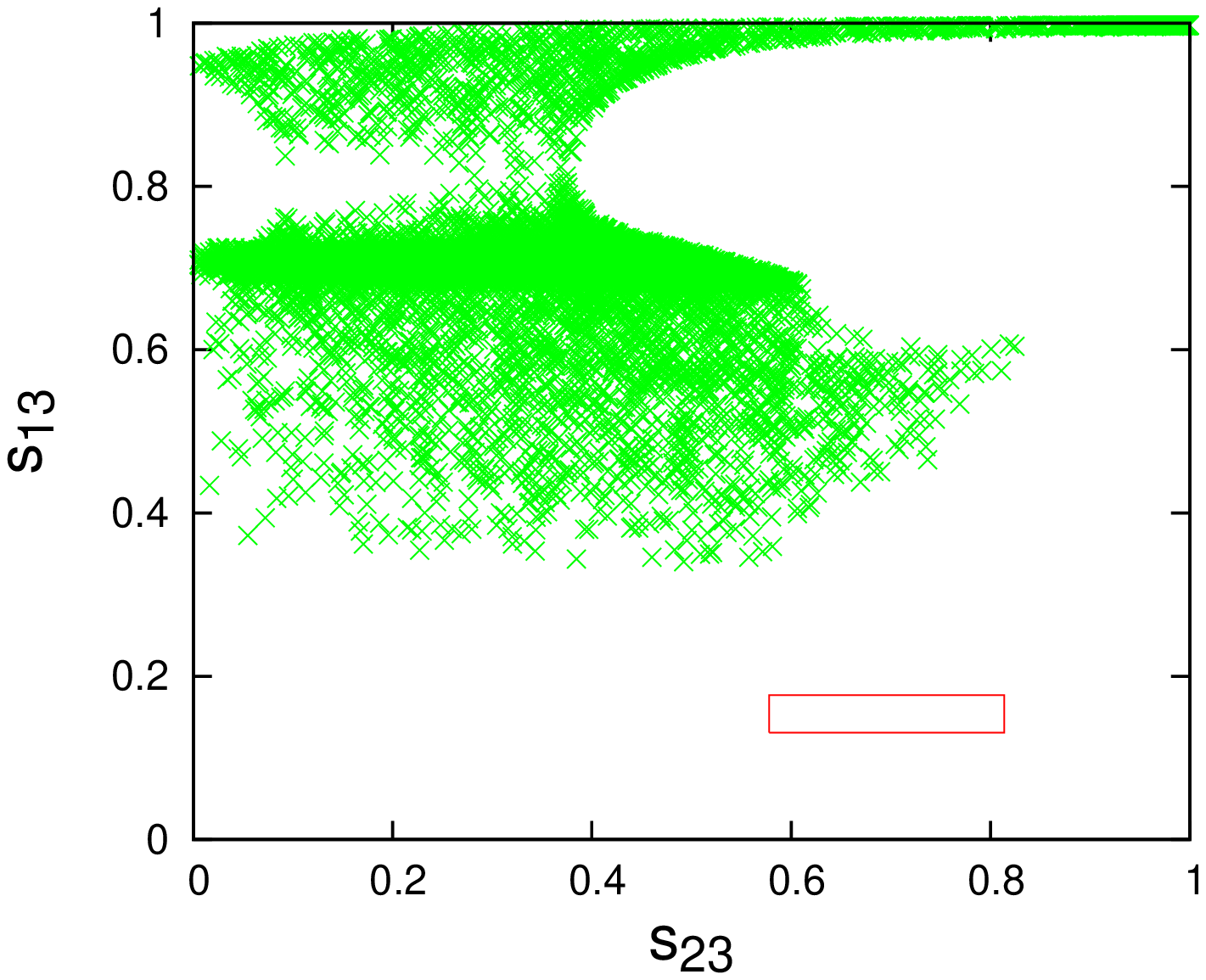}
\end{tabular}
\caption{Plots showing the parameter space for any two mixing
angles when the third angle is constrained by its  $3 \sigma$
range in the $D_l =0$ and $D_\nu\neq 0$ scenario for Class II
ansatz of texture five zero  Dirac mass matrices (inverted
hierarchy).} \label{t5cl2ih1}
\end{figure}

\begin{figure}
\begin{tabular}{cc}
  \includegraphics[width=0.2\paperwidth,height=0.2\paperheight,angle=-90]{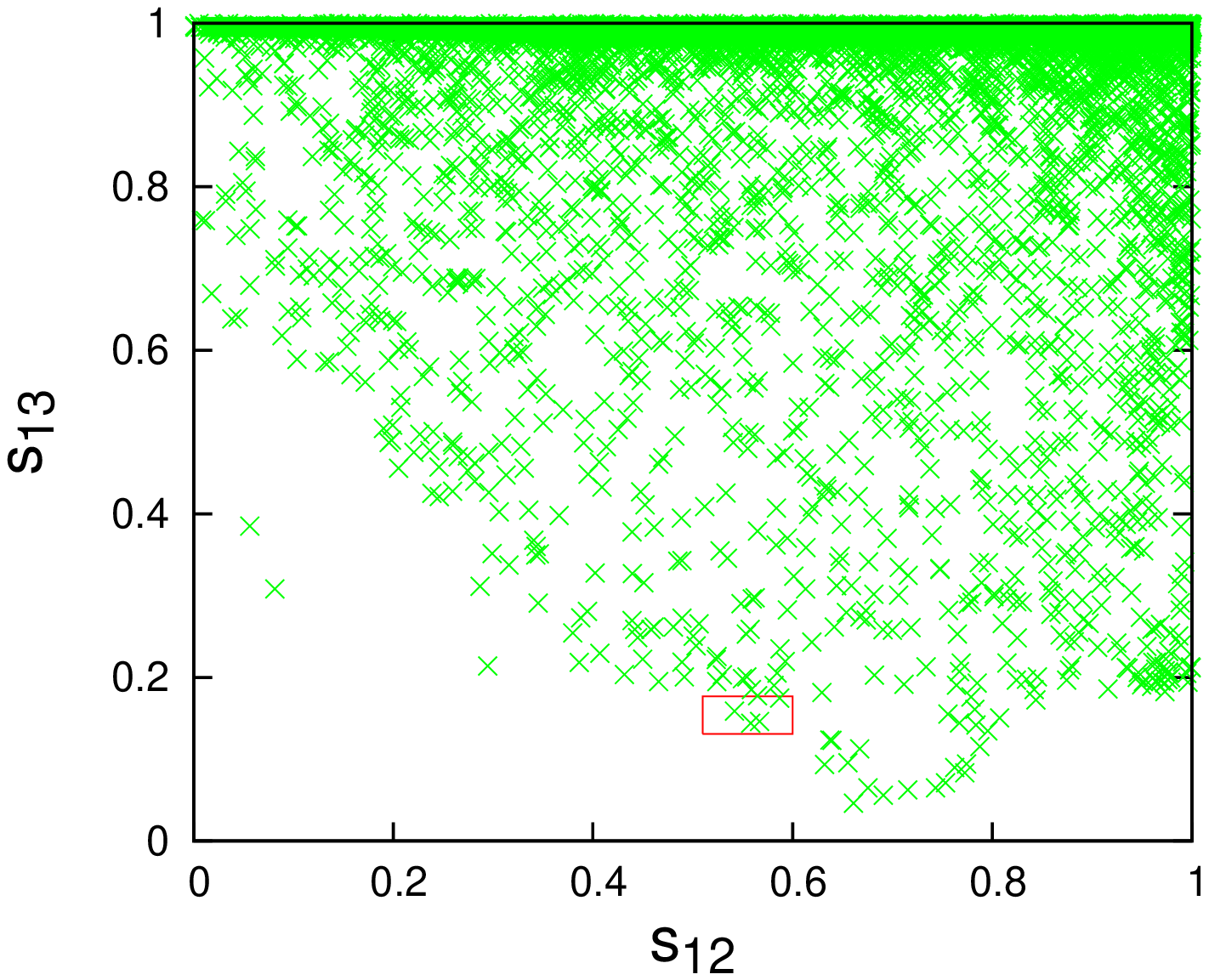}
  \includegraphics[width=0.2\paperwidth,height=0.2\paperheight,angle=-90]{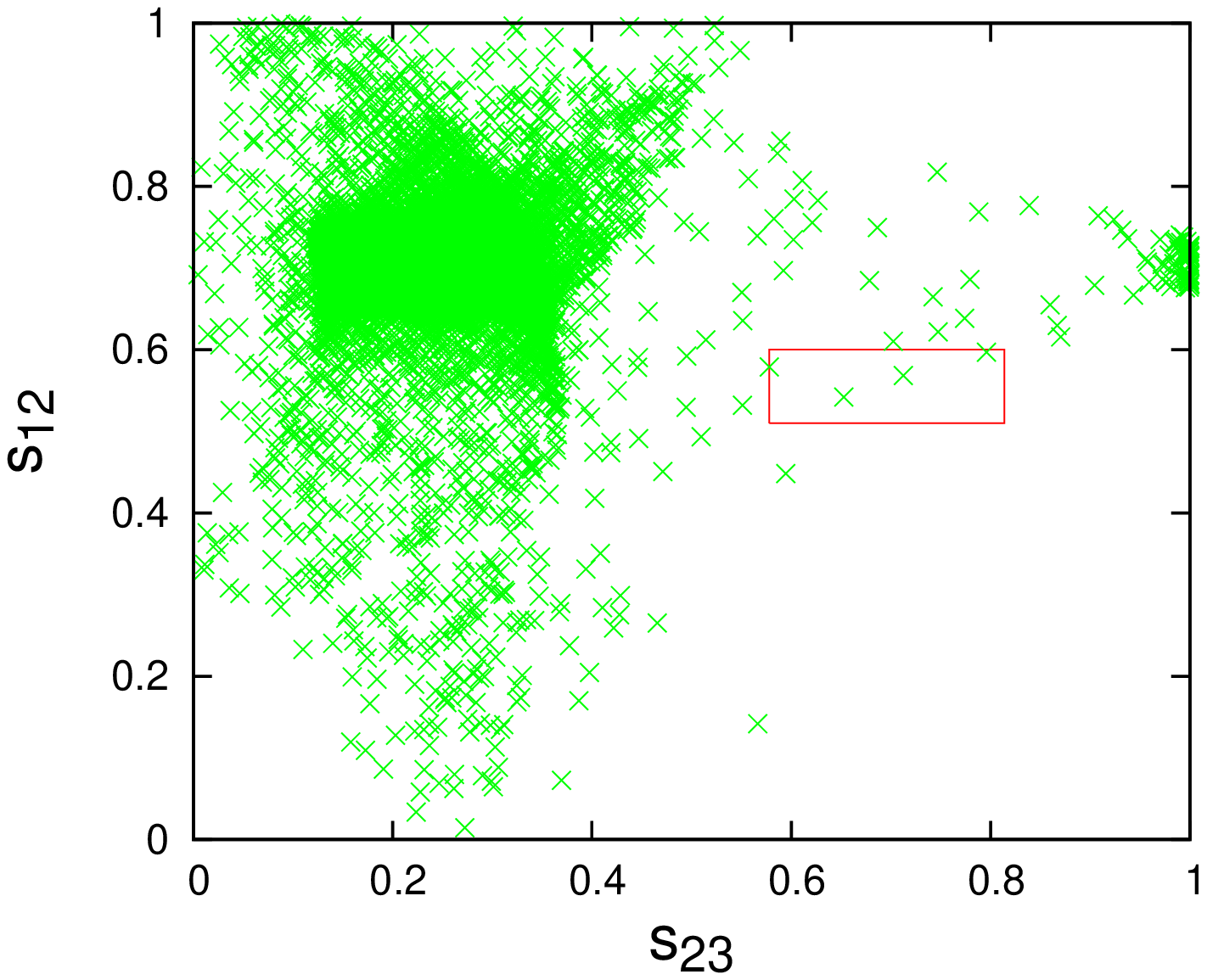}
  \includegraphics[width=0.2\paperwidth,height=0.2\paperheight,angle=-90]{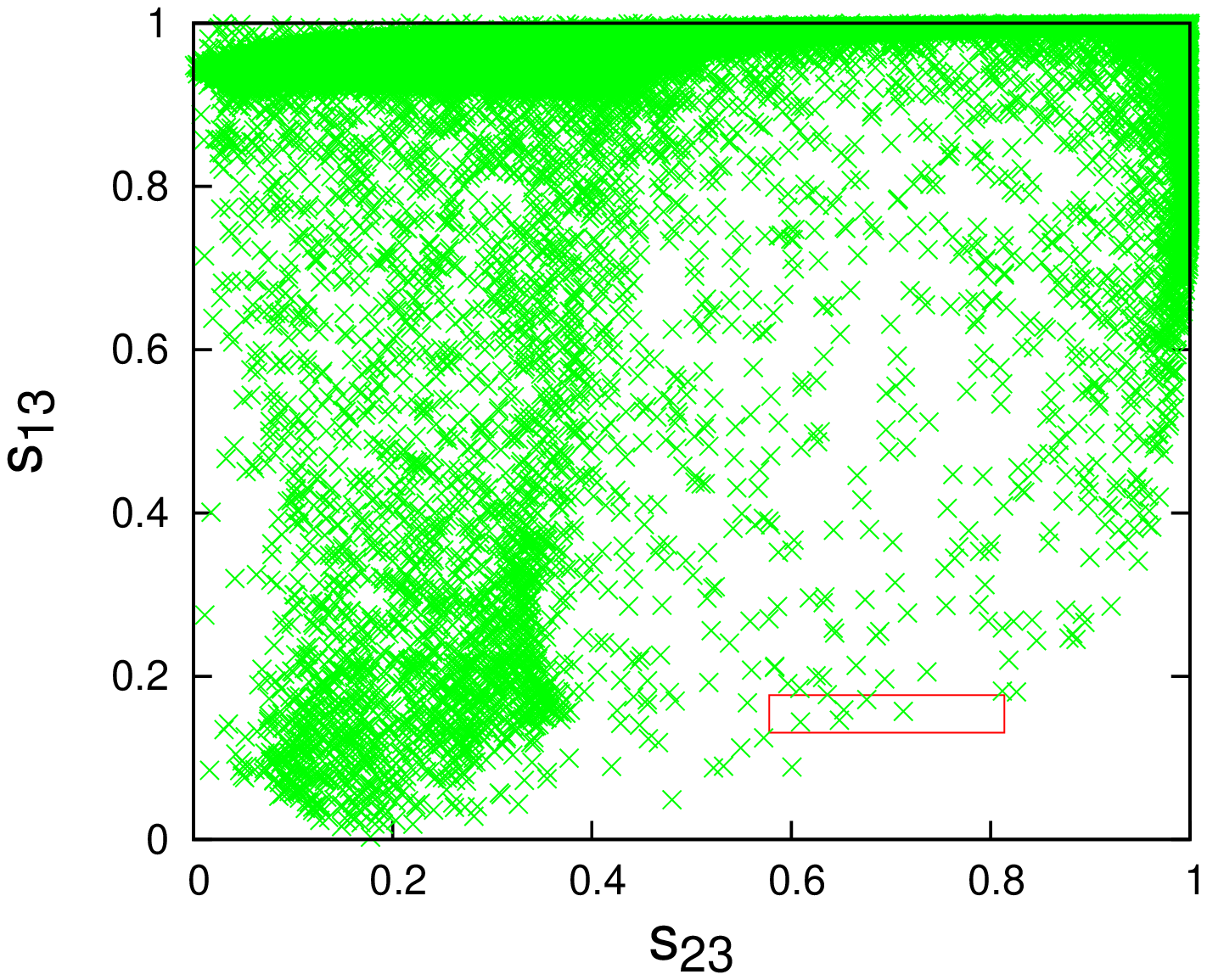}
\end{tabular}
\caption{Plots showing the parameter space for any two mixing
angles when the third angle is constrained by its  $1 \sigma$
range  in the $D_l \neq 0$ and $D_\nu = 0$ scenario for Class II
ansatz of texture five zero  Dirac mass matrices (inverted
hierarchy).} \label{t5cl2ih2}
\end{figure}

 We study both these possibilities in detail for all the neutrino mass orderings. Firstly, we examine
 the compatibility of matrices (\ref{cl2t51}) and (\ref{cl2t52}) with the inverted hierarchy
of neutrino masses.
For this purpose, in figures (\ref{t5cl2ih1}) and
 (\ref{t5cl2ih2}), we present the plots showing the parameter space allowed by this ansatz for any two mixing angles wherein
the third one  is constrained by its $3\sigma$ experimental bound for inverted hierarchy of neutrino masses.
The rectangular regions in these plots represent the $3\sigma$ ranges for the two mixing angles being considered.
Interestingly, one finds that for the case $D_l =0$ and $D_\nu\neq 0$ of texture five zero lepton mass matrices
inverted hierarchy is ruled out, whereas for the case $D_l \neq 0$ and $D_\nu=0$ of texture five zero lepton mass matrices
inverted hierarchy scenario seems to be viable.
\par
\begin{figure}
\begin{tabular}{cc}
  \includegraphics[width=0.2\paperwidth,height=0.2\paperheight,angle=-90]{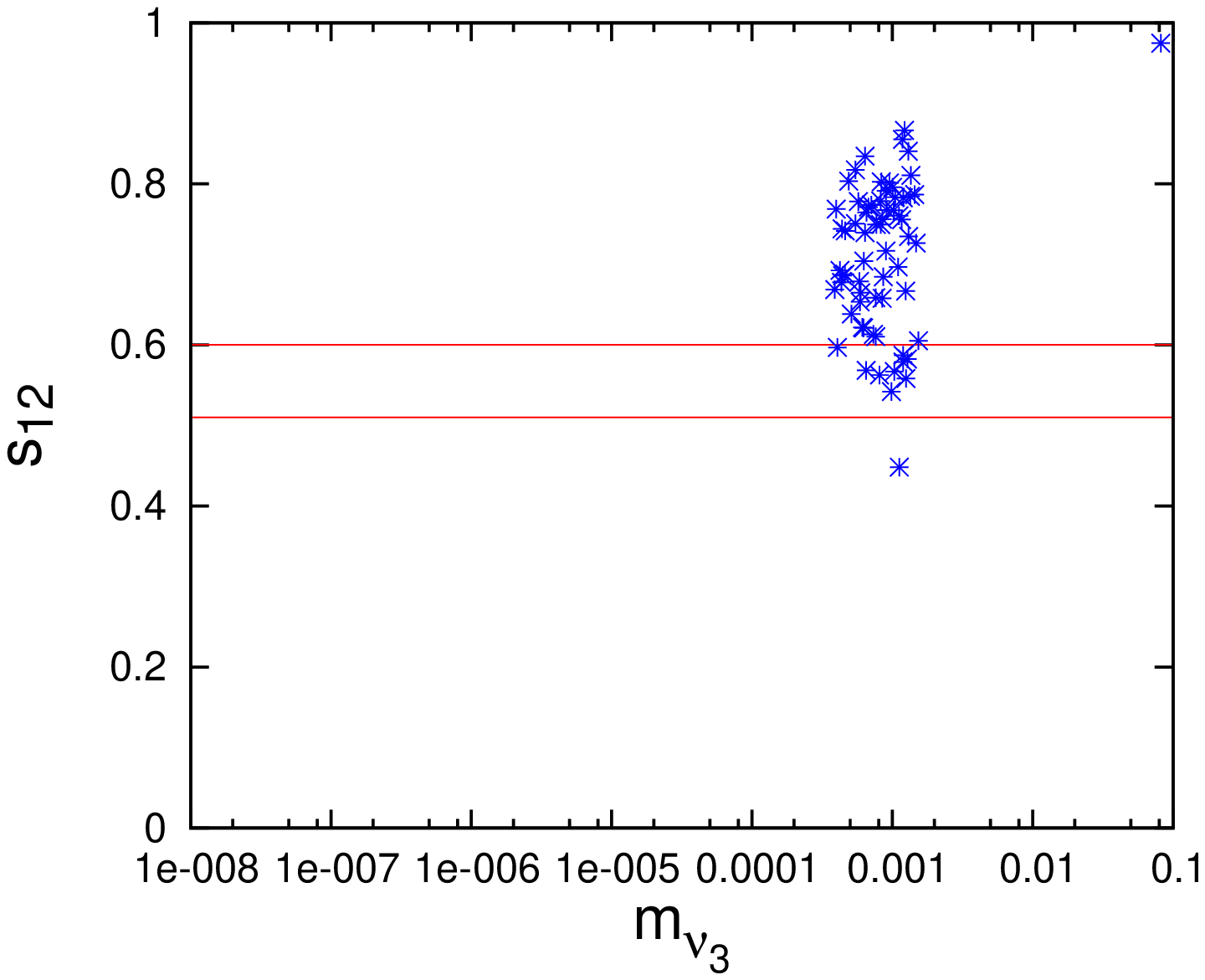}
  \includegraphics[width=0.2\paperwidth,height=0.2\paperheight,angle=-90]{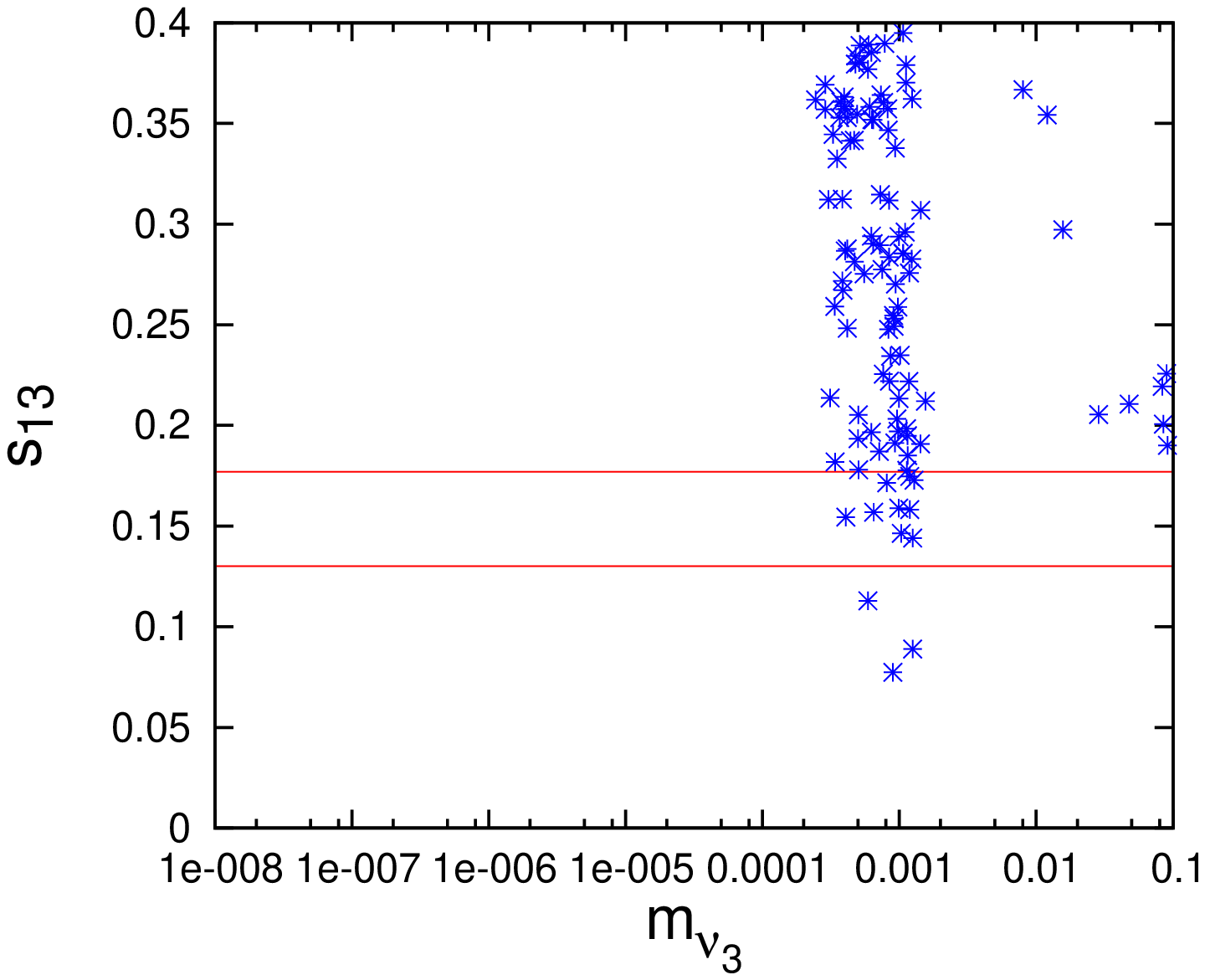}
  \includegraphics[width=0.2\paperwidth,height=0.2\paperheight,angle=-90]{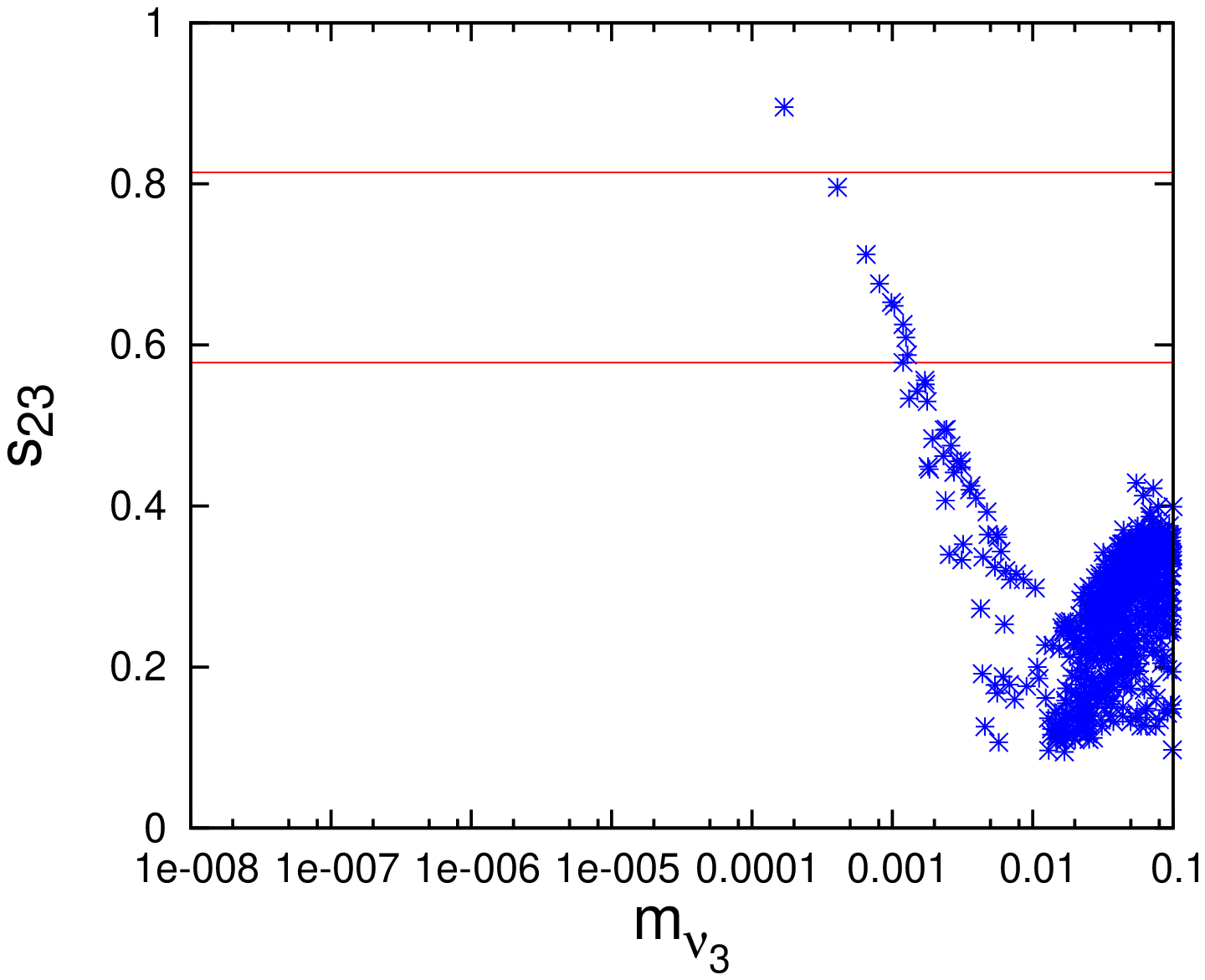}
\end{tabular}
\caption{Plots showing the lightest neutrino mass with mixing
angles when the other two angles are constrained by their $3
\sigma$ ranges  for $D_l\neq 0$ and $D_\nu = 0$ scenario for Class
II ansatz of texture five zero  Dirac mass matrices (inverted
hierarchy).} \label{t5cl2ih3}
\end{figure}

\begin{figure}
\begin{tabular}{cc}
  \includegraphics[width=0.2\paperwidth,height=0.2\paperheight,angle=-90]{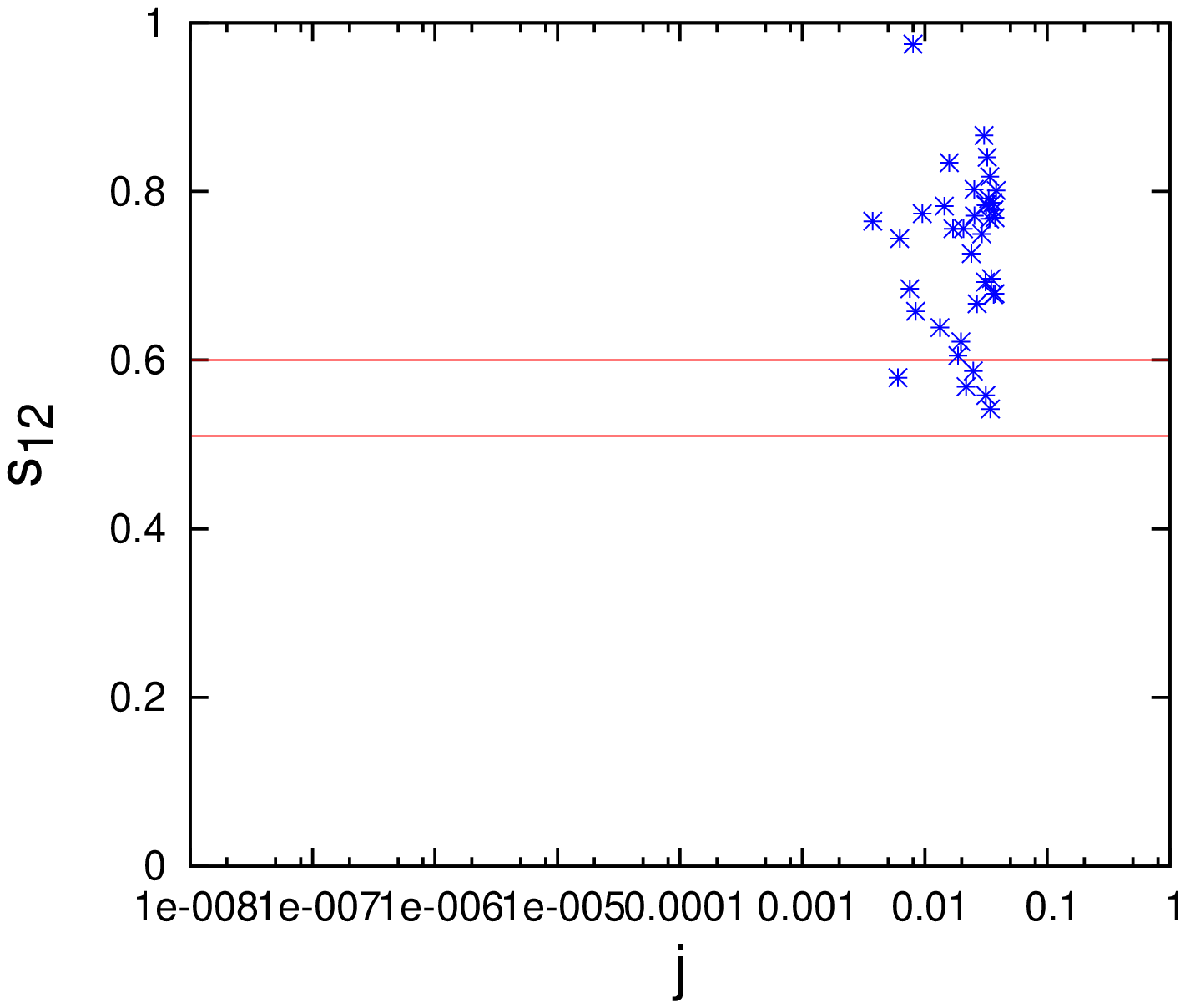}
  \includegraphics[width=0.2\paperwidth,height=0.2\paperheight,angle=-90]{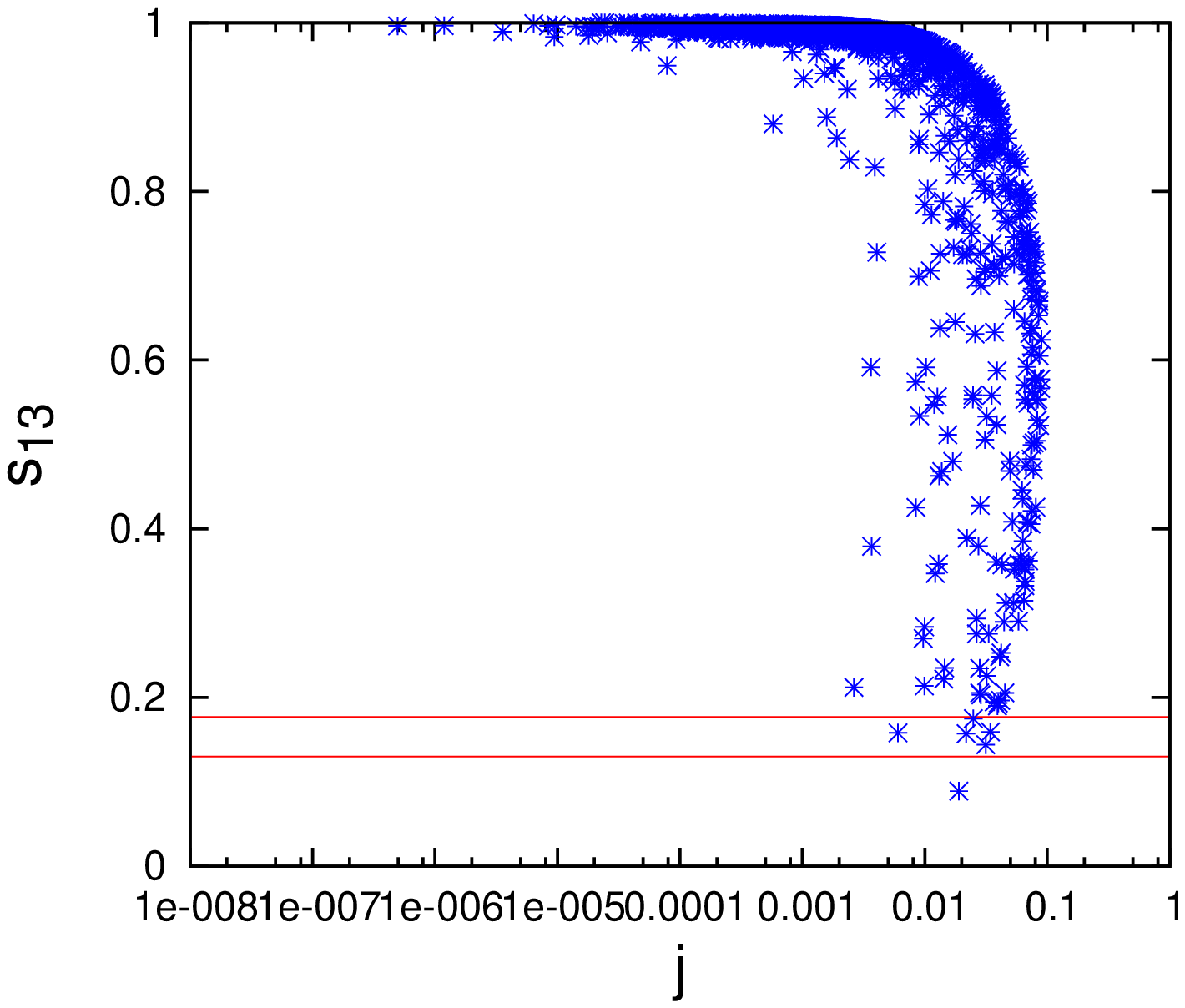}
  \includegraphics[width=0.2\paperwidth,height=0.2\paperheight,angle=-90]{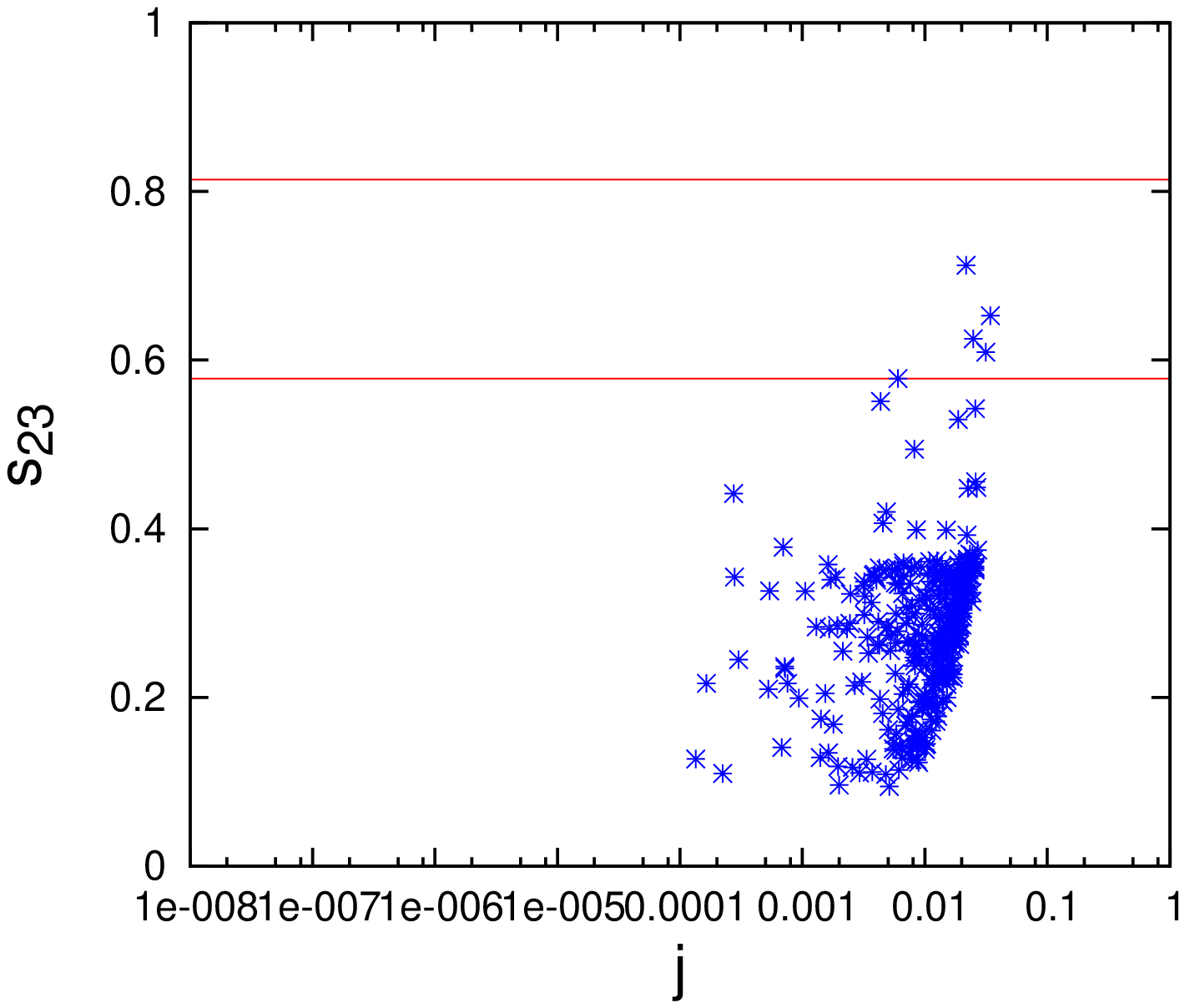}
\end{tabular}
\caption{Plots showing the variation of Jarlskog CP violating
parameter with mixing angles when the other two angles are
constrained by their $3 \sigma$ ranges   for $D_l\neq0$ and
$D_\nu= 0$ scenario for Class II ansatz of texture five zero Dirac
mass matrices (inverted hierarchy).} \label{t5cl2ih4}
\end{figure}

 For the $D_l \neq 0$ and $D_\nu = 0$ case of lepton mass matrices, wherein inverted hierarchy is shown to be viable, we
proceed next to study the the dependence of the lightest neutrino mass and Jarlskog's parameter on the the
leptonic mixing angles. To this end,  we present the plots showing variation of
the lightest neutrino mass and Jarlskog's parameter with the mixing angles in figures (\ref{t5cl2ih3}) and (\ref{t5cl2ih4})
respectively. While plotting these graphs, the other two mixing angles have been constrained by their $3\sigma$ ranges.
Interestingly, for this case one finds a very narrow range for both the lightest neutrino mass as well the Jarlskog's
parameter as can be seen from figures (\ref{t5cl2ih3}) and (\ref{t5cl2ih4}).

\begin{figure}
\begin{tabular}{cc}
  \includegraphics[width=0.2\paperwidth,height=0.2\paperheight,angle=-90]{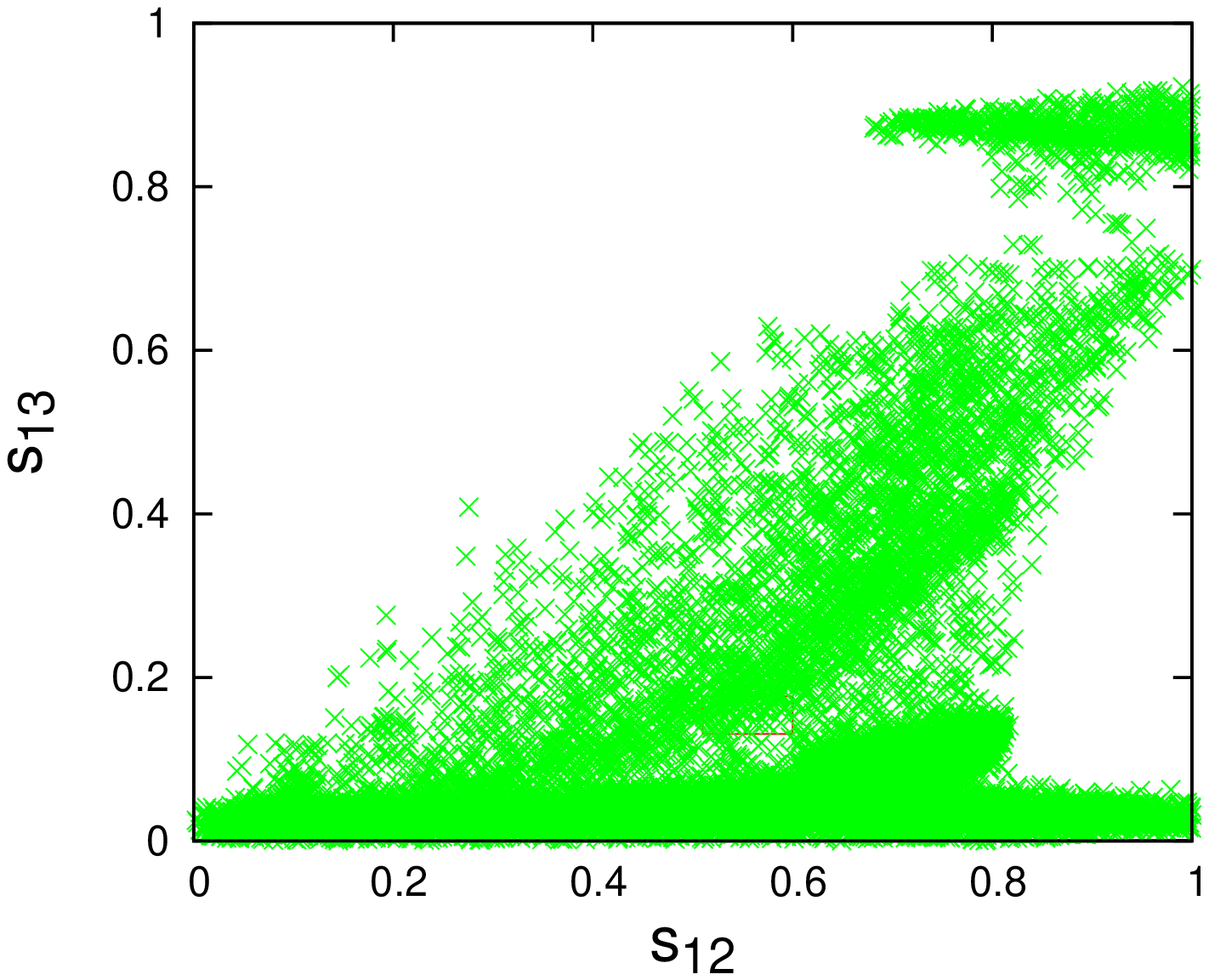}
  \includegraphics[width=0.2\paperwidth,height=0.2\paperheight,angle=-90]{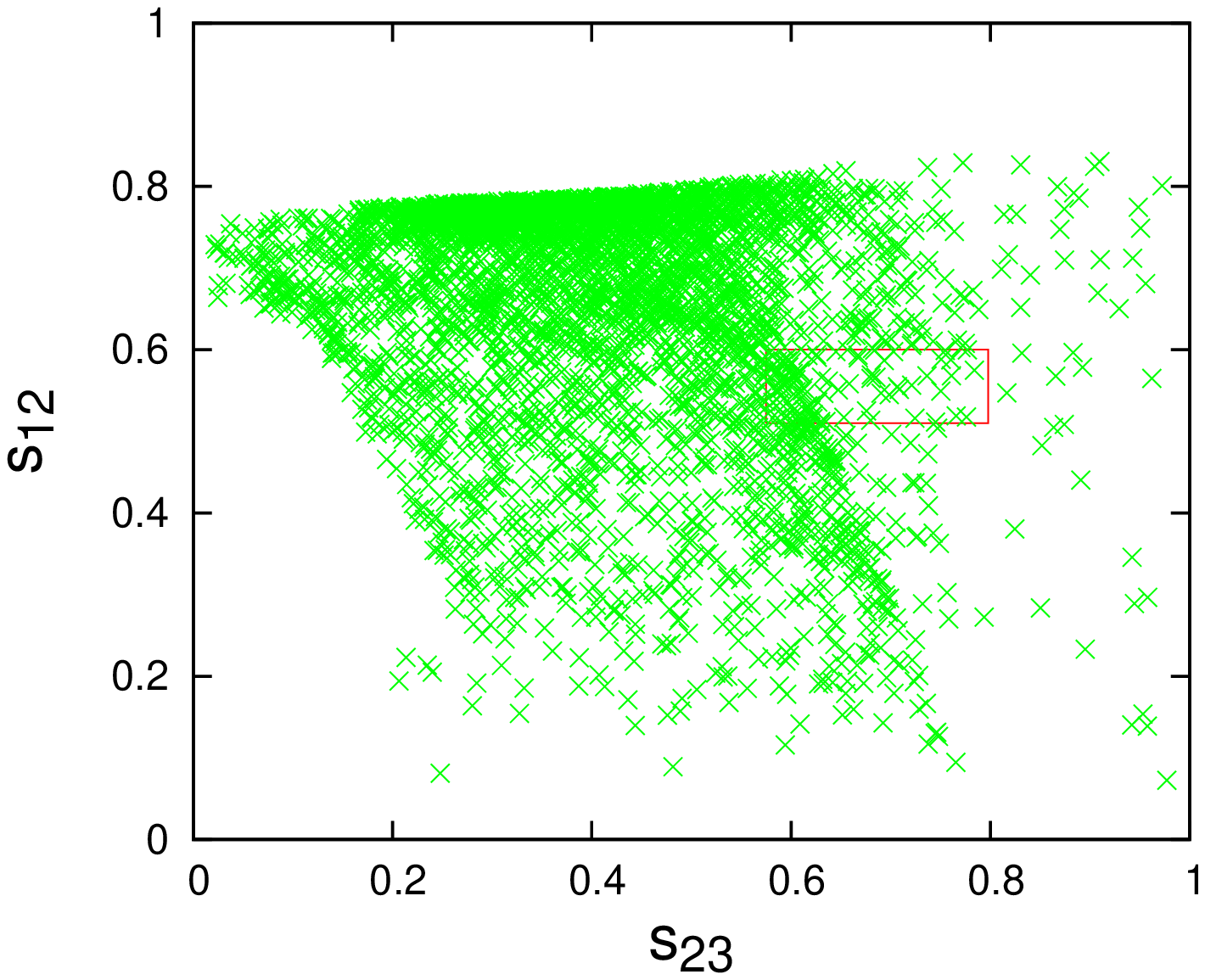}
  \includegraphics[width=0.2\paperwidth,height=0.2\paperheight,angle=-90]{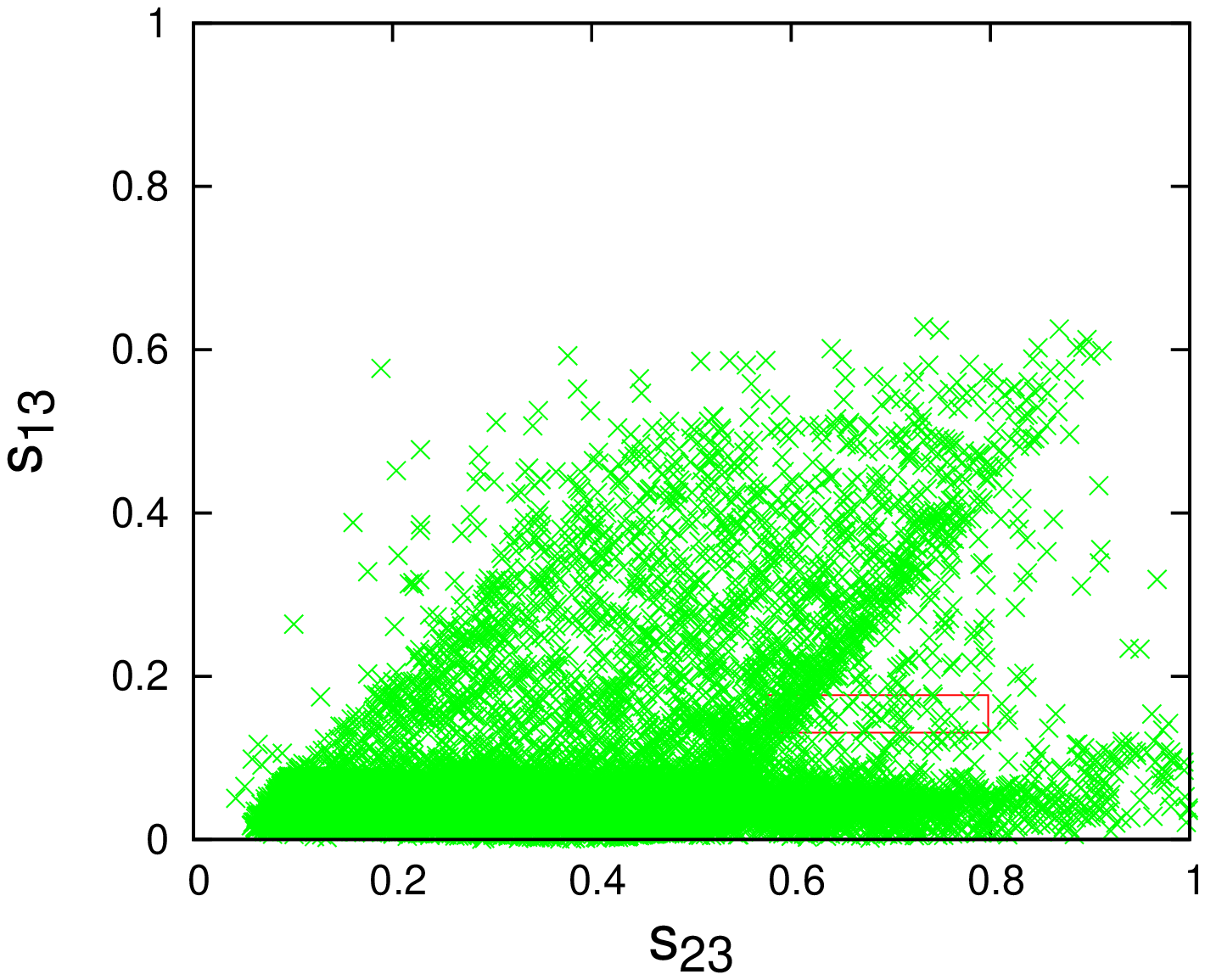}
\end{tabular}
\caption{Plots showing the parameter space for any two mixing
angles when the third angle is constrained by its  $1 \sigma$
range  in the $D_l =0$ and $D_\nu\neq 0$ scenario for Class II
ansatz of texture five zero  Dirac mass matrices (normal
hierarchy).} \label{t5cl2nh1}
\end{figure}
\begin{figure}
\begin{tabular}{cc}
  \includegraphics[width=0.2\paperwidth,height=0.2\paperheight,angle=-90]{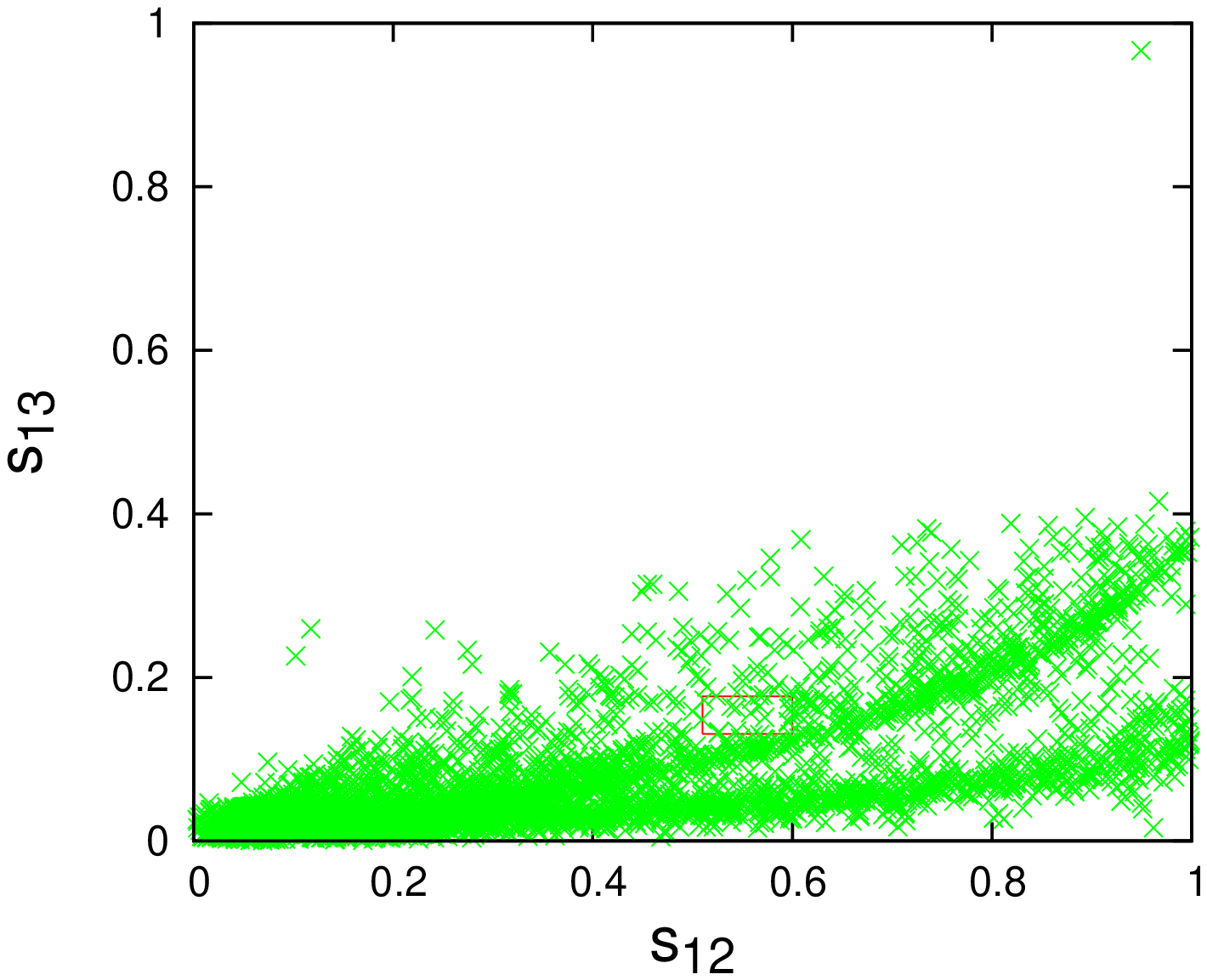}
  \includegraphics[width=0.2\paperwidth,height=0.2\paperheight,angle=-90]{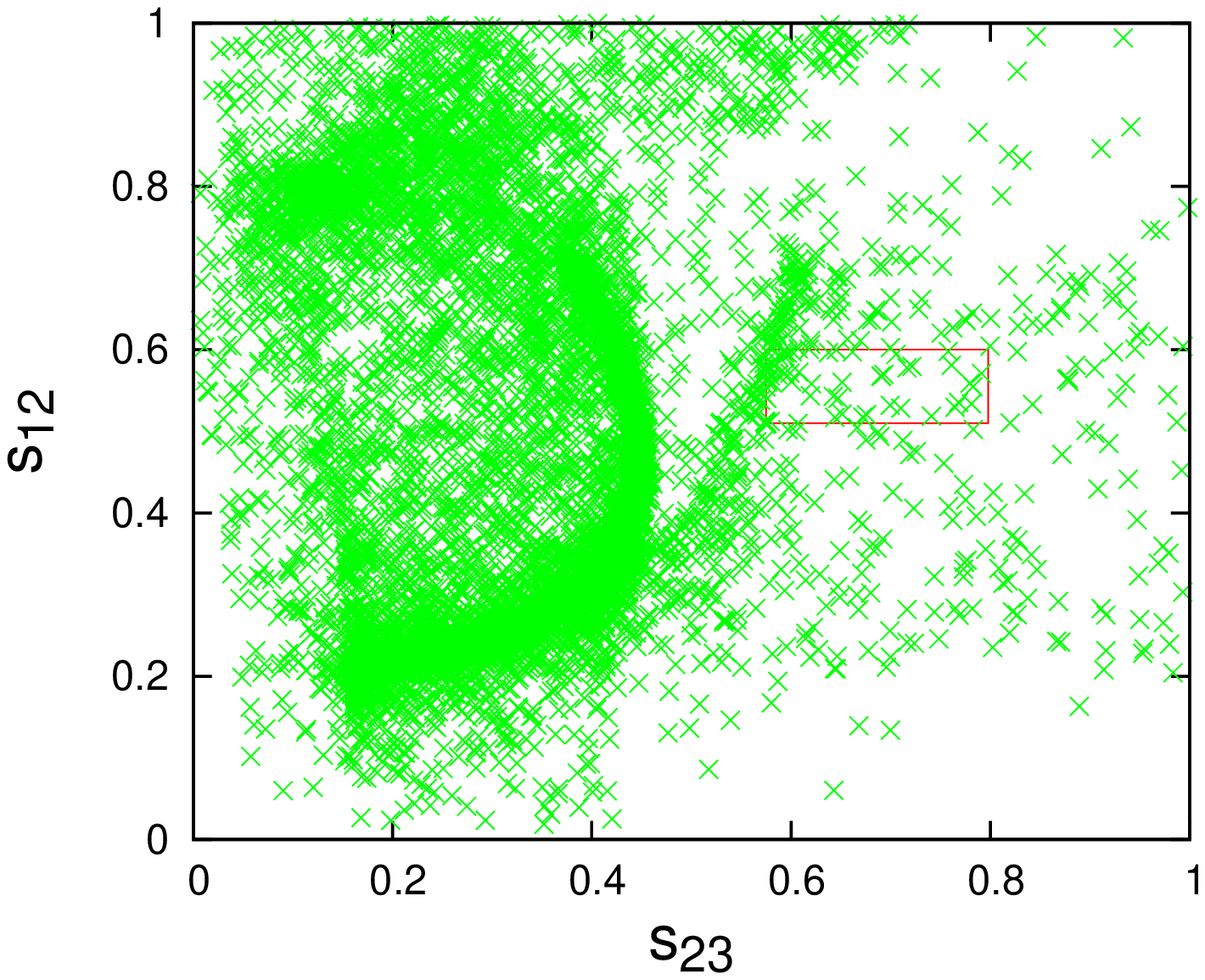}
  \includegraphics[width=0.2\paperwidth,height=0.2\paperheight,angle=-90]{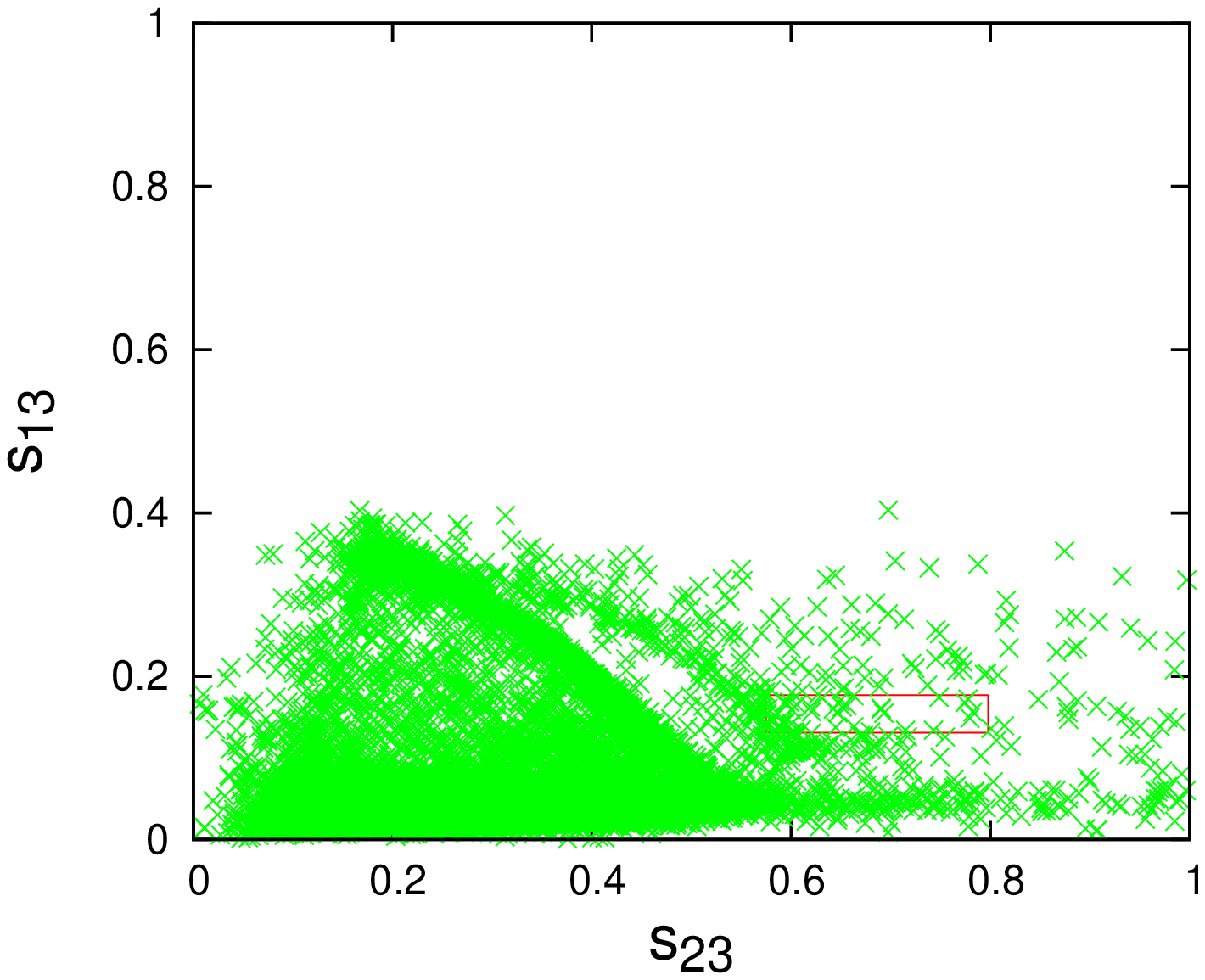}
\end{tabular}
\caption{Plots showing the parameter space for any two mixing
angles when the third angle is constrained by its  $3 \sigma$
range in the $D_l \neq 0$ and $D_\nu = 0$ scenario for Class II
ansatz of texture five zero  Dirac mass matrices (normal
hierarchy).} \label{t5cl2nh2}
\end{figure}

\begin{figure}
\begin{tabular}{cc}
  \includegraphics[width=0.2\paperwidth,height=0.2\paperheight,angle=-90]{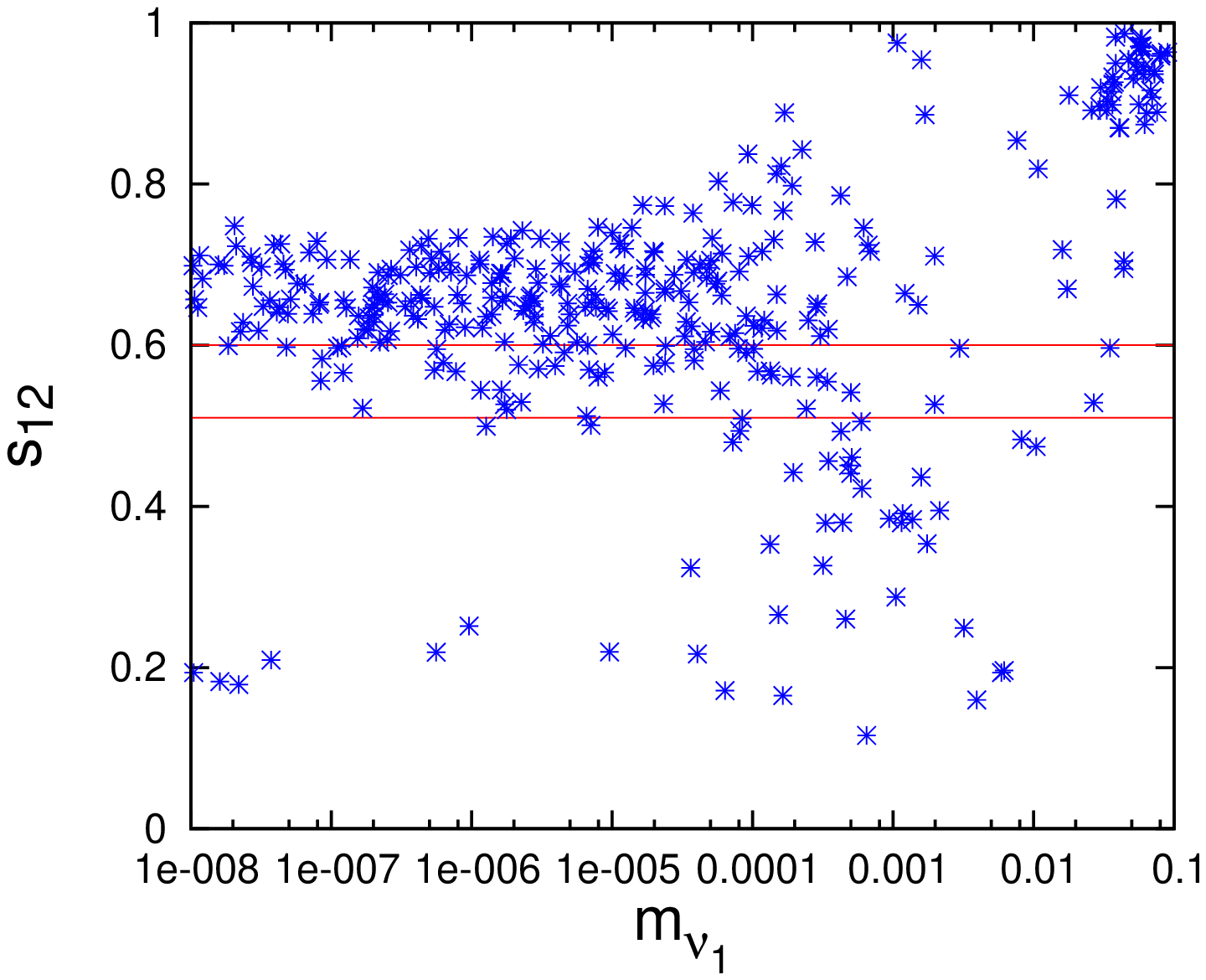}
  \includegraphics[width=0.2\paperwidth,height=0.2\paperheight,angle=-90]{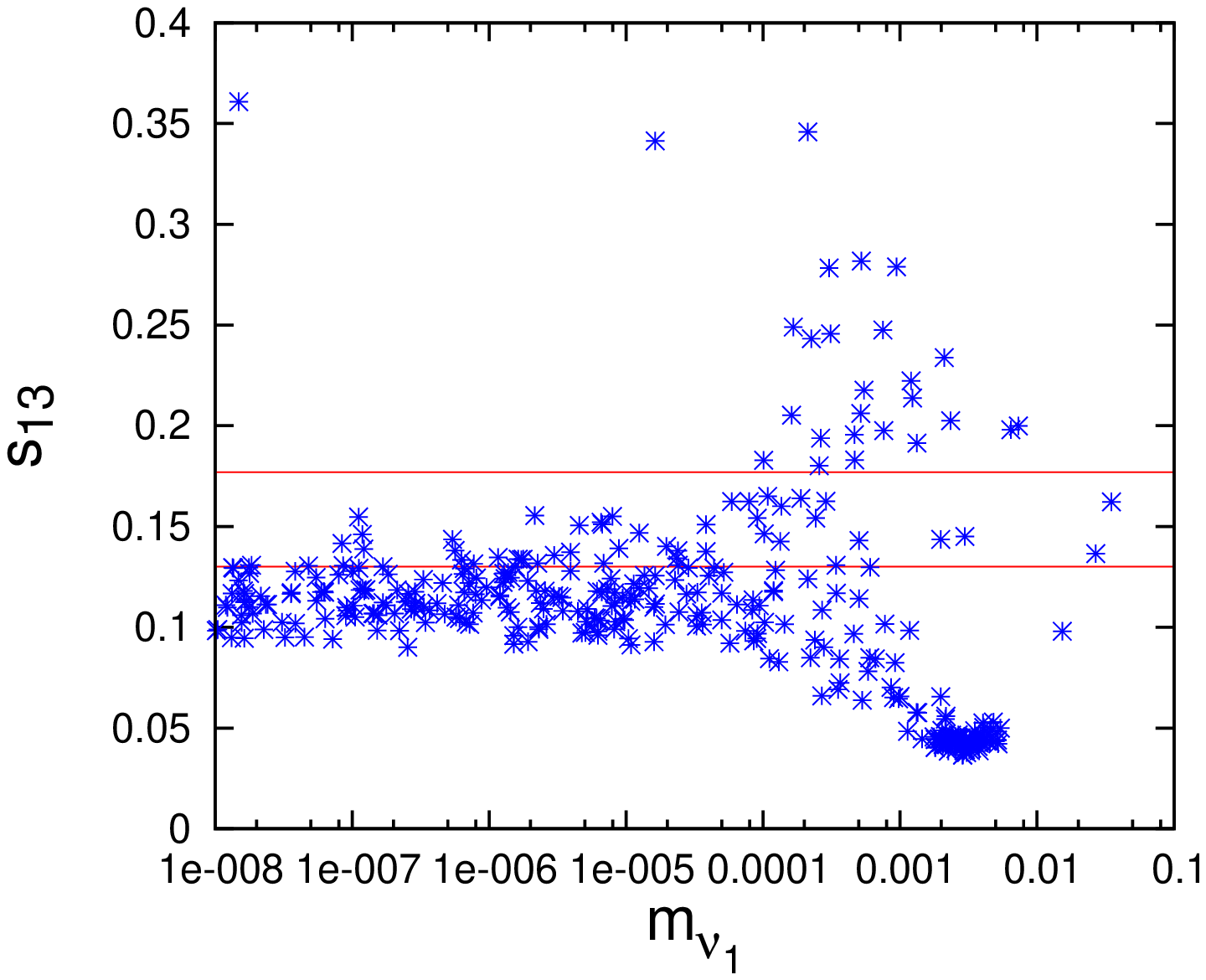}
  \includegraphics[width=0.2\paperwidth,height=0.2\paperheight,angle=-90]{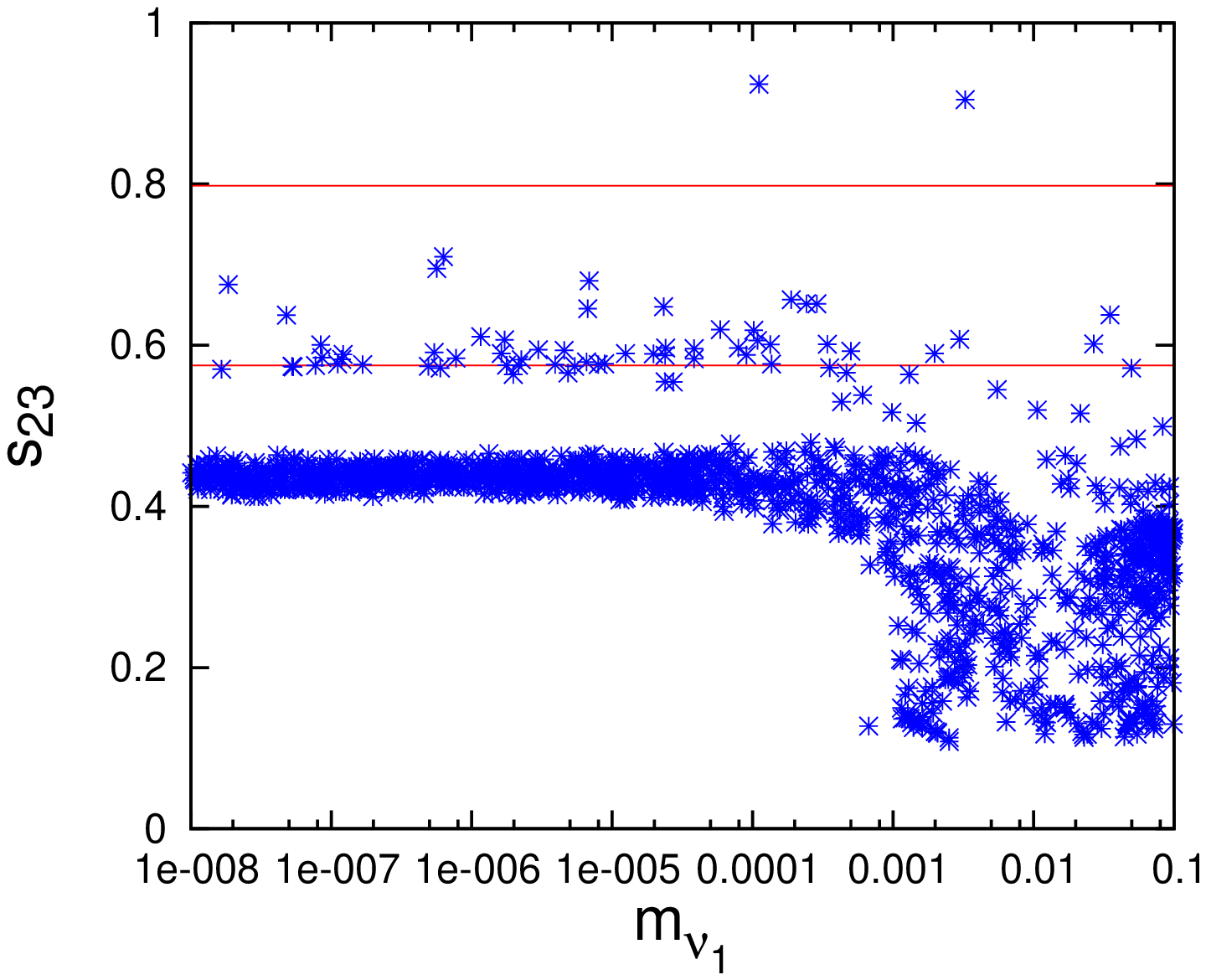}
\end{tabular}
\caption{Plots showing the lightest neutrino mass with mixing
angles when the other two angles are constrained by their $1
\sigma$ ranges  for $D_l\neq 0$ and $D_\nu = 0$ scenario for Class
II ansatz of texture five zero  Dirac mass matrices (normal
hierarchy).} \label{t5cl2nh3}
\end{figure}

\begin{figure}
\begin{tabular}{cc}
  \includegraphics[width=0.2\paperwidth,height=0.2\paperheight,angle=-90]{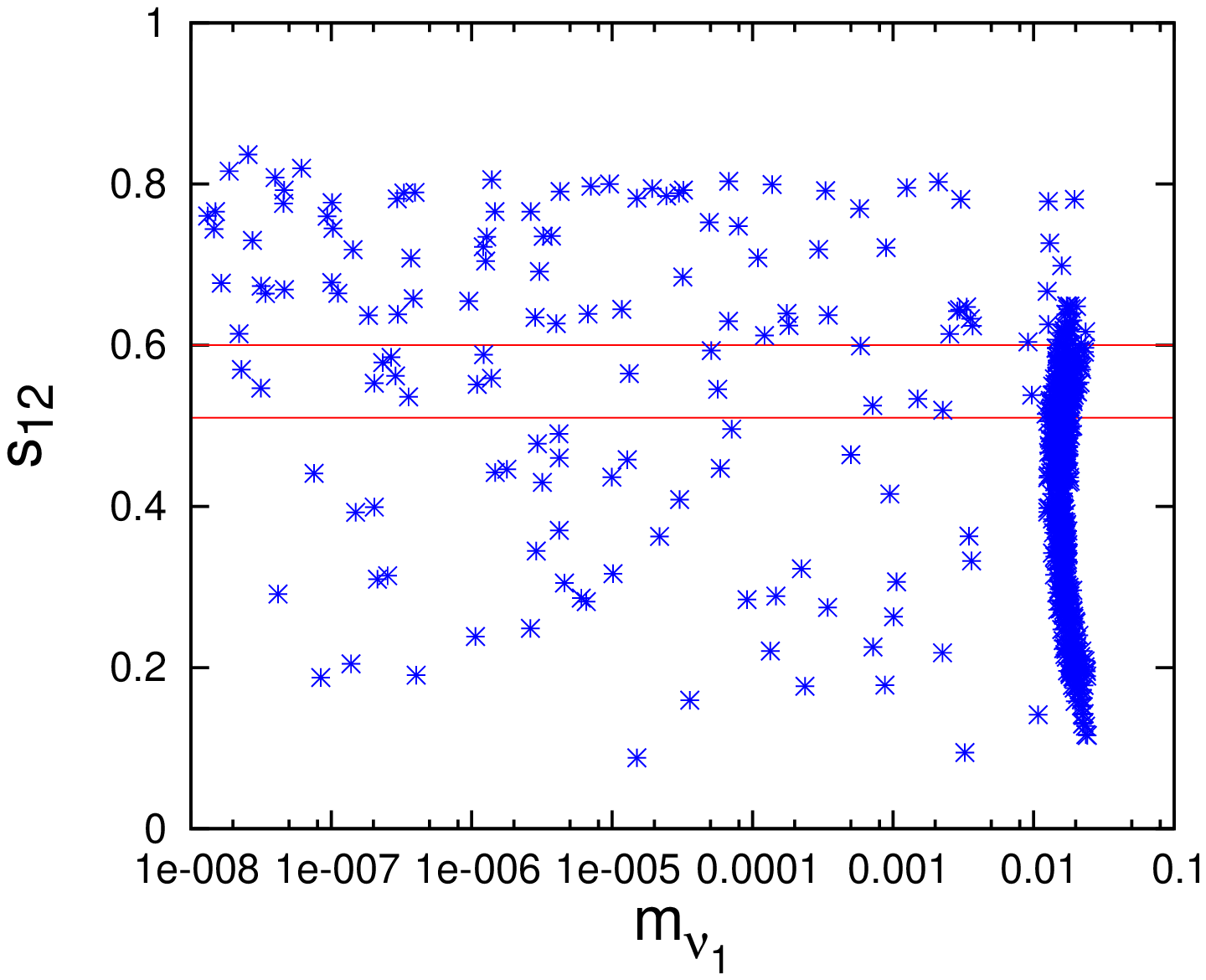}
  \includegraphics[width=0.2\paperwidth,height=0.2\paperheight,angle=-90]{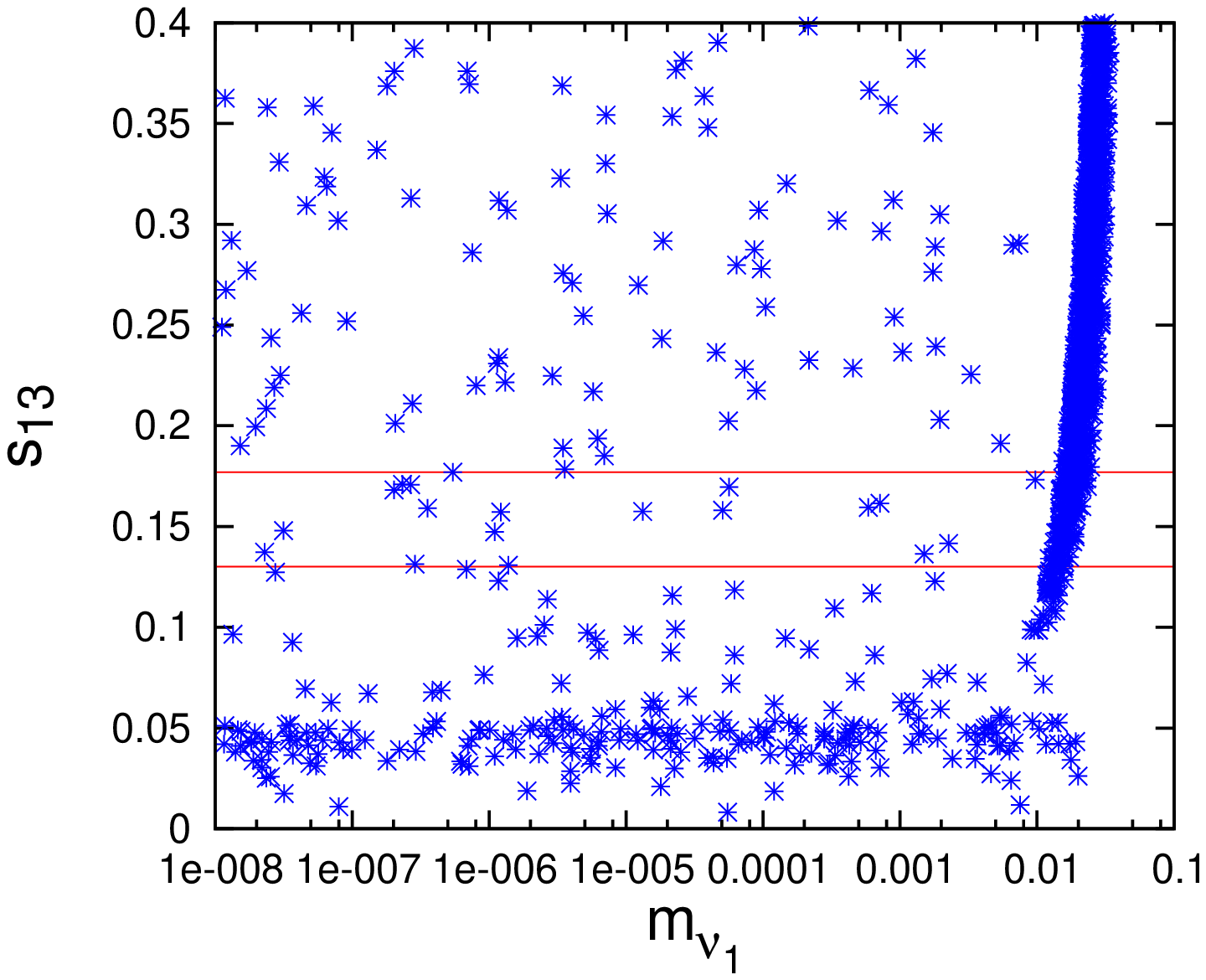}
  \includegraphics[width=0.2\paperwidth,height=0.2\paperheight,angle=-90]{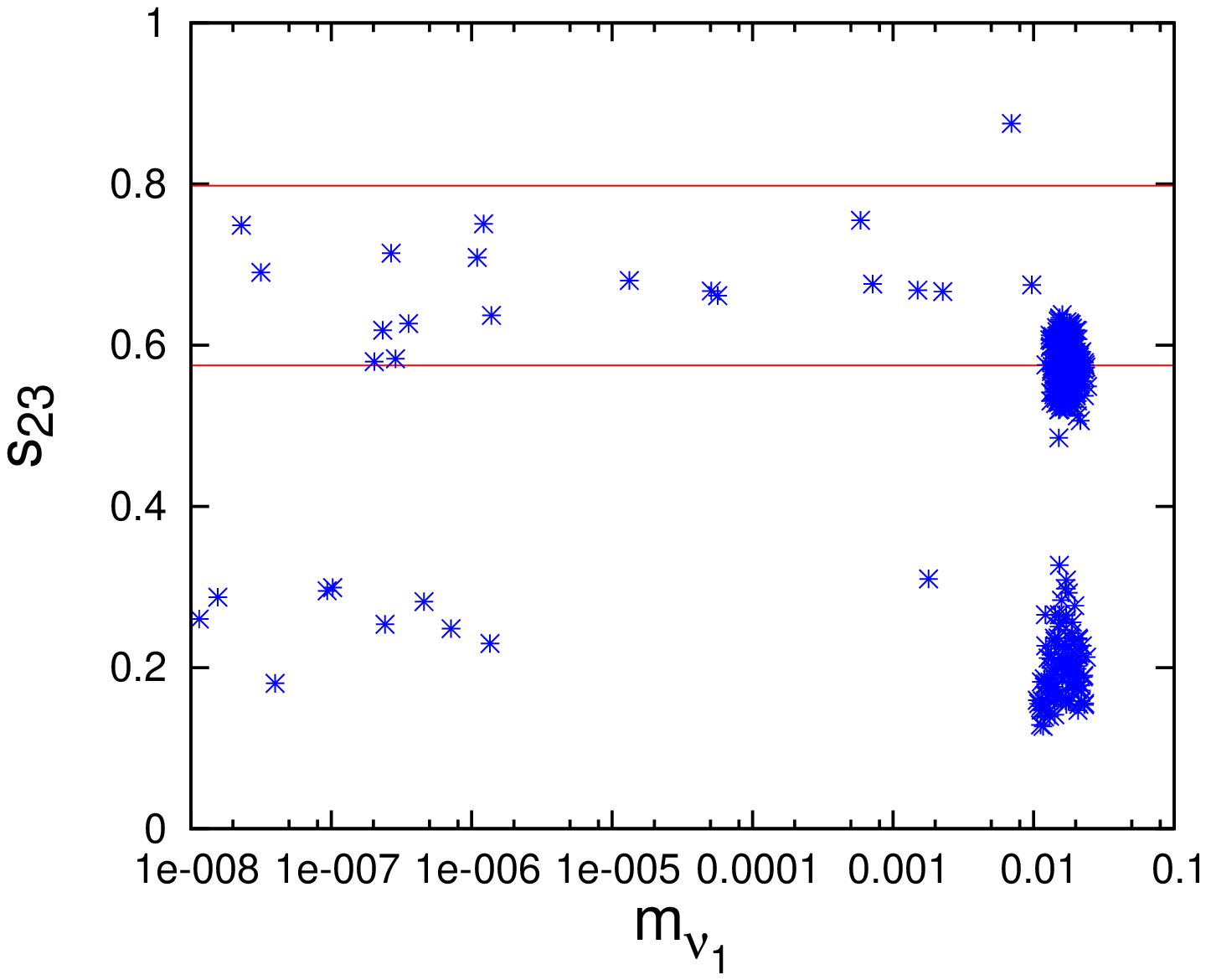}
\end{tabular}
\caption{Plots showing the lightest neutrino mass with mixing
angles when the other two angles are constrained by their $3
\sigma$ ranges   for $D_l= 0$ and $D_\nu \neq 0$ scenario for
Class II ansatz of texture five zero  Dirac mass matrices (normal
hierarchy).} \label{t5cl2nh4}
\end{figure}

\par After studying both the cases for texture five zero mass matrices pertaining to class II ansatz for inverted hierarchy,
we now carry out a similar
analysis pertaining to normal hierarchy. To this
end, in figures (\ref{t5cl2nh1}) and (\ref{t5cl2nh2}), we present the plots showing the parameter space corresponding
to any two mixing angles wherein the third one is constrained by its $3\sigma$ range. Interestingly, normal
hierarchy seems to be viable for both the cases, $D_\nu=0$ and $D_l \neq 0$ as well as $D_\nu \neq 0$ and $D_l=0$,
of texture five zero lepton mass matrices as can be seen from these
plots (\ref{t5cl2nh1}) and (\ref{t5cl2nh2}), wherein the parameter space shows significant overlap with the experimentally
allowed $3\sigma$ region shown by the rectangular boxes in each plot.
Next, we study the dependence of the lightest neutrino mass and
Jarlskog's parameter on the the leptonic mixing angles for this
case. To this end,  we present the plots showing variation of the
lightest neutrino mass with the mixing angles in figures
(\ref{t5cl2nh3}) and (\ref{t5cl2nh4}) respectively. While plotting
these graphs, the other two mixing angles have been constrained by
their $3\sigma$ ranges. Interestingly, for $D_l \neq 0$ and $D_\nu
=0$, the lightest neutrino mass is largely unrestricted, whereas
for the case $D_l=0$ and $D_\nu \neq 0$ one obtains an upper bound
$ \approx 0.01 eV$ for the lightest neutrino mass.

\begin{figure}
\begin{tabular}{cc}
  \includegraphics[width=0.2\paperwidth,height=0.2\paperheight,angle=-90]{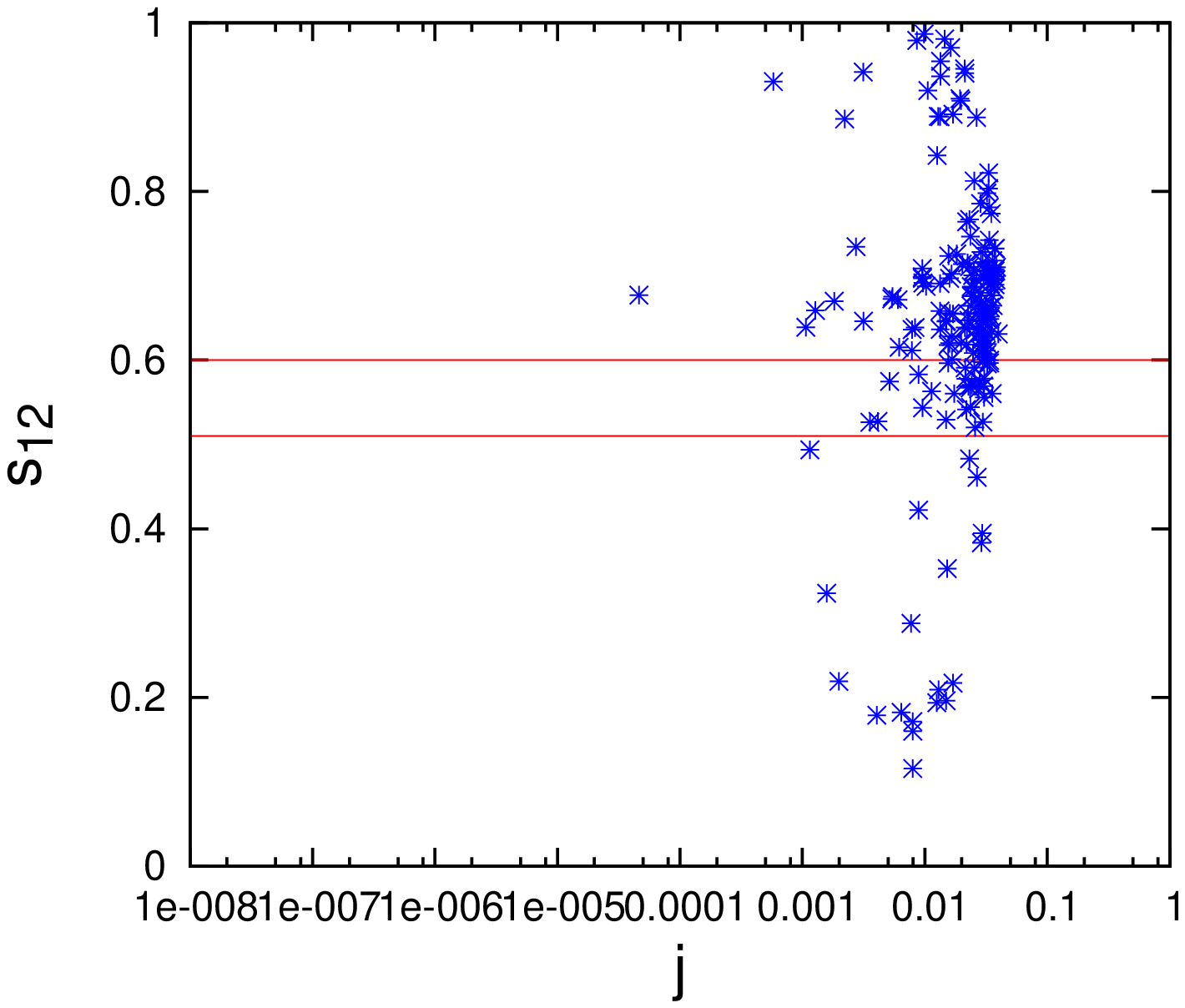}
  \includegraphics[width=0.2\paperwidth,height=0.2\paperheight,angle=-90]{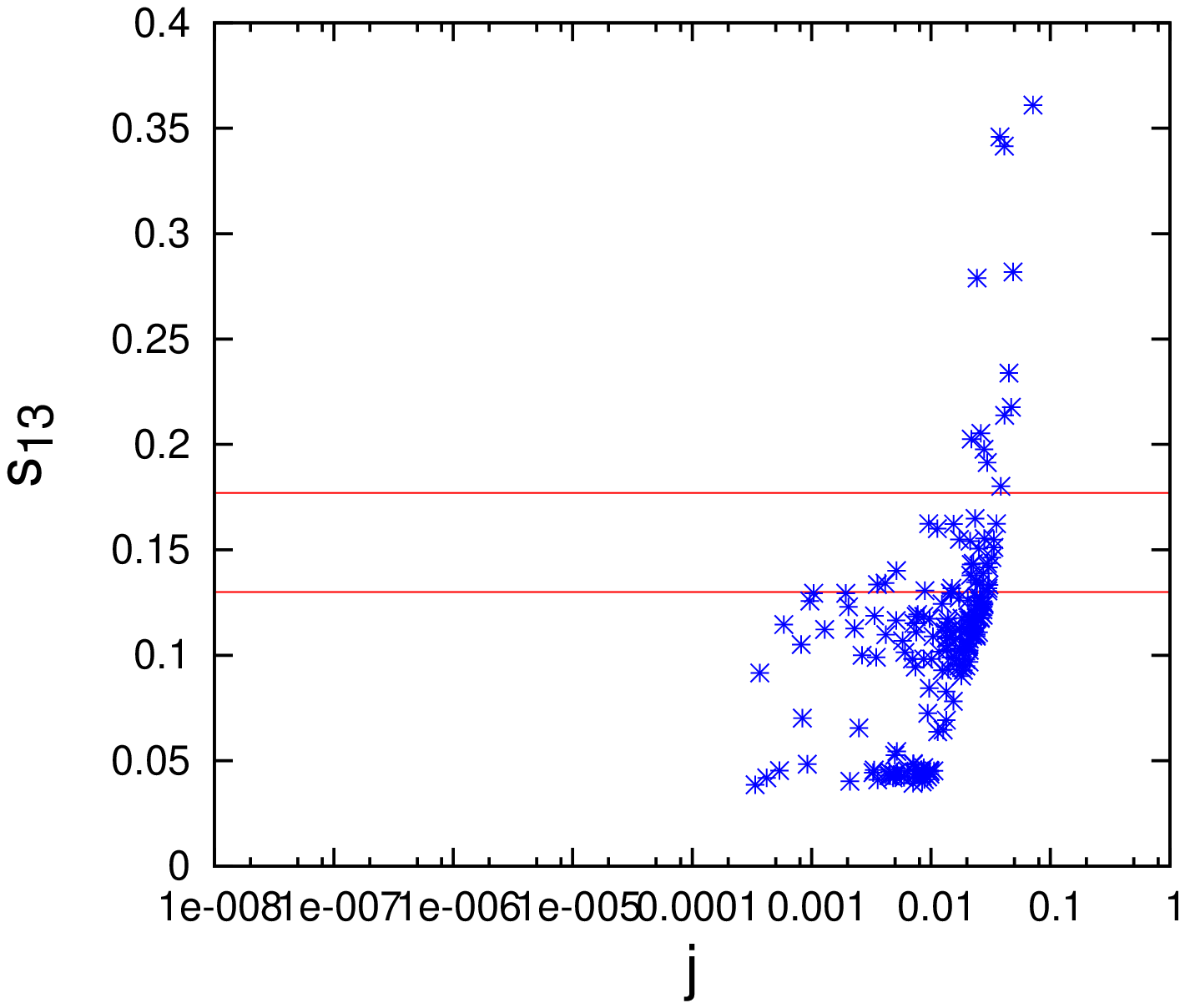}
  \includegraphics[width=0.2\paperwidth,height=0.2\paperheight,angle=-90]{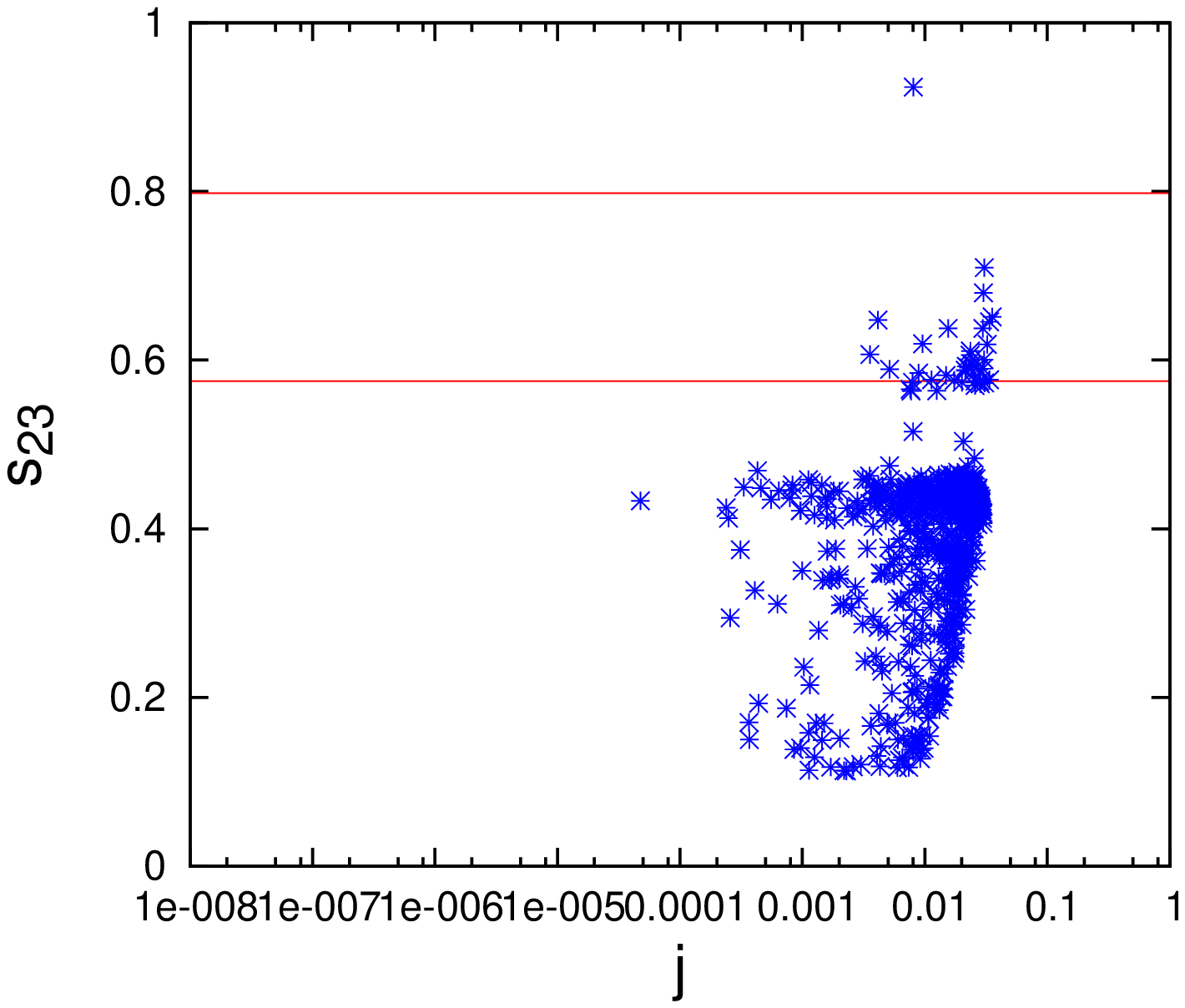}
\end{tabular}
\caption{Plots showing the variation of Jarlskog CP violating
parameter with mixing angles when the other two angles are
constrained by their $3 \sigma$ ranges  for $D_l\neq0$ and
$D_\nu=0$ scenario for Class II ansatz of texture five zero  Dirac
mass matrices (normal hierarchy).} \label{t5cl2nh5}
\end{figure}

\begin{figure}
\begin{tabular}{cc}
  \includegraphics[width=0.2\paperwidth,height=0.2\paperheight,angle=-90]{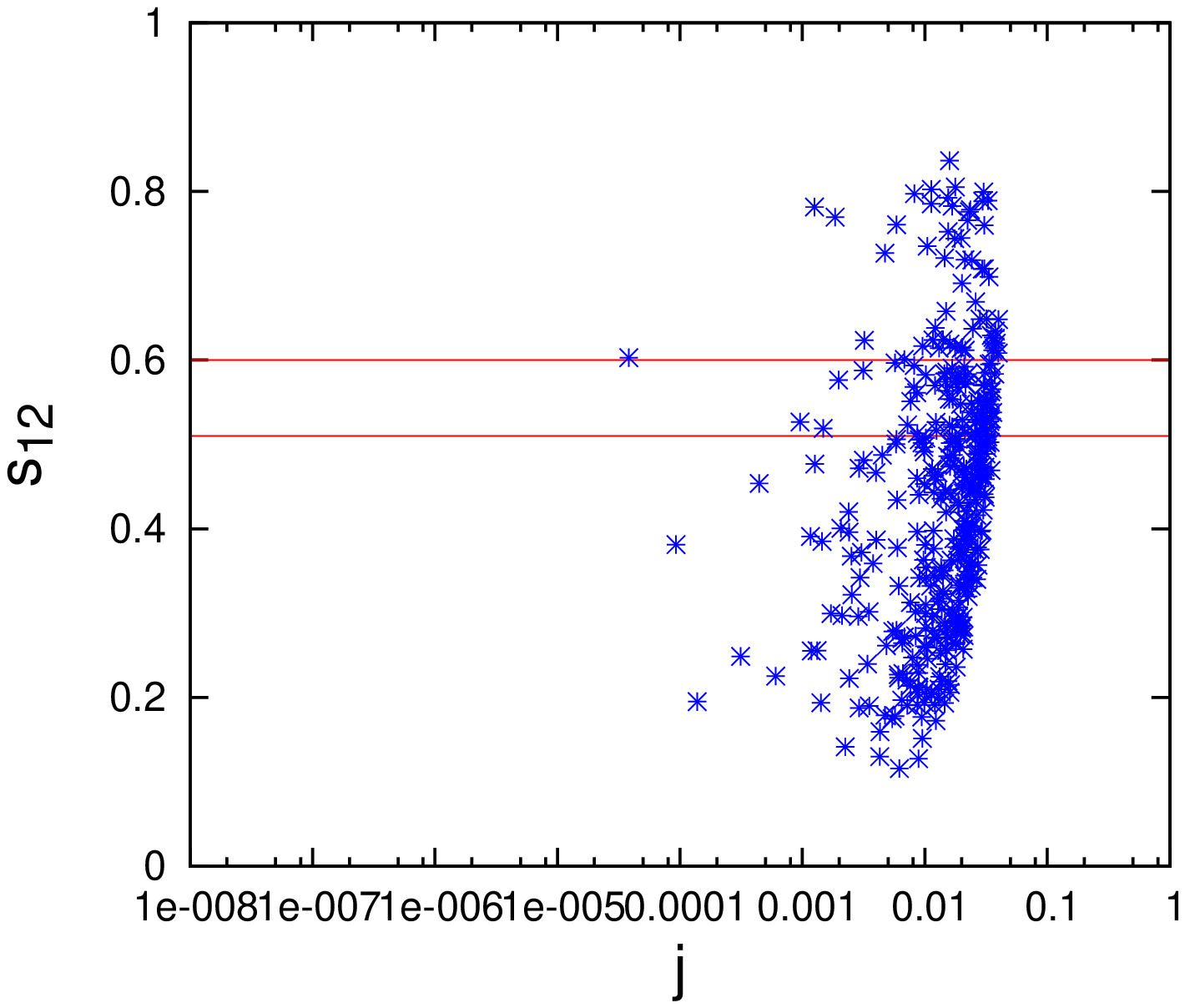}
  \includegraphics[width=0.2\paperwidth,height=0.2\paperheight,angle=-90]{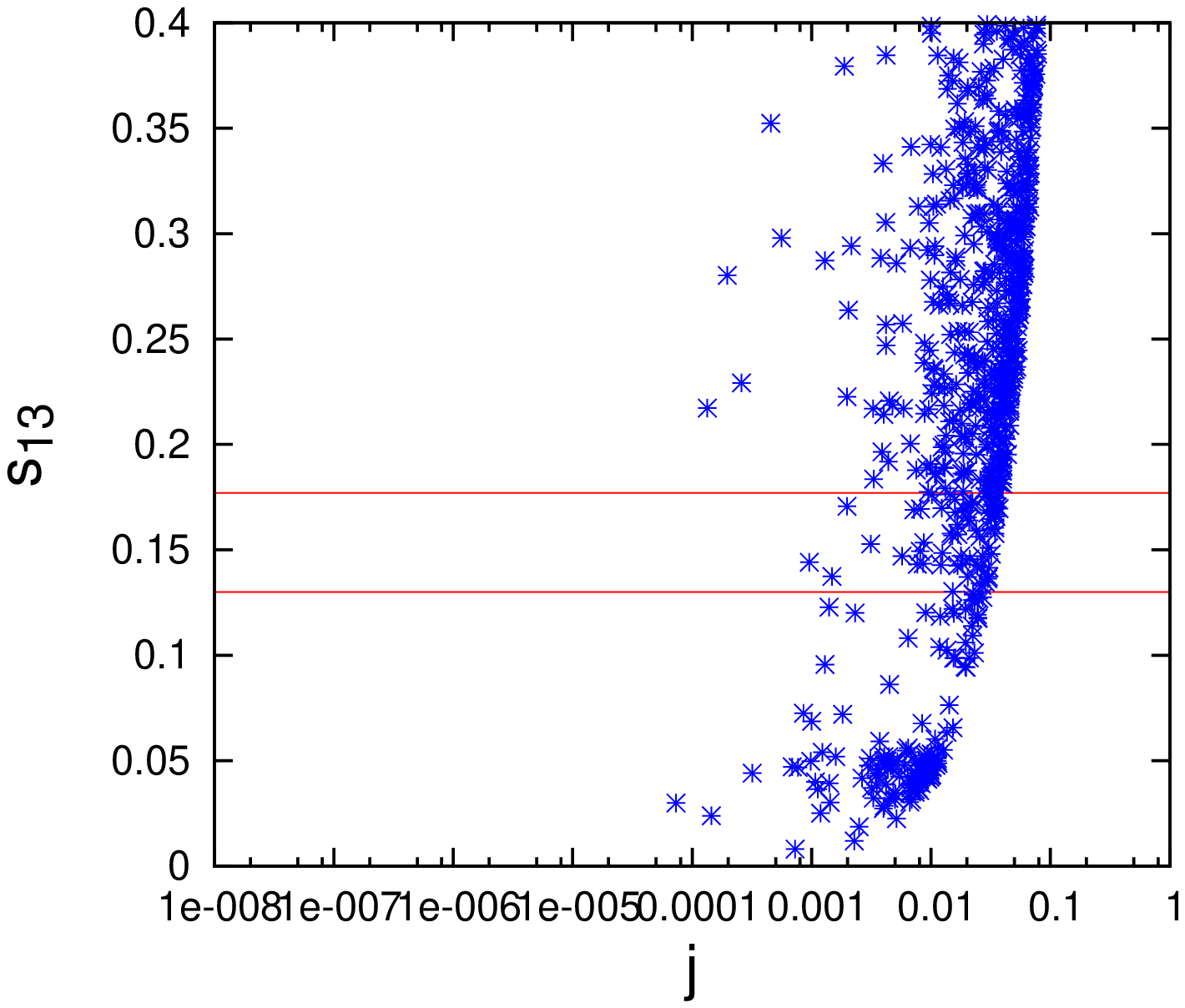}
  \includegraphics[width=0.2\paperwidth,height=0.2\paperheight,angle=-90]{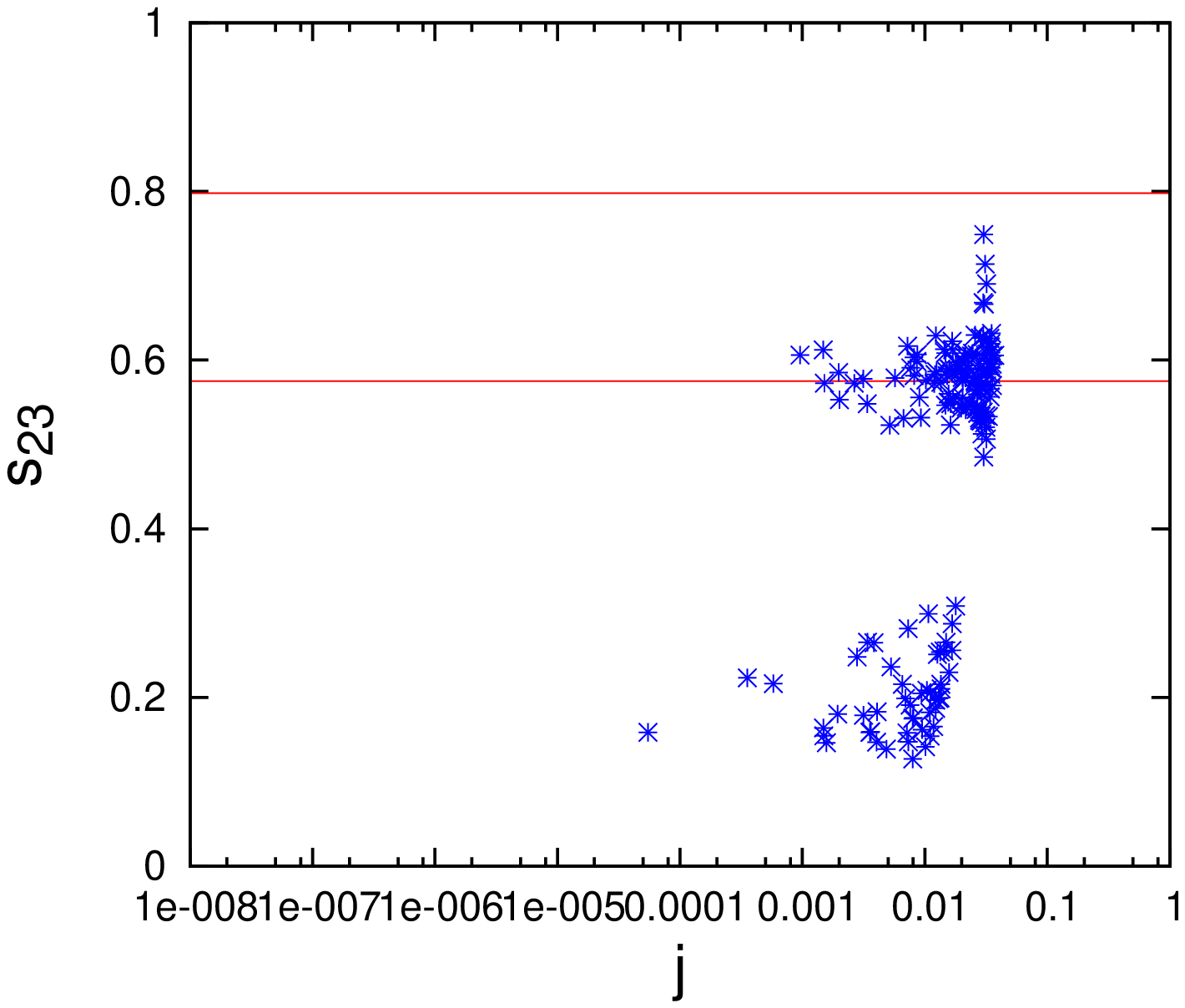}
\end{tabular}
\caption{Plots showing the variation of Jarlskog CP violating
parameter with mixing angles when the other two angles are
constrained by their $3 \sigma$ ranges   for $D_l= 0$ and $D_\nu
\neq 0$ scenario for Class II ansatz of texture five zero  Dirac
mass matrices (normal hierarchy).} \label{t5cl2nh6}
\end{figure}

Further, in figures (\ref{t5cl2nh5}) and (\ref{t5cl2nh6}) we
present the graphs showing the variation of the Jarlskog's
parameter with each of the mixing angles, keeping the other two
constrained by their $3\sigma$ ranges. Parallel lines in these
graphs show the $3\sigma$ ranges for the mixing angles being
considered. Interestingly, one finds that the range of J for
$D_l\neq0$ and $D_\nu=0$ is quite narrow as compared to the $D_l=
0$ and $D_\nu \neq 0$ case of texture five zero lepton mass
matrices for class II ansatz.

\subsubsection{Class III ansatz}

\begin{figure}
\begin{tabular}{cc}
  \includegraphics[width=0.2\paperwidth,height=0.2\paperheight,angle=-90]{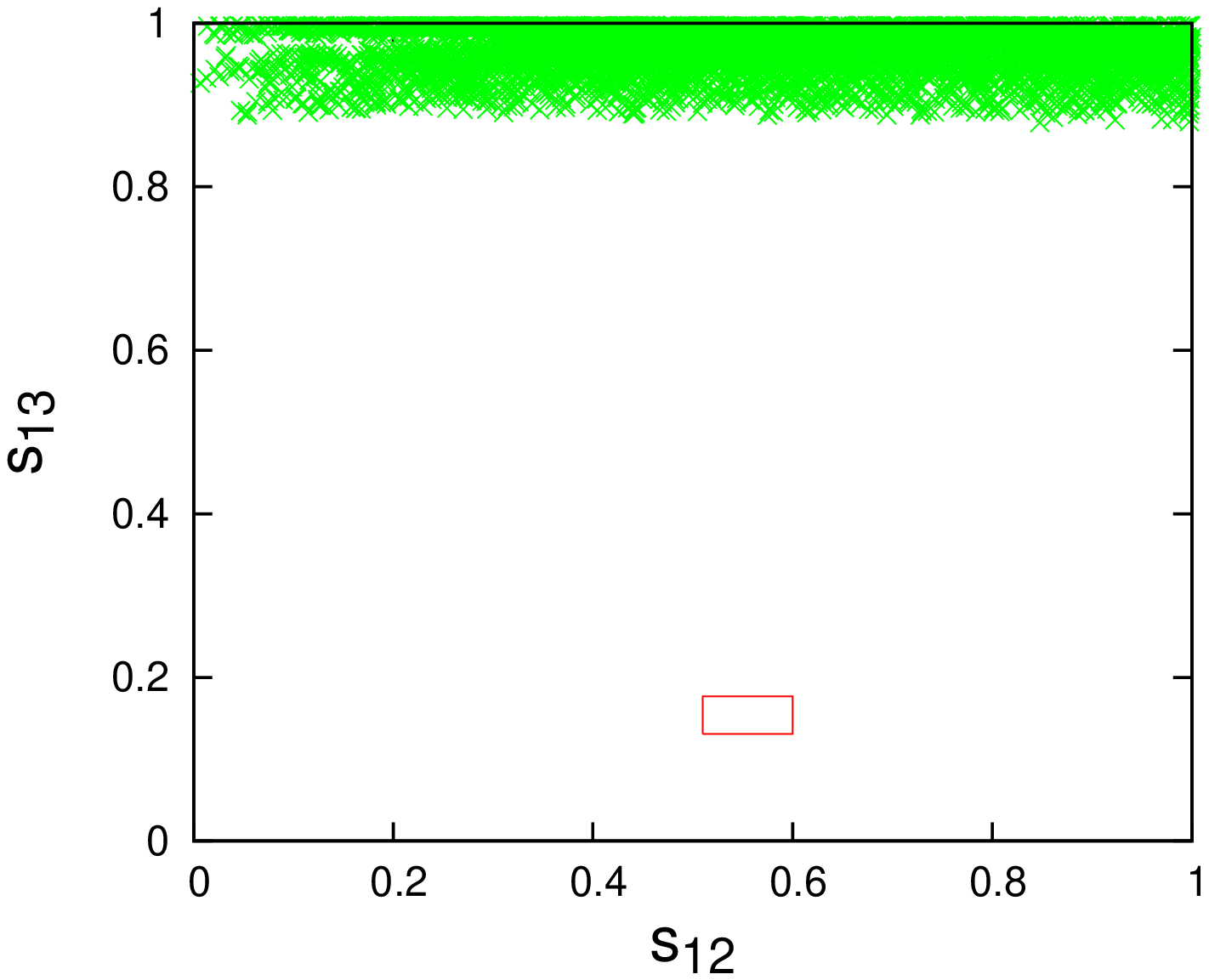}
  \includegraphics[width=0.2\paperwidth,height=0.2\paperheight,angle=-90]{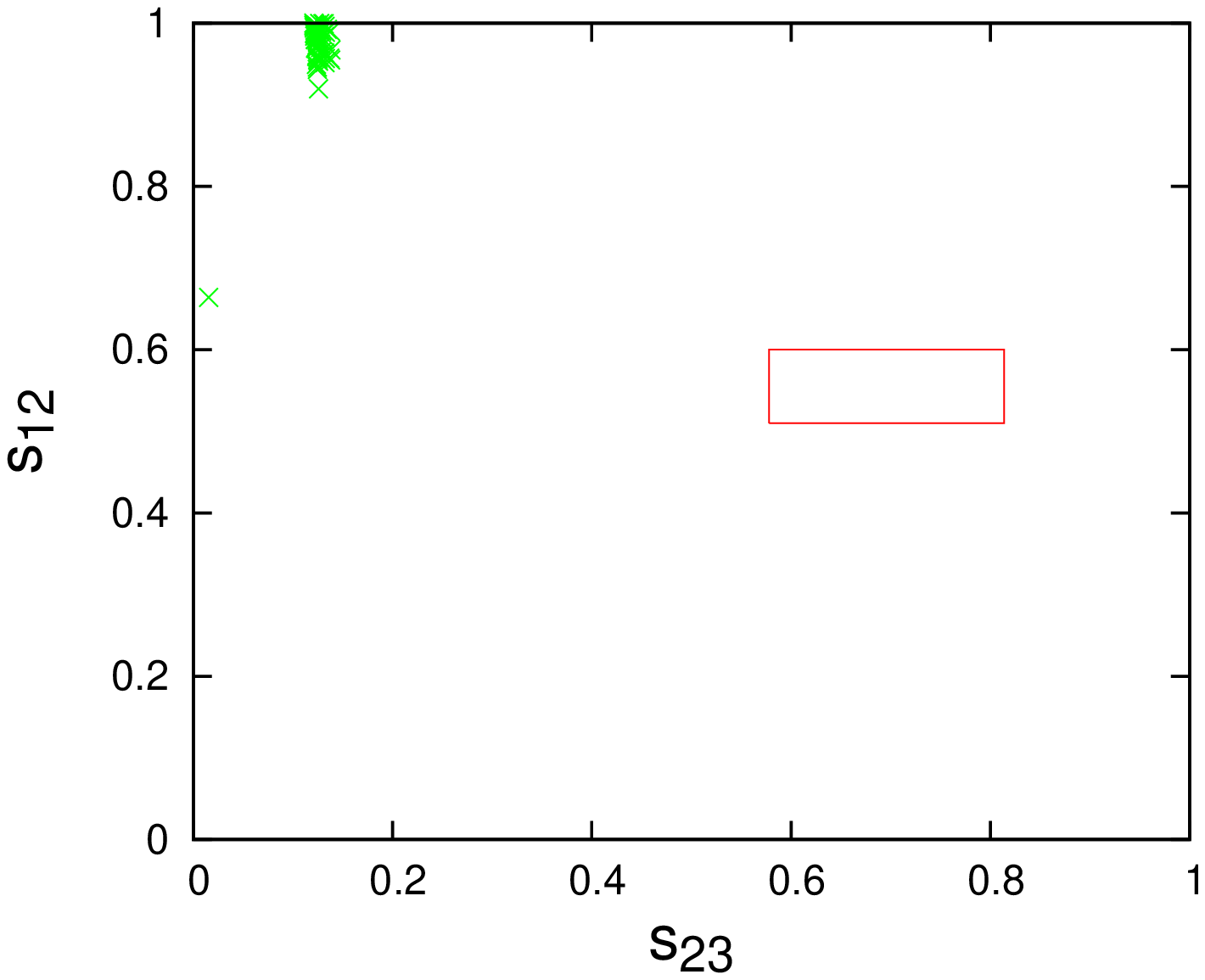}
  \includegraphics[width=0.2\paperwidth,height=0.2\paperheight,angle=-90]{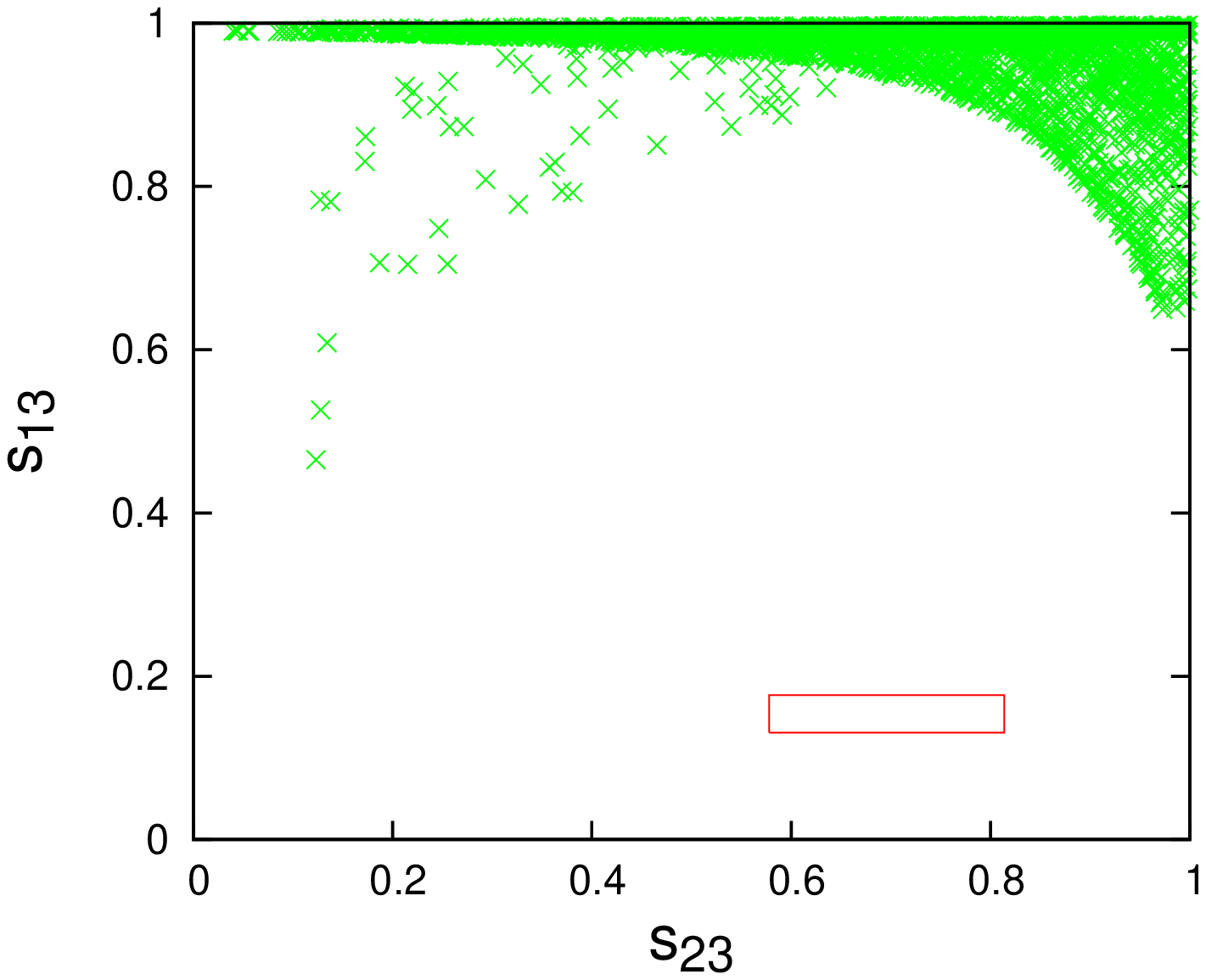}
\end{tabular}
\caption{Plots showing the parameter space for any two mixing
angles when the third angle is constrained by its  $3 \sigma$
range in the $D_l =0$ and $D_\nu\neq 0$ scenario for Class III
ansatz of texture five zero  Dirac mass matrices (inverted
hierarchy).} \label{t5cl3ih1}
\end{figure}

\begin{figure}
\begin{tabular}{cc}
  \includegraphics[width=0.2\paperwidth,height=0.2\paperheight,angle=-90]{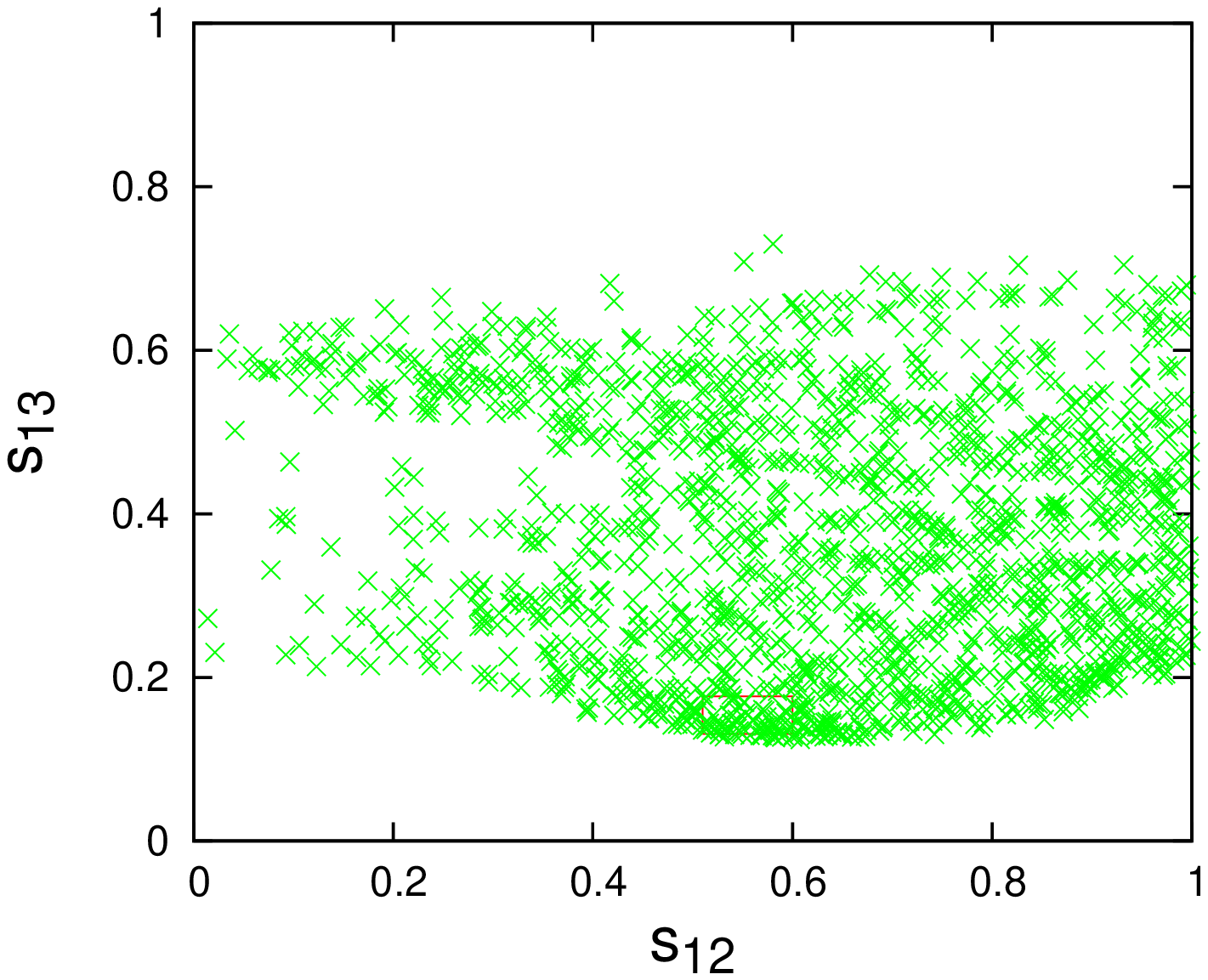}
  \includegraphics[width=0.2\paperwidth,height=0.2\paperheight,angle=-90]{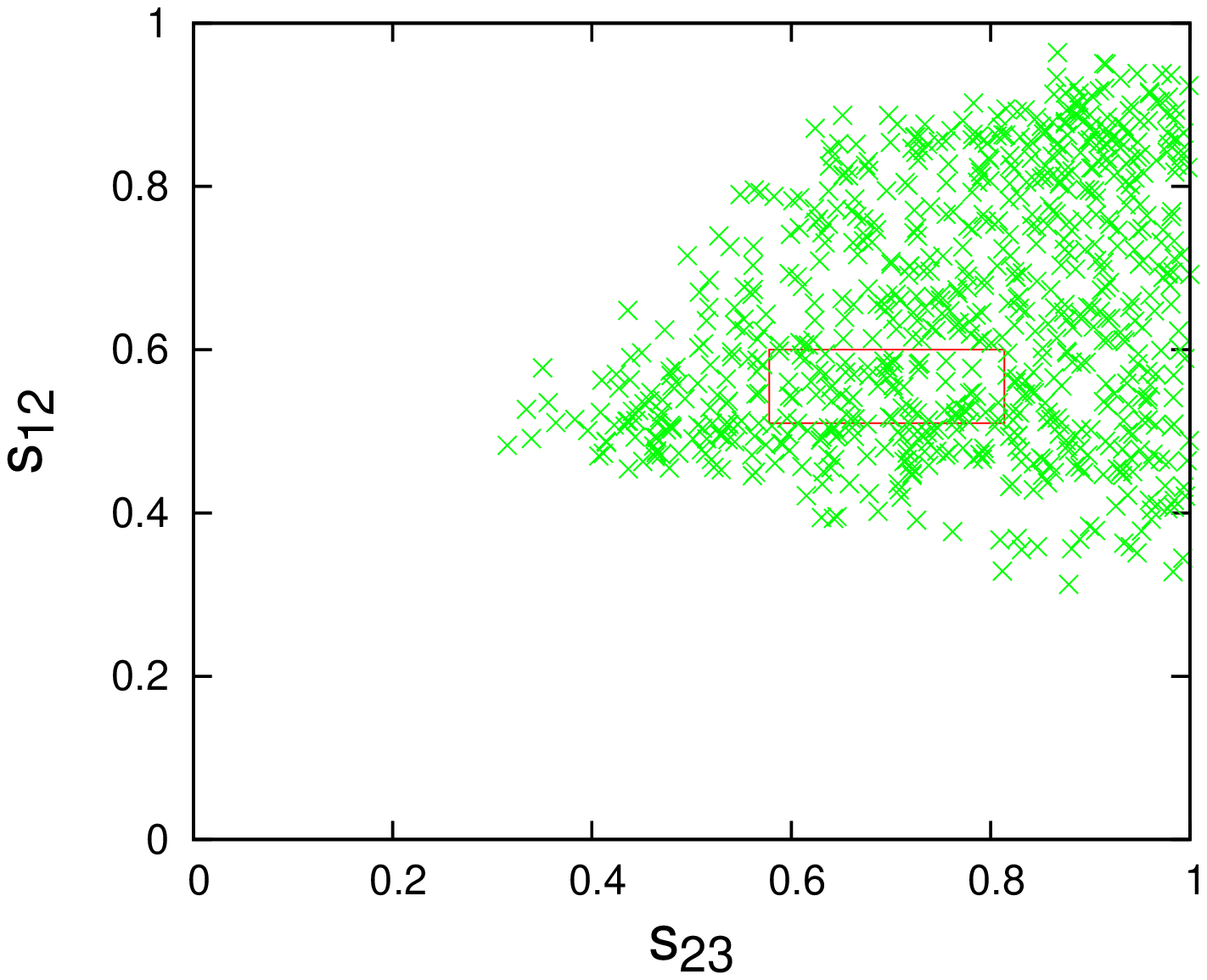}
  \includegraphics[width=0.2\paperwidth,height=0.2\paperheight,angle=-90]{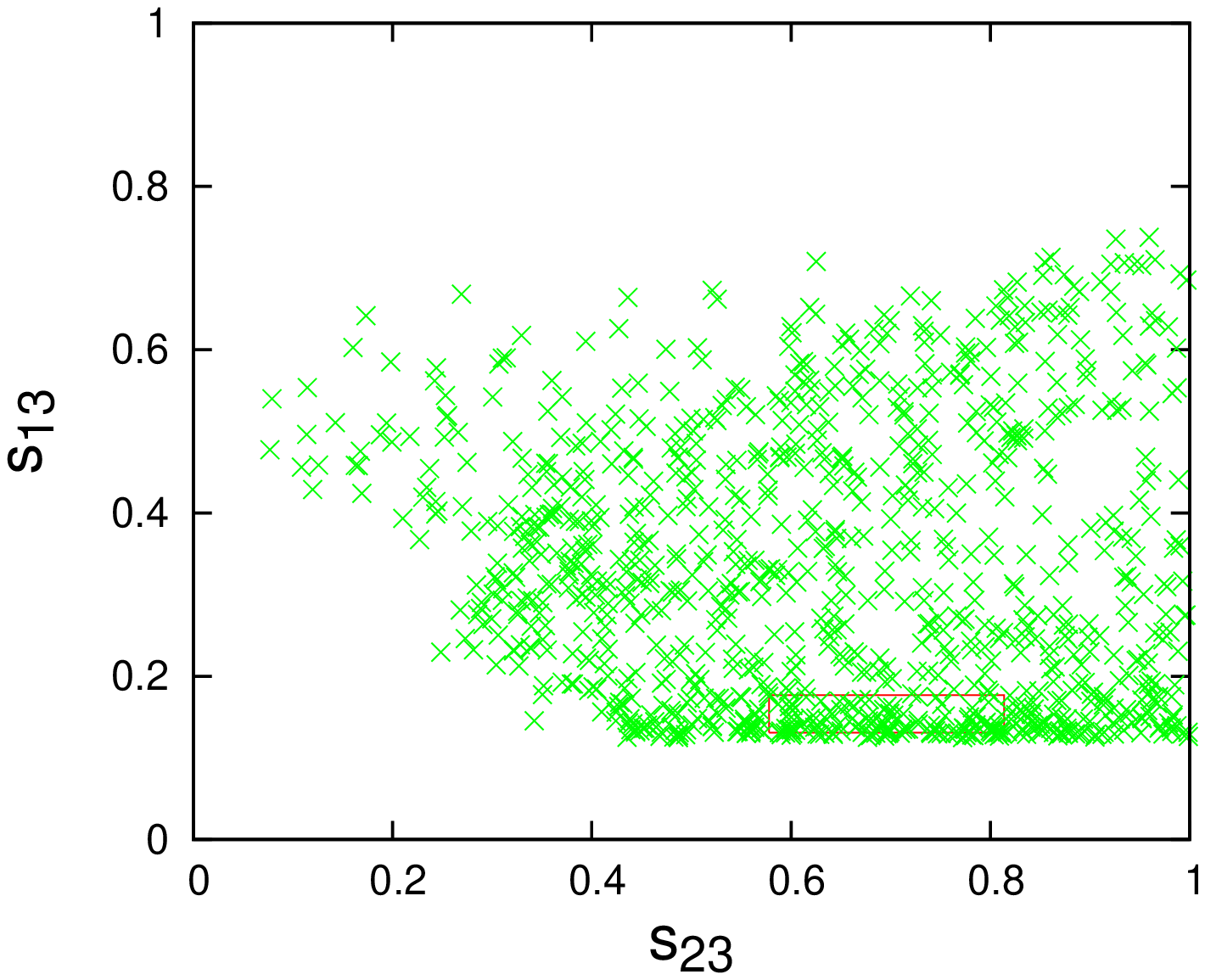}
\end{tabular}
\caption{Plots showing the parameter space for any two mixing angles when the third angle is constrained by
its  $3 \sigma$ range  in the $D_l \neq 0$ and $D_\nu = 0$ 
scenario for Class III ansatz of texture five zero  Dirac mass matrices (inverted hierarchy).}
\label{t5cl3ih2}
\end{figure}

The two possibilities for texture five zero lepton mass matrices for this class can be given as,

\be
 M_{l}=\left( \ba{ccc}
0 & A _{l}e^{i\alpha_l} & B_{l}     \\
A_{l}e^{-i\alpha_l} & 0 &  0     \\
 B_{l} & 0 &  E_{l} \ea \right), \qquad
M_{\nu}=\left( \ba{ccc}
0 & A _{\nu}e^{i\alpha_\nu} & B_{\nu}     \\
A_{\nu}e^{-i\alpha_\nu} & 0 &   D_{\nu}e^{i\beta_\nu}      \\
 B_{\nu}& D_{\nu}e^{-i\beta_\nu}  &  E_{\nu} \ea \right),
\label{cl3t51}
\ee
or \be
 M_{l}=\left( \ba{ccc}
0 & A _{l}e^{i\alpha_l} & B_{l}    \\
A_{l}e^{-i\alpha_l} & 0 &  D_{l}e^{i\beta_l}    \\
 B_{l} & D_{l}e^{-i\beta_l}  &  E_{l} \ea \right), \qquad
M_{\nu}=\left( \ba{ccc}
0 & A _{\nu}e^{i\alpha_\nu} & B_{\nu}     \\
A_{\nu}e^{-i\alpha_\nu} & 0 & 0        \\
 B_{\nu} & 0 &  E_{\nu} \ea \right),
\label{cl3t52}
\ee
We study both these possibilities in detail for all the neutrino mass orderings. Firstly, we examine
 the compatibility of matrices (\ref{cl3t51}) and (\ref{cl3t52}) with the inverted hierarchy
of neutrino masses.
For this purpose, in figures (\ref{t5cl3ih1}) and
 (\ref{t5cl3ih2}), we present the plots showing the parameter space allowed by this ansatz for any two mixing angles wherein
the third one  is constrained by its $3\sigma$ experimental bound for inverted hierarchy of neutrino masses.
The rectangular regions in these plots represent the $3\sigma$ ranges for the two mixing angles being considered.
Interestingly, one finds that for the case $D_l =0$ and $D_\nu\neq 0$ of texture five zero lepton mass matrices
inverted hierarchy is ruled out, whereas for the case $D_l \neq 0$ and $D_\nu=0$ of texture five zero lepton mass matrices
inverted hierarchy scenario seems to be viable.
\par
\begin{figure}
\begin{tabular}{cc}
  \includegraphics[width=0.2\paperwidth,height=0.2\paperheight,angle=-90]{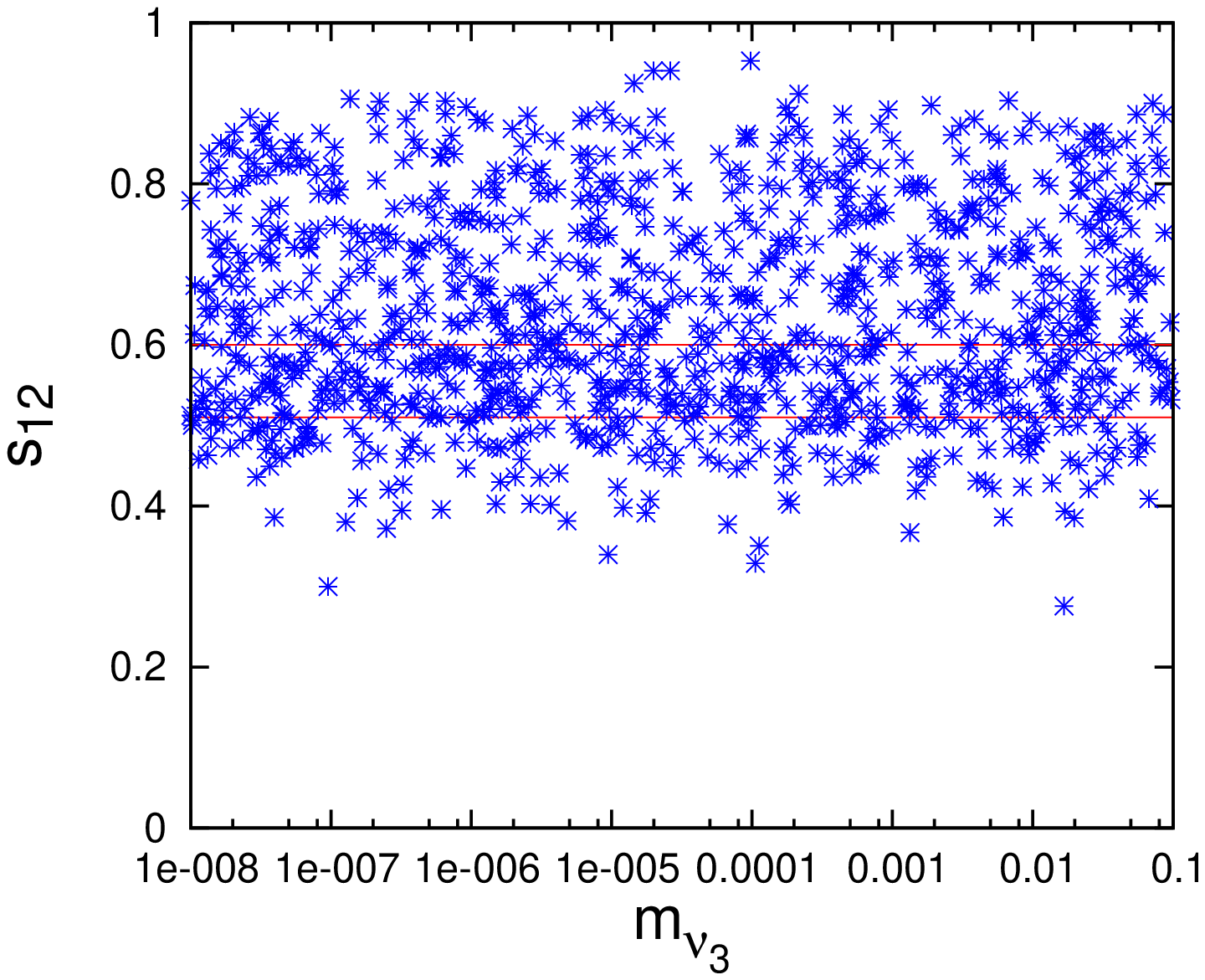}
  \includegraphics[width=0.2\paperwidth,height=0.2\paperheight,angle=-90]{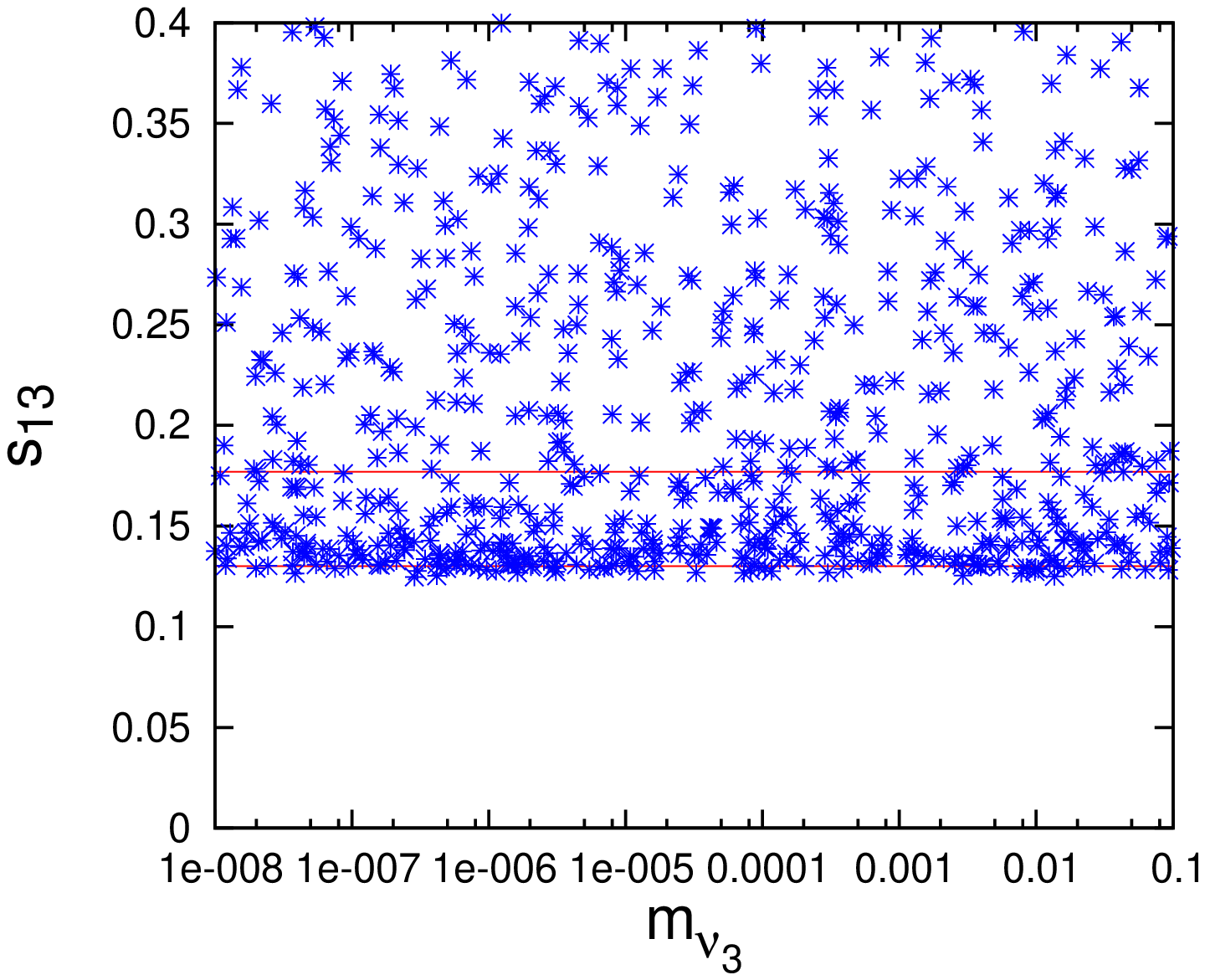}
  \includegraphics[width=0.2\paperwidth,height=0.2\paperheight,angle=-90]{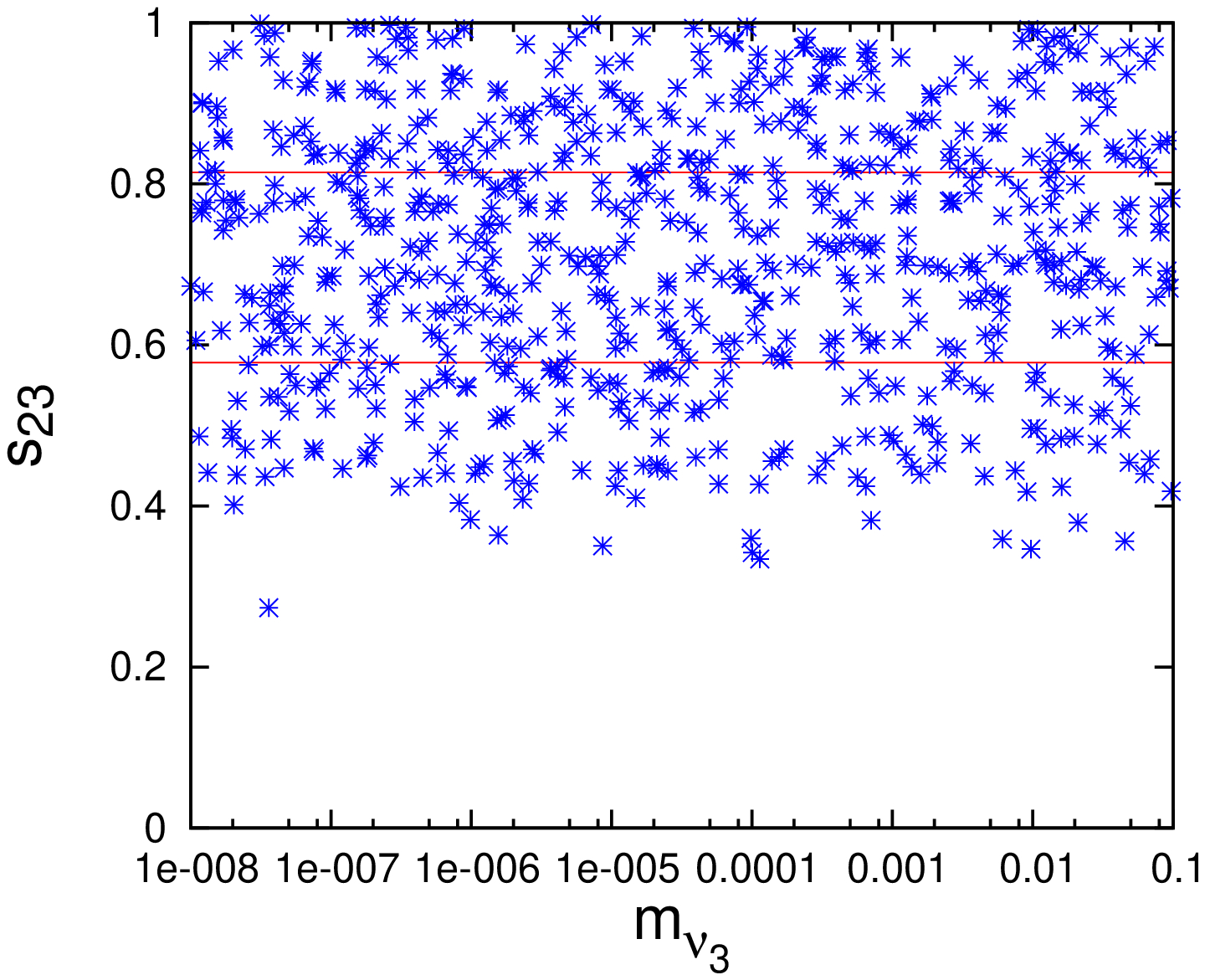}
\end{tabular}
\caption{Plots showing the lightest neutrino mass with mixing
angles when the other two angles are constrained by their $3
\sigma$ ranges   for $D_l\neq 0$ and $D_\nu = 0$ scenario for
Class III ansatz of texture five zero  Dirac mass matrices
(inverted hierarchy).} \label{t5cl3ih3}
\end{figure}

 For the $D_l \neq 0$ and $D_\nu = 0$ case of lepton mass matrices, wherein inverted hierarchy is shown to be viable, we
proceed next to study the the dependence of the lightest neutrino mass and Jarlskog's parameter on the the
leptonic mixing angles. To this end,  we present the plots showing variation of
the lightest neutrino mass with the mixing angles in figures (\ref{t5cl3ih3}). While
plotting these graphs, the other two mixing angles have been constrained by their $3\sigma$ ranges.
Interestingly, one finds that the lightest neutrino mass is unrestricted for this structure. Further,
on carrying out the calculations for the Jarlskog's parameter, one finds that the $3\sigma$ experimental bounds for
the mixing angles allow an exteremely narrow range $\sim$ 0.05 for Jarlskog's parameter.

\begin{figure}
\begin{tabular}{cc}
  \includegraphics[width=0.2\paperwidth,height=0.2\paperheight,angle=-90]{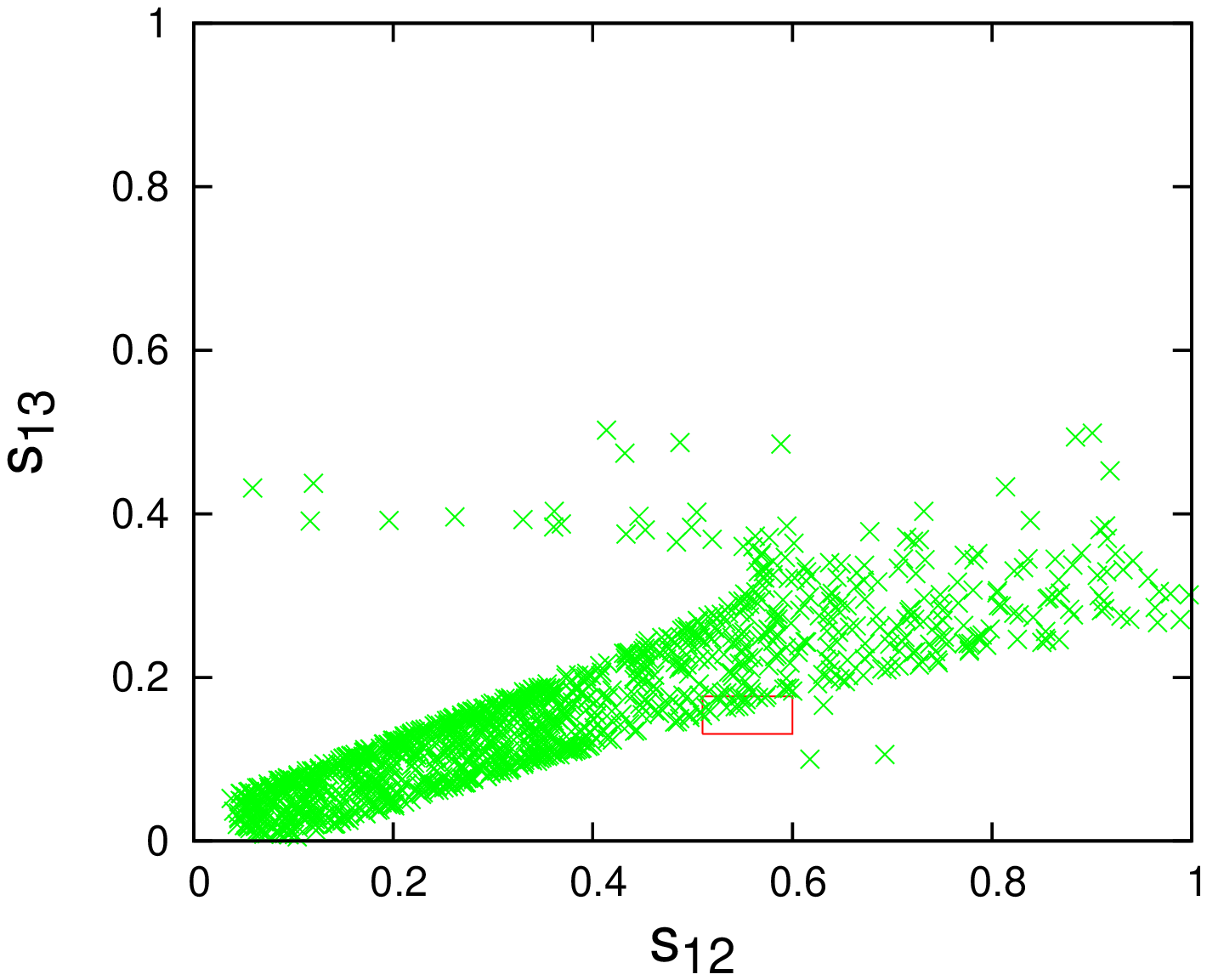}
  \includegraphics[width=0.2\paperwidth,height=0.2\paperheight,angle=-90]{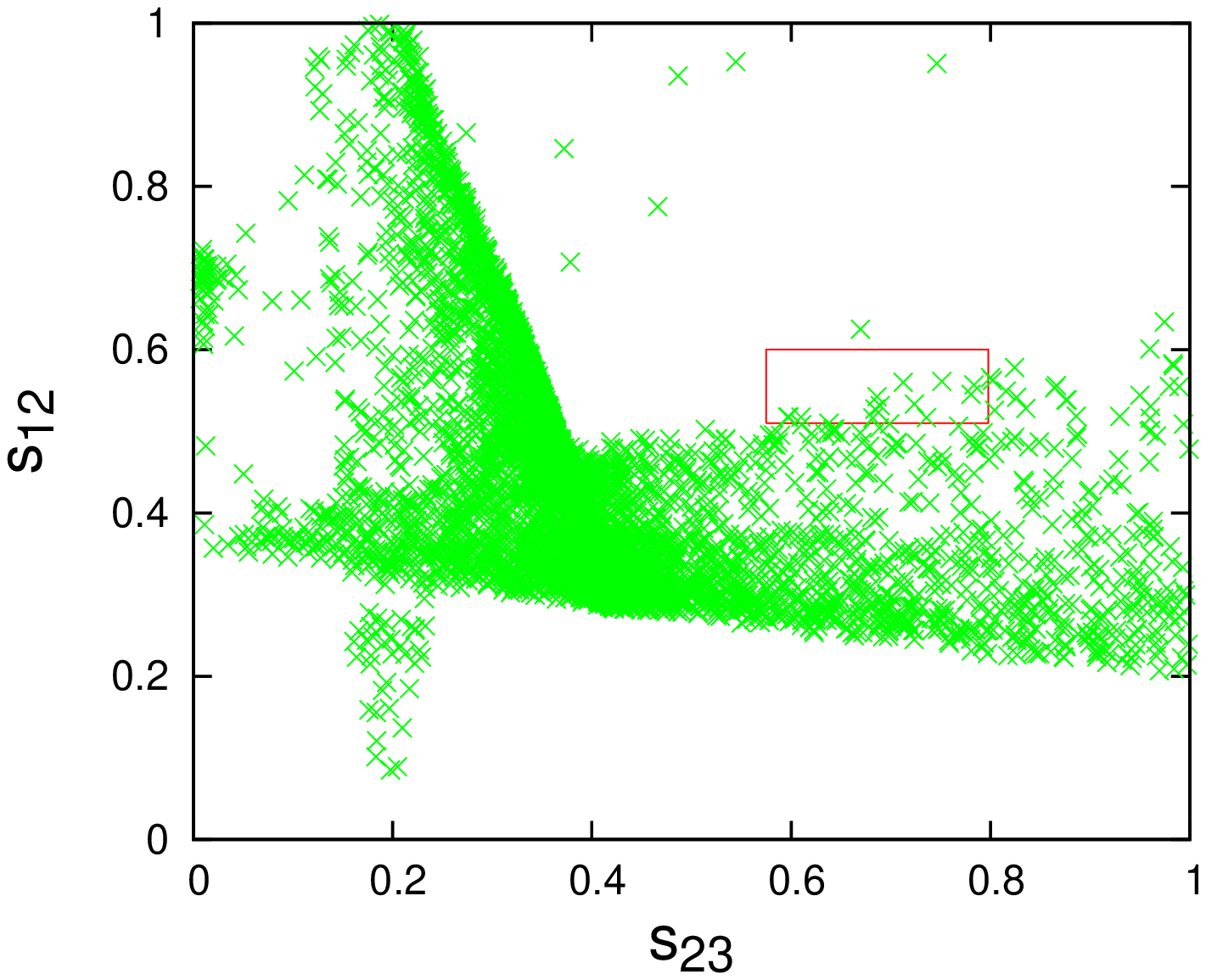}
  \includegraphics[width=0.2\paperwidth,height=0.2\paperheight,angle=-90]{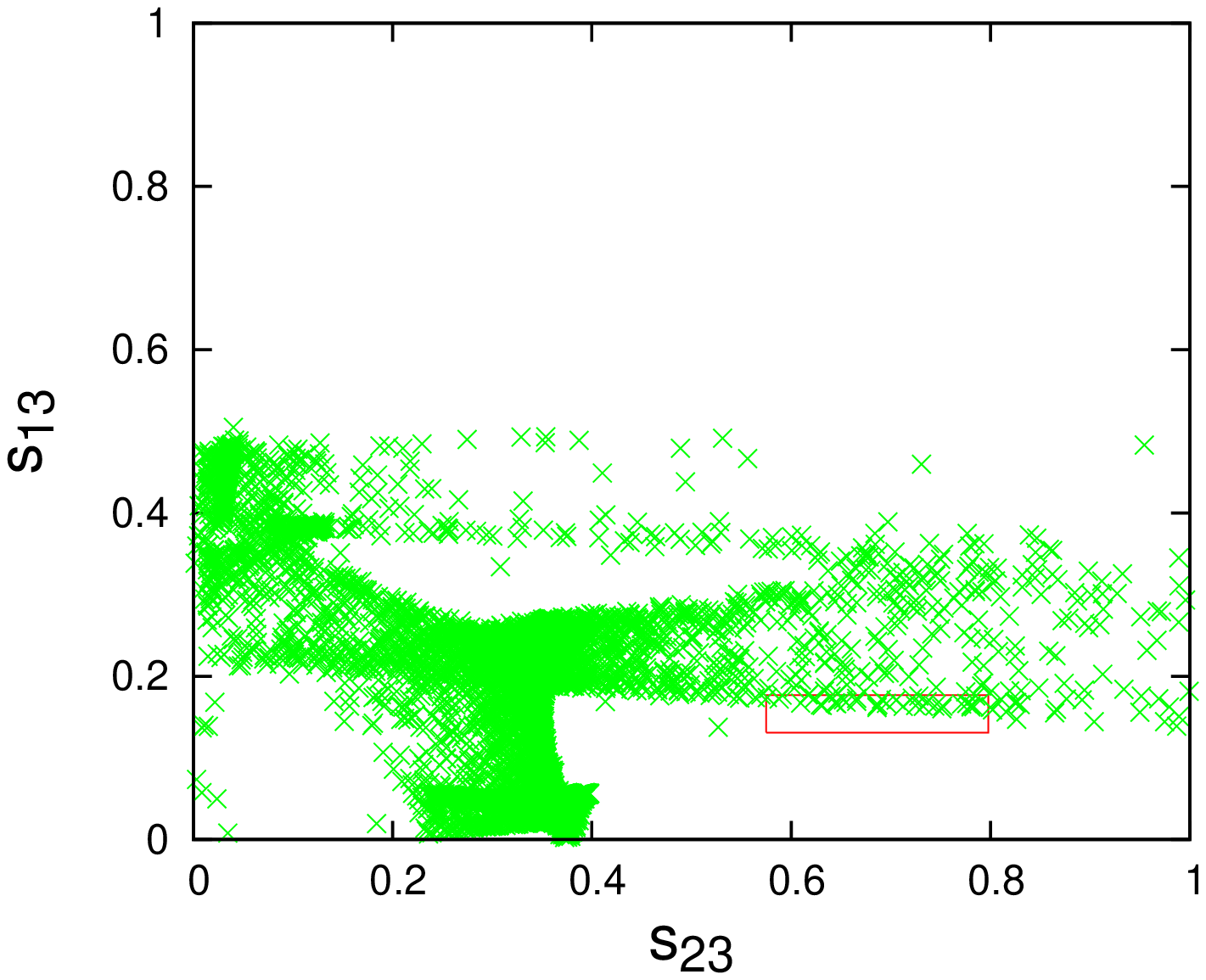}
\end{tabular}
\caption{Plots showing the parameter space for any two mixing
angles when the third angle is constrained by its  $1\sigma$ range
in the $D_l =0$ and $D_\nu\neq 0$ scenario for Class III ansatz of
texture five zero  Dirac mass matrices (normal hierarchy).}
\label{t5cl3nh1}
\end{figure}

\begin{figure}
\begin{tabular}{cc}
  \includegraphics[width=0.2\paperwidth,height=0.2\paperheight,angle=-90]{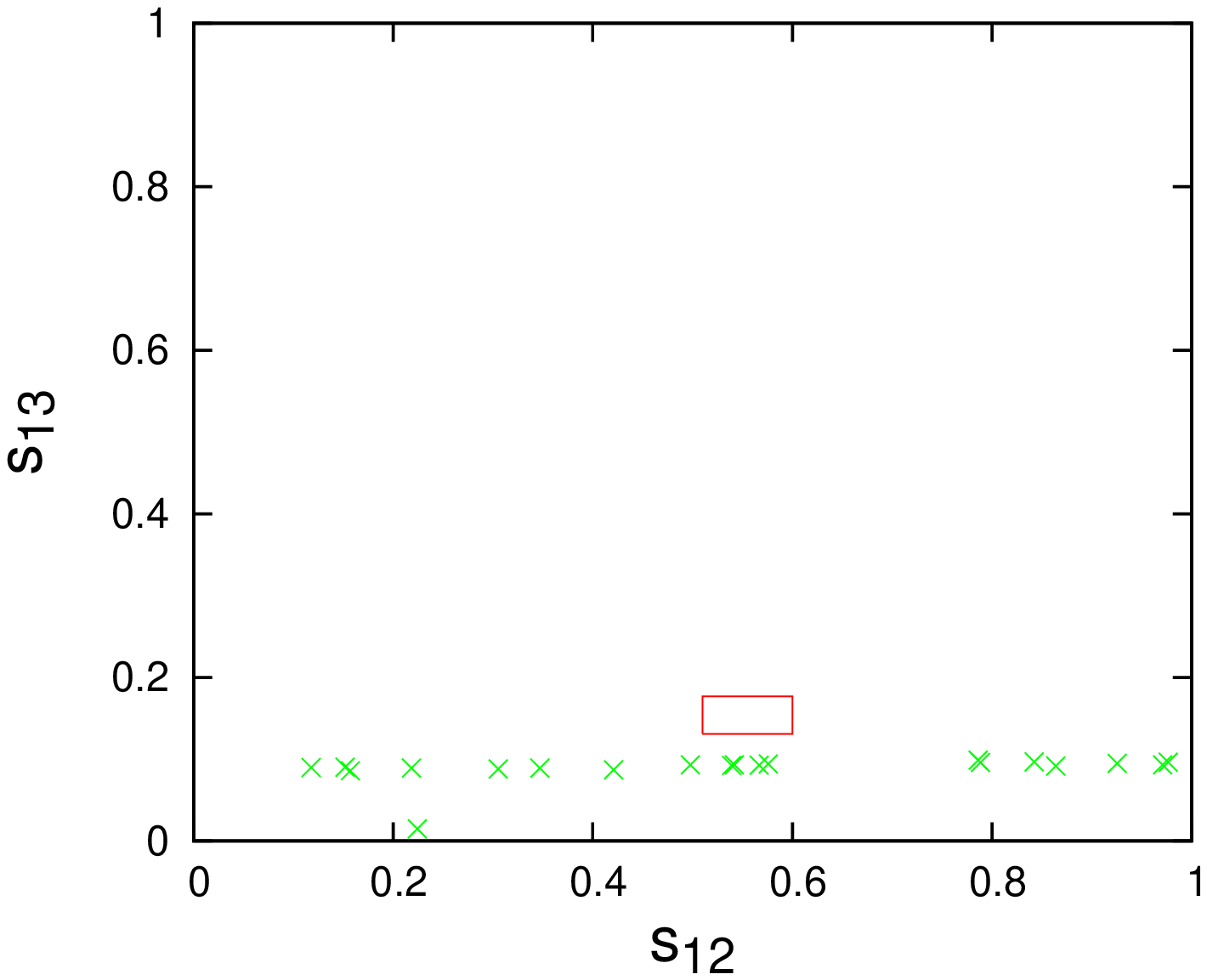}
  \includegraphics[width=0.2\paperwidth,height=0.2\paperheight,angle=-90]{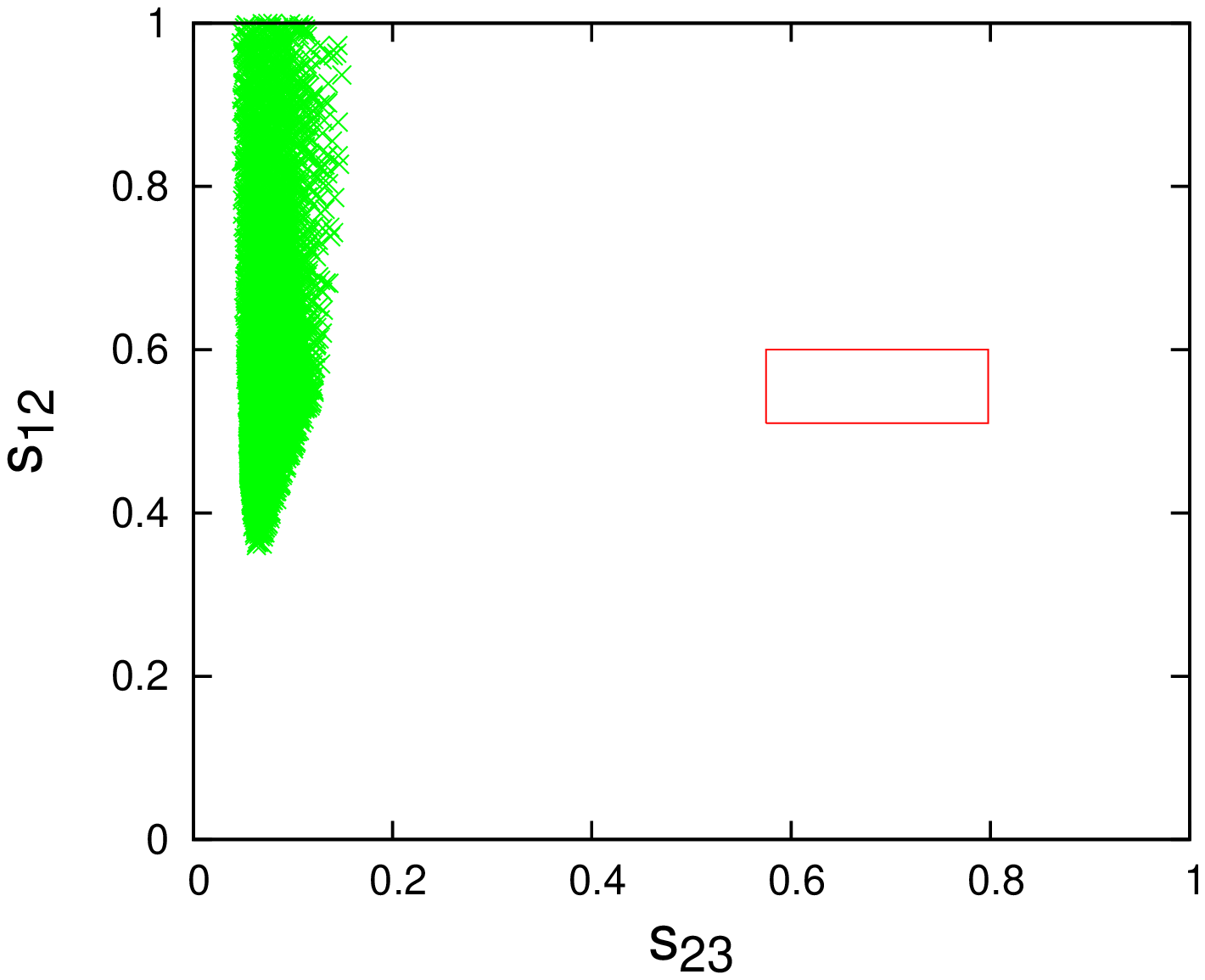}
  \includegraphics[width=0.2\paperwidth,height=0.2\paperheight,angle=-90]{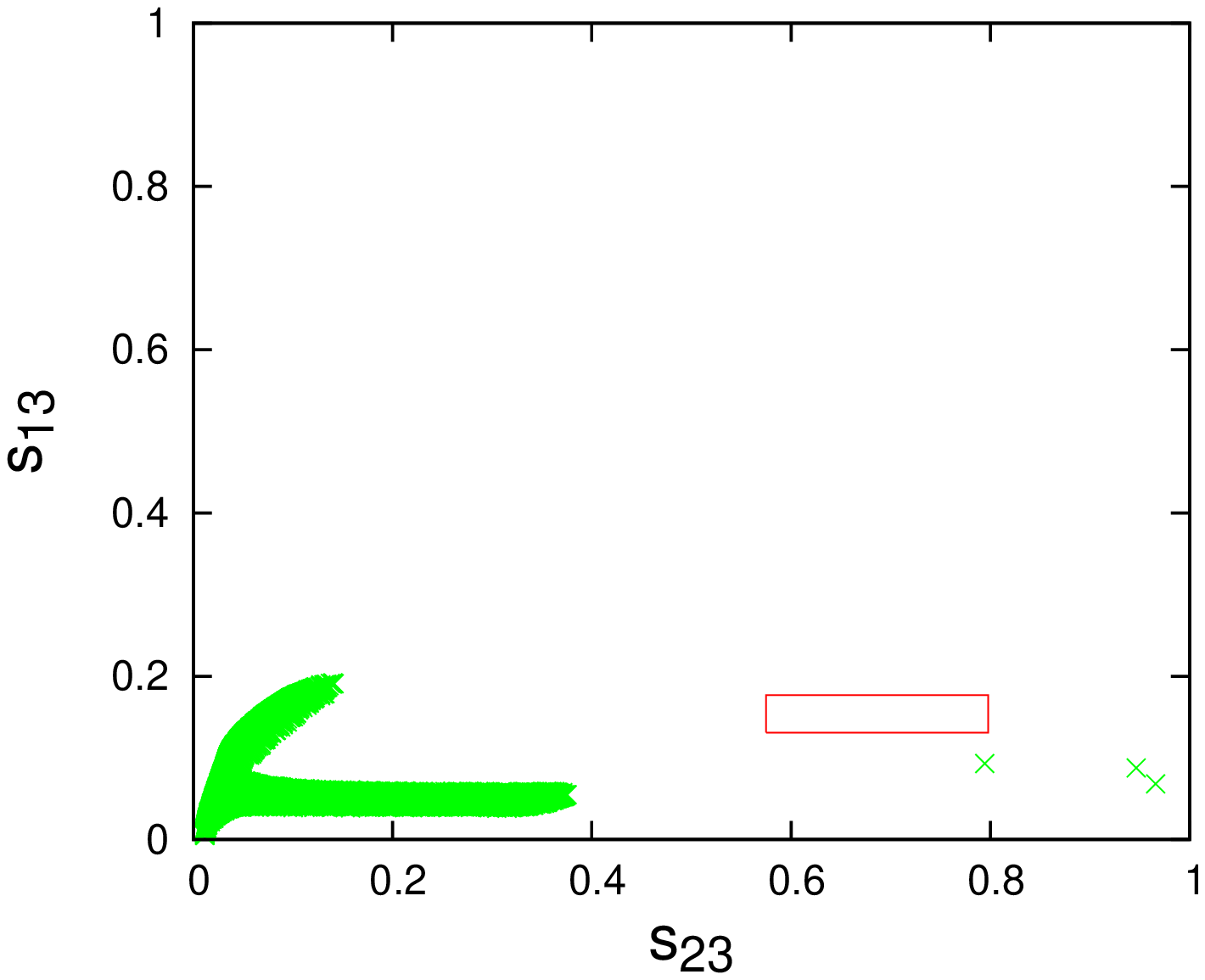}
\end{tabular}
\caption{Plots showing the parameter space for any two mixing
angles when the third angle is constrained by its  $3 \sigma$
range in the $D_l \neq 0$ and $D_\nu = 0$ scenario for Class III
ansatz of texture five zero  Dirac mass matrices (normal hierarchy).}
 \label{t5cl3nh2}
\end{figure}
\begin{figure}
\begin{tabular}{cc}
  \includegraphics[width=0.2\paperwidth,height=0.2\paperheight,angle=-90]{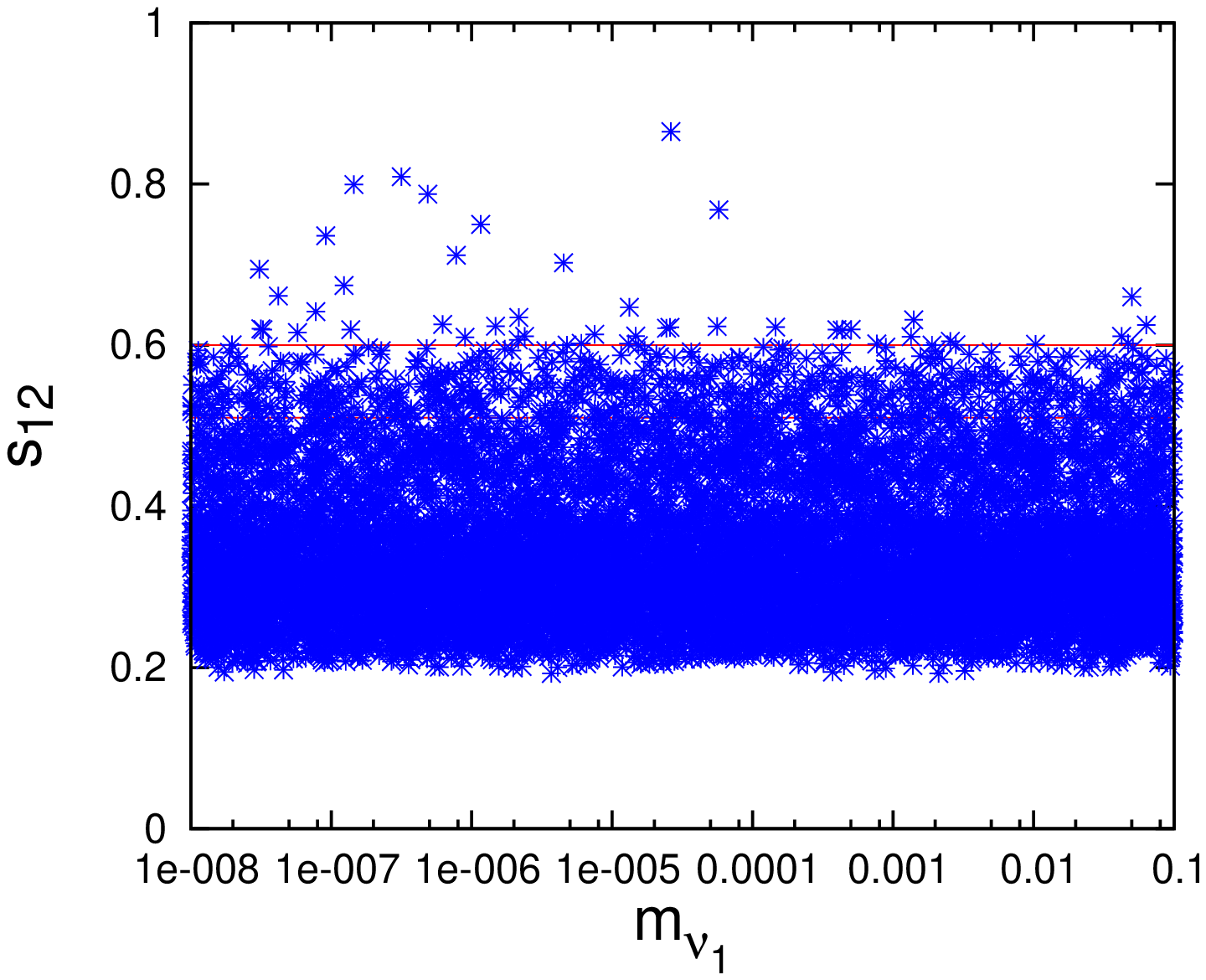}
  \includegraphics[width=0.2\paperwidth,height=0.2\paperheight,angle=-90]{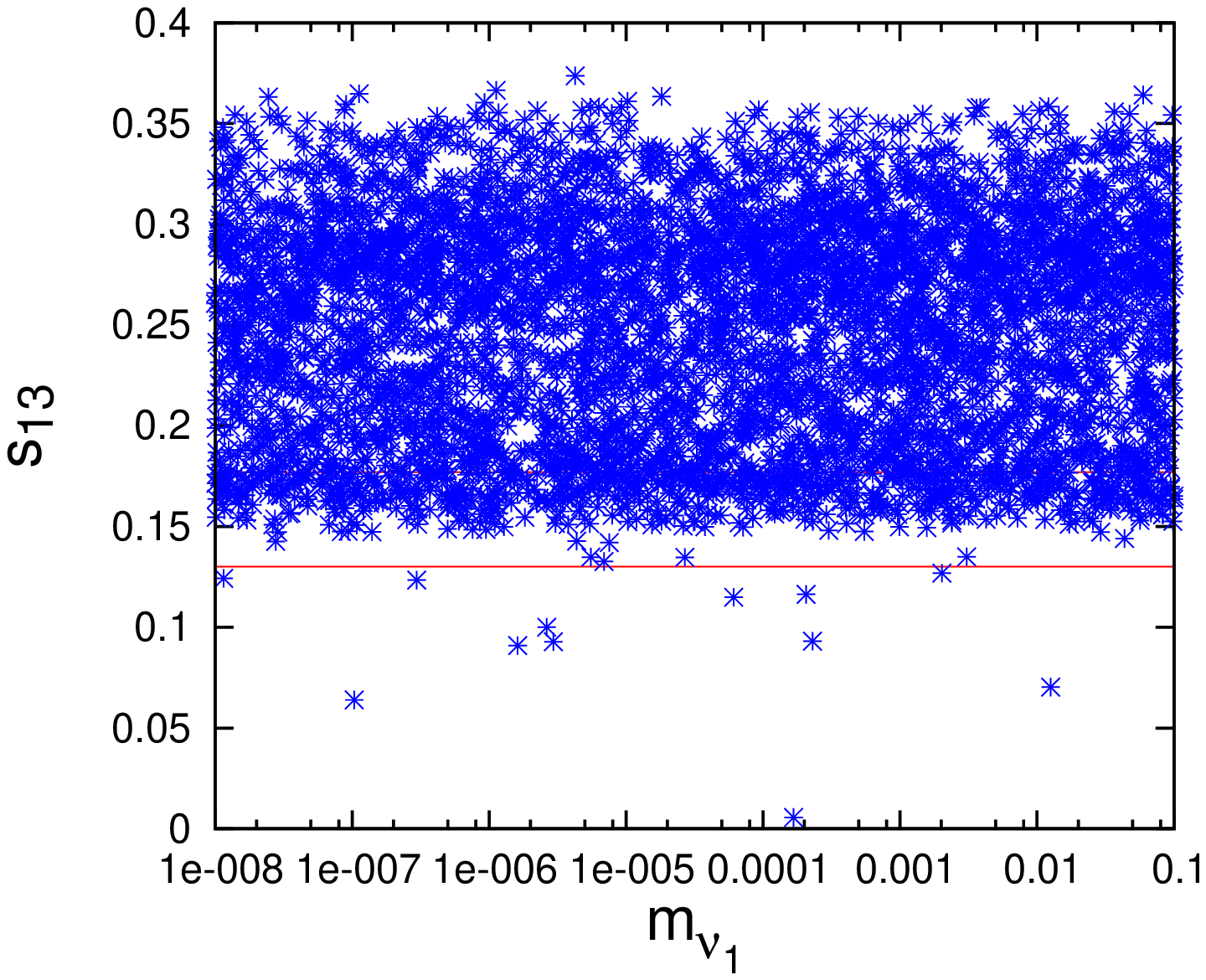}
  \includegraphics[width=0.2\paperwidth,height=0.2\paperheight,angle=-90]{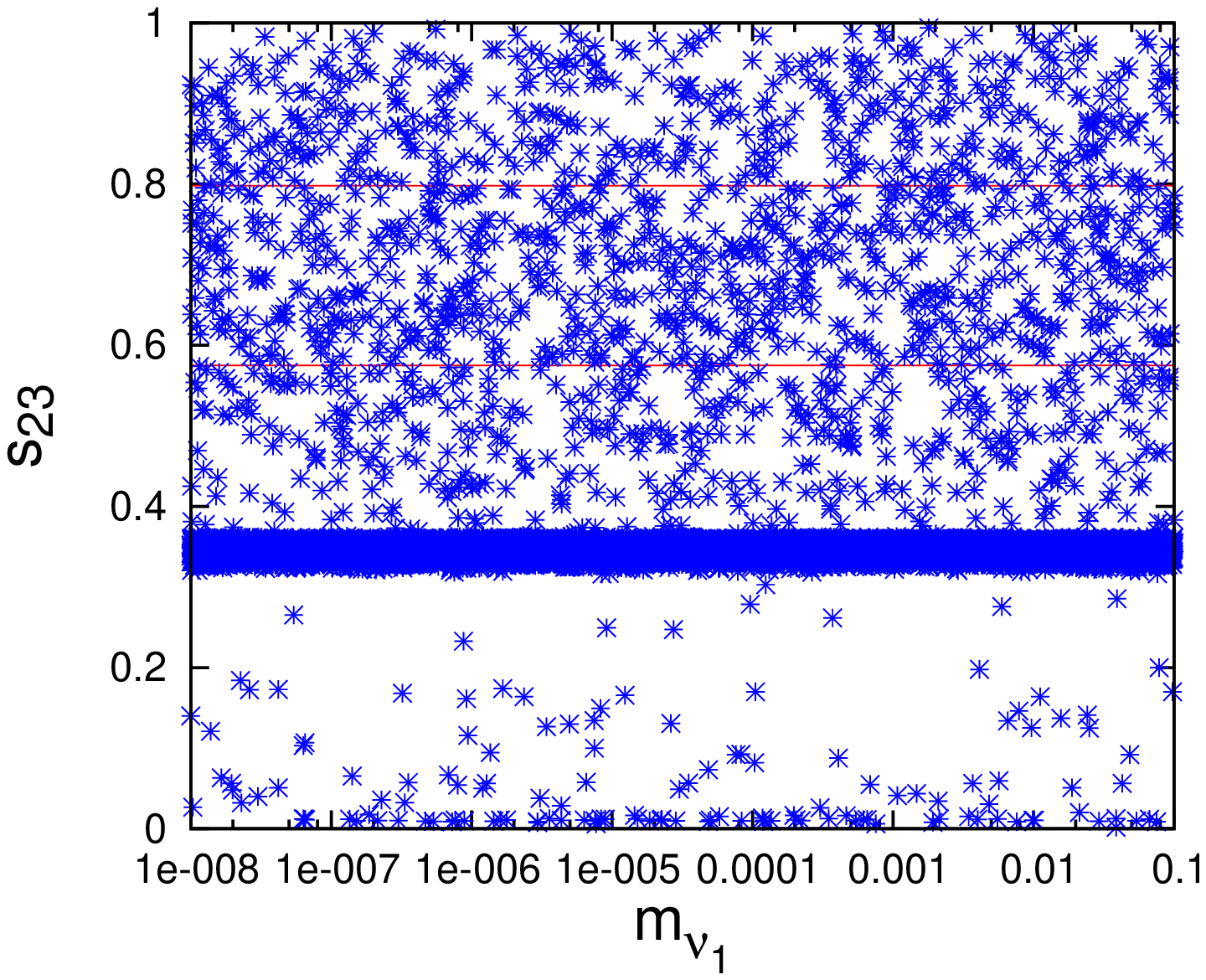}
\end{tabular}
\caption{Plots showing the variation of lightest neutrino mass
with mixing angles when the other two angles are constrained by
their $3 \sigma$ ranges  for $D_l=0$ and $D_\nu \neq0$ scenario
for Class III ansatz of texture five zero  Dirac mass matrices
(normal hierarchy).} \label{t5cl3nh3}
\end{figure}

\begin{figure}
\begin{tabular}{cc}
  \includegraphics[width=0.2\paperwidth,height=0.2\paperheight,angle=-90]{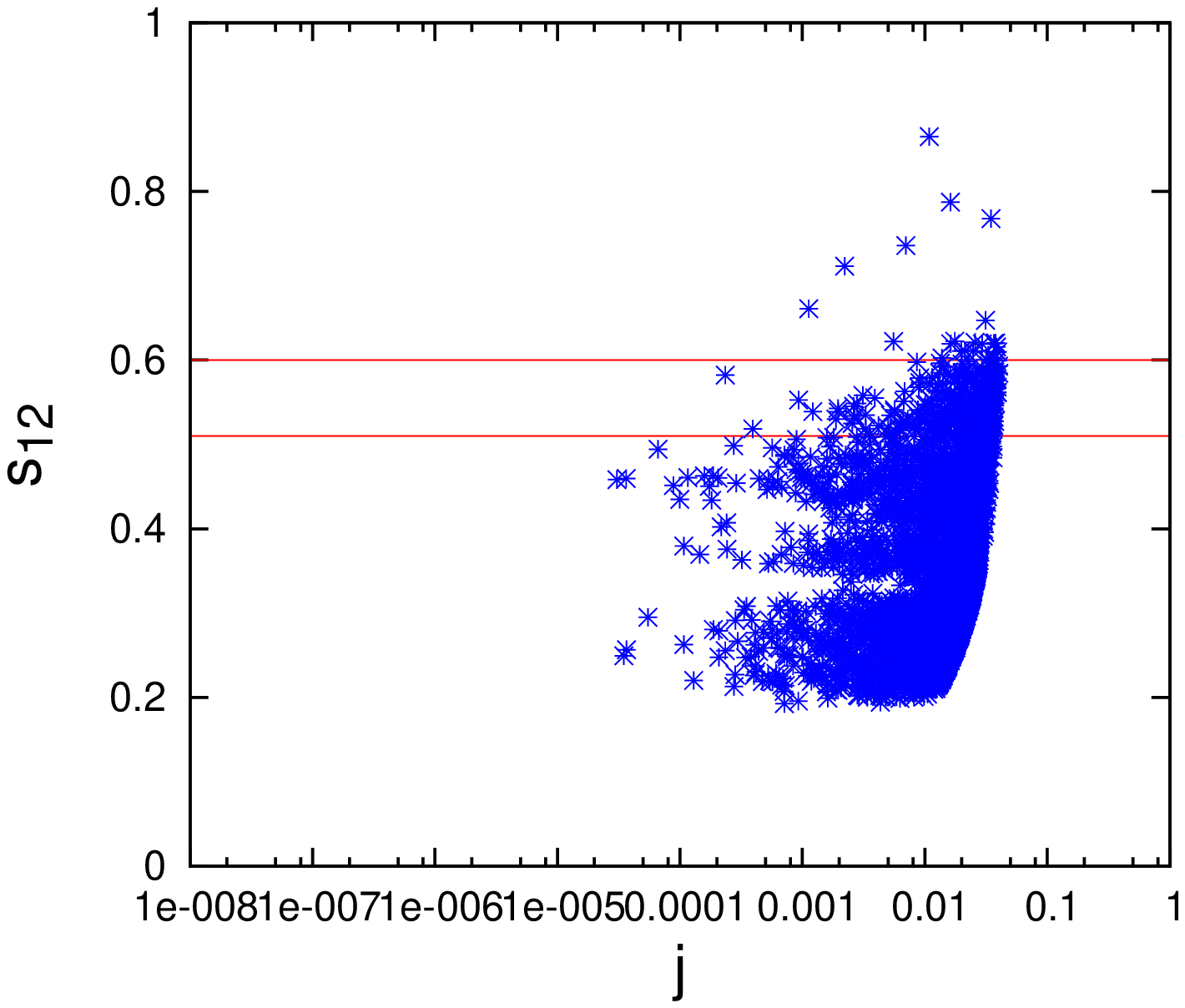}
  \includegraphics[width=0.2\paperwidth,height=0.2\paperheight,angle=-90]{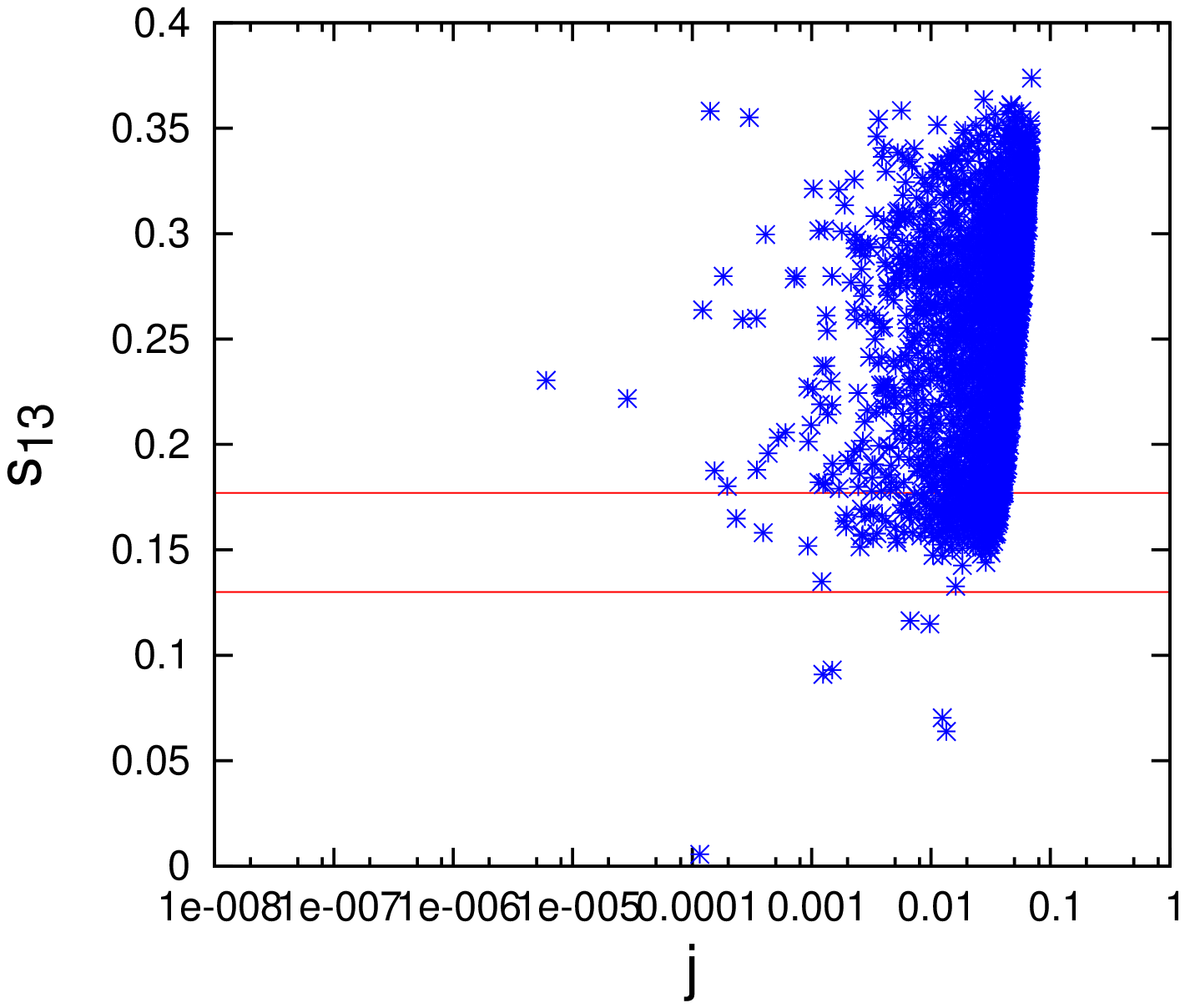}
  \includegraphics[width=0.2\paperwidth,height=0.2\paperheight,angle=-90]{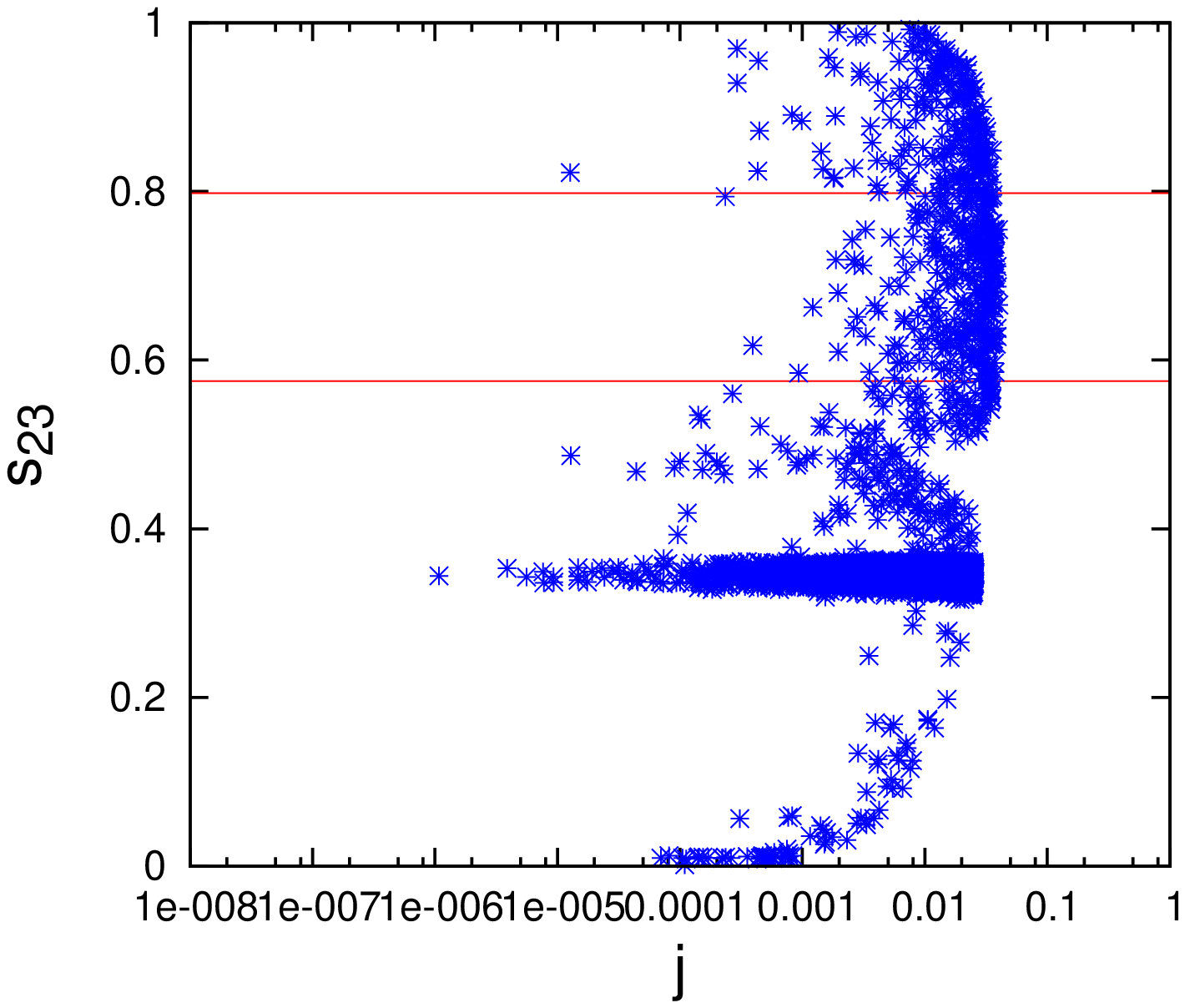}
\end{tabular}
\caption{Plots showing the variation of Jarlskog's parameter with
mixing angles when the other two angles are constrained by their
$3 \sigma$ ranges   for $D_l= 0$ and $D_\nu \neq 0$ scenario for
Class III ansatz of texture five zero  Dirac mass matrices
(normal hierarchy).} \label{t5cl3nh4}
\end{figure}
After studying both the cases for texture five zero mass matrices for inverted hierarchy pertaining to class II ansatz,
we now carry out a similar
analysis pertaining to normal hierarchy. To this
end, in figures (\ref{t5cl3nh1}) and (\ref{t5cl3nh2}), we present the plots showing the parameter space corresponding
to any two mixing angles wherein the third one is constrained by its $3\sigma$ range. Interestingly, normal
hierarchy seems to be ruled out for the case $D_l \neq 0$ and $D_\nu=0$, whereas for the case $D_l=0$ and $D_\nu \neq 0$
of texture five zero lepton mass matrices normal hierarchy
seems to be viable.
Next, we study the dependence of the lightest neutrino mass and Jarlskog's parameter on the the
leptonic mixing angles for the $D_l=0$ and $D_\nu \neq 0$ case of texture five zero mass matrices
corresponding to class III. To this end,  we present the plots showing variation of
the lightest neutrino mass and Jarlskog's parameter with the mixing angles in figures (\ref{t5cl3nh3}) and (\ref{t5cl3nh4})
respectively. While plotting these graphs, the other two mixing angles have been constrained by their $3\sigma$ ranges.
Interestingly, one finds that the lightest neutrino mass is largly unrestricted, whereas for Jarlskog's parameter
one approximately obtains a range, viz., $0.00001 \lesssim j \lesssim 0.05$. Further, since the lightest neutrino mass is unrestricted,
therefore the degenerate scenario pertaining to normal hierarchy can not be ruled out for this structure.

\section{Summary and conclusions}
To summarize, for Dirac neutrinos, we have carried out detailed calculations pertaining to non minimal
textures characterized by texture two zero Fritzsch-like structure as well as all possibilities for
texture four zero and five zero lepton mass matrices. Corresponding to these, we have considered all the
three possibilities for neutrino masses i.e. normal, inverted as well as degenerate scenarios. The
compatibility of these texture specific mass matrices has been examined by plotting the parameter space
corresponding to any two of the leptonic mixing angles. Further, for all the structures which seem to be
compatible with the recent lepton mixing data, the implications of the mixing angles on the lightest
neutrino mass as well as the Jarlskog parameter have also been studied.
\par The analysis reveals that the Fritzsch like texture two zero lepton mass matrices
are compatible with the recent lepton mixing data pertaining to normal as well as inverted neutrino mass
hierarchies. Interestingly, one finds that both the normal as well
as inverted neutrino mass hierarchies are compatible with  texture four zero mass matrices pertaining to
class II and III contrary to the case for texture four zero mass matrices pertaining to class I wherein inverted
hierarchy seems to be ruled out. None of the two possibilities pertaining to degenerate neutrino mass scenario
is compatible with texture four zero mass matrices in class I and II, whereas degenerate scenario can not be ruled out
for texture four zero mass matrices in class III. Mass matrices in class IV are phenomenologically excluded.
\par For texture five zero lepton mass matrices, we analyse both the cases, viz. $D_l=0,~D_\nu \neq 0 $ as well as
$D_l \neq 0,~D_\nu =0$ for all the three phenomenologically viable classes. For texture five zero matrices pertaining to
class I, inverted hierarchy is ruled out for both the cases, whereas normal hierarchy is viable for the $D_l=0,~D_\nu \neq 0$ case.
For class II, normal hierarchy is viable for both the cases while the inverted hierarchy is ruled out for the case $D_l=0,~D_\nu \neq 0$.
Finally, for texture five zero mass matrices pertaining to class III we find that inverted hierarchy
is viable for the case $D_l \neq 0,~D_\nu =0$, while the normal hierarchy is compatible with the $D_l=0,~D_\nu \neq 0$ case.

\vskip 0.5cm
{\bf Acknowledgements} \\S.S. would like to acknowledge UGC, Govt. of India, for financial support.
G.A. would like to acknowledge DST,
Government of India (Grant No: SR/FTP/PS-017/2012) for financial
support. S.S., P.F., G.A. acknowledge the Chairperson, Department
of Physics, P.U., for providing facilities to work.



\begin{thebibliography}{99}


\bibitem{t2k}T2K Collaboration,( K. Abe {\it et al.}), Phys. Rev. Lett.
{\bf 107 } (2011) 041801, [arXiv:1106.2822].

\bibitem{dc}DOUBLE-CHOOZ Collaboration, (Y. Abe {\it et al.), Indication for the Disappearance of Reactor
Electron Antineutrinos in the Double Chooz
Experiment},[arXiv:1112.6353].

\bibitem{db}DAYA-BAY Collaboration, (F. P. An {\it et al.), Observation of Electron-Antineutrino Disappearance at
Daya Bay}, [arXiv:1203.1669].

\bibitem{reno}RENO Collaboration, (J. K. Ahn {\it et al.) Observation of Reactor Electron Antineutrino Disappearance in
the Reno Experiment}, [arXiv:1204.0626].

\bibitem{topdown} A. Yu. Smirnov, hep-ph/0604213; C. D. Froggatt and H. B. Nielsen, Nucl. Phys. B {\bf 147}, 277 (1979);
Y. Nir and N. Seiberg, Phys. Lett. B {\bf 309}, 337 (1993);
M. Leurer, Y. Nir and N. Seiberg, Nucl. Phys. B {\bf 420}, 468 (1994);
L. E. Ibanez and G. G. Ross, Phys. Lett. B {\bf 332}, 100 (1994);
G. Altarelli and F. Feruglio, hep-ph/1002.0211;
A. J. Buras, C. Grojean, S. Pokorski and R. Ziegler, hep-ph/1105.3725 and references
therein; Z. Guo and B. Ma, JHEP {\bf 0909}, 091 (2009) and references therein.

\bibitem{tex} H. Fritzsch, Z. Z. Xing, Prog. Part. Nucl. Phys. { \bf 45} 1 (2000), and references therein;
Z. Z. Xing, Int. Jour. of Mod. Phys. A  {\bf 19} 1 (2004), and references therein,
hep-ph/9912358; M. Gupta, G. Ahuja, Int. Jour. of Mod. Phys. A {\bf 26} 2973 (2011);
N. G. Deshpande, M. Gupta and P. B. Pal, Phys. Rev. D {\bf 45}, 953 (1992); P. S. Gill
and M. Gupta, Pramana {\bf 45}, 333 (1995); {\it{ibid}}. J. Phys. G {\bf 23}, 335 (1997); {\it{ibid.}} Phys.
Rev. D {\bf 56}, 3143 (1997); M. Randhawa and M. Gupta, Phys. Rev. D {\bf 63}, 097301
(2001); M. Randhawa, G. Ahuja and M. Gupta, Phys. Lett. B 643, 175 (2006);
G. Ahuja, S. Kumar, M. Randhawa, M. Gupta and S. Dev, Phys. Rev. D {\bf 76}, 013006
(2007); G. Ahuja, M. Gupta, M. Randhawa and R. Verma, Phys. Rev. D {\bf 79}, 093006
(2009); R. Verma, G. Ahuja and M. Gupta, Phys. Lett. B {\bf 681}, 330 (2009).

\bibitem{singreview} M. Gupta and G. Ahuja, Int. J. Mod. Phys. A {\bf 27},
1230033 (2012).

\bibitem{ptep} P.Fakay, S.Sharma, G.Ahuja and M.Gupta, [arXiv:1401.8121].

\bibitem{so10} Stuart Raby, Phys.Lett. B {\bf 561} 119 (2003);
K. S. Babu, Jogesh C. Pati and Parul Rastogi Phys. Rev. D {\bf 71}, 015005 (2005);
K.S. Babu, Jogesh C. Pati and Frank Wilczek Nucl. Phys. B {\bf 566}, 33 (2000);
S.M.Barr Phys. Rev.Lett. {\bf{64}}, 353, 1990; K.S. Babu and S.M.Barr, Phys. Rev. Lett. {\bf{75}}, 11(1995);
W. Buchmüller and D. Wyler Phys. Lett. B {\bf 521} 291 (2001).

 \bibitem{st} H. Fritzsch, M. Gell-Mann and H. Leutwyler,
Phys. Lett. {\bf B 47}, 365 (1973). 

\bibitem{ewm1} S.L. Glashow, Nucl. Phys. {\bf 22}, 597 (1961);
S. Weinberg, Phys. Rev. Lett. {\bf 19}, 1264 (1967); A. Salam, in
{\it Elementary Particle Theory}, ed. N. Svartholm
(Almquist and Wiksells, Stockholm, 1969). 

\bibitem{sm} For excellent reviews on the Standard Model see,
J.F. Donoghue, E. Golowich and B.R. Holstein, {\it Dynamics of the
Standard Model}, (Cambridge University Press, 1992).

\bibitem{nmm} R. D. Peccei and K. Wang, Phys. Rev. D {\bf 53 },5 (1996).

\bibitem{branco} G.C. Branco, D. Emmanuel-Costa and R. G. Felipe, Phys. Lett. B {\bf  477}, 147 (2000);
G.C. Branco, D. Emmanuel-Costa, R. G. Felipe and H. Serodio, Phys. Lett. B {\bf 670}, 340 (2009).

\bibitem{frxing} H. Fritzsch and Z.Z, Xing, Phys. Lett. B {\bf  413}  396 (1997).




 \bibitem{pmns}B. Pontecorvo, Zh. Eksp. Theor. Fiz. (JETP) {\bf 33}, 549 (1957);
 {\it ibid.} {\bf 34}, 247 (1958); {\it ibid.} {\bf53}, 1771 (1967); Z.
Maki, M. Nakagawa, S. Sakata, Prog. Theor. Phys. {\bf 28}, 870
(1962).

\bibitem{fogli2012} G. L. Fogli, E. Lisi, A. Marrone, D. Montanino, A. Palazzo and
A. M. Rotunno, Phys. Rev. D. {\bf 86}, 013012 (2012).

\bibitem{t40lep} G. Ahuja, M. Gupta, M. Randhawa and Rohit Verma, Phys. Rev. D {\bf 79}, 093006 (2009);
G. Ahuja, S. Kumar, M. Randhawa, M. Gupta and S. Dev, Phys. Rev. D {\bf 76}, 013006 (2007);
M. Randhawa, G. Ahuja and M. Gupta, Phys. Rev. D {\bf 65}, 093016(2002).



























\end{thebibliography}
\end{document}